\documentclass[12pt]{article} 
\usepackage{citesort}
\usepackage{amssymb}
\usepackage{epsfig}
\usepackage{amsmath} 

\newcommand{\dal}{\mbox {$\sqcap$ \hspace{-0.42cm}$\sqcup\:$}}

\newcommand{\singlespace}{\renewcommand{\baselinestretch}{1}\large\normalsize}
\newcommand{\doublespace}{\renewcommand{\baselinestretch}{1.6}\large\normalsize}
\newcommand{\mc}{\multicolumn}
\newcommand{\lsim}{\mathrel{\mathop{\kern 0pt \rlap
  {\raise.2ex\hbox{$<$}}}
  \lower.9ex\hbox{\kern-.190em $\sim$}}}
\newcommand{\gsim}{\mathrel{\mathop{\kern 0pt \rlap
  {\raise.2ex\hbox{$>$}}}
  \lower.9ex\hbox{\kern-.190em $\sim$}}}
\def\dsla{\rlap{/}\partial}
\def\non{\rlap{\small /}\in}

\begin{document}

\title{\bf Weak decay of $\Lambda$--hypernuclei}
\author{W. M. Alberico$^\mathrm{a}$ and G. Garbarino$^\mathrm{b}$\\
\small \it $^\mathrm{a}$INFN, Sezione di Torino and Dipartimento di Fisica Teorica,\\
\small \it Universit\`a di Torino, 10125 Torino, Italy\\ 
\small \it $^\mathrm{b}$Departament d'Estructura i Constituents de la Mat\`{e}ria, \\
\small \it Universitat de Barcelona, 08028 Barcelona, Spain} 

\date{\empty}
\maketitle 
\thispagestyle{empty} 

\begin{abstract}
{\footnotesize
In this review we discuss the present status of strange nuclear
physics, with special attention to the weak decay of $\Lambda$--hypernuclei.
The models proposed for the evaluation of the $\Lambda$ decay widths are 
summarized and their results are compared with the data. The rates
$\Gamma_{\rm NM}=\Gamma_n+\Gamma_p\, (+\Gamma_2)$,
$\Gamma_{\pi^0}$ and $\Gamma_{\pi^-}$ are well explained by several 
calculations.
Despite the intensive investigations of the last years, the main open problem
remains a sound theoretical interpretation of the large experimental values 
of the ratio $\Gamma_n/\Gamma_p$. 
However, the large uncertainties involved in the experimental determination
of the ratio do not allow to reach any definitive conclusion. 
The $\Gamma_n/\Gamma_p$ puzzle is strongly related to the so--called
$\Delta I=1/2$ rule on the isospin change in the non--mesonic decay,
whose possible violation cannot be established at present, again due to
the insufficient precision of the data.
Although recent works offer a step forward in the solution of the
puzzle, further efforts (especially on the experimental side)
must be invested in order to understand the detailed dynamics
of the non--mesonic decay. 
Even if, by means of {\it single} nucleon spectra measurements,
the error bars on $\Gamma_n/\Gamma_p$ have been considerably reduced very 
recently at KEK (however, with central data compatible with older experiments),
a clean extraction of $\Gamma_n/\Gamma_p$ is needed.
What is missing at present, but planned for the next future, are measurements 
of 1) nucleon energy spectra in {\it double} coincidence and 
2) nucleon angular correlations: such observations allow to disentangle the 
nucleons produced in one-- and two--body induced decays and lead to a 
{\it direct} determination of $\Gamma_n/\Gamma_p$.
Notably, the two--body component of the non--mesonic decay rates has not been 
measured yet, due to the too low counting rates expected for a coincidence experiment.
For the asymmetric non--mesonic decay of polarized hypernuclei the situation 
is even more puzzling.
Indeed, strong inconsistencies appear already among data. A recent experiment
obtained a positive intrinsic $\Lambda$ asymmetry parameter, $a_{\Lambda}$, for
$^5_{\Lambda}\vec{\rm H}{\rm e}$.
This is in complete disagreement with a previous measurement, which obtained a
large and negative $a_{\Lambda}$ for $p$--shell hypernuclei, and with theory, 
which predicts a negative value moderately dependent on nuclear structure 
effects. Also in this case, improved experiments establishing with certainty 
the sign and magnitude of $a_{\Lambda}$ for $s$-- and
$p$--shell hypernuclei will provide a guidance for a deeper understanding of
hypernuclear dynamics and decay mechanisms.}

%

\vspace{0.6cm} 
\noindent {\footnotesize {\sl Keywords}:  
Production and structure of hypernuclei; mesonic and non--mesonic decay of $\Lambda$--hypernuclei;
$\Gamma_n/\Gamma_p$ puzzle; $\Delta I=1/2$ isospin rule; decay of polarized
$\Lambda$--hypernuclei. 

\vspace{0.15cm}
}
\end{abstract}  

\newpage{$\,$} 
\thispagestyle{empty} 

\newpage
\doublespace
\pagenumbering{Roman} 
\tableofcontents

\newpage
\singlespace
\pagenumbering{arabic}
\section{Hyperons and hypernuclei}
\label{hyp}
\subsection{Introduction}
\label{hyp1}
Hyperons (${\Lambda}$, ${\Sigma}$, $\Xi$, $\Omega$)
have lifetimes of the order of $10^{-10}$~sec (apart from the ${\Sigma}^0$, which 
decays into $\Lambda \gamma$). They decay weakly, with a mean free path 
$\approx c\tau ={\mathcal O}(10$ cm$)$. 
A hypernucleus is a bound system of neutrons, 
protons and one or more hyperons. We will denote with $^{A+1}_YZ$ a hypernucleus 
with $Z$ protons, $A-Z$ neutrons and a hyperon $Y$. A crucial point to 
describe the structure of these {\it strange nuclei} 
is the knowledge of the elementary 
hyperon--nucleon ($YN$) and hyperon--hyperon ($YY$) interactions. Hyperon masses
differ remarkably from the nucleonic mass, hence the flavour $SU(3)$ symmetry
is broken. The amount of this breaking is a fundamental question in order to 
understand the baryon--baryon interaction in the strange sector.

Among hyperons and nucleons the following esoenergetic strong reactions
($\Delta S=0$) are allowed:
\begin{equation}
\begin{array}{c c c c c c c l}
\Sigma^-p&\rightarrow & \Lambda n & \hspace{0.6mm}&
\Sigma^+n&\rightarrow & \Lambda p &\hspace{0.3mm}(Q\simeq 78 {\rm MeV}) \\
\Xi^-p&\rightarrow & \Lambda \Lambda & \hspace{0.6mm}&  
\Xi^0n&\rightarrow & \Lambda \Lambda &\hspace{0.3mm}(Q\simeq 26 {\rm MeV}) \\
\Omega^-p&\rightarrow & \Lambda \Xi^0 & \hspace{0.6mm}&
\Omega^-n&\rightarrow & \Lambda \Xi^- &\hspace{0.3mm}(Q\simeq 178 {\rm MeV}) 
\end{array} \nonumber
\end{equation}
(in parentheses are quoted the average released energies, the so--called $Q$--values),
hence, only the lightest hyperon (${\Lambda}$)
is generally stable with respect to the 
strong processes which occur in nuclear systems. In this review 
we shall be mainly concerned with $\Lambda$--hypernuclei.

The existence of hypernuclei is interesting since it gives a new dimension 
to the traditional world of nuclei (states with new symmetries, selection rules, 
etc). In fact, they represent the first kind of 
{\it flavoured nuclei}, in the direction of other exotic systems 
(charmed nuclei and so on).

Hypernuclear physics was born in 1952, when the first hypernucleus was 
observed through its decays
\cite{Da53}. Since then, it has known several phases of development and it 
has been characterized by more and more new challenging questions and answers.
However, this field has experienced great advances only in the last 10--15 years.
We can look at hypernuclear physics as a good tool to match nuclear and
particle physics.
Nowadays, the knowledge of hypernuclear phenomena is rather good, but some
open problems still remain. Actually, the study of this field
may help in understanding some crucial questions, related, to list a few, to:
\begin{itemize}
\item some aspects of the baryon--baryon weak interactions;
\item the $YN$ and $YY$ strong interactions in the $J^P=1/2^+$ baryon octet;
\item the possible existence of di--baryon particles;
\item the renormalization of hyperon and meson properties in the nuclear medium;
\item the nuclear structure: for instance, the long standing question 
of the origin of the spin--orbit interaction and other aspects of the many--body
nuclear dynamics;
\item the role played by quark degrees of freedom, flavour symmetry 
and chiral models in nuclear and hypernuclear phenomena.
\end{itemize}
In this review we will widely discuss a great deal of these problems.

\subsection{Hyperon--nucleon, hyperon--hyperon interactions and hypernuclear structure}
\label{hyp2}

We summarize here the phenomenological information available
nowadays on $YN$ and $YY$ interactions and on the structure 
of $\Lambda$--, $\Sigma$--, $\Xi$-- and $\Lambda \Lambda$--hypernuclei.

One of the main reasons of interest in hypernuclear physics lies 
in the characteristics of the $YN$ and $YY$ interactions which can be derived from.
Obviously, measurements of $YN$ and $YY$ cross sections would give more direct
information. However, such experiments are very difficult due to the short lifetime
of the hyperons, which gives flight paths limited to less than 10 cm: 
nowadays, no scattering data are available on the $YY$ interaction and very limited are the ones 
for the $\Lambda N$, $\Sigma N$ and $\Xi N$ interactions (especially in the last case).
Moreover, we remind the reader that the inverse reaction 
$pn\rightarrow p\Lambda$ in free space is under
investigation at COSY (J\"{u}lich) \cite{Ha95} and KEK \cite{Ki98}. Unfortunately, 
the experimental observation of this process is difficult because of its very low cross
section \cite{Pa99} with respect to the huge background.

The $NN$ interaction can be understood in terms of one--meson--exchange 
(OME) models, usually combined with
a proper parameterization of the repulsive component at short distance, which
originates from quark exchanges between the hadrons and has a range of about
$0.5$ fm, corresponding to a transferred momentum $q\gsim 400$
MeV. The extension of the OME description to strange particles of the $J^P=1/2^+$ baryon 
octet is still unsatisfactory.
Several models of the $YN$ and $YY$ interactions are available. For instance, 
with the help of the flavour $SU(3)$ symmetry, the
Bonn--J\"{u}lich \cite{Re94} and Nijmegen \cite{Ri96,Ri99,Ri99a} groups 
have developed several potentials using the
OME picture, also including, in some cases, two--meson--exchange. 
In addition to meson--exchange potentials, 
other groups (Tokyo \cite{Ya92}, T\"{u}bingen \cite{St90} and Kyoto--Niigata \cite{Fu96})
use quark cluster models to explain the short range interactions. 
Unfortunately, none of these potentials
is fully satisfactory and there are large discrepancies among the
different models (especially on the spin--isospin dependence). Since 
the data on $YN$ scattering are very limited (they consist almost exclusively
of spin--averaged cross sections),
it is impossible to fit the $YN$ interaction unambiguously: different $YN$ potentials
can reproduce the data equally well, but they exhibit differences on a more detailed
level, especially when the spin structure is concerned
(compare for example Refs.~\cite{Re94,Fu96,Ri99}). The measurement 
of spin observables in the $YN$ scattering as well as in the weak process
$pn\to p \Lambda$ could discriminate among the various
interaction models. On the other hand, the study of the hypernuclear structure 
and weak decays is
helpful in order to get useful information on the $YN$ and $YY$ interactions.

\subsubsection*{$\Lambda N$ interaction and $\Lambda$--hypernuclei}

\begin{figure}
\begin{center}
\mbox{\epsfig{file=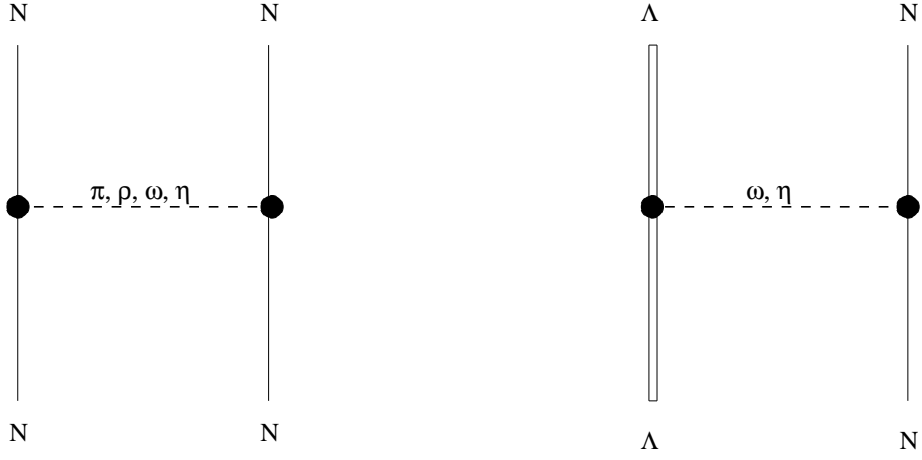,width=.8\textwidth}}
\vskip 2mm
\caption{$NN$ and $\Lambda N$ strong amplitudes in the one--meson--exchange model.} 
\label{potln}
\end{center}
\end{figure}

\begin{figure}
\begin{center}
\mbox{\epsfig{file=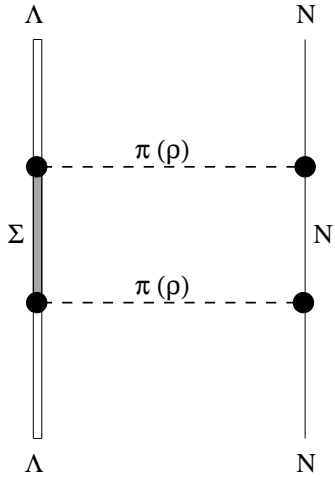,width=.29\textwidth}}
\vskip 2mm
\caption{Two pion ($\rho$) exchange contribution to the $\Lambda N$ potential.}
\label{pipi}
\end{center}
\end{figure}

The strong $\Lambda N$ interaction displays some 
different aspects with respect to the 
$NN$ one. For instance, due to isospin conservation in strong interactions,
the fact that the ${\Lambda}$ has isospin $I=0$ forbids the
emission of a pion (${\Lambda}{\rlap{/}{\to}} \Lambda \pi$).
In figure \ref{potln} we depict the $NN$
and $\Lambda N$ strong potentials in the OME model.
The strong $\Lambda \rightarrow \Sigma \pi$
and $\Sigma \rightarrow \Lambda \pi$ couplings are allowed, 
and the $\Lambda$ hyperon can
interact with a nucleon by exchanging an even number of pions and/or 
of $\rho$ mesons (see figure~\ref{pipi}).
The dominant part of the $\Lambda N$ interaction comes
from the two--pion--exchange, hence it has a shorter range than
the $NN$ one. Moreover, the $\Lambda N$ potential is weaker
than the $NN$ potential: roughly speaking, from the diagrams of 
figures~\ref{potln} and \ref{pipi},
for the tensorial components we have: $V^T_{\Lambda N}/V^T_{NN}\simeq 1/4$.

\begin{figure}
\begin{center}
\mbox{\epsfig{file=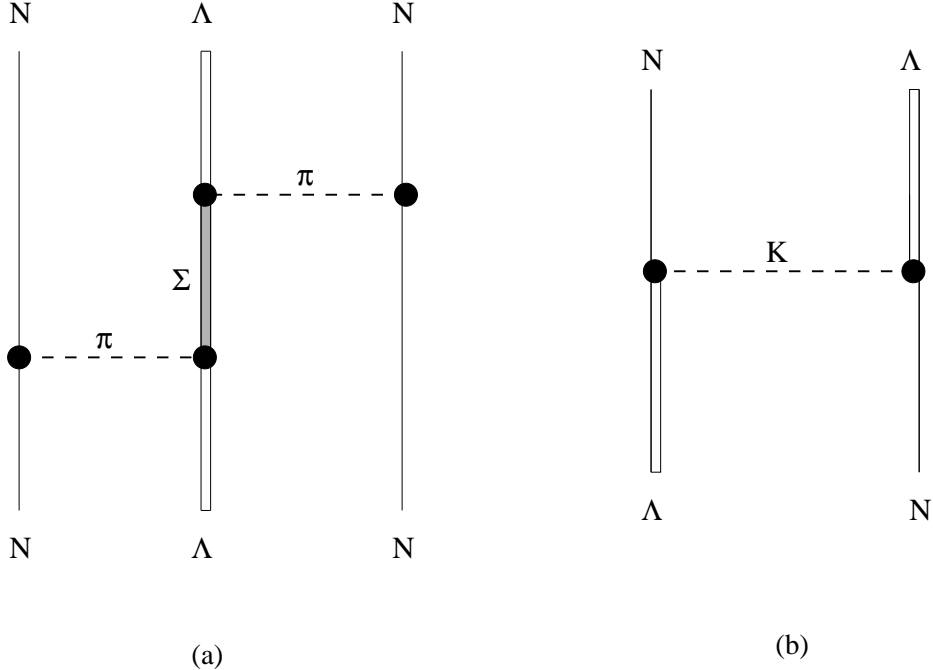,width=.8\textwidth}}
\vskip 2mm
\caption{$\Lambda$ three--body interaction (a) and two--body interaction 
with strangeness exchange (b).}
\label{kappa}
\end{center}
\end{figure}

Besides, three--body interactions 
and two--body interactions with strangeness exchange are also allowed (figure~\ref{kappa}). 
The $\Lambda NN$ three--body force, whose pionic component is depicted in 
Fig.~\ref{kappa}(a), is an important ingredient to investigate the 
structure of $\Lambda$--hypernuclei \cite{Us98,Sh99,Ak00}, especially in light 
systems. This is due to the $\Lambda N$--$\Sigma N$ strong coupling, which is 
sizeable in the nuclear medium \cite{Mo86,Gi88,Gi94,Gi95,Hi01}, and, on the other hand, 
leads to a non--negligible second order tensor force in the $\Lambda N$ interaction
(Fig.~\ref{pipi}). By assuming a repulsive $\Lambda NN$ potential, the small $\Lambda$
binding energies in light hypernuclei and the depth of the 
$\Lambda$--nucleus mean potential for heavy systems
can be reproduced \cite{Sh99} without requiring
a strong spin--dependence of the $\Lambda N$ interaction; 
the latter seems, at present, to be excluded
(see following discussion in this paragraph).
In particular, the $\Lambda NN$ interaction is
essential to explain the existence of the lightest hypernucleus, the
hypertriton ($^3_\Lambda $H$\equiv pn\Lambda$), which
is weakly bound. The $\Lambda$ binding energy, defined as:
\begin{equation}
B_{\Lambda}(^{A+1}_{\Lambda}Z)\equiv M(^{A+1}_{\Lambda}Z)-M(^AZ)-m_{\Lambda}<0 , \nonumber
\end{equation}
is as small as $\simeq 130$ KeV in hypertriton. In Ref.~\cite{Po96}, 
within a microscopic many--body scheme, the authors showed how the coupling to
intermediate $\Sigma N$ states in the $\Lambda N$ interaction (figure~\ref{pipi}) is crucial
for a correct evaluation of the $\Lambda$ binding energy in nuclear matter. 
In hypernuclei the $\Lambda N$--$\Sigma N$ coupling
is more important [especially because of the relatively small
$\Lambda$--$\Sigma$ mass difference ($\simeq 78$ MeV)] than the $NN$--$\Delta N$
coupling in conventional nuclei, where the latter plays a very small role
in binding few--nucleon systems ($m_{\Delta}-m_N\simeq 293$ MeV). Another signal
of the $\Lambda N$--$\Sigma N$ coupling comes from the observation 
that in $S$--wave relative 
states the $\Lambda p$ interaction is more attractive than the $\Lambda n$
interaction. This follows from a comparison of the
experimental $\Lambda$ binding energies in the $A=3$, $I=1/2$ doublet:
\begin{equation} 
\frac{B_{\Lambda}(^4_{\Lambda}{\rm He})-
B_{\Lambda}(^4_{\Lambda}{\rm H})}{B_{\Lambda}(^4_{\Lambda}{\rm He})}=
\frac{0.35 {\rm MeV}}{2.39 {\rm MeV}}\simeq 0.15 \nonumber
\end{equation}
($^4_{\Lambda}{\rm He}=ppn\Lambda$ and $^4_{\Lambda}{\rm H}=pnn\Lambda$
should differ only because of Coulomb effects, if the $\Lambda n$ and $\Lambda p$
interactions were of equal strength). The $\Lambda N$--$\Sigma N$ coupling 
gives a charge symmetry breaking more important than the one observed
in ordinary nuclei by comparing the neutron separation energies in
$^3$H and $^3$He (after correcting for the Coulomb interaction in $^3$He).
The Coulomb energy in $^4_{\Lambda}$He is expected to be only a little
more repulsive than in $^3$He: 
$E_C(^4_{\Lambda}{\rm He})-E_C(^3{\rm He})\simeq 0.02$ MeV.
Large part of the charge symmetry breaking observed in light
$\Lambda$--hypernuclei is due to the coupling between the $\Lambda N$ and 
$\Sigma N$ channels and turns out to be
quite sensitive to the mass difference between the initial and final state: 
\begin{equation}
\Delta m(\Lambda p\rightarrow \Sigma^+n)\simeq 75\: {\rm MeV} 
<\Delta m(\Lambda n\rightarrow \Sigma^-p)\simeq 80.5\: {\rm MeV} \nonumber
\end{equation}
(for transitions without charge exchange, $\Lambda p\to \Sigma^0 p$
and $\Lambda n\to \Sigma^0 n$, $\Delta m\simeq 77$ MeV).

Another important aspect of the $\Lambda N$ interaction is its spin--dependence.
A qualitative indication of the difference between the singlet ($J=0$) and
triplet ($J=1$) $\Lambda N$ interactions comes from the comparison of the $\Lambda$
binding energy in isobar nuclei not related by charge symmetry. For example,
$|B_{\Lambda}(^7_{\Lambda}{\rm Li})|$ is larger than 
$|B_{\Lambda}(^7_{\Lambda}{\rm Be})|$ by about $0.42$ MeV. 
The greater $|B_{\Lambda}|$ value 
corresponds to the hypernucleus whose nucleons' core has non--zero spin
(being an odd--odd system, while the core of $^7_{\Lambda}$Be is even--even);
it can be explained by
the effect of the spin--dependent $\Lambda$ interaction with
the unpaired nucleons, a proton and a neutron, in $^7_{\Lambda}$Li. However, the
$\Lambda N$ spin--spin interaction is weak. In ref \cite{Mill85}
the $\Lambda$--core spin doublet splittings ($J=|J_{\rm core}\pm 1/2|$),
which give the strength of the $\Lambda N$ spin--spin interaction, are
predicted to be of the order of $0.1$ MeV for $p$--shell hypernuclei,
and only for $^{7}_{\Lambda}$Li in the ground state
this splitting is sizeably larger ($\simeq 0.6$ MeV). The recent measurements,
at KEK and BNL \cite{Ta98,Tam01}, of the energy spacing of the $^7_{\Lambda}$Li
ground state doublet, $M1(\frac{3}{2}^+\to \frac{1}{2}^+)=691.7\pm 1.2$ KeV,
and of various $\gamma$--ray transitions in other $p$--shell hypernuclei confirmed
this prediction \cite{Mill01}. Experiments
with high energy resolution are then essential to study the spin--dependence
of the $\Lambda N$ interaction. From the analysis of the spins of ground and 
excited states in $\Lambda$--hypernuclei one expects the $\Lambda N$ interaction
to be more attractive in the spin--singlet state than in the spin--triplet state
\cite{Has95}.
In Ref.~\cite{St90} the authors found that the quark cluster model gives more
attraction in the triplet interaction; moreover, their meson--exchange potentials
are almost spin--independent. In the phenomenological OME models of the $\Lambda N$
interaction the situation is not clear \cite{Re94,Ri96,Ri99}: 
since there is no direct empirical information about the spin structure of the potential,
some versions favour the singlet interaction, while others favour the triplet one.
It has been found \cite{Ba86,Ba90} that the $\Lambda N$ effective interaction has 
repulsive character in the spin--parity $J^P=0^+$ channel, while for the $NN$ interaction an 
attractive $0^+$ pairing is well known. 
This anti--pairing effect originates from a delicate balance between the 
$\Lambda N$ inner repulsion and the attraction at intermediate distances. 

The spin--orbit component of the $\Lambda$--nucleus mean potential 
is rather small. The spin--orbit separation of the $\Lambda$ levels is at least one order
of magnitude smaller than the one typical of the $N$--nucleus interaction
\cite{Br78,Bo80,Ma81,Mill85,Mill01,Je90,Gi95,Ha99,Sa99}. Such effect could originate
from the weak tensor component of the 
$\Lambda N$ interaction, whose most important contributions come from the
exchange, forbidden at the lowest order, of pions and rhos. This supports
the hypothesis that the strong one--body spin--orbit potential experienced
in ordinary nuclei (central point in order to explain their exact
shell structure) originates from a two--body tensor force.    
However this point is not completely clear yet. In fact,
forces besides the spin--orbit one \cite{Mill01} as well as core excitations \cite{Hi98,Ha99,Ho01}
may contribute to the observed splittings as well. We will further discuss the problem of the
spin--orbit interaction in $\Lambda$--hypernuclei in the next section.

When a hyperon is embedded in the nucleus, one has to take into account the
influence of the medium on the hyperon. A simple mean field picture turns out
to be a good description of the bulk hypernuclear properties \cite{Do88}
(for example the hyperons binding energies and the excitation functions).
Within this approximation the hyperon maintains its single
particle behaviour in the medium, and it is well known that this occurs 
even for states well below the Fermi surface \cite{Pi91,Ha96}, a property which is
not observed for the nucleons. This is due to the fact that in hypernuclei 
the $\Lambda$ is a distinguishable 
particle, which weakly interacts with the nucleons' core. However, 
deviations from the independent particle description
can be produced, for instance, by three--body forces,
$\Lambda N$--$\Sigma N$ and $\Lambda \Lambda$--$\Xi N$ couplings,
QCD effects at the nuclear level
and non--localities due to relativistic effects. 

On the other hand, the presence of a hyperon influences the nuclear medium: hence, 
the Hartree--Fock approximation acquires a new self--consistency
requirement in strange nuclei. In spite of the relative weakness of the $\Lambda N$
interaction (with respect to the $NN$ interaction), for particular nucleon configurations
the single particle levels may be considerably lowered by the presence of a $\Lambda$:
for the deepest ones the energy shift
can reach $3\div 5$ MeV, while for the valence orbits a value of about $ 1$ MeV
is frequent \cite{Ba90}. For example,
the extra $1p$ neutron binding energy for a $p$--shell
hypernucleus $^{A+1}_{\Lambda}$Z due to the addition of the $\Lambda$
to the nucleus $^A$Z is calculable with the following relation:
\begin{equation}
B^{1p}_n(^{A+1}_{\Lambda}{\rm Z})-B^{1p}_n(^A{\rm Z})=
B_{\Lambda}(^{A+1}_{\Lambda}{\rm Z})-B_{\Lambda}(^A_{\Lambda}{\rm Z})<0 . \nonumber
\end{equation}
For a $1s$ neutron:
\begin{equation}
B^{1s}_n(^{A+1}_{\Lambda}{\rm Z})=B^{1p}_n(^{A+1}_{\Lambda}{\rm Z})+
M(^A_{\Lambda}{\rm Z})-M^*(^A_{\Lambda}{\rm Z}) , \nonumber
\end{equation}
where $M^*(^A_{\Lambda}{\rm Z})$ is the mass of the $1s$ neutron--hole
excited state of $^A_{\Lambda}{\rm Z}$, which can be produced by the 
$K^-n\rightarrow \Lambda \pi^-$ reaction on $^{A+1}Z$,
through the transformation of a $1s$
neutron into a $1s$ $\Lambda$--hyperon (see next section). Similar relations
hold for the proton levels.
The stability of the nucleons' core is increased by the presence of the 
$\Lambda$ particle, which plays then a ``glue--like'' role. Remarkable
examples are $^5$He and $^8$Be vs $^6_{\Lambda}$He and $^9_{\Lambda}$Be,
the former being unstable and the latter stable with respect to strong particle
emission. very recent $\gamma$--ray spectroscopy experiment at KEK \cite{Tan01} 
showed that the size of the $^6$Li core in $^7_{\Lambda}$Li is reduced
with respect to that of the loosely bound $^6$Li nucleus.
In a $^5_{\Lambda}$He--$d$ ($^4$He--$d$) cluster model for $^7_{\Lambda}$Li
($^6$Li) \cite{Hi99}, the rms distance between $^5_{\Lambda}$He and $d$ in 
$^7_{\Lambda}$Li is about 19\% smaller than the one between $^4$He and $d$ in $^6$Li.
The role of stabilizer of the $\Lambda$ in nuclei is due to its position
in the inner part of the nucleons' core, on single particle levels which are forbidden,
by the Pauli principle, to the nucleons.
On the other hand, the weak decay of the $\Lambda$ 
may cause the delayed fission of the 
host nucleus (because of the decreased stability and energy release of the
decay). This process has been used to measure the lifetime of heavy hypernuclei
\cite{Ar93,Ku98}.

We consider now the different behaviours, in nuclei, of
neutron and $\Lambda$ due to their decay modes.
In the free space neutron and $\Lambda$ are unstable; they decay
through the following weak channels:
\begin{equation}
\begin{array}{c c l l}
n &\rightarrow &  p\,e^-\,{\overline \nu}_e &(100\%) , \nonumber \\
\Lambda &\rightarrow & \pi^-\, p &(63.9\%) , \nonumber \\
        &            & \pi^0\, n &(35.8\%) . \nonumber
\end{array}
\end{equation}
The energy released in the neutron free decay is 
$Q^{\rm free}_n \equiv m_n-(m_p+m_e)\simeq 0.78$ MeV, 
while the binding energy of a nucleon in the nucleus is (in the average) 
$B_N\simeq -8$ MeV,
therefore a neutron in a nucleus is generally stable
(namely its decay is kinematically forbidden). On the other hand,
for the $\Lambda$ free decay the released energy is 
$Q^{\rm free}_{\Lambda} \equiv m_{\Lambda}-(m_{\pi}+m_N)\simeq 40$ MeV. This is
larger than the nuclear $\Lambda$ separation energy:
$|B_{\Lambda}|\lsim 27$ MeV (especially in
light hypernuclei), hence the $\Lambda$ is kinematically 
unstable even if embedded in nuclear systems.

The binding energies of nucleon and $\Lambda$ in nuclei, $B_N$ and 
$B_{\Lambda}$, have different behaviours as a function of the mass number:
$B_N$ saturates at about $-8$ MeV for nuclei with $A\gsim 10$, while 
$|B_{\Lambda}|$ monotonically increases with $A$ up to $\simeq 27$ MeV for
$^{208}_{\Lambda}$Pb. Indeed, the $\Lambda$ can
occupy whatever single particle state, the ground state of the
hypernucleus always corresponding to the hyperon in the $1s$ level.
It is then clear that a $\Lambda$
particle is a good probe of the inner part of nuclei. Actually, the Pauli
principle is active on the $u$ and $d$ quarks of nucleons and 
$\Lambda$ when they are very close to each other. For example, in the
case of  $^5_{\Lambda}$He$\equiv ppnn\Lambda$, if the constituent
baryons maintain their identity, both the hyperon and the nucleons
occupy $s$ levels, while at the
quark level an {\it up} quark in the $\Lambda p$ short range interaction
(and a {\it down} quark in $\Lambda n$)
has to occupy the $p$--level. The Pauli blocking effect 
at the quark level could be an important ingredient to explain
the anomalously small $^5_{\Lambda}$He binding energy with respect to 
calculations performed within the baryon picture.
A study of the role played by the quark
Pauli principle on the binding energies of single-- and double--$\Lambda$ $s$--shell 
hypernuclei can be found in Ref.~\cite{Su99}. The authors have found significant
effects when the assumed size of the baryons is of the order of the proton charge radius: 
$b\simeq 0.86$ fm.

With the exception of
hypernuclei of the $s$--shell, the depth of the $\Lambda$--nucleus mean field
is of about $30$ MeV \cite{Do88}, namely it is less attractive than 
the one typical for a nucleon 
($\simeq 50\div 55$ MeV). This characteristic reflects the smaller range and the 
weakness of the $\Lambda N$ interaction at intermediate distances
with respect to the $NN$ one.
It is possible to reproduce the experimental single particle
$\Lambda$ levels using Woods--Saxon wells with the above depth and appropriate radii. 
For $s$--shell hypernuclei the $\Lambda$ single particle potential displays a repulsive
soft core at short distances \cite{Mo91,St93,Ou98}. A measure of this effect is
given by the rms radii for a nucleon and a $\Lambda$ in these hypernuclei:
the hyperon rms radius is larger than the one for a nucleon.

\subsubsection*{$\Sigma N$ interaction and $\Sigma$--hypernuclei}

The investigation of the $\Sigma N$ interaction 
is richer but more difficult than that of the $\Lambda N$ interaction.
We remind the reader that it exhibits a long range OPE component, its central part is
weaker than the $\Lambda N$ one and it is very sensitive to 
spin and isospin \cite{Do89,Ha90,Ak97}. Very roughly, the strengths 
of the averaged
$NN$, $\Lambda N$ and $\Sigma N$ potentials are in the following
ratios: $NN/\Lambda N\simeq 3/2$, $NN/\Sigma N\simeq 3$. The strong
spin--isospin dependence in the $\Sigma N$ interaction is natural in OME models and it is
due to the exchange of both isoscalar ($\omega$, $\eta$) and
isovector mesons ($\pi$, $\rho$). The $\Sigma N$ spin--orbit strength is expected
to be about $0.5\div 1$ times the $NN$ one \cite{Do89}. 
Calculations and experimental observations have shown that both
the $\Lambda N$ and $\Sigma N$ effective potentials are strongly repulsive at 
short distances, and a repulsive core even remains in the
$\Lambda$--$\alpha$ and $\Sigma^0$--$\alpha$ folding potentials \cite{Ak97,Ou98}
(which describe the $\Lambda$
and $\Sigma$ dynamics in $^5_{\Lambda}$He and $^5_{\Sigma}$He, respectively);
however, differently from $^5_{\Lambda}$He ($B_{\Lambda}\simeq 3.12$ MeV),
because of the large repulsion in the inner region, the $\Sigma^0$--$\alpha$ potential
does not support bound states. 
On the contrary, for the $NN$ interaction the attraction at intermediate distances
is so strong that the $N$--$\alpha$ potential obtained by the folding procedure
does not contain the inner repulsive component.
For heavy nuclei a repulsive bump could appear on the surface of 
the $\Sigma$--nucleus potential because of the particular balance of
repulsion and attraction (which is less effective on the nuclear surface) in the
$\Sigma N$ interaction \cite{Ak97}.  

The first production signals interpreted as $\Sigma$--hypernuclear states, 
20 years ago at CERN, showed unexpected narrow peaks 
(less than $8$ MeV, instead of the $20\div 30$ MeV estimated for
nuclear matter \cite{Da81,Do89}) which were assigned to the formation of
$^9_{\Sigma}$Be, $^{12}_{\Sigma}$C, $^{12}_{\Sigma}$Be and
$^{16}_{\Sigma}$C \cite{Be85}. The first observation reported two narrow
peaks above the $\Sigma$ binding threshold separated by about 
$12$ MeV in the $^9$Be$(K^-,\pi^-)$ strong reaction and were ascribed to the
formation of $^9_{{\Sigma}^0}$Be. 
However, the measurements were carried out with very low statistics
and the identification of the peaks involved large ambiguities. Moreover,
none of the reported states could be assigned to hypernuclear ground states.
Recently, at BNL--AGS \cite{Ba99}, by employing ten times better statistic,
the existence of such narrow structures for $p$--shell hypernuclei has been excluded.
Due to the relevance of the $\Lambda N$--$\Sigma N$ coupling,
$\Sigma$--hypernuclei can also be regarded as resonant states of 
$\Lambda$--hypernuclei. On the other hand, in a $\Sigma$--hypernucleus the
$\Sigma N\rightarrow \Lambda N$ conversion creates a $\Lambda$ with a
kinetic energy of about $40$ MeV. Since the $\Lambda$--nucleus well depth is
smaller than this energy, the $\Lambda$--hyperon has a thick probability
to escape from the nucleus and decays after $\simeq 2$ cm.
From the theoretical point of view, the
existence of narrow $\Sigma$ state in nuclei cannot
be explained only in the (plausible)
hypothesis of a sizeable $\Sigma N\rightarrow \Lambda N$ strong converting process. 
Among the mechanisms introduced in order to suppress the
calculated widths of $\Sigma$--hypernuclei \cite{Br82,Do89,Os90,Ba90,Ya94}, 
the most relevant ones are the Pauli blocking effect on the final nucleon 
in $\Sigma N\rightarrow \Lambda N$, the suppression 
of particular spin--isospin transitions
and the medium polarization effect. The latter is 
accounted for in \cite{Br82,Os90} 
through the so--called {\it induced interaction approach}. Moreover, it is also possible
that the $\Sigma \rightarrow \Lambda$ conversion is less efficient in finite nuclei
because the $\Sigma$--nucleus potential has such a small depth that the
$\Sigma$ wave function is considerably pushed out of the nucleus. As already pointed
out, it has been established that for $s$--shell $\Sigma$--hypernuclei the hyperon
is pushed towards the nuclear surface by a central repulsion in the
$\Sigma$--nucleus potential.
The above effects can reduce the $\Sigma N\rightarrow \Lambda N$ width up to
$5\div 10$ MeV in $p$--shell hypernuclei \cite{Ba90,Os90}. 

There are many ambiguities in our knowledge of the properties of the 
$\Sigma$--nucleus potential, as obtained
from hypernuclear and $\Sigma N$ scattering data studies.
If this potential had small depth, in the production of 
heavy systems there should be the problem of resolving the 
small spacing among the single particle
levels. In fact, if the energy separation among the $\Sigma$--levels
is lower than their widths, these states cannot be resolved by the
experiment. The analysis of the few existing data
on $\Sigma^-$--atoms and of $(K^-,\pi^{\pm})$ production
indicate \cite{Be85,Do89,Ba90,Mar95,Ak97} a $\Sigma$ single particle potential depth
in the range $8\lsim |V^{\Sigma}_0| \lsim 15$ MeV for hypernuclei beyond the $s$--shell.
Very shallow depths ($-V^{\Sigma}_0 \simeq 10$ MeV) 
are consistent with the $(K^-,\pi^{\pm})$ analysis, which, in fact, has not proved
the existence of $\Sigma$ nuclear states beyond the $s$--shell. Instead,
$-V^{\Sigma}_0\simeq 20$ MeV is more consistent with
$\Sigma$--atoms data, which, however, are not sensitive to the interior part of the
nucleus: hence, $20$ MeV overestimates $|V^{\Sigma}_0|$.
Moreover, from the above cited analysis, the $\Sigma$--nucleus potential 
turns out to be strongly spin-- and isospin--dependent, with a 
spin--orbit part comparable with the $N$--nucleus one. 
From further theoretical speculations \cite{Ya94} and
experiments carried out in the last years at KEK \cite{Ha89,Ou94} and 
BNL \cite{Ba99,Na98}, the existence of 
$\Sigma$ bound states for nuclear mass numbers $A\geq 4$ seems to be strongly 
unlikely. On the other hand, the existence of the predicted \cite{Ak86} 
$^4_{\Sigma}$He bound state has been proved 
(with binding energy $-B_{\Sigma}\simeq 2.8\div 4.4$ MeV and width 
$\Gamma_{\Sigma}\simeq 7.0\div 12.1 $ MeV) both at KEK \cite{Ha89} and BNL \cite{Na98}. 
Actually, only for very light systems the
widths are expected to be narrower than the separation among the $\Sigma$--levels;
moreover, for hypernuclei other than the $s$--shell ones, 
the $\Sigma$--nucleus potential could not be deep
enough to accommodate $\Sigma$ bound states. Other investigations support the presence
of a substantial repulsive component in the $\Sigma$--nucleus potential also
in medium and heavy hypernuclei \cite{Ba94}. It is then clear that 
further experiments and theoretical work are needed to 
properly understand the existence of $\Sigma$--hypernuclei. 

\subsubsection*{Strangeness $S=-2$ hypernuclei and the $H$--dibaryon}

Some experiments have revealed the existence of 
$\Xi$--hypernuclei \cite{Do83,Ba90,Nak98,Ma98,Fuk98}.
They are produced through the $K^-p\rightarrow K^+ \Xi^-$, 
$K^-p\rightarrow K^0 \Xi^0$ and $K^-n\rightarrow K^0 \Xi^-$ strong reactions, which, 
because of the relatively large momentum transferred 
($\simeq 500$ MeV), preferentially excite high total spin hypernuclear states.
The measured $1s$ $\Xi^-$ binding energies (old emulsion data)
have been fitted by using a Woods--Saxon potential
with radius $R=1.1A^{1/3}$ fm, depth $20\lsim |V^{\Xi}_0| \lsim 28$ MeV
and surface diffuseness $a=0.65$ fm \cite{Do83}. The depth $V^{\Xi}_0$ compares
well with theoretical predictions based on Nijmegen OME
models and allows for the binding of several $\Xi$ levels.
More recent speculations favour smaller well depths, around 
$12\div 16$ MeV \cite{Fuk98,Ka00} for $^{12}_{\Xi^-}$C and $^{12}_{\Xi^-}$Be.
However, improved experiments are needed to extract precise information concerning
the $\Xi$--nucleus potential \cite{Mo01}; for example it is not yet
clear whether the potential depth exhibit a mass number dependence \cite{Ta95}. 
The authors of Ref.~\cite{Ak97} obtained a $\Xi^-$--$\alpha$ potential
characterized by a quite strong inner repulsion and a shallow attraction
at intermediate distances: the $\Xi^-$ wave function is pushed on the nuclear
surface and the small $\Xi^-$ binding energy has been reproduced.
They have also found that in the formation 
of $_{\Xi^-}^5$He an important role is played by the Coulomb interaction.

When a $\Xi^-$--hypernucleus (or a $\Xi^-$--atom) is formed,
the hyperon strongly interacts with a nucleon  
of the medium (exchanging a strange meson, $K$ or $K^*$) and produces two 
$\Lambda$'s with an energy release of only $28$ MeV
(further reduced by binding effects): $\Xi^- p\rightarrow \Lambda \Lambda$.
This offers the possibility of producing
double--$\Lambda$--hypernuclei \cite{Ao91,Ic01}, which were observed
for the first time during the 60's in emulsion experiments. The formation 
probability of a $\Lambda \Lambda$--hypernucleus is sizeable because
the $28$ MeV energy release in $\Xi^- p\rightarrow \Lambda \Lambda$ is only $0.1$ \% larger than the
separation energy in an $\alpha$--particle. Therefore, if an $\alpha$--cluster
is broken as a consequence of the $\Xi^-$ absorption, the final $\Lambda$'s
will not have enough energy to escape from the nucleus. The production probability
of a double--$\Lambda$ or twin--$\Lambda$ hypernucleus turned out to be
$(18\pm13)$\% in the experiment of Ref.~\cite{Ah00}, which used 
$\Xi^-$ atomic capture on $^{12}$C.

The strangeness $S=-2$ hypernuclei are quite interesting because they represent
the unique way of getting
information on the $\Xi N$ and $\Lambda \Lambda$ interactions.
In Ref.~\cite{Do83} the conversion width due to the process 
$\Xi^- p\rightarrow \Lambda \Lambda$ has
been estimated to be quite narrow (of the order of $5$ MeV or less), 
as one expects because of the small energy released in the process.
More recent calculations have found conversion widths narrower 
than the spacing among the $\Xi$ levels: typically 
$\Gamma_{\Xi}\simeq 1.6$ MeV for $s$--states and  $\Gamma_{\Xi}\simeq 0.9$ MeV
for $p$--states \cite{Fuk98}.  For $^5_{\Xi^-}$He the calculation of Ref.~\cite{Ku95}
obtained a very small width, $\Gamma_{\Xi}=0.76$ MeV, which results from a
small overlap between the $\Xi^-$ wave function and the nuclear core 
(the $\Xi^-$ binding energy being only $1.7$ MeV). Therefore, if the
experimental energy resolution is good enough, the smallness of 
$\Gamma_{\Xi}$ makes it feasible to perform spectroscopic studies.
Because of the small mass difference between initial and final states, the 
$\Lambda \Lambda$--$\Xi N$ coupling plays an important role in double $\Lambda$
hypernuclei. However, a suppression of this coupling coming from
the Pauli blocking on the nucleon becomes sizeable in medium--heavy
hypernuclei. On the contrary, the 
$\Lambda \Lambda$--$\Sigma \Sigma$ strong coupling will be
less important for the large mass difference ($\Delta m\simeq 155$ MeV).

The study of $\Xi$-- and $\Lambda \Lambda$--hypernuclei 
is closely related to the observation of hyperon mixed states due to the
$\Lambda \Lambda$--$\Xi N$--$\Sigma \Sigma$ couplings \cite{Gi94}
and, in particular, to the search for a stable $H$--particle. The latter 
is predicted to be a six quark state containing two $u$, two $d$ 
and two $s$ quarks coupled into a singlet $SU(3)$
state of both colour and flavour: 
it should have $J^P=0^+$, $I=0$ and it should be stable against strong decays 
(obviously, if its mass is smaller than twice the $\Lambda$ mass).
This object has baryonic number 2 but it is not an
ordinary nuclear state, namely the three quark clusters contained in $H$ are deconfined.
This kind of di--baryon was predicted by Jaffe in 1977 \cite{Ja77} within a quark bag model.
Nowadays, searching experiments are running \cite{Nak98,Ya98,Ch98,Ah00,Ya00,Al00}. 
From observations on double--$\Lambda$--hypernuclei, the expected mass is 
$m_H\equiv 2m_{\Lambda}+B_H\gsim 2m_{\Lambda}-28$ MeV 
\cite{St90,Ao91,Nak98}. The first calculation by 
Jaffe found a large value for $B_H$ ($-80$ MeV). Should 
the binding energy of the $H$--dibaryon
be more attractive than the binding energy of two $\Lambda$'s in nuclei, 
$B_H<B_{\Lambda \Lambda}\equiv M(^{A+2}_{\Lambda \Lambda}{\rm Z})-
M(^A{\rm Z})-2m_{\Lambda}$, then the di--baryon should
be strongly emitted from the nucleus
($^{A+2}_{\Lambda \Lambda}{\rm Z}\rightarrow {^A{\rm Z}} + H$), 
and the hypernucleus would have a very short
lifetime. On the contrary, if $B_H>B_{\Lambda \Lambda}$, successive decays 
of the two hyperons (weak processes) should be 
observed, but this would not necessarily imply the non--existence of the $H$ di--baryon:
the $\Lambda \Lambda$ interaction could also be attractive, although weaker
than $B_{\Lambda \Lambda}$. It is then clear that, in this sense, the
stability of double--$\Lambda$ hypernuclei may hinder the experimental detection of
the $H$--particle: the observation of the weak decay of a double--$\Lambda$ hypernucleus
only excludes the $H$ mass in the region $m_H<2m_{\Lambda}+B_{\Lambda \Lambda}$.
From the present experimental searches there is no unambiguous evidence 
which supports the existence of di--baryon resonances in the strange sector. 

Studies of the $\Lambda \Lambda$ contribution
to the experimental binding energy 
$B_{\Lambda \Lambda}$ are quite difficult because of the
few data available on double--$\Lambda$ hypernuclei and of the
density dependence of the $\Lambda \Lambda$ interaction 
($\Lambda \Lambda$--$\Xi N$ coupling, three body forces, etc). 
This interaction occurs by the exchange of $I=0$ mesons at lowest order, 
which favours an attractive character for $V_{\Lambda \Lambda}$. 
Nuclear emulsion experiments reported the
observation of three double--$\Lambda$ hypernuclei: $^6_{\Lambda \Lambda}$He,
$^{10}_{\Lambda \Lambda}$Be and $^{13}_{\Lambda \Lambda}$B. From these
events, an effective $\Lambda \Lambda$ matrix element
$-\langle V_{\Lambda \Lambda} \rangle\simeq \Delta B_{\Lambda \Lambda}
\equiv | B_{\Lambda \Lambda}| -2
| B_{\Lambda}| \simeq 4\div5$ MeV \cite{Gi94} was determined,
$| B_{\Lambda \Lambda}|$ being the separation energy 
of the $\Lambda$ pair from the
$^{A+2}_{\Lambda \Lambda}$Z hypernucleus and $| B_{\Lambda}|$ the hyperon separation
energy from the $^{A+1}_{\Lambda}$Z hypernucleus. 
However, a very recent counter--emulsion hybrid experiment, performed
at KEK \cite{Tak01}, favours a quite weaker $\Lambda \Lambda$ interaction:
$\Delta B_{\Lambda \Lambda}(^6_{\Lambda \Lambda}{\rm He})=1.01^{+0.27}_{-0.23}$ MeV.
The authors of Ref.~\cite{Tak01} advanced the hypothesis that in the previous
emulsion experiments, the single--$\Lambda$ hypernuclei were produced
in excited states. In this case, a value of $\Delta B_{\Lambda \Lambda}$ 
around 1 MeV is expected also from these experiments.
The production of $^4_{\Lambda \Lambda}$H hypernuclei has been reported, very recently,
in a counter experiment at BNL \cite{Ah01}. Unfortunately, due to the limited statistics,
the authors have not determined $\Delta B_{\Lambda \Lambda}$.
The quantity $\Delta B_{\Lambda \Lambda}$, called $\Lambda \Lambda$ bond energy, 
is expected to decrease as the nuclear mass number
$A$ increases and goes to zero in the limit $A\rightarrow \infty $: 
for increasing $A$ the attraction between the $\Lambda$'s
becomes weaker because of the larger $\Lambda \Lambda$ average distance.
We note here that in Ref.~\cite{La98} the
author, by using the Skyrme--Hartree--Fock approach, found the approximation
$-\langle V_{\Lambda \Lambda}\rangle \simeq \Delta B_{\Lambda \Lambda}$ between
$\Lambda \Lambda$ interaction strength and bond energy 
to be questionable. Indeed these quantities seem to be sizeably affected by the interplay
of several factors, such as the spatial distributions of the $\Lambda$'s and
the core polarization: for $^{13}_{\Lambda \Lambda}$B the author evaluated
$-\langle V_{\Lambda \Lambda}\rangle \simeq 2.1\div 5.6$ MeV (depending 
on the various parameterizations used for the $\Lambda \Lambda$ Skyrme potential)
once the $\Lambda \Lambda$ potential was adjusted to reproduce the value
$\Delta B_{\Lambda \Lambda}= 4.8$ MeV of the old emulsion experiment.
In a double--$\Lambda$ hypernucleus the two hyperons are in the
${^1S_0}$ relative state (the ${^3S_0}$ is not allowed by Pauli principle).
We can then compare the $\Lambda \Lambda$ interaction matrix elements with the
${^1S_0}$ ones for $\Lambda n$ and neutron--neutron interactions in light systems:
$-\langle V_{\Lambda n}\rangle \simeq 2\div 3$ MeV, 
$-\langle V_{nn}\rangle \simeq 6\div 7$ MeV. 
We know that the ${^1S_0}$ $nn$ system is not bound.
However, a ${^1S_0}$ $\Lambda \Lambda$ bound system, 
which has a smaller matrix element than $nn$, cannot be excluded on this basis
because of the unknown balance between the short range 
repulsion and the intermediate distance attraction in the $\Lambda \Lambda$ interaction.
On the other hand, also the $\Lambda \Lambda$--$\Xi N$ coupling must be taken into
account \cite{Af00}. Measurements of $B_{\Lambda \Lambda}$ in medium and
heavy double--$\Lambda$--hypernuclei are expected, too.

We conclude this section by recalling
that hypernuclei are always unstable with
respect to weak decay. A variety of processes are in principle
accessible (which do not have counterpart in the non--strange sector).
Limiting ourself to $\Delta S=1$ transitions we have:
\begin{itemize}
\item for $\Lambda$--hypernuclei:
\begin{equation}
\begin{array}{c c c l l}
\Lambda N&\rightarrow &nN & & (Q\simeq 176\: {\rm MeV}) ; \nonumber
\end{array} 
\end{equation}
\item for $\Sigma$--hypernuclei:
\begin{equation}
\begin{array}{c c c l l}
\Sigma N&\rightarrow &NN & & (Q\simeq 255\: {\rm MeV}) ; \nonumber
\end{array}
\end{equation}
\item for $\Xi$--hypernuclei:
\begin{equation}
\begin{array}{c c c l l}
\Xi N &\rightarrow &\Lambda N & & (Q\simeq 202\: {\rm MeV}) , \nonumber \\ 
& \rightarrow & \Sigma N & & (Q\simeq 123\: {\rm MeV}) ; \nonumber
\end{array}
\end{equation}
\item for $\Lambda \Lambda$--hypernuclei:
\begin{equation}
\label{lamlam}
\begin{array}{c c c l l}
\Lambda \Lambda &\rightarrow & \Lambda n & & (Q\simeq 176\: {\rm MeV}) , \\ 
&\rightarrow & \Sigma N & & (Q\simeq 97\: {\rm MeV}) ;
\end{array}
\end{equation}
\item and many other processes for multi strangeness systems ($S\leq -3$), for example:
\begin{equation}
\begin{array}{c c c l l}
\Xi \Lambda &\rightarrow &\Xi N & & (Q\simeq 174\: {\rm MeV}) , \nonumber \\
& \rightarrow & \Lambda \Lambda & & (Q\simeq 199\: {\rm MeV}) , \nonumber \\
\Xi \Xi &\rightarrow &\Xi \Lambda & & (Q\simeq 199\: {\rm MeV}) . \nonumber
\end{array}
\end{equation}
\end{itemize}
These decays are expected to have lifetimes of the order of $10^{-10}$ sec or less. 
However, when hyperons other than the $\Lambda$
are embedded in a nucleus, strong processes, which have very short lifetimes
($\tau\simeq 10^{-22}\div 10^{-21}$ sec), dominate
over the quoted weak decays, preventing them to occur.
For double--$\Lambda$ hypernuclei, the $\Lambda$--induced weak decay rates 
[Eq.~(\ref{lamlam})] of $s$--shell systems are estimated \cite{It01,Pa01a} 
to be suppressed by a factor $25\div 70$
with respect to the free $\Lambda$ width, and are impossible to 
detect at present.

\newpage
\section{Production of hypernuclei}
\label{prod}
\subsection{Introduction}
\label{prod1}
The new development (in the last 15 years) of counter experiments
have opened a new phase of hypernuclear physics. In fact, the old
experiments in the 60's used emulsion and 
bubble chamber techniques and, practically, they only
measured the hyperons binding energies. Through
counter techniques, the experiments have discovered new and interesting features
of the hypernuclear structure, although several questions still remain
unsolved.

Hypernuclei can be produced by using strong processes in which a particle 
(generally a pion or a kaon) hits a nucleus. 
Since strangeness has to be conserved, one can use the following 
production reactions: 
\begin{enumerate}
\item Processes with strangeness exchange:
\begin{equation}
\begin{array}{l l l}
K^-n &\rightarrow &\Lambda \pi^- , \nonumber \\ 
& & \Sigma^0 \pi^- , \nonumber \\
& & \Sigma^- \pi^0 , \nonumber \\
K^-p &\rightarrow &\Sigma^- \pi^+ , \nonumber \\
& & \Sigma^+ \pi^- , \nonumber \\
& & \Lambda \pi^0 , \nonumber \\
& & \Sigma^0 \pi^0 ;
\end{array}
\end{equation}

\item Processes with associated production of strange hadrons:
\begin{equation}
\begin{array}{l l l l l}
\pi^+n&\rightarrow &\Lambda K^+ , & & \nonumber \\
      & & \Sigma^0 K^+ , & & \nonumber \\
      & & \Sigma^+ K^0 , & & \nonumber \\
\pi^+p&\rightarrow &\Sigma^+ K^+ , & & \nonumber \\
\gamma p&\rightarrow &\Lambda K^+ & & {\rm (photoproduction)} , \nonumber \\
e^- p&\rightarrow & e^-\Lambda K^+  & & {\rm (electroproduction)} , \nonumber \\
pN&\rightarrow &\Lambda K^+N & & {\rm (proton-induced)} ;
\end{array}
\end{equation}

\item Reactions in which strangeness exchange and associated production of
strangeness are combined (used for the production of $S=-2,-3$ hypernuclei):
\begin{equation}
\begin{array}{l l l}
K^-p&\rightarrow &\Xi^- K^+ , \nonumber \\
& & \Xi^0 K^0 , \nonumber \\
& & K^+K^0\Omega^- , \nonumber \\
K^-n&\rightarrow &\Xi^- K^0 ,  \nonumber \\
K^-p&\rightarrow &\Lambda \pi^0 , 
\hspace{0.1 in}{\rm followed \: by} \hspace{0.1 in} \pi^0 p \rightarrow \Lambda K^+ , \nonumber \\
pp&\rightarrow & \Xi^0 K^+K^+n . 
\end{array}
\end{equation}
\end{enumerate}
In the following, we denote with $N(a,b)Y$, or simply with $(a,b)$, the process:
\begin{equation}
aN\rightarrow Yb , \nonumber
\end{equation}
where $N$ is a nucleon and $Y$ a hyperon. The considered reactions have different 
characteristics depending on their kinematics. In the following we shall see that,
because of the complementarity of the reactions,
the combined use of various production modes is important for exhaustive 
spectroscopic studies.

\begin{figure}
\begin{center}
\mbox{\epsfig{file=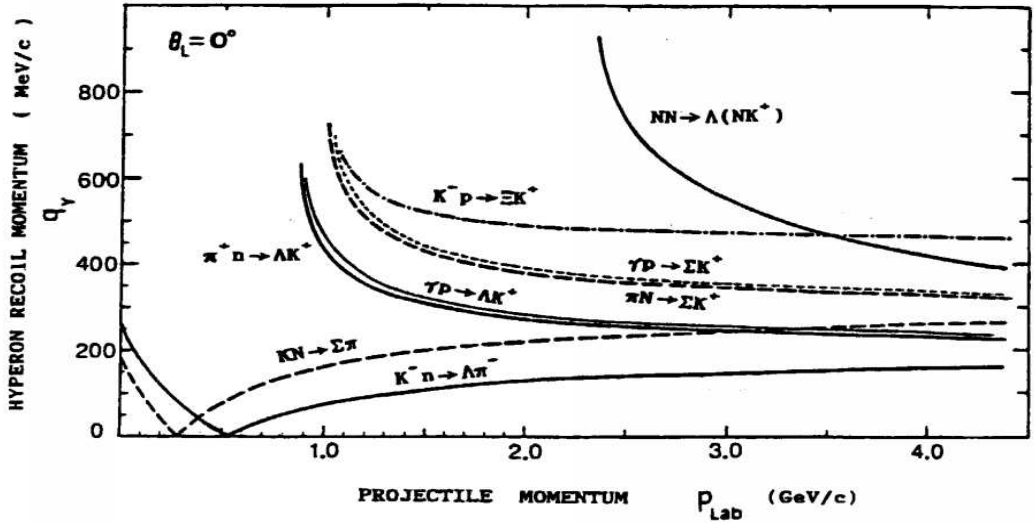,width=1.0\textwidth}}
\vskip 2mm
\caption{Momentum $q_Y$ transferred to the hyperon $Y$ as a function of
the projectile momentum in the laboratory frame $p_{Lab}$ for the reaction
$aN\protect\to Yb$ at $\theta_{b,Lab}=0^{\circ}$ (taken from Ref.~\cite{Ba90}).}
\label{momtrans}
\end{center}
\end{figure}

In order to produce a hypernucleus, the hyperon emerging from the reaction
has to remain inside the nuclear system. The formation
probability of a hypernucleus depends on the energy transferred in the
production. When the momentum transferred to the hyperon, $q_Y$, is much larger
than the nuclear Fermi momentum $k_F$, the hyperon has a very small 
{\it sticking probability}
and it leaves the nucleus. Instead, when $q_Y\lsim k_F$, the hyperon is created,
with a high probability, in a bound state. In figure~\ref{momtrans}  
are shown the momenta transferred to the hyperon $Y$ in the reactions $N(a,b)Y$ as a 
function of the projectile momentum $p_{a}$ at $\theta_{b}=0^{\circ}$ 
in the laboratory frame. 
With the exception of the $(K^-,\pi^{\pm})$ reactions, the other ones reported 
in the figure are endoenergetic, therefore the
hyperon cannot be produced at rest: $q_Y$ decreases
as the projectile momentum increases but it remains finite for high $p_{a}$. 
In this situation the hyperon is produced with a non--negligible probability 
above its emission threshold, namely with $B_{\Lambda}>0$ (quasi--free production).
Some hypernuclear states in the continuum may 
be quasi--bound state: they do not emit the hyperon but nucleons and/or 
cluster of nucleons. 
\subsection{The $(K^-,\pi^{\pm})$ strangeness exchange reactions}
\label{prod2}
In the $(K^-,\pi^{\pm})$ production reactions the incident $K^-$ transforms 
the struck neutron (proton) into a $\Lambda$ or $\Sigma^0$ ($\Sigma^-$) and a $\pi^-$
($\pi^+$) is emitted with an energy spectrum which is directly related to the 
populated hypernuclear level.
The reactions $n(K^-,\pi^-)\Lambda$, $p(K^-,\pi^+)\Sigma^-$
(used for the first time at CERN \cite{Be81} and BNL \cite{Ma81}
to produce $\Lambda$-- and $\Sigma$--hypernuclei)
are esoenergetic and can create the
hyperon at rest ($q_Y=0$). By considering, as an approximation, the 
initial neutron in $n(K^-,\pi^-)\Lambda$ at rest, the
transferred momentum is zero, $\vec p_K=\vec p_{\pi}\equiv \vec p$, and the
pion is emitted at $\theta=0^{\circ}$ in the laboratory frame. 
Thus, from energy--momentum conservation:
\begin{equation}
\sqrt{\vec p\,^2+m^2_K}+m_N=m_{\Lambda}+\sqrt{\vec p\,^2+m^2_{\pi}} , \nonumber
\end{equation}
and the momentum for the production of the $\Lambda$ 
at rest (called magic momentum) can be derived as follows:
\begin{equation}
E_K\equiv \sqrt{\vec p\,^2+m^2_K}=\frac{m^2_K-m^2_{\pi}+(m_{\Lambda}-m_N)^2}
{2(m_{\Lambda}-m_N)} \hspace{0.12 in}\Rightarrow \hspace{0.12 in} 
p \simeq 530\: {\rm MeV} . \nonumber
\end{equation}
If the production reaction is $p(K^-,\pi^+)\Sigma^-$, the kaon magic momentum is 
$p\simeq 280$ MeV. 

Since both the initial $K^-$ and the final pion are strongly absorbed in the nucleus
(they have a small mean free path), the kaon induced
reactions preferentially populate less bound $\Lambda$--levels and they have been 
only employed for $s$-- and $p$--shell hypernuclear studies.
Moreover, the low intensity and poor resolution of the kaon beams hinder the use of the
$(K^-,\pi^{\pm})$ reactions.

By using the strangeness exchange reaction at $\theta=0^{\circ}$,
the hyperon is predominantly produced in a state with the same 
quantum numbers of the struck nucleon,
namely the neutron hole and the $\Lambda$ are coupled to $J^P=0^+$ and $\Delta l=0$
(substitutional reaction). By increasing $\theta$ the relative importance 
of $\Delta l=1, 2,$ etc transitions 
increases, and hypernuclear states with higher spin can be produced.
From measures at both $\theta\simeq 0^{\circ}$ and $\theta > 10\div 15^{\circ}$ 
it has been possible to study a large part of the level structure of light hypernuclei
\cite{Ma81,Sa99}.

Spectroscopic studies with the reaction $n(K^-,\pi^-)\Lambda$ in a few hypernuclei
($^{13}_{\Lambda}$C, $^{16}_{\Lambda}$O and others) 
have shown that the spin--orbit part of the
$\Lambda$--nucleus mean potential is very small compared to 
the one of a nucleon \cite{Br78,Bo80,Ma81}, although the exact magnitude is not 
known yet. Taken, for instance, the case of $^{16}_{\Lambda}$O, 
the measured $\Lambda$ and nucleon
$p_{1/2}-p_{3/2}$ spin--orbit shifts are \cite{Br78}:
\begin{equation}
\Delta E_{\Lambda}(^{16}_{\Lambda}{\rm O}; 1p_{1/2}-1p_{3/2})\leq 0.3\: {\rm MeV} \ll
\Delta E_N(^{16}_{\Lambda}{\rm O}; 1p_{1/2}-1p_{3/2})\simeq 6\: {\rm MeV} . \nonumber
\end{equation}
This estimate comes from the observed peaks in the excitation spectrum, which are
reported in table~\ref{spinorb} with the relative $(N^{-1},\Lambda)$ configurations.
We see that the $p_{1/2}-p_{3/2}$ 
spin--orbit separation for the nucleon is obtained by subtracting the 
energies of peaks \#3 and \#4. From the observation that almost
the same separation exists between the peaks \#1 and \#2, we can infer that
the analogous spin--orbit separation for the $\Lambda$ is compatible with zero.
Subsequent $(\pi^+,K^+)$ experiments have confirmed small $\Lambda$ spin--orbit splittings. 
Very recently, the hyperon $1p_{1/2}-1p_{3/2}$ splitting of $^{13}_{\Lambda}$C
hypernuclei, produced by the $^{13}{\rm C}(K^-,\pi^-)^{13}_{\Lambda}$C reaction, has been
measured at BNL \cite{Aj01}, with the result:
\begin{eqnarray}
\Delta E_{\Lambda}(^{13}_{\Lambda}{\rm C}; 1p_{1/2}-1p_{3/2})&=&0.152\pm 0.065\, {\rm MeV} \nonumber \\
&<<& \Delta E_N(^{13}_{\Lambda}{\rm C}; 1p_{1/2}-1p_{3/2})\simeq 3\div 5\, {\rm MeV}. \nonumber
\end{eqnarray}
To our best knowledge, only the analysis of Dalitz
{\em et al.}~\cite{Da97} of old emulsion data on $^{16}_{\Lambda}$O found larger effects:
$\Delta E_{\Lambda}(^{16}_{\Lambda}{\rm O}; 1p_{1/2}-1p_{3/2})= 1.30\div 1.45$ MeV. 
On the other hand, the smallness of the $\Lambda$--nucleus spin--orbit interaction
arises naturally in a relativistic mean field description \cite{Je90}.
\begin{table}
\begin{center}
\caption{Peaks observed at CERN in $^{16}_{\Lambda}$O production 
(taken from Ref.~\protect\cite{Br78}).}
\label{spinorb}
\vspace{0.5cm}
\begin{tabular}{|c|c c|} \hline
\mc {1}{|c|}{Peak} &
\mc {1}{c}{$B_{\Lambda}$ (MeV)} &
\mc {1}{c|}{Configuration} \\ \hline\hline
 \#1 & 3.5 &   $(1p^{-1}_{3/2},1p^{\Lambda}_{3/2})_{J^P=0^+}$  \\
 \#2 &  -2.5&  $(1p^{-1}_{1/2},1p^{\Lambda}_{1/2})_{0^+}$ \\
 \#3 &  -7 &   $(1p^{-1}_{3/2},1s^{\Lambda}_{1/2})_{1^-}$  \\
 \#4 &  -13&   $(1p^{-1}_{1/2},1s^{\Lambda}_{1/2})_{1^-}$  \\ \hline
\end{tabular}
\end{center}
\end{table}
The $\Lambda$  $p_{1/2}-p_{3/2}$ splitting is generally considered to 
be originated predominantly from the $\Lambda N$ spin--orbit
force acting on the $\Lambda$ in the $p_{1/2}$ and $p_{3/2}$ levels of the 
$\Lambda - ^{15}$O system. However, excitations of the core
may contribute to the spin--orbit splitting as well \cite{Hi98,Ha99},
especially in heavy hypernuclei \cite{Mo89,Ho01}. Hence,
the smallness of the $\Lambda$ spin--orbit splittings does not necessarily imply a weak 
$\Lambda N$ spin--orbit interaction\footnote{Yet, we know that it is 
smaller than the $NN$ spin--orbit force, 
with a ratio $V^{\it l-s}_{NN}/V^{\it l-s}_{\Lambda N}\simeq 3\div 10$
between the strengths expected from phenomenological
studies of the baryon--baryon potentials \cite{Re94,Ri99,Fu96}. 
Very recent results from hypernuclear high--precision $\gamma$--ray
spectroscopy experiments \cite{Ta98,Ta01} seem to suggest an even smaller $\Lambda N$
spin--orbit interaction: $V^{\it l-s}_{NN}/V^{\it l-s}_{\Lambda N}>> 10$ \cite{Tam01}.}: 
we also have to take into account the response of the nucleons' core
to the added $\Lambda$, which can modify the mentioned shifts itself.
Indeed, there is evidence \cite{Hi98} that the core
response is able to reduce significantly the $\Lambda$ spin--orbit splitting,
already in $^{13}_{\Lambda}$C. Hypernuclear structure calculations with
core--excited states \cite{Mill85,Mill01} will be important in future analysis.

In the last 15 years the strangeness exchange reaction has 
been used at BNL \cite{Gr85,Sz91} 
for production and decay studies of hypernuclei from $^4_{\Lambda}$H to 
$^{12}_{\Lambda}$C.
However, because of the small momentum transfer and the large background
coming from the in--flight kaon decays, the measurement could not be extended
to heavy hypernuclei. At BNL \cite{Na98,Ba99} and KEK \cite{Ha89,Ou94}, 
the $(K^-,\pi^-)$ reaction confirmed the existence of the
$^4_{\Sigma}$He bound state, which was under discussion for about ten years.

The $({\rm stopped}\: K^-,\pi^-)$ reaction has been used at KEK  \cite{Ya86,Sa91,Ou98},
and, in the near future, will be employed at Da$\Phi$ne \cite{FI98}, 
the Frascati $\phi$--factory.
Moreover, this process was the standard method to produce $\Lambda$--hypernuclei
in emulsion and bubble chamber experiments during the 60's.
When the $K^-$ is stopped in the target, it is captured into an atomic level
and then, after cascade down to inner levels, it is absorbed in 
the nuclear surface, converting a nucleon into a $\Lambda$ or $\Sigma$. 
The momentum transferred to the produced $\Lambda$ is close to $k_F$ 
(for $0^{\circ}$ scattering--angle, $q_{\Lambda}\simeq 250$ MeV), while when a
$\Sigma$ is produced, $q_{\Sigma}\simeq 180$ MeV. 
The process with absorption of a kaon 
at rest in nuclei has the good feature of a large production yields, 
especially for $\Sigma$--hypernuclei \cite{Ya86}, and a large
number of hypernuclear states is accessible.
Moreover, differently from the in--flight reaction, it allows a clean separation of
the quasi--free hypernuclear production (because of the larger transferred momentum), 
resulting in a better determination of the weak decay rates,
especially in light systems \cite{Ou98}. 

At KEK \cite{Ou95}, the mesonic decay widths ($\Lambda\rightarrow \pi N$) 
for $^4_{\Lambda}$H and $^4_{\Lambda}$He, produced by the in--flight reaction, 
have been measured quite accurately. 
The $\Lambda$ decay into $\pi^0n$ has been directly identified for
the first time in $^4_{\Lambda}$He. Similarly, a measurement of the 
$\pi^0n$ decay channel for $^{11}_{\Lambda}$B and $^{12}_{\Lambda}$C
is presented in Ref.~\cite{Sa91}. Observations of this kind are 
of great importance also in connection 
with a proper parameterization of the $\pi^0$--nucleus optical potential.
\subsection{The $n(\pi^+,K^+)\Lambda$ strangeness associated production reaction}
\label{prod3}
When a $\pi^+$ hits a neutron, by the creation of a $s\overline{s}$
quark pair one has the associated production of two strange hadrons
in the final state: the $s$--quark becomes a constituent of a $\Lambda$
and the $\overline{s}$ is transferred to the meson, which becomes
a $K^+$. The $n(\pi^+,K^+)\Lambda$ reaction 
is complementary to the $n(K^-,\pi^-)\Lambda$ one. In fact, differently from the
latter, the former is best suited for studying deeply bound states in medium and
heavy hypernuclei \cite{Pi91,Ha96}.  It allows almost background free spectra and
it has the advantages of using good
quality and large intensity pion beams. In addition, the final $K^+$
is moderately distorted by the nucleus
($\lambda_{K^+}\simeq 2\lambda_{K^-}\simeq 2\lambda_{\pi^\pm}\simeq 4$ fm).
The reaction $n(\pi^+,K^+)\Lambda$ thus preferentially populates bound states with high 
$(n^{-1},\Lambda)$ spin configurations. 
Since the mass of the final strange hadrons pair is sizeably larger that the
mass of the initial particles, the $n(\pi^+,K^+)\Lambda$ reaction is endothermic,
with a quite large momentum transferred
to the hyperon: $q_Y\simeq 300\div 400$ MeV at $0^{\circ}$ scattering--angle 
(see figure~\ref{momtrans}). 
Hence, this reaction is able to populate all possible
$\Lambda$ levels, from the deepest one up to the quasi--free region.
We note that when the $\Lambda$ is produced above its emission threshold,
namely in the quasi--free region, it may leave the nucleus or spread its energy
inside the nucleus. In the latter case, by the emission of nucleons and/or
photons, a variety of hypernuclear states are accessible.
Because of the relatively large momenta transferred to the hyperon,
the relevant cross section for the associated production reaction
is one/two orders of magnitude smaller
than the one typical of the strangeness exchange reaction.
However, this defect is overcompensated by the high intensity of the 
available pion beams. From experiments using this reaction we have high quality
information about the spectroscopy of
many light to heavy $\Lambda$--hypernuclei \cite{Do88,Pi91,Ha96,Ha98,Ho01}. 

The associated production reaction has been
used for the first time at BNL \cite{Mi85,Pe86} for $^{12}_{\Lambda}$C,
while more recently it has been employed
at KEK \cite{No95,Ha96,Ho01}. Here, it allowed to accurately measure the lifetime
of $\Lambda$--hypernuclei over a broad range of mass numbers \cite{Bh98}
(from $^{12}_{\Lambda}$C to $^{56}_{\Lambda}$Fe, and data
on $^{89}_{\Lambda}$Y are under analysis now), with the explicit identification of 
the produced hypernuclei. Moreover, $(\pi^+,K^+)$ spectroscopy experiments 
at KEK \cite{Na01,Ho01} observed double--peak structures in $^{12}_{\Lambda}$C, 
$^{16}_{\Lambda}$O, $^{51}_{\Lambda}$V and $^{89}_{\Lambda}$Y, interpretable
as $\Lambda$ spin--orbit splittings. The magnitude of the shifts suggests a 
$\Lambda$--nucleus spin--orbit interaction stronger than the one extracted from
$(K^-,\pi^-)$ experiments. However, the interpretation of the measured spectra is 
still under discussion.

At KEK \cite{Sa98} the $n(\pi^+,K^+)\Lambda$ 
reaction has been also utilized to measure the weak decay width for
$\Lambda\rightarrow \pi^- p$ in $^{12}_{\Lambda}$C. 
This measurement has been carried out with 
a relatively small error and allowed a quite precise 
determination of the medium distortion 
acting on the pion coming out from the decay, a useful point for a better
understanding of the pion--nucleus interactions.

The KEK Superconducting Kaon Spectrometer worked with an energetic 
resolution of $1.5\div 2$ MeV FWHM\footnote{Full Width at Half Maximum}. 
Nowadays, there is an effort for sub--MeV resolution
spectroscopy (and pion beams with high statistics and intensity), again
by using the $(\pi^+,K^+)$ reaction, at the Japan Hadron
Facility (JHF) \cite{No98}. The use of high resolutions is important,
in particular, for the observation of the hypernuclear fine--structure
and, in turn, for a better understanding of the $\Lambda N$ spin--isospin 
dependent interactions.

\subsection{The $p(e,e^{\prime},K^+)\Lambda$ reaction}
\label{electroprod}
The electroproduction reaction is characterized by large momentum transfer
($\simeq 350$ MeV) and by the dominance of the spin--flip amplitudes in the elementary
process $p(\gamma,K^+)\Lambda$.
Thus, the electroproduction cross sections are small and the reaction
mainly populates stretched and unnatural parity hypernuclear states. 
The smallness of the $(e,e^{\prime},K^+)$ reaction cross section
is partially compensated by the high intensity of the initial electron beam
relatively to that of the final kaon beam.
This reaction could complement our knowledge
of hypernuclear spectroscopy derived from studies performed with meson beams.
Indeed, the high precision of electron beams can considerably improve 
the quality of experimental data. Moreover, the $(K^-,\pi^\pm)$ and 
$(\pi^+,K^+)$ reactions hardly produce ground states and deep--hole states
in heavy hypernuclei, because of the strong pion and kaon absorption
in the nuclear medium. Unnatural parity states are also difficult to
excite in $(K^-,\pi^\pm)$ and $(\pi^+,K^+)$ experiments, due to their 
moderate spin--flip amplitudes.

At TJNAF laboratories \cite{Hu00}, by using the
electroproduction reaction, hypernuclear levels will be observed
with high--resolution ($\simeq 0.6$ MeV FWHM) and, through fission fragment
detection techniques, the lifetimes of heavy hypernuclei will be measured
with great accuracy and precise identification of the decayed system \cite{Ta99}.

\newpage
\section{Weak decay modes of $\Lambda$--hypernuclei}
\label{decay}
\subsection{Introduction}
\label{decay1}
In the production of hypernuclei, the populated state may be highly excited, 
above one or more threshold energies for particle decays. These states are 
unstable with respect to the emission of the hyperon, of photons and nucleons. 
The spectroscopic studies of strong and electromagnetic de--excitations give
information on the hypernuclear structure which are complementary
to those we can extract from excitation functions and angular distributions
studies. Once the hypernucleus is stable with respect to
electromagnetic and strong processes, it is in the ground state, with
the hyperon in the $1s$ level, and can only 
decay via a strangeness--changing weak interaction, through the 
disappearance of the hyperon.
This is the most important decay mechanism, because
it opens the possibility to study some very interesting questions,
which have been quoted in the introduction of section \ref{hyp}.

Now we come to the main subject of the review, 
the study of the weak decay of 
$\Lambda$--hypernuclei. In the next two subsections we briefly discuss the
main characteristics of the decay channels for these systems. 
In subsection~\ref{mes} we will introduce the mesonic mode 
($\Lambda \rightarrow \pi N$), which
resembles what happens to the $\Lambda$ in free space, and in
subsection~\ref{nonmes} the so--called non--mesonic modes ($\Lambda N\rightarrow NN$, 
$\Lambda NN\rightarrow NNN$, etc), which can only occur in nuclear systems.
Semi--leptonic and weak radiative $\Lambda$ decay modes:
\begin{equation}
\begin{array}{l l l l l}
\Lambda &\rightarrow & n\gamma        & & ({\rm B.R.}= 1.75\cdot 10^{-3}) , \nonumber \\
     & &   p\pi^-\gamma               & & ({\rm B.R.}= 8.4\cdot 10^{-4}) , \nonumber \\
     & &   pe^-\overline{\nu}_e       & & ({\rm B.R.}= 8.32\cdot 10^{-4}) , \nonumber \\
     & &   p\mu^-\overline{\nu}_{\mu} & & ({\rm B.R.}= 1.57\cdot 10^{-4}) ,  \nonumber  
\end{array}
\end{equation} 
are neglected in the following, being, in free space, orders of magnitude less important 
than the mesonic decay (${\rm B.R.}=0.997$).
\subsection{Mesonic decay} 
\label{mes}
The free $\Lambda$ decays via the pionic channels:
\begin{equation}
\begin{array}{l l l l l}
\label{lambdadec}
\Lambda &\rightarrow & \pi^- p & & (\Gamma^{\rm free}_{\pi^-}/
\Gamma^{\rm free}_{\Lambda}=0.639) \nonumber \\
&   & \pi^0 n & & (\Gamma^{\rm free}_{\pi^0}/
\Gamma^{\rm free}_{\Lambda}=0.358) , \nonumber
\end{array}
\end{equation}
with a lifetime $\tau^{\rm free}_{\Lambda}\equiv \hbar/\Gamma^{\rm free}_{\Lambda}=
2.632\cdot 10^{-10}$ sec.

The experimental ratio of the relevant widths,
$\Gamma^{\rm free}_{\pi^-}/\Gamma^{\rm free}_{\pi^0}\simeq 1.78$,
and the $\Lambda$ polarization observables are
compatible with the $\Delta I=1/2$ rule on the isospin change
(for $\Gamma^{\rm free}_{\pi^-}/\Gamma^{\rm free}_{\pi^0}$ 
this follows from a simple Clebsch--Gordan analysis), 
which is also valid for the
decay of the $\Sigma$ hyperon and for pionic kaon decays (namely in non--leptonic
strangeness changing processes). Actually, this rule is slightly violated 
in the $\Lambda$ free decay, since 
it predicts $\Gamma^{\rm free}_{\pi^-}/\Gamma^{\rm free}_{\pi^0}=2$ 
(taking the same phase space for the two channels and neglecting
the final state interactions). Nevertheless, the ratio $A_{1/2}/A_{3/2}$ between 
the $\Delta I=1/2$ and the $\Delta I=3/2$ transition amplitudes is very large
(of the order of 30). This isospin rule is based on experimental observations
but its dynamical origin is not yet understood on theoretical grounds. 
On the other hand, it is not clear whether it is a universal characteristic of all 
non--leptonic processes with $\Delta S\neq 0$. The $\Lambda$ free
decay in the {\it Standard Model} can occur 
through both $\Delta I=1/2$ and $\Delta I=3/2$
transitions, with comparable strengths: an $s$ quark converts into a $u$ quark
through the exchange of a $W$ boson. Moreover, the effective
4--quark weak interaction derived from the {\it Standard Model} 
including perturbative QCD corrections (box and penguin quark diagrams, namely
one--loop gluon radiative corrections) 
gives too small $A_{1/2}/A_{3/2}$ ratios ($\simeq 3\div 4$, as calculated at
the hadronic scale of about $1$ GeV by using renormalization 
group techniques \cite{Gi79,Pa90}). Therefore, non--perturbative
QCD effects at low energy (such as hadron structure and reaction mechanism), which are more
difficult to handle, and/or final state interactions could be responsible
for the enhancement of the $\Delta I=1/2$ amplitude and/or the suppression of the
$\Delta I=3/2$ amplitude. In the low energy regime, chiral perturbation
theory is the effective theory which is usually employed for describing 
hadronic phenomena \cite{Ok98a}. However, it is well known that, when used in 
connection with perturbative QCD corrections, it is not able to reproduce
the rates for hyperon non--leptonic weak decays. 

Taking into account energy--momentum conservation in the mesonic decay,  
$m_{\Lambda}$ is equal to $\sqrt{\vec p\,^2+m_{\pi}^2}+\sqrt{\vec p\,^2+m_N^2}$ in the
center--of--mass system, thus the momentum of the final nucleon is $p\simeq 100$ MeV and
corresponds to an energy release $Q_{\Lambda}\simeq m_{\Lambda}-m_N-m_{\pi}\simeq 40$ MeV.
We have neglected the binding energies of the recoil 
nucleon and $\Lambda$, which tend to decrease $p$.
Hence, in nuclei the mesonic decay is disfavoured by the Pauli principle,
particularly in heavy systems. It is strictly forbidden in normal infinite nuclear matter 
(where the Fermi momentum is $k_F^0\simeq$ 270 MeV), while in finite nuclei it can 
occur because of three important effects: 
\begin{itemize}
\item In nuclei the hyperon 
has a momentum distribution (being confined in a limited spatial region) 
that allows larger momenta to be available to the final nucleon; \\

\item The final pion feels an attraction by the medium, such that for fixed
momentum $\vec q$ it has an energy smaller than the free one 
[$\omega(\vec q)<\sqrt{\vec q\,^2+m_{\pi}^2}$], and consequently, due to
energy conservation, the final nucleon again has more chance to come out above 
the Fermi surface. It has been shown \cite{Os93,Mo94} that the pion distortion 
increases the mesonic width by one or two 
orders of magnitude for very heavy hypernuclei
($A\simeq 200$) with respect to the value obtained without the medium distortion. 
For light and medium hypernuclei this enhancement factor is smaller,
being about $2$ for $A\simeq 16$. \\

\item At the nuclear surface the local Fermi momentum is considerably
smaller than $k_F^0$ (namely the Pauli blocking is less effective) and 
favours the decay. 
\end{itemize}

Nevertheless, the mesonic width rapidly decreases as 
the nuclear mass number $A$ of the hypernucleus increases \cite{Os93,Mo94}.
A further (but very small) effect which reduces the mesonic rate, 
especially in medium and
heavy hypernuclei, is the absorption of the final pion in the medium.
Actually, while energy--momentum conservation forbids the absorption
of a on--shell pion by a free nucleon, the absorption by a correlated pair
of nucleons is allowed for both on-- and off--shell pions, and the corresponding 
$\Lambda$ decay is observed as non--mesonic, resulting in a
final state with 3 nucleons: $\Lambda NN\rightarrow NNN$. Hence,
the mesonic channel is strictly related to the three--body non--mesonic decay.

From the study of the mesonic channel it is possible to extract
important information on the pion--nucleus optical potential, which we do not
know nowadays in a complete and unequivocal form from pionic 
atoms and low energy pion--nucleus scattering
experiments (on the other hand, no data are available for neutral pions). 
The nuclear mesonic rate, $\Gamma_{\rm M}=\Gamma_{\pi^-}+\Gamma_{\pi^0}$, is very sensitive to 
the pion self--energy in the medium \cite{It88,Os93,Mo94}: it is significantly
enhanced by the attractive $P$--wave part of the optical potential, but 
exclusive decays to closed shell nuclei mainly select the repulsive $P$--wave interaction 
and reduce the mesonic rate with respect to the calculation using non--distorted
(free) pion waves. 

The mesonic width is also extremely sensitive to the $Q$--value of the process, 
$Q_{\Lambda}\simeq 40$ MeV $+B_{\Lambda}-B_N$, 
which is in fact very small and decreases with the nuclear mass number. 
This implies a great sensitivity of the available phase space
to the mass of the final light particle, i.e. the pion (in analogy with
the problem of determining the neutrino mass from the nuclear $\beta$ decay),
and to the $\Lambda$ and final nucleon binding energies.
It is then clear that a systematic measurement of the
mesonic decays in medium--heavy systems is 
strongly advisable. Unfortunately, no data are nowadays available on the mesonic decay
for $A\geq 56$ hypernuclei, apart from some old emulsion and bubble chamber 
limits for $40<A<100$ \cite{Co90}.

From calculations and experiments on mesonic decays of 
$s$--shell hypernuclei we have
evidence for a central repulsion in the $\Lambda$--nucleus mean potential 
\cite{Ku95a,Ak97,Ou98}
(named, for this reason, ``Isle'' potential). This is an indication for a particular
balance between the strongly repulsive $\Lambda N$ interaction at short range,
which automatically appears in quark based models \cite{St90,St93}, 
and the weak (with respect to the $NN$ one) $\Lambda N$ attraction 
at intermediate distances. 

The following consideration about the Pauli principle in very light systems is
interesting. We have discussed how the Pauli exclusion principle suppresses the nuclear
mesonic decay. However, in $A=3$ hypernuclei the mesonic decays 
into two--body final states are enhanced (with respect to the corresponding free $\Lambda$ decays) 
as a result of the anti--symmetrization of the nucleons in the particular final states
\cite{Da59,Ku95a}. 
For example, the experimental rate for the two--body process 
$^4_{\Lambda}$H$\rightarrow ^4$He$+ \pi^-$ 
($\simeq 0.69\; \Gamma^{\rm free}_{\Lambda}$) is almost equal to the 
$\Lambda \rightarrow \pi^-p$ free rate ($\simeq 0.64\; \Gamma^{\rm free}_{\Lambda}$). 
Adding also the contribution of three--body mesonic decays with a $\pi^-$ in the final state, 
the rate is about the total $\Lambda$ free width: 
$\Gamma(^4_{\Lambda}{\rm H}\rightarrow \pi^-+{\rm all})>\Gamma^{\rm free}_{\pi^-}$. 
Moreover, again from data,
$\Gamma(^4_{\Lambda}$He$\rightarrow \pi^0+{\rm all})\gsim \Gamma^{\rm free}_{\pi^0}$.

From theoretical calculations \cite{Os93,Mo94}
and experimental measurements \cite{Os98} there is evidence that
the $\Gamma_{\pi^-}/\Gamma_{\pi^0}$ ratio in nuclei strongly 
oscillates around the value 2, predicted by the $\Delta I=1/2$ rule for a nucleus
with an equal number of neutrons and protons and closed shells. 
However, this is essentially due to nuclear shell 
effects and might not be directly related to the weak process itself. 
On the other hand, in the calculation of Refs.~\cite{Os93,Mo94} the
$\Delta I=1/2$ rule is enforced in the $\Lambda \rightarrow \pi N$ free vertex;
however, shell effects, also related to the Pauli blocking for the available final nuclear
states, make $\Gamma_{\pi^-}/\Gamma_{\pi^0}$ strongly dependent
on the hypernuclear structure. We remind the reader that $\Gamma_{\pi^-}/\Gamma_{\pi^0}$
is also sensitive to final state interactions and Coulomb effects.
\subsection{Non--mesonic decay} 
\label{nonmes}
\begin{figure}
\begin{center}
\mbox{\epsfig{file=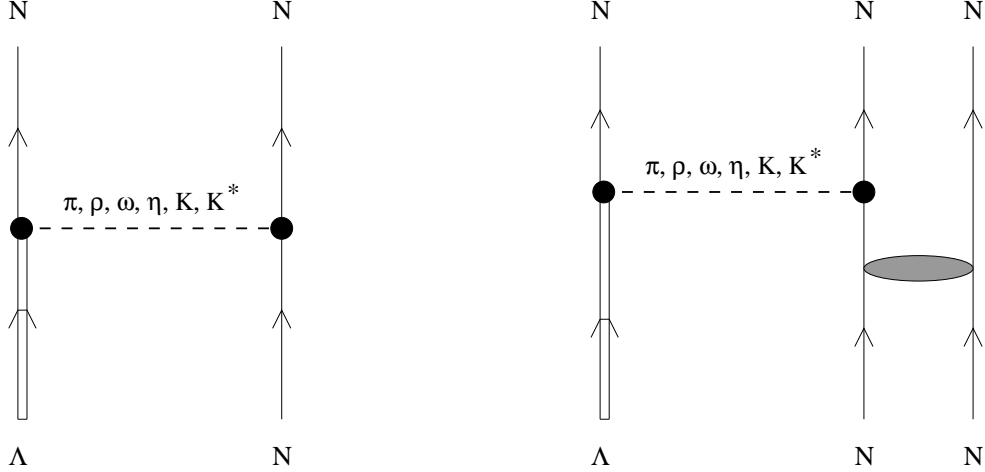,width=.85\textwidth}}
\vskip 2mm
\caption{One--nucleon (a) and two--nucleon (b) induced $\Lambda$ decay in nuclei.}
\label{nm12}
\end{center}
\end{figure}

When the pion emitted from the weak hadronic vertex $\Lambda \rightarrow \pi N$ is virtual,
then it will be absorbed by one or more nucleons of the medium, resulting in 
a non--mesonic process of the following type:
\begin{eqnarray}
\label{gn}
\Lambda n & \rightarrow & nn \hspace{4mm} \left(\Gamma_n\right) , \\
\label{gp}
\Lambda p & \rightarrow & np \hspace{4mm} \left(\Gamma_p\right) , \\
\label{g2}
\Lambda NN & \rightarrow & nNN \hspace{4mm} \left(\Gamma_2\right) .
\end{eqnarray}
The total weak decay rate of a $\Lambda$--hypernucleus is then:
\begin{equation}
\Gamma_{\rm T}=\Gamma_{\rm M}+\Gamma_{\rm NM} , \nonumber
\end{equation}
where:
\begin{equation}
\Gamma_{\rm M}=\Gamma_{\pi^-}+\Gamma_{\pi^0} , \hspace{0.15 in}
\Gamma_{\rm NM}=\Gamma_1+\Gamma_2 , \hspace{0.15 in}
\Gamma_1=\Gamma_n+\Gamma_p , \nonumber
\end{equation}
and the lifetime is $\tau=\hbar/\Gamma_{\rm T}$.
The channel (\ref{g2}) can be interpreted by assuming that the pion 
is absorbed by a pair of nucleons
correlated by the strong interaction. Obviously, the non--mesonic
processes can also be mediated by the exchange of more massive mesons than
the pion (see figure \ref{nm12}). 

The non--mesonic mode is only possible in nuclei and,
nowadays, the systematic study of the
hypernuclear decay is the only practical way to get information on the weak process
${\Lambda}N \rightarrow NN$ (which provides the first extension of the weak 
$\Delta S=0$ $NN\rightarrow NN$ interaction to strange baryons), 
especially on its parity--conserving part, which
is masked by the strong interaction in the weak $NN\rightarrow NN$ reaction.
In fact, there are not experimental observations for the process
$\Lambda N\rightarrow NN$ using 
$\Lambda$ beams: it is, however, under study (at COSY \cite{Ha95} and KEK \cite{Ki98}) 
the measurement of the (low) cross section for the inverse reaction
$pn\rightarrow p{\Lambda}$, which could give much cleaner information.

The precise measurement of $\Gamma_n$ and $\Gamma_p$ in
$s$--shell hypernuclei is very important for the study of
the spin--isospin dependence and of the validity of the
$\Delta I=1/2$ rule in the non--mesonic processes
(see the analysis presented in subsection \ref{passh}); 
on the other side, it is relevant in connection with the hypernuclear structure
dependence, which is rather important in these very light systems. 
In $s$--shell hypernuclei all
nucleons are confined (as the hyperon) into the $s$--level, while complications arise
with increasing mass number, due to the appearance of more initial
$\Lambda N$ states and of the nucleons' rescattering inside the residual nucleus, 
which entangles the kinematics of the measured nucleons.

The final nucleons in the non--mesonic processes emerge with large
momenta: disregarding the $\Lambda$ and nucleon binding energies
and assuming the available energy $Q=m_{\Lambda}-m_N\simeq 176$ MeV 
to be equally splitted among the final nucleons, it turns out that
$p_N\simeq 420$ MeV for the one--nucleon induced channels 
[Eqs.~(\ref{gn}), (\ref{gp})] and 
$p_N\simeq 340$ MeV in the case of the two--nucleon induced mechanism 
(\ref{g2}). Therefore, the non--mesonic decay mode is not forbidden 
by the Pauli principle: on the contrary, the final nucleons escape
from the nucleus with great probability and the non--mesonic mechanism dominates 
over the mesonic mode for all but the $s$--shell hypernuclei. For very light
systems the two
decay modes are competitive, the smallest value for the non--mesonic width 
corresponding to hypertriton, where it is evaluated to be 
1.7\% of the $\Lambda$ free decay rate \cite{Go97}.
The non--mesonic channel is characterized by large momentum transfer, thus,
apart from very light hypernuclei, the details of the hypernuclear structure do not 
have a substantial influence (then providing useful information 
directly on the hadronic weak interaction).
On the other hand, the $NN$ and $\Lambda N$ short 
range correlations turn out to be very important. 

\begin{figure}
\begin{center}
\mbox{\epsfig{file=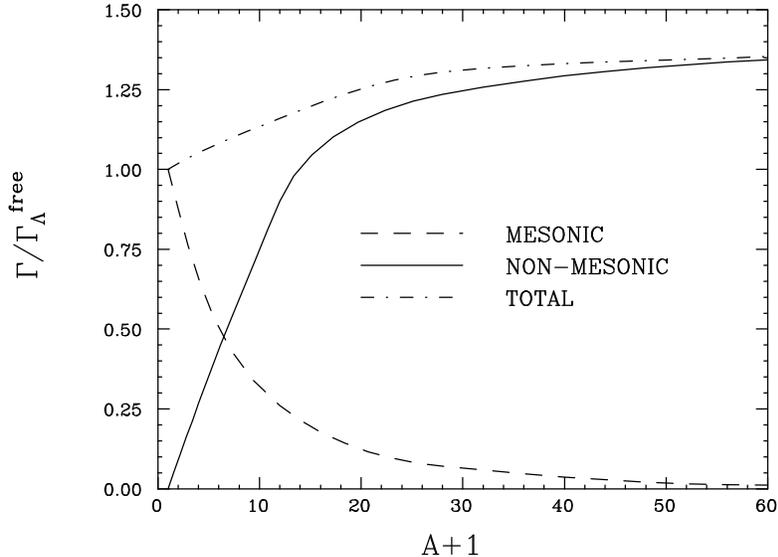, width=.7\textwidth}}
\vskip 2mm
\caption{Qualitative behaviour of mesonic, non--mesonic and total decay widths
as a function of the baryonic number $A+1$.}
\label{qual}
\end{center}
\end{figure}

There is an anticorrelation between mesonic and non--mesonic 
decay modes such that the experimental
lifetime is quite stable from light to heavy hypernuclei \cite{Co90,Os98}, apart
from some fluctuation in light systems because of shell structure effects: 
${\tau}_{\Lambda}=(0.5\div 1)\,{\tau}^{\rm free}_{\Lambda}$.
Since the mesonic width is less than 1\% of the total width
for $A>100$, the above consideration implies that the non--mesonic rate is rather
constant in the region of heavy hypernuclei. 
This can be simply understood from the following consideration.
If one naively assumes a zero range approximation for the non--mesonic process 
$\Lambda N\rightarrow NN$ (actually, the
range is not zero, but very small, due to the large transferred momenta),
$\Gamma_1$ is proportional to the overlap between the $\Lambda$
and nuclear densities:
\begin{equation}
\Gamma_1(A)\propto \int d{\vec r}\,|\psi_{\Lambda}(\vec r)|^2\rho_A(\vec r) , \nonumber
\end{equation}
where the $\Lambda$ wave function $\psi_{\Lambda}$ (nuclear density $\rho_A$) 
is normalized to unity (to the nuclear mass number $A$). 
This overlap integral increases with the 
mass number and reaches a constant value: 
by using, for simplicity, $\Lambda$ harmonic oscillator wave functions 
(with frequency $\omega$ adjusted to the experimental hyperon levels in 
hypernuclei) and Fermi distributions for the nuclear densities, we find 
$\Gamma_1(^{12}_{\Lambda}{\rm C})/\Gamma_1(^{208}_{\Lambda}{\rm Pb})\simeq 0.56$,
while $\Gamma_1$ is $90$ \% of the saturation value for $A\simeq 65$. In figure
\ref{qual} the qualitative behaviour of mesonic, non--mesonic and total widths as a
function of the nuclear mass number $A$ is shown.
For $A\leq 11$ the experimental 
data are quite well fitted by $\Gamma_{\rm NM}/\Gamma^{\rm free}_{\Lambda}\simeq 0.1A$:
$\Gamma_1$ (namely the probability of the
$\Lambda N\rightarrow NN$ process) is proportional to the number of $\Lambda N$ pairs, 
$A$, as it is expected from the above simple description, where we neglect 
the contribution of $\Gamma_2$.
Actually, the observed saturation of the $\Lambda N\rightarrow NN$ interaction
is strictly related to its range: the saturation occurs when the radius
of the hypernucleus becomes sensitively larger than the range of the interaction. 
By inspecting the experimental data of Refs.~\cite{Sz91,No95,Bh98,Ku98} 
we can conclude that the decaying $\Lambda$
can interact at most with about $15\div 20$ neighbouring nucleons,
namely almost exclusively with $s$-- and $p$--shell nucleons.
However, for a more quantitative explanation it will be important to collect data 
(with good precision, like in the KEK experiment \cite{Bh98} or in the planned
FINUDA \cite{FI98}) for hypernuclei
between $^{12}_{\Lambda}$C and $^{28}_{\Lambda}$Si and in the
region  $A=100\div 200$. Yet, from the available data 
one can say, very roughly, that the long distance 
component of the $\Lambda N\rightarrow NN$ interaction 
has a range of about 1.5 fm and corresponds, as we expect, to the OPE component
of the interaction.
\subsubsection{The $\Gamma_n/\Gamma_p$ puzzle} 
\label{puzz}
Nowadays, the main problem concerning the weak decay rates 
is to reproduce the experimental value for the ratio ${\Gamma}_n/{\Gamma}_p$
between the neutron-- and the proton--induced widths 
${\Lambda}n\rightarrow nn$ and ${\Lambda}p\rightarrow np$. The theoretical
calculations underestimate the central data for all considered 
hypernuclei (see tables~\ref{tab6}--\ref{tab8}):
\begin{equation}
\left\{\frac{{\Gamma}_n}{{\Gamma}_p}\right\}^{\rm Th}\ll
\left\{\frac{{\Gamma}_n}{{\Gamma}_p}\right\}^{\rm Exp} ,
\hspace{0.2in}
0.5\lsim \left\{\frac{{\Gamma}_n}{{\Gamma}_p}\right\}^{\rm Exp}\lsim 2 \nonumber
\end{equation}
\noindent (only for $^4_{\Lambda}$He this ratio seems to be less than 0.5),
although the large experimental error bars do not allow to reach any definitive conclusion.
The data are quite limited and not precise because of the difficulty in
detecting the products of the non--mesonic decays, especially for the neutron--induced one.
Moreover, the present experimental energy resolution for the detection of the outgoing
nucleons do not allow to identify the final state of the
residual nuclei in the processes 
$^A_{\Lambda}{\rm Z}\rightarrow {^{A-2}{\rm Z}} + nn$ and
$^A_{\Lambda}{\rm Z}\rightarrow {^{A-2}({\rm Z-1})} + np$. As a consequence,
the measurements supply decay rates averaged over several nuclear final states.

In the one--pion--exchange approximation, by assuming the $\Delta I=1/2$ rule
in the $\Lambda \rightarrow \pi^-p$ and $\Lambda \rightarrow \pi^0n$ 
free couplings, the calculations (discussed in the next section) 
give small ratios, in the range $0.05\div 0.20$.
This is due to the $\Delta I=1/2$ rule, which fixes the vertex ratio 
$V_{\Lambda \pi^-p}/V_{\Lambda \pi^0n}=-\sqrt{2}$ (both in $S$-- and $P$--wave
interactions), and to the particular form of
the OPE potential, which has a strong tensor and weak central and parity--violating
forces: the large tensor transition $\Lambda N(^3S_1)\rightarrow NN(^3D_1)$
requires, in fact, $I=0$ $np$ pairs in the anti--symmetric final state.
In $p$--shell and heavier hypernuclei the relative $\Lambda N$ $L=1$ state is found
to give only a small contribution to tensor transitions for the neutron--induced
decay, so it cannot improve the OPE ratio.
The contribution of the $\Lambda N$ $L=1$ relative state to $\Gamma_{\rm NM}$ 
seems to be of about $5\div 15$\% in $p$--shell hypernuclei \cite{Be92,It98,Pa01}. 
For these systems we expect the dominance of the $S$--wave interaction in the initial state,
due to the small $\Lambda N$ relative momentum. By using a 
simple argument about the isospin structure of the transition $\Lambda N\rightarrow NN$
in OPE, it is possible to estimate that for pure $\Delta I=3/2$ transitions
($V_{\Lambda \pi^-p}/V_{\Lambda \pi^0n}=1/\sqrt{2}$) the OPE ratio
can increase up to about $0.5$.
However, the OPE model with $\Delta I=1/2$ couplings
has been able to reproduce the one--body stimulated non--mesonic rates 
$\Gamma_{1}=\Gamma_n+\Gamma_p$ for light and medium hypernuclei
\cite{It95,Pa97,Ok98,It98,Pa01}. Hence, the problem seems to consist in overestimating the
proton--induced rate and underestimating the neutron--induced one.

In order to solve this puzzle (namely to explain
both $\Gamma_n+\Gamma_p$ and $\Gamma_n/\Gamma_p$), many attempts have
been made up to now, but without success. We recall the 
inclusion in the ${\Lambda}N\rightarrow NN$
transition potential of mesons heavier than the pion (also including the exchange
of correlated or uncorrelated two--pions) 
\cite{Du86,Sh94,It95,Du96,Pa97,It98,Pa01,Os01}, the inclusion of interaction terms that
explicitly violate the ${\Delta}I=1/2$ rule 
\cite{Ma94,Ma95,Pa98} and the description of the 
short range baryon--baryon interaction in terms of quark degrees of freedom
(by using a hybrid quark model in \cite{Ch83} and a direct quark 
mechanism in \cite{Ok96,Ok98,Ok99}), 
which automatically introduces $\Delta I=3/2$ contributions.
The calculations of Refs.~\cite{It98,Ok99,Pa01,Os01,Ju01} are the only 
ones which have found a sizeable increase of the neutron 
to proton ratio with respect to the OPE value. We shall come back to the
problem of the $\Gamma_n/\Gamma_p$ ratio more extensively in subsection
\ref{sumth} and section \ref{newpuzzle}.

\newpage
\section{Present status of experiment and theory} 
\label{sumexpth}
\subsection{Experiment}
\label{sumexp}
We shortly summarize here the main experiments which have 
observed the weak decay of $\Lambda$--hypernuclei.

The decay of a hypernucleus was observed 
for the first time in 1952 \cite{Da53} in a nuclear
emulsion used for cosmic--ray observations. The experiments on the
weak decays started in the first 60's and employed
negative kaons stopped in emulsions and
bubble chambers \cite{Te62}. They were mostly based on the
detection of the emitted negative pions,
and only established rough limits on the total 
lifetimes of $s$--shell $\Lambda$--hypernuclei. 

In the following years \cite{Po76}, until the first 70's,
although with great difficulties 
(the identification of hypernuclei was hard, statistics and precision 
were very low, etc),
the experiments succeeded in separating the mesonic and non--mesonic
$\Lambda$ decays and established the first limits on the 
partial rates. In these experiments hypernuclei were produced by using 
kaon or pion beams, as explained in section \ref{prod}. 
They showed \cite{Mo74,Co90,Sz91} that:
\begin{itemize}
\item for $s$--shell hypernuclei the mesonic and non--mesonic
widths were comparable, $\Gamma_{\rm NM}/\Gamma_{\pi^-}\simeq 0.3\div 1.5$, 
and $0.3\lsim \Gamma_n/\Gamma_p \lsim 2$; \\
\item for $p$--shell hypernuclei: 
$\Gamma_{\rm NM}/\Gamma_{\pi^-}\simeq 2\div 7$ and $0.6\lsim \Gamma_n/\Gamma_p\lsim 2$; \\
\item for medium and heavy hypernuclei ($40<A<100$)
the non--mesonic processes were dominant,
$\Gamma_{\rm NM}/\Gamma_{\pi^-}\simeq 100\div 200$, and
$1.5\lsim \Gamma_n/\Gamma_p\lsim 9$; \\
\item the total lifetimes for light hypernuclei ($A\le 15$) oscillated in the interval 
$\tau/\tau^{\rm free}_{\Lambda}\simeq 0.3\div 1$. 
\end{itemize}

The interest in the detection of hypernuclear decays seems
out of stock in the first 70's, until the first half of the 80's \cite{Ma81,Mi85}, 
when at the Brookhaven synchrotron, by using modern techniques
(scintillators, proportional chambers, etc, which allow direct timing
observations), it was measured the
lifetime of $^{11}_{\Lambda}$B and $^{12}_{\Lambda}$C \cite{Gr85}, 
produced by the $(K^-,\pi^-)$ reaction. 
After some years, through the detection of protons,
neutrons (from non--mesonic decays) and negative pions 
(from mesonic decays), the partial rates for 
$^5_{\Lambda}$He and $^{12}_{\Lambda}$C have been measured \cite{Sz91}. 
The total lifetime was measured directly, and the mesonic 
rate into $\pi^0n$ obtained by subtraction:
$\Gamma_{\pi^0}=\Gamma_{\rm T}-\Gamma_n-\Gamma_p-\Gamma_{\pi^-}$.
It must be noted that lifetime measurements are free from nuclear final
state interactions and material effects, which, on the contrary, affects very much
the measurement of the partial rates $\Gamma_n$ and $\Gamma_p$.
The so--called ``modern era'' of hypernuclear physics 
starts with counter experiment like these, which
very much improved the quality of data. More recently, with the same techniques,
$^4_{\Lambda}$He and $^9_{\Lambda}$Be hypernuclei have been studied at BNL \cite{Ze98}.

In the middle of the 80's, at CERN LEAR, the lifetimes of $^{209}_{\Lambda}$Bi and
$^{238}_{\Lambda}$U (produced by anti--proton annihilation) were
measured \cite{Bo87}, although with very large error bars, with results comparable with the
lifetimes of light hypernuclei. More recent results, obtained with an improved 
apparatus, are published in Ref.~\cite{Ar93}: large uncertainties remain because
of the limited precision of the recoil shadow method.
The experiment measures the fission fragments
of the produced hypernucleus, with a delay time which is equal to the
hypernuclear lifetime. In fact, the fission events are 
mainly induced by the energy released in the non--mesonic decay
(the probability of a time delayed fission due to the $\Lambda$ decay
is more than up to 2 orders of magnitude higher than the one of
prompt fission due to other sources \cite{Ar93}). Experiments of this kind are very
difficult to perform (the produced hypernucleus cannot be 
unambiguously identified) and, as already mentioned,
the lifetimes are generally measured with large errors.
Only very recently \cite{Ku98,Ka01,Ku01}, more accurate results have been obtained from 
delayed fission experiments. Nevertheless, there is a certain disagreement among these
new data, and the saturation value of the lifetime for very heavy
hypernuclei is not established with precision.
It is important to remind the reader that, for the decay of very heavy hypernuclei,
the application of more accurate techniques, employing
direct timing methods (used, for instance, in the BNL experiment of Ref.~\cite{Sz91}), 
is practically impossible due to the large background of light particles.

In the last 15 years there has been a rapid development of various experiments,
which have led to a more systematic investigation of hypernuclei, although
no experiment has been able to measure directly the whole set of partial decay rates.
At Brookhaven \cite{Mi85,Pi91} (starting since 1983) and KEK \cite{Has89,No95,Ha96} 
(since 1989), the $(\pi^+,K^+)$ strangeness
associated production reaction has been used. At J\"{u}lich (COSY) \cite{Ku98,Ku01}, 
by using proton--nucleus scattering processes, the total lifetime
of very heavy hypernuclei (in the region of bismuth and uranium: 
$A\simeq 200\div 240$) has been measured by delayed fission observations.
By using the same techniques
\cite{Ka01}, again at COSY, the lifetime for hypernuclei with mass numbers
$A= 180\pm 5$, produced in $p$--Au collisions, has been obtained.
At BNL AGS \cite{Ru98}, the $^{12}_{\Lambda}$B hypernucleus has been produced by the
$({\rm stopped}\: K^-,\pi^0)$ reaction, the $\Lambda$ being created on
a proton, with the final $\pi^0$ detected by a high energy resolution 
(less than $1$ MeV FWHM) neutral meson spectrometer. 
At BNL \cite{Ma98} and KEK \cite{Nak98} the process $(K^-,K^+)$
produces strangeness $-2$ hypernuclei, which are important for the study of the
$\Lambda \Lambda$ and $\Xi N$ interactions.

Several experiments are planned for the future.
At TJNAF laboratories \cite{Hu00}, by using the
electroproduction reaction $(e,e^{\prime},K^+)$, hypernuclear levels will be observed
with high--resolution ($\simeq 0.6$ MeV FWHM) and, through fission fragment 
detection techniques, the lifetimes of heavy hypernuclei will be measured
with great accuracy and precise identification of the decayed system \cite{Ta99}.
In the near future, within the Japan Hadron Facility (JHF) project, 
at KEK a germanium detector system will measure the hypernuclear 
$\gamma$--ray transitions with an energy resolution around 300 KeV FWHM \cite{No98}. 
Germanium detectors with a few KeV resolution are already collecting
$\gamma$--spectroscopy data at BNL and KEK \cite{Ta98,Ta01}. Experiments of this kind will be
essential for a better understanding of the $\Lambda N$ spin--dependent interactions.
Finally, the FINUDA facility \cite{FI98} will make 
use of very thin targets ($\simeq 0.1$ gr$/$cm$^2$) and
large detector acceptance ($\simeq 2\pi$ sr). The $({\rm stopped}\; K^-,\pi^-)$ 
production reaction, already employed at KEK \cite{Ya86,Sa91,Ou98}, 
will be used, with low energy $K^-$ ($\simeq 16$ MeV) 
coming from the decay of the $\phi$ mesons ($\phi \rightarrow K^+K^-$, 
B.R.$=49.1$\%). This mesons will be 
created at the Da$\Phi$ne $e^+e^-$ collider at a center--of--mass energy of $1.02$ GeV.
The experiment has been designed to work with high production rate (about 
80 hypernuclei produced per hour at the $e^+e^-$ 
luminosity of $10^{32}$cm$^{-2}$s$^{-1}$), high--resolution spectroscopy ($\simeq 0.7$ MeV FWHM) 
and high precision in the measurements of the
weak decay rates (2\% statistical error on the total lifetimes for one week of data
taking at ${\mathcal L}=10^{32}$cm$^{-2}$s$^{-1}$). 
The $\pi^-$ coming from the hypernuclear production could be detected
by the FINUDA spectrometer
in coincidence with all the particles emitted in the subsequent decay. 
It will be possible to measure the $\Gamma_n/\Gamma_p$ ratio with
precision better than the one of the other running experiments and to use about 10 
different targets, covering the whole mass range,
for a systematic study of production and decay of hypernuclei.
We think that the wide program of FINUDA could represent a new step forward 
in understanding the hypernuclear phenomena.

The main results obtained with the above listed experiments will be
quoted at the end of the next subsection, for a comparison with
theoretical predictions.
\subsection{Theory}
\label{sumth}
We summarize here the historical evolution of the various theoretical 
approaches utilized for the evaluation
of the weak decay of $\Lambda$--hypernuclei. Some details of the formalisms
employed in the calculations are given in the next section. The numerical results
of the different calculations are reported and discussed at the end of this 
subsection, in tables~\ref{tab1}--\ref{tab8}.

The first calculations of the mesonic rate for light hypernuclei
date at the end of the 50's \cite{Da59}. The Pauli blocking effect
for nuclear decay was estimated and used
in order to assign the spin to the ground state of $s$--shell hypernuclei.

The possibility of non--mesonic hypernuclear decay was suggested  for the first time 
in 1953 \cite{Ch53} and interpreted in terms of the free space 
$\Lambda \rightarrow \pi N$ decay, where the pion was considered as virtual and 
then absorbed by a bound nucleon.

In the 60's Block and Dalitz 
\cite{Bl62,Bl63,Da73} developed a \underline{phenomenological model},
which has been more recently updated \cite{Do87,Sc92,Al99b}. Within this approach,
some important characteristics of the non--mesonic decays 
(for instance the validity of the $\Delta I=1/2$ rule) of $s$--shell hypernuclei
can be reproduced in terms of elementary spin--dependent branching 
ratios for the $\Lambda n\rightarrow nn$ and $\Lambda p\rightarrow np$ processes, 
by fitting the available experimental data (see the discussion in subsection 
\ref{passh}). Although this kind of analysis makes use 
of several delicate assumptions, it has the good feature
that it does not need the knowledge of the effective Hamiltonian for the
reaction mechanism. An interesting empirical conclusion of 
Ref.~\cite{Bl63}, never explained on theoretical ground, is the dominance of the
$\Lambda N(^3S_1)\rightarrow NN(^3P_1)$ transition, which leads to large 
$\Gamma_n/\Gamma_p$ ratios ($\simeq 1\div 2$) for $^5_{\Lambda}$He. 
Following the phenomenological approach, it emerges that
in order to establish the degree of violation of the $\Delta I=1/2$ rule in the non--mesonic
decays of $s$--shell hypernuclei, we need more precise measurements
of $\Gamma_n$ and $\Gamma_p$, especially for 
$^4_{\Lambda}$H. With the present data one cannot exclude a large
violation of the $\Delta I=1/2$ rule \cite{Co90a,Sc92,Ru99,Al99b}.
In Ref.~\cite{Ru99} the authors came to the conclusion 
that the $\Delta I=1/2$ rule is strongly violated by observing that the experimental 
lifetimes of heavy hypernuclei (in the region $A\simeq 180\div 220$)
seem to favour $\Gamma_n/\Gamma_p$ ratios larger than 2, while $\Gamma_n/\Gamma_p$
should be $\leq 2$ if the $\Delta I=1/2$ rule is not violated.
However, we point out that, in the phenomenological analysis,
the inequality $\Gamma_n/\Gamma_p\leq 2$ for nuclear matter is
valid if the $\Lambda$ decays by interacting only with
$s$--shell nucleons, while in heavy systems $p$--shell nucleons are expected to contribute
too [see Eqs.~(\ref{uno2}), (\ref{ggtot}) and the discussion of paragraph \ref{blda} 
for $^{12}_{\Lambda}$C]. 

After the first analysis by Block and Dalitz, 
microscopic models of the $\Lambda N\to NN$ interaction began to
be developed. The first paper, for nuclear matter, including only
$L=0$ $\Lambda N$ relative states, is due to Adams \cite{Ad67}
(his numerical results are quoted in tables~\ref{tab1}, \ref{tab6}).
He used the \underline{OPE description} 
with $\Delta I=1/2$ $\Lambda N\pi$ couplings within a Fermi gas model and
found a large sensitivity of the decay widths
to the $NN$ and $\Lambda N$ short range repulsive 
correlations. For $\Lambda N$ they were described through the arbitrary insertion, 
in the two--body transition matrix element, of an analytical function which was an
approximation to the exact solution of the Bethe--Goldstone equation with
a hard--core potential ($r_{\rm core}\simeq 0.4$ fm). The obtained results 
were not realistic also because the employed $\Lambda N\pi$ coupling was too
small to reproduce the $\Lambda$ free lifetime. Taking this into account, 
the Adams' results for $\Gamma_{\rm NM}$ should be multiplied by 6.81, as it is done in 
table~\ref{tab1}.

Afterwards, in order to improve the OPE model, mesons heavier than
the pion have been introduced as mediators of the $\Lambda N\rightarrow NN$
interaction. McKellar and Gibson \cite{Mc84} evaluated the width for
a $\Lambda$ in nuclear matter, adding the \underline{exchange of the}
\underline{$\rho$--meson} and
taking into account $\Lambda N$ relative $S$ states only. They
calculated the $\Lambda N\rho$ weak vertex (experimentally not accessible) 
by using the factorization approximation (which, however, 
contains many ambiguities) and a pole model. The authors
assumed the $\Delta I=1/2$ rule and made the calculation by
using the two possible relative signs (being at the time unknown and not fixed by
their model) between the pion and rho potentials,
$V_{\pi}+|V_{\rho}|$ and $V_{\pi}-|V_{\rho}|$, with
very different results in the two cases. In table~\ref{tab1} the listed results
are the ones with the (nowadays fixed) right sign, namely $V_{\pi}-|V_{\rho}|$.
It is important to note that for mesons
heavier than the pion, no experimental indication supports the validity of the 
$\Delta I=1/2$ rule for their couplings with baryons (for example, see Ref.~\cite{Ma95} 
for an evaluation of the violation of the $\Delta I=1/2$ rule in the 
$\Lambda \rightarrow \rho N$ vertex). Some years later, Nardulli \cite{Na88}
determined the relative sign ($-$) between $\pi$ and $\rho$ exchange by using a somewhat
different pole model, also implementing the available information 
from weak non--leptonic and radiative decays. Refs.~\cite{Mc84,Na88} obtained
a non--mesonic width in the ($\pi +\rho$)--exchange model smaller than the OPE one. 
This characteristic resulted from a destructive interference
between the two mesons and would have been confirmed in the future. 
In \cite{Na88} the $\Gamma_n/\Gamma_p$ ratio in
($\pi +\rho$)--exchange resulted sizeably increased with respect to the OPE value
(see table~\ref{tab6}). 
However, more recent calculations \cite{Pa95-96,Pa97} showed a small effect of the 
$\rho$--exchange on $\Gamma_n/\Gamma_p$. Takeuki, Takaki and Band$\overline{\rm o}$ \cite{Ta85,Ba85}
applied a model with $(\pi +\rho)$--exchange to $^4_{\Lambda}$H, 
$^4_{\Lambda}$He and $^5_{\Lambda}$He, finding quite small
non--mesonic rates when pion and rho have negative relative phase (see table~\ref{tab3}).
The same result was obtained in Ref.~\cite{Mc84} for nuclear matter.
More recently \cite{Pa95-96}, a $(\pi +\rho)$ model has been applied to $^{12}_{\Lambda}$C. 
The authors found the central potential from $\rho$--exchange (omitted in the
previous calculations) to be more important than the tensor part. Moreover, the
$\Gamma_n/\Gamma_p$ ratio remained unchanged when the rho--meson was included
(see tables~\ref{tab2}, \ref{tab7}).
The conclusion we can draw from the calculations that include 
the $\rho$--exchange is that the results 
strongly depend on the model used for the evaluation of the $\Lambda N\rho$ vertex.
Nevertheless, today there is a general consent that the inclusion of the $\rho$
cannot improve the calculation of $\Gamma_n/\Gamma_p$ \cite{Du96,Pa97}.

In 1986 Dubach {\em et al.}~\cite{Du86} introduced a \underline{OME model} with
\underline{$\pi, \rho, K, K^*, \omega$ and $\eta$}, for a calculation in nuclear
matter. The $\Gamma_n/\Gamma_p$ is expected to be
sensitive to the isospin structure of the transition potential. 
Therefore, the inclusion
of mesons heavier than the pion could give ratios in better agreement with the
data. In order to evaluate the meson--baryon--baryon vertices which are not accessible to
the experiment, a quite large number of different models 
(pole model, $SU(6)_w$ symmetry, PCAC, Goldberger--Treiman relations, etc)
have been used, also enforcing the $\Delta I=1/2$ rule.
The calculation of the above vertices is strongly model--dependent, and this
makes the use of potentials with mesons other than the pion
almost unpraticable. The above cited paper only reports preliminary results, 
while the final ones are published in
Ref.~\cite{Du96} (see tables~\ref{tab1}--\ref{tab3}, \ref{tab6}--\ref{tab8}).
Here the model is also extended, within the extreme single particle shell model,
to finite nuclei ($^5_{\Lambda}$He and $^{12}_{\Lambda}$C). 
The controversial results presented in \cite{Du96}
caused debate and critical discussions in the literature. Because of the few details available
from Ref.~\cite{Du96} (which, on the other hand, does not take into account
the hadronic form factors and quotes some different results from the preliminary 
ones of Ref.~\cite{Du86}),
it is not possible to compare the model used in this work for the OME potential
with other ones proposed afterwards, for example by Ramos and Bennhold \cite{Ra94}
and Parre\~{n}o {\em et al.}~\cite{Pa97}, where
the decay is again treated in a shell model framework. In Ref.~\cite{Pa97},
differently from \cite{Du96}, all the 
possible $\Lambda N$ and $NN$ relative states have been included.
The unknown hadronic vertices have been obtained from pole model, soft meson theorems
and $SU(6)_w$ symmetry. The repulsive baryon--baryon correlation were based on 
Nijmegen and J\"{u}lich $\Lambda N$ and $NN$ interactions.
The calculation of the unknown hadronic vertices turned out to be
model--dependent and the obtained decay rates were very different from the 
ones of Ref.~\cite{Du96}. Ref.~\cite{Pa97} calculated the non--mesonic widths in the OME 
picture to be different at most by 15\% from the OPE ones for $^{12}_{\Lambda}$C and
$^5_{\Lambda}$He (see tables~\ref{tab2}, \ref{tab3}, \ref{tab7}, \ref{tab8}). The 
$\Gamma_n/\Gamma_p$ ratio in the full OME turned out to be 30\% smaller than the
OPE value for $^{12}_{\Lambda}$C, in contrast with the improved ratio of
Ref.~\cite{Du96}, even if it was quite sensitive to the isospin structure of the exchanged mesons 
(the largest changes corresponding to the inclusion of the strange meson $K$). 
This was mainly due to the destructive interference between the exchange of mesons with
the same isospin [($\pi, \rho$), ($K, K^*$), ($\omega, \eta$)]. Moreover,
the contribution of mesons heavier than the pion were suppressed by form
factors and short range correlations. Very recently, in Ref.~\cite{Pa01},
the authors of Ref.~\cite{Pa97} corrected a mistake they made in the inclusion of the
$K$-- and $K^*$--exchange. This had the effect of increasing the $\Gamma_n/\Gamma_p$ ratio: 
$(\Gamma_n/\Gamma_p)^{\rm OME}\simeq 4(\Gamma_n/\Gamma_p)^{\rm OPE}$.
The only inclusion of the $K$--meson in addition to the pion
leads to a smaller $\Gamma_n +\Gamma_p$ and significantly enhances the 
$\Gamma_n/\Gamma_p$ ratio (see tables~\ref{tab2}, \ref{tab3}, \ref{tab7}, \ref{tab8}). 
This behaviour has been confirmed by other
recent calculations \cite{Ok99,Os01} (even if the different numerical 
results are not always compatible), which are discussed in the following. 
In \cite{Pa01} the authors also presented a detailed ($T$--matrix) study of the final state
interactions acting between the nucleons emitted in the non--mesonic decay.

Recently, the authors of Ref.~\cite{Go97} studied the decay of the
hypertriton ($^3_{\Lambda}$H) in the full OME picture of Ref.~\cite{Pa97}.
They worked in the framework of the Faddeev equation, which allows
to exactly calculate (at least in principle)
wave functions and final scattering states for three--body systems. 
They reproduced the experimental $\Lambda$ separation 
energy and the total lifetime of $^3_{\Lambda}$H,
obtaining $\Gamma_{T}/\Gamma^{\rm free}_{\Lambda}=1.03$. The non--mesonic width was
found to be 1.7\% of the free $\Lambda$ decay rate, only a little less
than the calculation with pion--exchange alone. 

The OME model of Ref.~\cite{Pa97} has also been employed, in 
Ref.~\cite{HP01}, to discuss the effects of the nuclear deformation on 
the non--mesonic decay of $p$--shell hypernuclei. By using the Nilsson model with
realistic values of the deformation parameter, the authors found that, due to 
nuclear deformation, both $\Gamma_n+\Gamma_p$ and $\Gamma_n/\Gamma_p$ can change at most by 
about 10\% with respect to the spherical limit. In view of the 
corrections made in Ref.~\cite{Pa01}
to the OME model of Ref.~\cite{Pa97}, the results
obtained in the above discussed Refs.~\cite{Go97,HP01} should be
updated.

In addition to the exchange of one pion, other authors
have included in the non--mesonic transition potential the
\underline{exchange of two pions} 
(correlated \cite{Sh94,It95,It98} or not \cite{Ba88,Sh94a,Sh95} into 
$\sigma$ and $\rho$ resonances). In Ref.~\cite{Ba88} the two--pion--exchange 
mechanism contains the $\Sigma N$ intermediate state,
and the $\Delta I=1/2$ rule is enforced in the $\Lambda N\pi$ and $\Sigma N\pi$
vertices. The intermediate $NN$ state has to be excluded in order to avoid a
double counting when the uncorrelated $2\pi$--exchange is employed in
connection with short range correlations.
The authors of \cite{Ba88} found that the $\Sigma$ component further reduces 
(with respect to the OPE value) the $\Gamma_n/\Gamma_p$ ratio.
On the contrary, a 15\% increase of the OPE ratio was found in 
Ref.~\cite{Sh95}, due to the $\Sigma N$ intermediate 
state in uncorrelated $2\pi$--exchange (see tables~\ref{tab1}, \ref{tab6}).
From the results of Ref.~\cite{Sh94a} on the $\Lambda N\to NN$ matrix elements
we can point out that the inclusion of both the $\Sigma N$ and $\Delta N$ 
intermediate states in the $2\pi$--exchange could sizeably increase 
the OPE $\Gamma_n/\Gamma_p$ ratio. This conclusion comes from the observation that
large ${^1S_0}\to {^1S_0}$ transitions were obtained for uncorrelated $2\pi$--exchange.
The same finding about the importance of ${^1S_0}\to {^1S_0}$ transitions comes
from Ref.~\cite{Sh94} for correlated $2\pi$--exchange.
In Ref.~\cite{It98} an improvement of $\Gamma_n/\Gamma_p$ for $^{12}_{\Lambda}$C
and $\Delta I=3/2$ contributions 
(introduced by the boson--boson coupling model)
less important than the $\Delta I=1/2$ ones were found by employing
correlated two--pion--exchange ($2\pi/\rho + 2\pi/\sigma$) in addition to the OPE
(see tables~\ref{tab2}, \ref{tab3}, \ref{tab7}). 
The authors also studied the $A$--dependence of
the non--mesonic decay rate and reproduced the data for light hypernuclei
but not the saturation of $\Gamma_{\rm NM}$ at large $A$. 

The baryon--baryon short range interactions have been studied by Cheung, Heddle 
and Kisslinger \cite{Ch83} by using an \underline{hybrid model}, 
through which the decay is described by two separate mechanisms: the long range interactions 
($r\geq 0.8$ fm) are treated in terms of hadronic degrees of freedom 
(OPE with $\Delta I=1/2$ rule), while the short range interactions (which cannot be explained 
in terms of meson exchange) are described by a 6--quark cluster model, which includes both
$\Delta I=1/2$ and $\Delta I=3/2$ components (see tables~\ref{tab1}, \ref{tab2}). 

In Ref.~\cite{Ma94}, Maltman and Shmatikov combined
a OME potential containing \underline{($\pi + K$)--exchange} 
for long distance interactions
and a \underline{quark model} picture at short distances again
in a hybrid model. By employing also the effective weak Hamiltonian 
modified by perturbative QCD effects of Ref.~\cite{Pa90}, the authors
obtained significant violations of the $\Delta I=1/2$ rule in the 
$J=0$ $\Lambda N\rightarrow NN$ amplitudes. 
As they pointed out, this should sizeably modify the value of $\Gamma_n/\Gamma_p$
in nuclei. In Ref.~\cite{Ma95} the same authors evaluated the $\Delta I=3/2$ contribution 
to the $\Lambda N\rho$ coupling by using the factorization approximation and
obtained for it a magnitude comparable with the $\Delta I=1/2$ contribution.

More recently, Inoue {\em et al.}~\cite{Ok96,Ok98} 
treated the non--mesonic $\Lambda$ decay in $s$--shell 
hypernuclei within a \underline{{\it direct quark model}} 
combined with the \underline{OPE} description (enforcing here the $\Delta I=1/2$ rule).
In their model the $NN$ and $\Lambda N$ repulsion at short 
distance originates from quark exchange
between baryons (induced by the quark anti--symmetrization) and gluon exchange between quarks.
The main uncertainties in this kind of approach come from the parameterization
of the effective weak Hamiltonian for quarks, obtained through the so--called 
{\it Operator Product Expansion} \cite{Pa90},
which contains perturbative QCD effects and, by construction, 
terms associated to both $\Delta I=1/2$ and
$\Delta I=3/2$ transitions. The authors found that the direct quark (DQ)
mechanism was significant, giving sizeable $\Delta I=3/2$ 
contributions in the $J=0$ channel, in agreement with Ref.~\cite{Ma94}. The
results on the $\Gamma_n/\Gamma_p$ ratio
are more consistent (even if sizeable discrepancies still remain)
with the experiment, because of a large increase (with respect to the OPE)
of the neutron--induced decay rate (see tables~\ref{tab3}, \ref{tab8}). 
Unfortunately, the calculation is only made for $s$--shell hypernuclei 
(and, as we will mention just below, for symmetric nuclear matter in \cite{Ok99}), 
and the employed quark Hamiltonian
is not able to reproduce the large ratio between the $\Delta I=1/2$ and $\Delta I=3/2$
amplitudes observed in the free $\Lambda$ decay. At present, the data on
hypernuclei do not allow the extraction of the $\Delta I=3/2$ amplitude in 
$\Lambda N \rightarrow NN$ (see the discussion of subsection \ref{passh}). 
Notice, however, that in Ref.~\cite{Ok99a} Oka showed
that the $\Delta I=3/2$ component is probed, in a clean way, by
the soft $\pi^+$ emission observed in the $\Lambda$ nuclear decay
($\Lambda \rightarrow \pi^0 n$ followed by $\pi^0 p\rightarrow \pi^+ n$), because
of the absence of the $\Delta I=1/2$ component.
Another delicate point in models with a direct quark contribution
is related to double counting and superposition problems, which 
could arise when both the quark and hadronic description are employed together.
However, this does not seem to be a problem in the calculations by Inoue {\em et al.}, 
because their relativistic formalism does not allow the exchange of quark--antiquark pairs in 
the direct quark baryon--baryon interaction. Moreover, in the soft pion limit, they 
determined the relative phase between the OPE and the direct quark contributions. 
Apart from the quoted problems, it would be interesting 
to establish the connection between the quark effective weak Hamiltonian
and the phenomenological $\Lambda$ weak vertex. On the other hand, 
the possibility that the subnucleonic degrees of freedom play a role in
nuclear systems remains an interesting field of research, and the non--mesonic
decay of $\Lambda$--hypernuclei could reveal itself as
the only good tool to study such effects.
The systematic measurement of the partial non--mesonic rates will be useful
in distinguishing the different decay mechanisms (meson exchange and direct quark
interactions). 

Very recently \cite{Ok99}, the \underline{direct quark mechanism} has 
been combined with a full \underline{OME potential} 
(\underline{$\pi, \rho, K, K^*, \eta, \omega$}), for calculations
in nuclear matter and in $s$--shell hypernuclei (see tables~\ref{tab1}, \ref{tab3},
\ref{tab6}, \ref{tab8}). The authors compared the {\it OME} model with
the {\it Light Mesons ($\pi, K$) $+$ Direct Quark} model: the short range repulsion
is given, respectively, by heavy--meson--exchange and direct quark mechanism.
Heavy--meson--exchange and direct quark
contributions employed together could cause double counting problems: in any case, 
the authors obtained that the OME $+$ DQ description does 
not improve the results. Both the previous pictures, namely OME 
and $\pi+ K$ $+$ DQ, gave the best results of the calculation: also
the $\Gamma_n/\Gamma_p$ ratio is significantly improved when the 
$\pi+ K$ $+$ DQ model is employed, both for
$^5_{\Lambda}$He hypernuclei and nuclear matter (considered as an approximate 
description of heavy hypernuclei).  

Effects of a \underline{violation of the $\Delta I=1/2$ rule} in 
$\Lambda N\rightarrow NN$ have been studied in Ref.~\cite{Pa98} by Parre\~{n}o 
{\em et al.} The employed OME model is the same of Ref.~\cite{Pa97}, with hadronic
couplings evaluated in the factorization approximation. The conclusion reached
by the authors is different from the one of Inoue {\em et al.}~\cite{Ok98}: even by
introducing large $\Delta I=3/2$ contributions 
(of the order of the $\Delta I=1/2$ ones), to take into account the
ambiguities in the factorization approximation, the calculated $\Gamma_n/\Gamma_p$
ratio for $^{12}_{\Lambda}$C remains abundantly below the experimental values
(see tables~\ref{tab2}, \ref{tab7}). We recall that the OME model employed in
Refs.~\cite{Pa97,Pa98} has been recently corrected in Ref.~\cite{Pa01}.
A similar correction could affect the conclusions obtained with a 
$\Delta I=3/2$ contribution.

Concerning the $\Gamma_n/\Gamma_p$ puzzle we quote here a different
mechanism, which has been suggested by the 
authors of Ref.~\cite{Ju01}. In order to describe the process $\Lambda N\rightarrow NN$,
they employed, in addition to the \underline{OPE} at large distances,
a phenomenological \underline{4--baryon point interaction} for short range interactions,
including $\Delta I=3/2$ contributions as well. Such a 4--baryon interaction was
initially considered by Block and Dalitz \cite{Bl63} as an approximation of the
short range interactions mediated by heavy mesons.
By properly fixing the different phenomenological
coupling constants of the problem (in particular, by using a small
$\Delta I=3/2$ component), the authors of Ref.~\cite{Ju01} could fit fairly well both
the experimental $\Gamma_n+\Gamma_p$ and $\Gamma_n/\Gamma_p$
in $^4_{\Lambda}$He, $^5_{\Lambda}$He and $^{12}_{\Lambda}$C
(see tables~\ref{tab2}, \ref{tab3}, \ref{tab7}, \ref{tab8}).

Up to this point we have essentially discussed theoretical evaluations of the
non--mesonic hypernuclear decay, which is the dominant channel for all
hypernuclei, but for the $s$--shell ones. Concerning the \underline{mesonic decay},
Refs.~\cite{Mo91,St93} presented a study of 
the $^5_{\Lambda}$He decay using a quark model based hypernuclear wave 
function. The authors have shown how the short range $\Lambda N$ repulsion
(which naturally arises in a quark model) is relevant to reproduce the observed  
mesonic rates in $s$--shell hypernuclei (see table~\ref{tab5}). 
In 1993, Nieves and Oset \cite{Os93} calculated
the mesonic widths for a broad range of $\Lambda$--hypernuclei
(from $^{12}_{\Lambda}$C to $^{209}_{\Lambda}$Pb). They
used a shell model picture and distorted
pionic wave functions, solutions of a pion--nucleus optical potential. The
results showed wild oscillations of $\Gamma_{\pi^-}/\Gamma_{\pi^0}$
around the value (equal to $2$) predicted by the $\Delta I=1/2$ rule for
$N=Z$ closed shell hypernuclei. This was due to effects of the hypernuclear shell structure. Similar
calculations have been carried out by Itonaga, Motoba, and Band$\overline{\rm o}$ in Refs.~\cite{Ba85,It88}.
With respect to Ref.~\cite{Os93}, they use different optical potentials and
descriptions of the energy balance in the decays (more accurate in \cite{Os93}),
and obtain somewhat dissimilar results (especially in very heavy systems).
Motoba and Itonaga updated the calculations in Ref.~\cite{Mo94}
by using an improved optical potential (see tables~\ref{tab4}, \ref{tab5}).
Both the evaluations of Nieves--Oset and Motoba--Itonaga--Band$\overline{\rm o}$ showed how the
mesonic rate strongly depends on the competition between the Pauli blocking,
which suppresses the decay, and the enhancement due to the pion wave distortion
in the medium. When the pion wave is distorted by the optical potential, for 
$A>100$ the mesonic width is enhanced by one/two orders of magnitude with
respect to the calculation without pion distortion. For the decay into $\pi^-p$
the Coulomb distortion alone gives rise to a non--negligible enhancement.
The results of the above calculations for light to heavy hypernuclei 
are shown in Fig.~\ref{mesonic} of subsection \ref{pheres}.

A different approach, which allows a unified treatment of mesonic and non--mesonic
channels and automatically includes all the partial waves of the relative
$\Lambda N$ motion, has been suggested by Oset and Salcedo \cite{Os85}
(see tables~\ref{tab1}--\ref{tab5}) and utilizes
the Random Phase Approximation (RPA) within the framework of the
\underline{polarization} \underline{propagators}.  
We shall discuss in detail this method in the next section.
Here we only remind the reader that the crucial point for the evaluation of the decay rates
is a realistic description of the pion self--energy in the medium and 
(especially for the evaluation of
$\Gamma_{\rm NM}$) of the baryon--baryon short range correlations. More recently,
this model has been applied to the calculation, in nuclear matter, of the three--body
decay $\Lambda NN\rightarrow NNN$ \cite{Al91} (see table~\ref{tab1}), 
through a purely phenomenological parameterization (by means of data on deep 
inelastic $(e,e^{\prime})$ scattering and pionic atoms)
of the {\sl 2p--2h} excitations in the pion self--energy.
A more detailed analysis of $\Gamma_2$, also implemented in finite nuclei via
the local density approximation, has been made in Ref.~\cite{Ra95}
(see tables~\ref{tab2}, \ref{tab4}). Here, the authors employed
a more realistic {\sl 2p--2h} polarization propagator, based again on an empirical
analysis of pionic atoms but also extended to kinematical regions not accessible by
this phenomenology. The introduction of a new non--negligible 
(as found in the above mentioned calculations) 
two--nucleon induced non--mesonic channel requires a reanalysis of the $\Gamma_n/\Gamma_p$ ratio. 
The most recent calculations performed within the 
polarization propagator method can be found in
Refs.~\cite{Al99,Al99a,Os01}. They reproduce quite well
both the mesonic and non--mesonic rates for light to heavy hypernuclei, although
problems related to the $\Gamma_n/\Gamma_p$ ratio still remain.
The results obtained in \cite{Al99,Al99a} will be discussed in detail in subsections
\ref{pheres} and \ref{micres}; 
those of Ref.~\cite{Os01} are listed in tables~\ref{tab2}, \ref{tab7}.
In Ref.~\cite{Os01} the one--nucleon induced non--mesonic decay
has been studied within a meson--exchange approach including one pion, one kaon,
correlated and uncorrelated $2\pi$--exchange and $\omega$--exchange
(\underline{$\pi+K+2\pi/\sigma+$ unc $2\pi+\omega$}). The correlated 
$2\pi$--exchange (in the $\sigma$ channel) has been treated in terms of a chiral unitary approach 
to the $\pi \pi$ scattering. For the $NN$ interaction this approach
leads to a $\sigma$--meson--exchange potential with a moderate attraction at 
$r\gsim 0.9$~fm and a repulsion at shorter distances, in contrast with the
attraction of the conventional $\sigma$--exchange. Once the correlated and
uncorrelated $2\pi$--exchange are added, a net $NN$ attraction is obtained
for all distances. In order to restore the behaviour of realistic 
$NN$ potentials, which present a moderate attraction only at intermediate distances,
the authors of Ref.~\cite{Os01} introduced the exchange of the $\omega$--meson
to produce the required repulsion. A large cancellation between 
$\sigma$--exchange and uncorrelated $2\pi$--exchange has been found
for momenta around the relevant value $420$ MeV. Consequently,
the total $2\pi$--exchange contribution to the decay turned out to be small
(around 10\% on $\Gamma_n$ and $\Gamma_p$). The $\omega$--exchange 
gave a contribution of the same order of magnitude. On the contrary,
the $K$--exchange, also constrained by chiral unitary theory, has been of
primary importance to reproduce the experimental non--mesonic rate
$\Gamma_{\rm NM}=\Gamma_n+\Gamma_p$ and to improve the OPE 
$\Gamma_n/\Gamma_p$ ratio for $^{12}_{\Lambda}$C: $(\Gamma_n/\Gamma_p)^{\rm Full}\simeq
4.5\, (\Gamma_n/\Gamma_p)^{\rm OPE}$. In Ref.~\cite{Zh99}, still within the
polarization propagators framework, by using a relativistic mean--field
approximation to the Walecka model, the authors evaluated the ring OPE non--mesonic
decay widths to be considerably smaller than the non--relativistic 
ones of Refs.~\cite{Al99,Al99a,Os01} (see tables \ref{tab2}, \ref{tab4} and
subsections \ref{pheres}, \ref{micres}).
This also seems to be unrealistic when compared with the findings of
Ref.~\cite{Sh93}. Here, by employing the Walecka model within the wave function
formalism, the relativistic OPE calculation gave nuclear matter non--mesonic rates
larger (by about 40\%) than the non--relativistic ones
(see table~\ref{tab1}, \ref{tab6}). However, we remind the reader that
this calculation does not include the effects of vertex form factors and 
short range correlations, which significantly reduce the non--mesonic rates,
both in the relativistic and non--relativistic descriptions.

In tables~\ref{tab1}--\ref{tab8} the numerical results obtained within
the above discussed models are summarized and compared with experimental data.
The decay widths are in units of the free $\Lambda$ width.

\subsection*{\normalsize \underline{Tab.~\ref{tab1}: Decay width for a $\Lambda$ in
nuclear matter}}

The results of Adams are corrected for the small $\Lambda \pi N$ coupling 
constant he used, as explained above. All the uncorrelated
OPE decay widths are compatible with a value of about 4. The result by Cheung 
{\em et al.}~is sizeably smaller than 4, but we recall that in their calculation the
pion--exchange is only active for $r>0.8$ fm, while a large OPE contribution comes
from smaller distances. This is equivalent to use very strong
short range correlations (SRC), which prevent the process for $r<0.8$ fm.
Differences among the various calculations are observed when the effects
of SRC and form factors (FF) are included in the OPE models. They reduce
the uncorrelated widths by a factor $\gsim 2$. Adams used an inappropriate
(too strong) correlation for the tensorial transition ${^3S_1}\to {^3D_1}$.
Neglecting the tensorial SRC, his correlated result (1.57) is more realistic. 
The differences among the other calculations may be understood taking into account the
parameterizations used for SRC and FF. For example, in the polarization
propagator method (PPM) of Ref.~\cite{Os85}, 
a monopole FF with cut--off $\Lambda_{\pi}=1.3$ GeV is used, while in Ref.~\cite{Mc84}
a softer FF is employed ($\Lambda_{\pi}\simeq 0.6$ GeV). This is responsible
for the ratio 2.1 between the results of Refs.~\cite{Os85} and \cite{Mc84}.
The inclusion of the $\rho$--exchange in the transition potential
decreases the decay rate (this characteristic has been confirmed in finite 
nucleus calculations): in Ref.~\cite{Mc84} the $\rho$--meson leads to an 
unrealistic almost complete cancellation of the OPE contribution. The results of
Nardulli refer to different choices for the FF. Also the one--meson--exchange
(OME) models (we refer, here and in the following, to OME models when the transition
potential contains the exchange of $\pi, \rho, K, K^*, \omega,$ and $\eta$ mesons) tends 
to reduce the rate with respect to the pure OPE calculation. This is also true, as we
shall see in the next tables, for finite hypernuclei. In particular, the
$K$--meson--exchange considerably cancels the OPE contribution \cite{Ok99}.
From inspection of the experimental data on heavy hypernuclei one concludes that realistic
values of the $\Lambda$ decay rate in nuclear matter lie in the range $1.5\div 2$.

\begin{table}
\begin{center}
\caption{Decay width for a $\Lambda$ in nuclear matter ($\Gamma_{\rm T}\equiv \Gamma_{\rm NM}$).}
\label{tab1}
\begin{tabular}{|c|c|c|c|c|} \hline
\mc {1}{|c|}{Ref.} &
\mc {1}{c|}{Unc} &
\mc {1}{c|}{+ SRC} &
\mc {1}{c|}{+ SRC + FF} &
\mc {1}{c|}{Model} \\ \hline\hline
Adams 1967 \cite{Ad67}           & 3.47 & 0.38  &              & OPE \\
                                 &      & 1.57  &              & no tensor SRC \\   
\hline
Dalitz 1973 \cite{Da73}          &      &       & 2.0          & Contact int. \\ \hline
Cheung {\em et al.}~1983--86 \cite{Ch83}   & 0.99 &       & 0.77         & OPE $> 0.8$ fm\\
                            &      &       & 3.0          & Hybrid \\ \hline
McKellar--Gibson 1984 \cite{Mc84} & 4.13 & 2.31  & 1.06         & OPE\\
                            & 1.13 & 0.72  & 0.10         & $\pi + \rho$\\ \hline
Oset--Salcedo 1985 \cite{Os85}    & 4.3  &       & 2.2          & PPM \\ \hline
Dubach {\em et al.}~1986 \cite{Du86}   & 3.89 & 1.82  &              & OPE\\
                            &      & 1.55  &              & $\pi + \rho$\\
                            &      & 1.23  &              & OME\\ \hline
Nardulli 1988 \cite{Na88}        &      &       &$0.7\div2.1$  & $\pi + \rho$\\ \hline
Alberico {\em et al.}~1991 \cite{Al91} &      &       & 1.74         & PPM with 2B \\  \hline
Shinmura 1993 \cite{Sh93}                 &   2.92     &      &        & OPE \\
                                          &   3.97     &      &        & Rel OPE\\
\hline
Shinmura 1995 \cite{Sh95} &        &       &  1.73    & OPE \\
                                          &        &       &  1.85      & $\pi +$ unc $2\pi$\\ \hline
Dubach {\em et al.}~1996 \cite{Du96}   & 4.66 & 1.85  &              & OPE\\
                            &      & 1.38  &              & OME\\ \hline
Sasaki {\em et al.}~2000 \cite{Ok99}      &      & 2.819 & 1.850        & OPE\\
                                 &      & 2.068 & 1.216        & $\pi +K$\\ 
                                 &      & 2.863 & 1.906        & OME\\
                                 &      &       & 2.456        & $\pi + K$ + DQ\\ \hline
\end{tabular}
\end{center}
\end{table} 

\newpage

\subsection*{\normalsize \underline{Tab.~\ref{tab2}: Non--mesonic decay width for
$^{12}_{\Lambda}$C.}}

The OPE results of Cheung {\em et al.}~underestimate the experiment for the same reason
explained in connection with the calculation in nuclear matter. Note that here the
reduction obtained in going from the uncorrelated case to the correlated one is 
even smaller than what occurred in nuclear matter. In this calculation,
the SRC in OPE plays a little role because
the $\pi$--exchange is only active for distances $r>0.8$ fm.
However, the complete result (hybrid model) of Ref.~\cite{Ch83} is realistic. 
The relativistic polarization propagator method (Rel PPM) of Ref.~\cite{Zh99}
predicts a too small decay rate. On the contrary, the non--relativistic
PPM's of Refs.~\cite{Os85,Ra95} overestimate the data because of the
use of unrealistic SRC and $\Lambda$ wave functions. The calculation
by Dubach {\em et al.}~\cite{Du96} provides a too large uncorrelated OPE rate
[too small is the reduction with respect to their calculation (3.89 in Ref.~\cite{Du86}
and 4.66 in Ref.~\cite{Du96}) for nuclear matter] and
too small correlated results, both in OPE and in the full OME. The 
available computational details of Refs.~\cite{Du86,Du96} do not allow
to explain these controversial results. All the other correlated OPE calculations 
(apart from the case of Ref.~\cite{Ju01})
are compatible with the experiment and give rates reduced
with respect to the uncorrelated ones by a factor $1.5\div 2$. The
$\pi+\rho$, $\pi + 2\pi/\rho + 2\pi/\sigma$ and OME rates 
are quite similar to the OPE estimates. 
In Ref.~\cite{Pa98}, $\Delta I=3/2$ contributions to the
$\Lambda \to \pi N$ transition are evaluated in the factorization 
approximation: their effect on the non--mesonic rate seems very small.
In Ref.~\cite{Os01}, the authors used the polarization
propagator method with $(\pi +K+2\pi/\sigma +$ unc $2\pi +\omega)$--exchange. 
The result for the one--nucleon induced non--mesonic rate of the full calculation 
is reduced with respect to the OPE value of about 30\%. 
This is due, almost completely, to $K$--exchange. The full result
including the two--body induced contribution (2B) has been obtained 
by adding the value $\Gamma_2=0.27$ obtained in Ref.~\cite{Ra95}.
Realistic calculations supply non--mesonic widths in $^{12}_{\Lambda}$C 
reduced by a factor $1.5\div 2$ with respect to the values for nuclear matter. 
The results of Parre\~{n}o and Ramos of Ref.~\cite{Pa01} correct those of
Ref.~\cite{Pa97} (due to a mistake in the inclusion of the $K$ and $K^*$
contributions) and correspond to the use of different Nijmegen models 
\cite{Ri99,Ri99a} for the hadronic coupling constants. The authors also
made an accurate evaluation of the final state interactions
between the outgoing nucleons, by using the scattering $NN$ wave function
from the Lippmann--Schwinger ($T$--matrix) equation obtained with the Nijmegen $NN$ 
potentials. The $K$--exchange decreases the rate $\Gamma_n +\Gamma_p$ with respect to  
the one calculated in OPE by about 26\% in Ref.~\cite{Os01} and $37\div 45$\%
in Ref.~\cite{Pa01}.

\begin{table}
\begin{center}
\caption{Non--mesonic decay width for $^{12}_{\Lambda}$C.}
\label{tab2}
\begin{tabular}{|c|c|c|c|c|} \hline
\mc {1}{|c|}{Ref.} &
\mc {1}{c|}{Unc} &
\mc {1}{c|}{+ SRC} &
\mc {1}{c|}{+ SRC + FF} &
\mc {1}{c|}{Model} \\ \hline\hline
Cheung {\em et al.}~1983--86 \cite{Ch83}           & 0.48  &       & 0.41             & OPE $>0.8$ fm \\
                                    &       &       & 1.28             & Hybrid \\  \hline
Oset--Salcedo 1985 \cite{Os85}            &       &       & 1.5              & PPM \\ \hline
Ramos--Bennhold 1994 \cite{Ra94}            & 1.58  &       & 0.87             & OPE\\
                                    & 4.30  &       & 0.98             & OME\\ \hline
Ramos {\em et al.}~1995 \cite{Ra95}            &       &       & 1.72             & PPM with 2B\\ \hline
Parre\~{n}o {\em et al.}~1995 \cite{Pa95}      & 1.641 & 1.186 & 0.964            & OPE\\ \hline
Parre\~{n}o {\em et al.}~1995--96 \cite{Pa95-96}   & 1.716 & 1.239 & 1.110            & OPE\\
                                    &       &       & 0.991            & $\pi+\rho$\\ \hline
Dubach {\em et al.}~1996 \cite{Du96}            & 3.4   & 0.5   &                  & OPE\\
                                    &       & 0.2   &                  & OME\\ \hline
Parre\~{n}o {\em et al.}~1997 \cite{Pa97}             & 1.682 & 1.232 & 0.885            & OPE\\
                                    & 2.055 &       & 0.859            & $\pi+\rho$\\
                                    & 2.301 &       & 0.753            & OME\\ \hline
Parre\~{n}o {\em et al.}~1998 \cite{Pa98}             &       &       & 0.753            & OME\\
         &       &       & $0.753\div0.796$ & OME + $\Delta I=3/2$\\ \hline
Itonaga {\em et al.}~1998 \cite{It98}                     &       &       & 1.05
& $\pi + 2\pi/\rho + 2\pi/\sigma$ \\ \hline
Zhou--Piekarewicz 1999 \cite{Zh99}           &       &       & 0.413            & Rel PPM \\ \hline
Jun {\em et al.}~2001 \cite{Ju01}       &       &       & 0.468      & OPE\\ 
                                  &       &       & 1.174     & OPE + 4BPI\\ \hline
Jido {\em et al.}~2001 \cite{Os01}   &    &    & 1.075 & OPE \\
                               &    &    & 0.795 & $\pi +K$ \\
                               &    &    & 0.769 & $\pi +K+2\pi +\omega$ \\
                               &    &    & 1.039 &  Full with 2B \\ \hline
Parre\~{n}o--Ramos~2001 \cite{Pa01}       &  &  & $0.751\div 0.762$  & OPE\\
(correction of \cite{Pa97})               &  &  & $0.413\div 0.485$  & $\pi +K$\\
                                          &  &  & $0.554\div 0.726$  & OME \\ \hline\hline   
Exp BNL 1991 \cite{Sz91}                 &       &       & $1.14\pm0.20$    & \\ \hline
Exp KEK 1995 \cite{No95}                 &       &       & $0.89\pm0.18$    & \\ \hline
Exp KEK 2000 \cite{Ou00}                 &       &       & $0.83\pm0.11$    & \\ \hline 
\end{tabular}
\end{center}
\end{table}

\newpage

\subsection*{\normalsize \underline{Tab.~\ref{tab3}: Non--mesonic decay width for
$^5_{\Lambda}$He}}

In this and in the following tables, 
only the results obtained including FF and SRC are 
listed. Unrealistic rates are predicted by Refs.~\cite{Ta85,Os85}. The result
of Ref.~\cite{Ta85} presents a strong cancellation between $\pi$-- and 
$\rho$--exchange. In \cite{Os85} the authors overestimated $\Gamma_{\rm NM}$
because they employed a wave function for the hyperon too much superimposed with
the nuclear core. We remind the reader that
$\Lambda-^4$He potentials consistent with experimental observations
have a repulsive core. By using the same model, with a more realistic $\Lambda$ wave
function (calculated from a variational method)
the same authors obtained \cite{Os86a} a non--mesonic width
compatible with the experiment. There are remarkable differences among
the several OPE estimates, ranging from 0.144 (Takeuchi {\em et al.}) to 0.9
(Dubach {\em et al.}). Because of the lack of technical details, the calculation of Dubach {\em et
al.}~cannot be easily compared with the other ones. We remark that they do not
take into account of the FF, which reduce the non--mesonic width, especially the OPE one.
The large difference between their OPE and 
OME results could originate form a double counting between heavy--meson--exchange and SRC.
It is also rather strange that the uncorrelated OPE result of Dubach {\em et al.}~(0.6,
not shown in the table) is smaller than the correlated one (0.9).
Another point to recall is that in Ref.~\cite{Du96}
the correlated OPE and OME non--mesonic rates
for $^{12}_{\Lambda}$C are smaller than the corresponding rates for $^5_{\Lambda}$He
of table~\ref{tab2}, while, from experiment, we know that
$\Gamma_{\rm NM}(^{12}_{\Lambda}{\rm C})\simeq 2\Gamma_{\rm NM}(^5_{\Lambda}{\rm He})$.
The calculations by Inoue {\em et al.}~\cite{Ok96,Ok98} and Sasaki {\em et al.}~\cite{Ok99} show different 
OPE results. They can be understood in terms of the different FF and SRC employed.
The calculation or Ref.~\cite{Pa01} is an updating of that presented in \cite{Pa97}:
the intervals shown correspond to the use of different Nijmegen models for
the hadronic coupling constant.
We note that for $(\pi +K)$--exchange the results of Ref.~\cite{Pa01} are 
substantially compatible with the value of Ref.~\cite{Ok99}. The reduction of the 
$\pi+K$ rate with respect to the OPE one is larger in Ref.~\cite{Pa01} ($36\div 45$\%) than in
Ref.~\cite{Ok99} (26\%).

\begin{table}
\begin{center}
\caption{Non--mesonic decay width for $^5_{\Lambda}$He.}
\label{tab3}                      
\begin{tabular}{|c|c|c|} \hline
\mc {1}{|c|}{Ref.} &
\mc {1}{c|}{$\Gamma_{\rm NM}$} &
\mc {1}{c|}{Model} \\ \hline\hline
Dalitz 1973 \cite{Da73}                & 0.5   & Contact int. \\ \hline   
Takeuchi {\em et al.}~1985 \cite{Ta85}        & 0.144 & OPE \\
                                  & 0.033 & $\pi+\rho$ \\ \hline
Oset--Salcedo 1985 \cite{Os85}          & 1.15  & PPM \\ \hline
Oset--Salcedo--Usmani 1986 \cite{Os86a}       & 0.54  & PPM \\ \hline
Itonaga {\em et al.}~1995 \cite{It95}        & 0.20  & OPE \\  
                                  & 0.30  & $\pi+2\pi/\sigma$ \\ \hline
Parre\~{n}o {\em et al.}~1995 \cite{Pa95}    & 0.56  & OPE \\ \hline
Dubach {\em et al.}~1996 \cite{Du96}         & 0.9   & OPE \\   
                                  & 0.5   & OME \\ \hline
Inoue {\em et al.}~1996 \cite{Ok96}            & 0.333 & OPE \\
                                       & 0.381 & DQ only \\ \hline
Parre\~{n}o {\em et al.}~1997 \cite{Pa97}    & 0.414 & OME \\ \hline
Itonaga {\em et al.}~1998 \cite{It98}            & 0.39 & $\pi + 2\pi/\rho + 2\pi/\sigma$ \\ \hline
Inoue {\em et al.}~1998 \cite{Ok98}            & 0.216 & OPE \\
                                  & 0.627 & OPE + DQ \\ \hline
Sasaki {\em et al.}~2000 \cite{Ok99}            & 0.370 & OPE \\
                                       & 0.302 & $\pi+K$ \\ 
                                       & 0.519 & $\pi+K+$ DQ \\ \hline
Jun {\em et al.}~2001 \cite{Ju01}            & 0.158  & OPE \\ 
                                       & 0.426 & OPE + 4BPI \\ \hline
Parre\~{n}o--Ramos~2001 \cite{Pa01}    & $0.424\div 0.425$ & OPE \\
(correction of \cite{Pa97})            & $0.235\div 0.272$ & $\pi +K$ \\
                                       & $0.317\div 0.425$ & OME \\ \hline\hline
Exp BNL 1991 \cite{Sz91}               & $0.41\pm0.14$ &  \\ \hline
Exp KEK 1995 \cite{No95a}              & $0.50\pm0.07$ &  \\    \hline
\end{tabular} 
\end{center}
\end{table}

\newpage

\subsection*{\normalsize \underline{Tab.~\ref{tab4}: Mesonic decay rate for
$^{12}_{\Lambda}$C.}}

The results reported in the table are all
compatible with the data, which, however, have very large error bars.
The only exception is the calculation of Ref.~\cite{Zh99}, supplying a 
decay rate which underestimates the recent KEK data \cite{Sa00}.
The estimates obtained with the wave function method (WFM) of Refs.~\cite{It88,Os93,Mo94}
are consistent with the experimental ratio $\Gamma_{\pi^0}/\Gamma_{\pi^-}\simeq 1\div
2>\left(\Gamma_{\pi^0}/\Gamma_{\pi^-}\right)^{\rm free}=1/2$,
which reflects the particular nuclear shell structure of
$^{12}_{\Lambda}$C.

\vspace{1cm}

\begin{table}[thb]
\begin{center}
\caption{Mesonic decay rate for $^{12}_{\Lambda}$C.}
\label{tab4}
\begin{tabular}{|c|c|c|} \hline   
\mc {1}{|c|}{Ref.} &
\mc {1}{c|}{$\Gamma_{M}$} &
\mc {1}{c|}{Model} \\ \hline\hline
Oset--Salcedo 1985 \cite{Os85}             & 0.41             & PPM \\ \hline
Itonaga--Motoba--Band$\overline{\rm o}$ 1988 \cite{It88}            & $0.233\div0.303$ & WFM \\ \hline  
Ericson--Band$\overline{\rm o}$ 1990 \cite{Er90}       & 0.229 & WFM \\ \hline
Nieves--Oset 1993 \cite{Os93}              & 0.245            & WFM \\ \hline
Itonaga--Motoba 1994 \cite{Mo94}            & 0.228            & WFM \\ \hline
Ramos {\em et al.}~1995 \cite{Ra95}             & 0.31             & PPM \\ \hline
Zhou--Piekarewicz 1999 \cite{Zh99}      & 0.112             & Rel PPM \\ \hline\hline
Exp BNL 1991 \cite{Sz91}                        & $0.11\pm0.27$    &   \\ \hline
Exp KEK 1995 \cite{No95}                        & $0.36\pm0.13$    &   \\ \hline
Exp KEK 2001 \cite{Sa00}:                   &                     &   \\ 
$\Gamma_{\pi^-}$                            & $0.113\pm 0.014$    &   \\ 
$\Gamma_{\rm M}$ (with $\Gamma_{\pi^0}$ from \cite{Sa91})      & $0.31\pm 0.07$    & \\ \hline
\end{tabular} 
\end{center}
\end{table} 

\newpage

\subsection*{\normalsize \underline{Tab.~\ref{tab5}: Mesonic decay rate for
$^{5}_{\Lambda}$He.}}

The theoretical results agree with the experimental data. This is also
true for $\Gamma_{\pi^-}/\Gamma_{\pi^0}$, which does not
deviate much from the $\Delta I=1/2$ value ($=2$) for free decays. We expect this
result, since $^5_{\Lambda}$He has a closed shell core with $N=Z$. A repulsive core in the 
$\Lambda - \alpha$ mean potential (used in all but the calculation of 
Ref.~\cite{Os85}) is favoured. Moreover, it comes out naturally
in the quark model descriptions of Refs.~\cite{Mo91,St93}. The results of
Refs.~\cite{It88,Mo92} refer to the use of different pion--nucleus
optical potentials.

\vspace{1cm}

\begin{table}[thb]
\begin{center}  
\caption{Mesonic decay rate for $^{5}_{\Lambda}$He.}
\label{tab5}
\begin{tabular}{|c|c|c|} \hline
\mc {1}{|c|}{Ref.} &
\mc {1}{c|}{$\Gamma_{M}$} &
\mc {1}{c|}{Model} \\ \hline\hline
Oset--Salcedo 1985 \cite{Os85}             & 0.65             & PPM \\ \hline
Oset--Salcedo--Usmani 1986 \cite{Os86a}             & 0.54             & PPM \\ \hline
Itonaga--Motoba--Band$\overline{\rm o}$ 1988 \cite{It88}            & $0.331\div0.472$ & WFM \\ \hline
Motoba {\em et al.}~1991 \cite{Mo91}            & 0.608            & WFM + Quark Model \\ \hline
Motoba        1992 \cite{Mo92}            & 0.61             & WFM   \\ \hline
Straub {\em et al.}~1993 \cite{St93}              & 0.670             & WFM + Quark Model   \\ \hline
Kumagai--Fuse {\em et al.}~1995  \cite{Ku95a}           & 0.60             & WFM  \\ \hline\hline
Exp BNL 1991 \cite{Sz91}                  & $0.59^{+0.44}_{-0.31}$     &    \\ \hline   
\end{tabular}
\end{center} 
\end{table}  

\newpage

\subsection*{\normalsize \underline{Tab.~\ref{tab6}: $\Gamma_n/\Gamma_p$ ratio for
nuclear matter.}}

The OPE ratios of Adams \cite{Ad67} and Shinmura \cite{Sh93} seem unrealistic: in fact,
they are considerably larger than the other OPE estimates. We note, however, that
Adams' (Shinmura's) calculation did not include hadronic FF (SRC and FF).
The $(\pi +\rho)$ calculation by Nardulli supplies values of $\Gamma_n/\Gamma_p$
(the interval in the table corresponds to the use of
different FF) close to the experimental indication for
$^{12}_{\Lambda}$C. However, no other estimate that employed a 
$(\pi +\rho)$--exchange potential has confirmed an important role of the 
$\rho$--meson in the calculation of $\Gamma_n/\Gamma_p$. In Refs.~\cite{Du86,Du96,Ok99} 
the introduction of heavier mesons supplies improved ratios: 
a great improvement, due to both the exchange of the $K$--meson and the DQ process, 
has been found by Sasaki {\em et al.}~\cite{Ok99}. 

\vspace{1cm}

\begin{table}[h]
\begin{center}  
\caption{$\Gamma_n/\Gamma_p$ ratio for nuclear matter.}
\label{tab6}
\begin{tabular}{|c|c|c|} \hline
\mc {1}{|c|}{Ref.} &
\mc {1}{c|}{$\Gamma_n/\Gamma_p$} &
\mc {1}{c|}{Model} \\ \hline\hline
Adams 1967 \cite{Ad67}               & 0.35             & OPE   \\ \hline
Dubach {\em et al.}~1986--96 \cite{Du86,Du96}  & 0.06   &  OPE  \\
                                & 0.08   &  $\pi+\rho$ \\  
                                & 0.34   &  OME  \\ \hline 
Nardulli 1988 \cite{Na88}            & $0.67\div1.25$   & $\pi+\rho$   \\ \hline
Shinmura 1993 \cite{Sh93}  & 0.255   &  Rel OPE  \\ \hline
Shinmura 1995 \cite{Sh95}  & 0.07    &  OPE  \\
                                           & 0.08   &  $\pi +$unc $2\pi$ \\ \hline
Sasaki {\em et al.}~2000 \cite{Ok99}          & 0.087  & OPE \\
                                     & 0.430  & $\pi+K$ \\
                                     & 0.398  & OME \\
                                     & 0.716  & $\pi+K+$ DQ \\ \hline
\end{tabular}
\end{center} 
\end{table}  

\newpage

\subsection*{\normalsize \underline{Tab.~\ref{tab7}: $\Gamma_n/\Gamma_p$ ratio for
$^{12}_{\Lambda}$C.}}

All the calculations but the ones of Refs.~\cite{Du96,It98,Ju01,Os01,Pa01} 
strongly underestimate the observed ratios. However, we must notice that
the various data have very large error bars and
there are still problems about the methods employed by the experiments to extract 
$\Gamma_n/\Gamma_p$ (see the discussion of section \ref{newpuzzle}).
In Ref.~\cite{Ju01}, in addition to the OPE at large distances, a 4--baryon point interaction 
(4BPI), including $\Delta I=3/2$ contributions as well, is employed
to describe the short range interactions through a purely phenomenological 
model which fits the partial
non--mesonic rates for light hypernuclei. However, the values of some of 
the parameters used in this model are questionable. 
The large $\Gamma_n/\Gamma_p$ ratio obtained by Dubach {\em et al.}
in OME is not confirmed by the calculations of Refs.~\cite{Ra94,Pa97,Pa01}. 
Moreover, we note that the calculation of Dubach {\em et al.}~obtains a realistic
$\Gamma_n/\Gamma_p$ but strongly underestimate $\Gamma_n+\Gamma_p$ (see table~\ref{tab2}). 
Also surprising is the large difference between the results of Ref.~\cite{Du96} 
for $^{12}_{\Lambda}$C and nuclear matter (see table~\ref{tab6}). 
The OME calculation in Ref.~\cite{Pa97} overestimates $\Gamma_p$ and
underestimates $\Gamma_n$: $\Gamma_p\simeq 2\Gamma_p^{exp}$, 
$\Gamma_n\simeq 0.1 \Gamma_n^{exp}$ (we refer, here, to the data of 
Ref.~\cite{Sz91}). In Ref.~\cite{Pa01}, the results of
\cite{Pa97} have been corrected for a mistake made in the inclusion of the
strange mesons exchange (a sign error in certain transitions mediated by
$K$-- and $K^*$--exchange). The new calculation shows an 
improvement of the OME $\Gamma_n/\Gamma_p$ ratio,
mainly due to $K$--exchange. The results quoted in the table has been
obtained by means of different models for the calculation of the unknown hadronic
vertices and by using the Lippmann--Schwinger equation to obtain the
scattering wave function for the final $NN$ states.
In Ref.~\cite{Pa98}, by introducing $\Delta I=3/2$ contributions in the OME $\Lambda N\to NN$
transition amplitude (OME + $\Delta I=3/2$) of Ref.~\cite{Pa97}
(which, we remind the reader, contains the above discussed error),
variations of $\Gamma_n$ only have been obtained. The inclusion of correlated 
$2\pi$--exchange in \cite{It98} (both in the $\sigma$ and $\rho$ channels) improves the 
calculated ratio. In Ref.~\cite{Os01}, thanks to the $K$--exchange,
a significant improvement of the OPE ratio has been obtained. The 
two--pion--exchange (correlated in the $\sigma$ channel
and uncorrelated) as well as the $\omega$--exchange
turned out to have small effects on the decay rates. 
The $(\pi +K)$ calculation of Ref.~\cite{Os01} provides a ratio about 52\% larger
than the maximum value obtained in Ref.~\cite{Pa01}.

\begin{table}
\begin{center}
\caption{$\Gamma_n/\Gamma_p$ ratio for $^{12}_{\Lambda}$C.}
\label{tab7}
\begin{tabular}{|c|c|c|} \hline
\mc {1}{|c|}{Ref.} &
\mc {1}{c|}{$\Gamma_n/\Gamma_p$} &
\mc {1}{c|}{Model} \\ \hline\hline
Ramos--Bennhold 1994 \cite{Ra94}            & 0.19   & OPE   \\
                                    & 0.27   &  OME  \\ \hline
Parre\~{n}o {\em et al.}~1995--96 \cite{Pa95-96}   & 0.12   & OPE   \\
                                    & 0.12   & $\pi+\rho$   \\ \hline
Dubach {\em et al.}~1996 \cite{Du96}           & 0.20   &  OPE  \\
                                    & 0.83   &   OME \\ \hline
Parre\~{n}o {\em et al.}~1997 \cite{Pa97}      & 0.104  & OPE   \\
                                    & 0.095  & $\pi+\rho$   \\
                                    & 0.068  & OME   \\ \hline
Parre\~{n}o {\em et al.}~1998 \cite{Pa98}      & 0.068  &  OME   \\
                                    & $0.034\div0.136$   &  OME + $\Delta I=3/2$ \\ \hline
Itonaga {\em et al.}~1998 \cite{It98}          & 0.10   & OPE   \\
                                    & 0.36  &  $\pi+2\pi/\rho+2\pi/\sigma$  \\ \hline
Jun {\em et al.}~2001 \cite{Ju01}         & 0.08  & OPE \\ 
                                    & 1.14 & OPE + 4BPI \\ \hline 
Jido {\em et al.}~2001 \cite{Os01}        & 0.12     & OPE \\
                                    & 0.52     & $\pi +K$ \\
                                    & 0.53     & $\pi +K+2\pi +\omega$ \\ \hline
Parre\~{n}o--Ramos~2001 \cite{Pa01} & $0.078\div 0.079$    & OPE \\
(correction of \cite{Pa97})         & $0.205\div 0.343$    & $\pi +K$ \\
                                    & $0.288\div 0.341$    & OME \\ \hline\hline
Exp 1974 \cite{Mo74}                      & $0.59\pm0.15$   &    \\ \hline
Exp BNL 1991 \cite{Sz91}                 & $1.33^{+1.12}_{-0.81}$   &    \\ \hline
Exp KEK 1995 \cite{No95}                 & $1.87^{+0.67}_{-1.16}$   &    \\ \hline
Exp KEK 2001 \cite{Ha01}                 & $1.17^{+0.22}_{-0.20}$  & \\ \hline
\end{tabular}   
\end{center}
\end{table} 

\newpage

\subsection*{\normalsize \underline{Tab.~\ref{tab8}: $\Gamma_n/\Gamma_p$ ratio for
$^{5}_{\Lambda}$He.}}

Also for $^{5}_{\Lambda}$He, apart from the phenomenological fit of 
Ref.~\cite{Ju01} and the $(\pi+K+ {\rm DQ})$
calculation of Sasaki {\em et al.}~\cite{Ok99}, the theory underestimates the experiment. 
In Refs.~\cite{Ok96,Ok98} Inoue {\em et al.}~showed how the direct quark (DQ) mechanism
is an important ingredient in the evaluation of $\Gamma_n/\Gamma_p$.
The calculation of Sasaki {\em et al.}~\cite{Ok99} found a large
improvement of the ratio, due to the combined effects of $K$--exchange
and DQ mechanism. However, this model tends to overestimate the observed
total non--mesonic rates for heavy hypernuclei (see results for nuclear matter in
table~\ref{tab1}). As explained by the authors, this effect could be originated
from the fact that the short range baryon--baryon correlations used in the
calculation were not sufficiently strong. The results of Ref.~\cite{Pa97}
have been revisited in Ref.~\cite{Pa01}: here, in addition to a correction of 
an error in the previous OME calculation, the authors made a detailed analysis
of the final state $NN$ interactions and found a considerable improvement of the
ratio. The $(\pi +K)$ calculation of this paper agrees with that
of Ref.~\cite{Ok99}. 

\vspace{1cm}

\begin{table}[thb]
\begin{center}
\caption{$\Gamma_n/\Gamma_p$ ratio for $^{5}_{\Lambda}$He.}
\label{tab8} 
\begin{tabular}{|c|c|c|} \hline
\mc {1}{|c|}{Ref.} &
\mc {1}{c|}{$\Gamma_n/\Gamma_p$} &
\mc {1}{c|}{Model} \\ \hline\hline
Itonaga {\em et al.}~1995 \cite{It95}        & 0.13   & OPE \\
                                  & 0.17   & $\pi+2\pi/\sigma$ \\ \hline
Inoue {\em et al.}~1996 \cite{Ok96}            & 0.12   & OPE \\
                                  & 0.95   & DQ only \\ \hline
Dubach {\em et al.}~1996 \cite{Du96}         & 0.05   & OPE \\  
                                  & 0.48   & OME \\ \hline
Parre\~{n}o {\em et al.}~1997 \cite{Pa97}    & 0.073  & OME \\ \hline
Inoue {\em et al.}~1998 \cite{Ok98}            & 0.132  & OPE \\
                                       & 0.489  & OPE + DQ \\ \hline
Sasaki {\em et al.}~2000 \cite{Ok99}            & 0.133  & OPE \\
                                       & 0.450  & $\pi+K$ \\ 
                                       & 0.701  & $\pi+K+$ DQ \\ \hline
Jun {\em et al.}~2001 \cite{Ju01}            & 0.10   & OPE \\ 
                                       & 1.30   & OPE + 4BPI \\ \hline
Parre\~{n}o--Ramos~2001 \cite{Pa01}    & 0.086 & OPE \\
(correction of \cite{Pa97})            & $0.288\div 0.498$ & $\pi +K$ \\
                                       & $0.343\div 0.457$ & OME \\ \hline\hline
Exp BNL 1991 \cite{Sz91}               & $0.93\pm0.55$  &  \\ \hline
Exp KEK 1995 \cite{No95a}              & $1.97\pm0.67$  &  \\ \hline
\end{tabular}
\end{center} 
\end{table}

\newpage

The theoretical calculations quoted in the
tables for the non--mesonic decay show that further efforts
(both on the theoretical and experimental side) must be
focused on a better understanding of the detailed dynamics of this channel.
Some models find an overall agreement with the experimental
total non--mesonic rates, but for the partial rates, neutron-- and proton--induced,
there are large discrepancies. Only the calculations of 
Refs.~\cite{It98,Ok99,Pa01,Os01,Ju01} obtained improved $\Gamma_n/\Gamma_p$ ratios
as well as realistic total rates. 
Recent calculations showed the importance of both the $K$--meson--exchange 
and the direct quark mechanism \cite{Ok99,Os01,Pa01} for a considerable improvement 
of $\Gamma_n/\Gamma_p$. On the other hand, the mesonic widths are 
well explained by the proposed models.

\newpage
\section{Models for calculation}
\label{model}
\subsection{Introduction}
\label{model1}
In this section we present the frameworks utilized in the literature
for the formal derivation of $\Lambda$ decay rates in nuclei. 
In subsections \ref{m-gen} and \ref{nm-gen} we discuss the general features 
of the approach used for direct finite nucleus calculations. 
It is usually called Wave Function Method
(WFM) and it has been employed by large part of the authors 
\cite{Os93,Mo94,Du96,Pa97,Ok99,Pa01}.
This method makes use of shell model nuclear and hypernuclear wave functions 
(both at hadronic and quark level) as well as pion wave functions 
generated by pion--nucleus optical potentials. In subsection \ref{pm}
the Polarization Propagator Method (PPM), applied for the first time
to hypernuclear decay in Ref.~\cite{Os85} and subsequently in 
Refs.~\cite{Al91,Ra95,Zh99,Al99,Al99a,Os01}, is summarized.  We shall see how the
decay widths can be evaluated, in nuclear matter, by means of a 
many--body description of the hyperon self--energy. The
Local Density Approximation (LDA) allows then one to implement the calculation
in finite nuclei. 
Finally a microscopic approach, based again on the PPM, is presented in 
subsection \ref{func}: here the Feynman diagrams contributing to the
$\Lambda$ self--energy are classified by means of a functional integral approach,
according to the prescriptions of the so--called bosonic loop expansion.

The numerical results of the literature obtained with WFM 
and PPM calculations have already been discussed in the previous section. 
Those obtained by the authors of the present review by
applying the formalism of subsections \ref{pm}, \ref{func}
are the subject of subsections \ref{pheres}, \ref{micres}. 

\subsection{Wave Function Method: mesonic decay}
\label{m-gen}

The weak effective Hamiltonian for the ${\Lambda}\to \pi N$ decay can
be parameterized in the form:
\begin{equation}
\label{lagran}
{\mathcal H}^W_{{\Lambda}\pi N}=iG m_{\pi}^2\overline{\psi}_N(A+B\gamma_5)
{\vec \tau} \cdot {\vec \phi}_{\pi}{\psi}_{\Lambda} ,
\end{equation}
where the values of the weak coupling constants $G= 2.211\cdot 10^{-7}/m_{\pi}^2$, $A=1.06$ and
$B=-7.10$ are fixed on the free ${\Lambda}$ decay. The constants $A$ and $B$
determine the strengths of the parity
violating and parity conserving ${\Lambda}\rightarrow \pi N$ amplitudes, respectively.
In order to enforce the $\Delta I=1/2$ rule (which fixes
$\Gamma^{\rm free}_{\pi^-}/\Gamma^{\rm free}_{\pi^0}=2$), in Eq.~(\ref{lagran})
the hyperon is assumed to be an isospin spurion with $I=1/2$, $I_z=-1/2$.

In the non--relativistic approximation, the free $\Lambda$ decay width 
$\Gamma^{\rm free}_{\Lambda}=\Gamma^{\rm free}_{\pi^-}+\Gamma^{\rm free}_{\pi^0}$
is given by:
\begin{equation}
\Gamma^{\rm free}_{\alpha}=c_{\alpha}(G m^2_{\pi})^2\int
\frac{d\vec q}{(2\pi)^3\,2\omega({\vec q})}\,2\pi\,
\delta[m_{\Lambda}-\omega({\vec q})-E_N] \left(S^2+\frac{P^2}{m^2_{\pi}}{\vec q}^2\right) ,
\nonumber
\end{equation}
where $c_{\alpha}=1$ for $\Gamma_{\pi^0}$ and $c_{\alpha}=2$  for $\Gamma_{\pi^-}$
(expressing the $\Delta I=1/2$ rule),
$S=A$, $P=m_{\pi}B/(2m_N)$, whereas $E_N$ and $\omega(\vec q)$ 
are the total energies of nucleon and pion, respectively.
One then easily finds the well known result:
\begin{equation}
\Gamma^{\rm free}_{\alpha}=c_{\alpha}(G m^2_{\pi})^2\frac{1}{2\pi}\frac{m_Nq_{\rm c.m.}}{m_{\Lambda}}
\left(S^2+\frac{P^2}{m^2_{\pi}}q^2_{\rm c.m.}\right) ,\nonumber 
\end{equation}
which reproduces the observed rates. In the previous equation,
$q_{\rm c.m.}\simeq 100$ MeV is the pion momentum in the center--of--mass frame.

In a finite nucleus approach, the mesonic width
$\Gamma_{\rm M}=\Gamma_{\pi^-}+\Gamma_{\pi^0}$ is calculable by means
of the following formula:
\begin{eqnarray}
\hspace{-0.6cm} \Gamma_{\alpha}&=&c_{\alpha}(G m^2_{\pi})^2\sum_{N\non F}\int
\frac{d\vec q}{(2\pi)^3\,2\omega({\vec q})}\,2\pi\,
\delta[E_{\Lambda}-\omega({\vec q})-E_N] \nonumber \\
&&\times \left\{S^2\left|\int d{\vec r} \phi_{\Lambda}(\vec r)
\phi_{\pi}({\vec q}, {\vec r})\phi^*_N(\vec r)\right|^2
+\frac{P^2}{m^2_{\pi}}
\left|\int d{\vec r} \phi_{\Lambda}(\vec r){\vec \triangledown}
\phi_{\pi}({\vec q}, {\vec r}) \phi^*_N(\vec r)\right|^2 \right\} , \nonumber
\end{eqnarray}
where the sum runs over non--occupied nucleonic states, and $E_{\Lambda}$ is the
hyperon total energy. The $\Lambda$ and nucleon wave functions $\phi_{\Lambda}$ and $\phi_N$
are obtainable within a shell model. The pion wave function $\phi_{\pi}$ 
corresponds to an outgoing wave, solution of the Klein--Gordon equation with proper
pion--nucleus optical potential $V_{\rm opt}$:
\begin{equation}
\left\{{\vec \triangledown}^2-m^2_{\pi}-2\omega V_{\rm opt}(\vec r)
+\left[\omega-V_C(\vec r)\right]^2\right\}
\phi_{\pi}({\vec q}, {\vec r})=0 , \nonumber
\end{equation}
where $V_C(\vec r)$ is the nuclear Coulomb potential and the energy eigenvalue
$\omega$ depends on $\vec q$.

Different calculations
\cite{It88,Os93,Mo94} have shown how strongly the mesonic decay is sensitive to
the pion--nucleus optical potential, which can be parameterized in terms of the
nuclear density, as discussed in Refs.~\cite{It88,Mo94}, or evaluated microscopically,
as in Ref.~\cite{Os93}.

\subsection{Wave Function Method: non--mesonic decay}
\label{nm-gen}
Within the meson--exchange--mechanism, the weak transition $\Lambda N\to NN$
is assumed to proceed via the mediation of virtual mesons of the pseudoscalar
($\pi$, $\eta$ and $K$) and vector ($\rho$, $\omega$ and $K^*$) octets
\cite{Du96,Pa97,Ok99,Pa01} (see Fig~\ref{nm12}).
Two--pion--exchange has been considered in the literature as well 
\cite{Ba88,Sh94,Sh95,It98}.

The fundamental ingredients for the calculation of the $\Lambda N\to NN$ transition
within a OME model are the weak and strong hadronic vertices.
The ${\Lambda}\pi N$ weak Hamiltonian is given in Eq.~(\ref{lagran}).
For the strong $NN\pi$ Hamiltonian one has the usual pseudoscalar coupling:
\begin{equation}
{\mathcal H}^S_{NN\pi}=ig_{NN\pi}\overline{\psi}_N\gamma_5
{\vec \tau} \cdot {\vec \phi}_{\pi}{\psi}_N , \nonumber
\end{equation}
$g_{NN\pi}$ being the strong coupling constant for the $NN\pi$ vertex.
In momentum space, the non--relativistic transition potential in the OPE approximation
is then:
\begin{equation}
V_{\pi}({\vec q})=-G m_{\pi}^2\frac{g_{NN\pi}}{2m_N}\left(
A+\frac{B}{2\bar{m}}{\vec \sigma_1} \cdot {\vec q}\right)
\frac{{\vec \sigma_2} \cdot {\vec q}}{{\vec q}^2+m_{\pi}^2} 
{\vec \tau_1} \cdot {\vec \tau_2}, \nonumber
\end{equation}
where $\bar{m}=(m_{\Lambda}+m_N)/2$ and
${\vec q}$ is the momentum of the exchanged pion 
(directed towards the strong vertex), whose static free propagator is
$-({\vec q}^2+m^2_{\pi})^{-1}$. One can ignore relativistic effects and
use for calculations the above non--relativistic potential~\cite{Os98}.

Given the large momentum ($\simeq 420$ MeV) exchanged in the $\Lambda N\to NN$
transition, the OPE mechanism describes the long range part of the interaction,
and more massive mesons are expected to contribute at shorter distances.
A difficulty appears when one wants to
include other mesons in the exchange potential. In fact, for mesons $m$ other than the
pion, the weak and strong vertices $\Lambda Nm$ and $NNm$ are experimentally 
unknown; moreover, their theoretical evaluation resulted quite model--dependent,
as explained in the previous section. For example, if
one includes in the calculation the contribution of the $\rho$--meson,
the weak $\Lambda N\rho$ and strong $NN\rho$ Hamiltonians:
\begin{eqnarray} 
{\mathcal H}^W_{\Lambda N\rho}&=&G m_{\pi}^2\overline{\psi}_N\left(
\alpha \gamma^{\mu}-i\beta \frac{\sigma^{\mu \nu}q_{\nu}}{2\bar{m}}
+\epsilon \gamma^{\mu} \gamma_5\right){\vec \tau} \cdot {\vec \rho}_{\mu}
\psi_{\Lambda} , \nonumber \\
{\mathcal H}^S_{NN\rho}&=&\overline{\psi}_N\left(
g^V_{NN\rho}\gamma^{\mu}+i\frac{g^T_{NN\rho}}{2\bar{m}}\sigma^{\mu \nu}q_{\nu}\right)
{\vec \tau} \cdot {\vec \rho}_{\mu} \psi_N , \nonumber
\end{eqnarray}
are needed \cite{Pa97}. 
They give the following $\rho$--meson transition potential:
\begin{eqnarray}
V_{\rho}({\vec q})&=&G m_{\pi}^2\left[g^V_{NN\rho}\alpha -
\frac{(\alpha+\beta)(g^V_{NN\rho}+g^T_{NN\rho})}{4m_n m}
({\vec \sigma}_1 \times {\vec q})\cdot ({\vec \sigma}_2 \times {\vec q}) \right. \nonumber \\
&&\left. +i\frac{\epsilon (g^V_{NN\rho}+g^T_{NN\rho})}{2m_m}({\vec \sigma}_1 \times {\vec \sigma}_2)
\cdot {\vec q}\right]
\frac{{\vec \tau}_1 \cdot {\vec \tau}_2}{{\vec q}^2+m^2_{\rho}}\,, 
\nonumber
\end{eqnarray}
where the weak coupling constants $\alpha$, $\beta$ and $\epsilon$ 
must be evaluated theoretically.

The potential for a OME calculation accounting for the exchange of pseudoscalar
and vector mesons can be expressed through the following decomposition:
\begin{equation}
\label{pot-ome}
V(\vec r)=\sum_m V_m(\vec r)=
\sum_m \sum_{\alpha}V^{\alpha}_m(r)\hat{O}^{\alpha}(\hat{{\vec r}})\hat{I}_m ,
\end{equation}
where $m=\pi$, $\rho$, $K$, $K^*$, $\omega$, $\eta$; the spin operators
$\hat{O}^{\alpha}$ are (PV stands for parity--violating):  
\begin{equation} 
\hat{O}^{\alpha}(\hat{{\vec r}})=
\begin{cases}
\hat{1} & \text{central spin--independent} , \\
{\vec \sigma}_1 \cdot {\vec \sigma}_2 & \text{central spin--dependent } , \\
S_{12}(\hat{{\vec r}})=3({\vec \sigma}_1 \cdot \hat{{\vec r}})
({\vec \sigma}_2 \cdot \hat{{\vec r}})-{\vec \sigma}_1 \cdot {\vec \sigma}_2 & \text{tensor} , \\
{\vec \sigma}_2 \cdot \hat{{\vec r}} & \text{PV for pseudoscalar mesons} , \\
({\vec \sigma}_1 \times {\vec \sigma}_2)\cdot \hat{{\vec r}} &
\text{PV for vector mesons},
\end{cases} \nonumber
\end{equation}
whereas the isospin operators $\hat{I}_m$ are:
\begin{equation}
\hat{I}_m=
\begin{cases}
\hat{1} & \text{isoscalars mesons ($\eta$, $\omega$)} , \\
{\vec \tau}_1 \cdot {\vec \tau}_2 & \text{isovector mesons ($\pi$, $\rho$)} ,\\
\text{linear combination of $\hat{1}$ and ${\vec \tau}_1 \cdot {\vec \tau}_2$} &
\text{isodoublet mesons ($K$, $K^*$)} .
\end{cases} \nonumber
\end{equation}  
For details concerning the potential (\ref{pot-ome}), see Refs.~\cite{Pa97,Du96}.

Assuming the initial hypernucleus to be at rest,
the one--body induced non--mesonic decay rate can then be written as:
\begin{equation}
\label{nm-wfm}
\Gamma_1=\int \frac{d{\vec p}_1}{(2\pi)^3}\int \frac{d{\vec p}_2}{(2\pi)^3}
\,2\pi\, \delta({\rm E.C.}) \overline{{\sum}}\left|{\mathcal M}({\vec p}_1,{\vec p}_2)\right|^2 ,
\end{equation} 
where $\delta({\rm E.C.})$ stands for the energy conserving delta function:
\begin{equation}
\delta({\rm E.C.})=\delta\left(m_H-E_{\it R}-2m_N-\frac{{\vec p}^2_1}{2m_N}-
\frac{{\vec p}^2_2}{2m_N}\right) ; \nonumber
\end{equation}
moreover:
\begin{equation}
{\mathcal M}({\vec p}_1,{\vec p}_2)
\equiv \langle \Psi_R; N({\vec p}_1)N({\vec p}_2)|
\hat{T}_{\Lambda N\to NN}|\Psi_H\rangle  \nonumber
\end{equation}
is the amplitude for the transition of the initial hypernuclear state $\Psi_H$
of mass $m_H$ into a final state composed by a
residual nucleus $\Psi_R$ with energy $E_R$ and an antisymmetrized
two nucleon state $N({\vec p}_1)N({\vec p}_2)$, 
${\vec p}_1$ and ${\vec p}_2$ being the nucleon momenta. The sum $\overline{\sum}$
in Eq.~(\ref{nm-wfm}) indicates an average over the third component of the
hypernuclear total spin and a sum over the quantum numbers of the residual
system and over the spin and isospin third components of the outgoing
nucleons. Customarily, in shell model calculations
the weak--coupling scheme is used to describe the
hypernuclear wave function $\Psi_H$, the nuclear core wave function
being obtained through the technique
of fractional parentage coefficients \cite{Pa97}. The many--body transition amplitude 
${\mathcal M}({\vec p}_1,{\vec p}_2)$ is then expressed in terms of
two--body amplitudes $\langle NN|V|\Lambda N \rangle$ of the OME potential of
Eq.~(\ref{pot-ome}).

Since the $\Lambda$ decays from an orbital angular momentum $l=0$ state,
in the non--mesonic decay rate one can easily isolate 
the contributions of neutron-- and proton--induced transitions \cite{Pa97}, and 
the $\Gamma_n/\Gamma_p$ ratio can be directly evaluated. 
The $NN$ final state interactions and the $\Lambda N$ correlations
(which are absent in an independent particle shell model)
can also be implemented in the calculation \cite{Pa95,Pa97,Ok99,Pa01}.

\subsection{Polarization Propagator Method and Local Density Approximation}
\label{pm}
The ${\Lambda}$ decay in nuclear systems can be studied by using the 
Polarization Propagator Method~\cite{Os82}, which is usually employed 
within the Random Phase Approximation (RPA). 
The calculation of the widths is performed in nuclear matter and
then it is extended to finite nuclei via the LDA.
This many--body technique has been applied for the first time to hypernuclear decays in 
Ref.~\cite{Os85}. It provides a unified picture
of the different decay channels and it is equivalent to the
WFM \cite{Os94} (in the sense that it is a semiclassical
approximation of the exact quantum mechanical problem).
For the calculation of the mesonic rates the WFM is more reliable than the
PPM in LDA, this channel being rather sensitive to the 
shell structure of the hypernucleus, due to the small energies involved.
In general it is advisable to avoid the use of the LDA to describe very
light systems. On the other
hand, the propagator method in LDA offers the possibility of calculations 
over a broad range of mass numbers, while the WFM is hardly exploitable for medium
and heavy hypernuclei. 

\subsubsection{Nuclear matter}

To calculate the ${\Lambda}$ width one needs the 
imaginary part of the ${\Lambda}$ self--energy:
\begin{equation}
\label{Gamma}
{\Gamma}_{\Lambda}=-2\;{\rm Im}\,{\Sigma}_{\Lambda} .
\end{equation}
By using the customary Feynman rules, from Fig.~\ref{self}
the ${\Lambda}$ self--energy in the non--relativistic limit is obtained as:
\begin{figure}
\begin{center}
\mbox{\epsfig{file=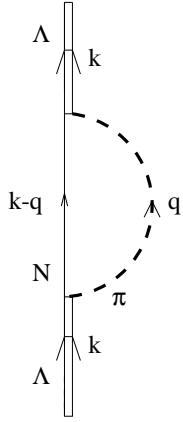,width=.15\textwidth}}
\vskip 2mm
\caption{${\Lambda}$ self energy in nuclear matter.}
\label{self}
\end{center}
\end{figure}
\begin{equation}
\label{Sigma1}
{\Sigma}_{\Lambda}(k)=3i(G m_{\pi}^2)^2\int \frac{d^4q}{(2\pi)^4}
\left(S^2+\frac{P^2}{m_{\pi}^2}\vec q\,^2\right)F_{\pi}^2(q)
G_N(k-q)G_{\pi}(q) ,
\end{equation}
the factor 3 being a consequence of the $\Delta I=1/2$ rule.
The nucleon and pion propagators in nuclear matter are, respectively:
\begin{equation}
G_N(p)=\frac{{\theta}(\mid \vec p \mid-k_F)}{p_0-E_N(\vec p)-V_N+i{\epsilon}}+
\frac{{\theta}(k_F-\mid \vec p \mid)}{p_0-E_N(\vec p)-V_N-i{\epsilon}} , 
\end{equation}
and:
\begin{equation}
\label{proppion}
G_{\pi}(q)=\frac{1}{q_0^2-\vec q\,^2-m_{\pi}^2-{\Sigma}_{\pi}^*(q)} .
\end{equation}
The above form of the non--relativistic nucleon propagator refers to a
non--interacting Fermi system but includes corrections due to Pauli principle
and an average binding. Other effects of the nucleon renormalization 
in the medium are found to be negligible in the processes we are treating \cite{Os91}.
In the previous equations, $p=(p_0,\vec p)$ and $q=(q_0,\vec q)$ denote 
four--vectors, $k_F$ is the Fermi momentum, $E_N$ is the nucleon total free
energy, $V_N$  the nucleon binding energy (which is density--dependent), 
and ${\Sigma}_{\pi}^*$ is the pion proper self--energy in nuclear matter. 
Moreover, in Eq.~(\ref{Sigma1}) we have included a monopole form factor describing
the hadronic structure of the $\pi\Lambda N $ vertex:
\begin{equation}
F_{\pi}(q)=\frac {{\Lambda}_{\pi}^2-m_{\pi}^2}{{\Lambda}_{\pi}^2-q_0^2+\vec q\,^2} , \nonumber
\end{equation}
which is normalized to unity for on--shell pions.
Since at present there is no reason to introduce a different form factor 
in the weak vertex, one utilizes here the same expression usually employed for 
the $\pi NN$ strong vertex. For instance, in the pole dominance description
of the parity--conserving weak vertex, a form factor identical to the strong
one is assigned. From empirical studies on the $NN$ interaction it follows that
$\Lambda_{\pi NN}\simeq 1.3$ GeV, and the same value can be used for $\pi\Lambda N $. 
We note here that the parity--conserving
term ($l=1$ term) in Eq.~(\ref{Sigma1}) contributes only about 12\% of the total
free decay width. However, the $P$--wave interaction becomes dominant 
in the nuclear non--mesonic decay, because of the larger exchanged momenta. 

\begin{figure}
\begin{center}
\mbox{\epsfig{file=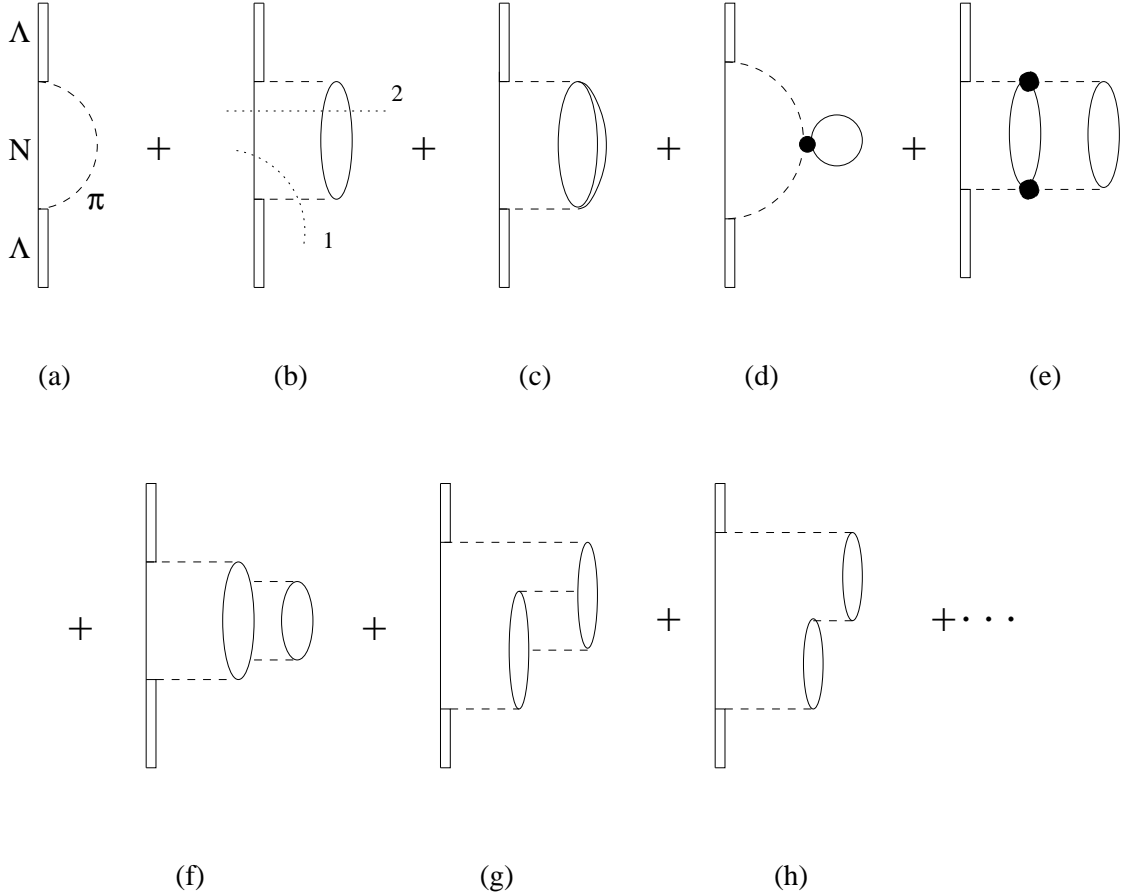,width=0.97\textwidth}}
\vskip 2mm
\caption{Lowest order terms for the ${\Lambda}$ self--energy in 
nuclear matter. The meaning of the various diagrams is explained in the text.}
\label{self1}
\end{center}
\end{figure}

In Fig.~\ref{self1} we show the lowest order Feynman diagrams for the ${\Lambda}$ 
self--energy in nuclear matter. Diagram (a) represents the bare
self--energy term, including the effects of the Pauli principle and of
binding on the intermediate nucleon.
In (b) and (c) the pion couples to a particle--hole ({\sl p--h}) and a 
{\sl ${\Delta}$--h} pair, respectively. Diagram (d) is an insertion of $S$--wave
pion self--energy at lowest order. In diagram (e) we show a {\sl 2p--2h} excitation coupled to the
pion through $S$--wave $\pi N$ interactions. 
Other {\sl 2p--2h} excitations, coupled in $P$--wave, are 
shown in (f) and (g), while (h) is a RPA iteration of diagram (b). 

In Eq.~(\ref{Sigma1}) there are two different sources of imaginary part. The analytical
structure of the integrand allows the integration over $q_0$ \cite{Os85}. After 
performing this integration, an imaginary part 
is obtained from the (renormalized) pion--nucleon
pole and physically corresponds to the mesonic decay of the hyperon. Moreover, the
pion proper self--energy $\Sigma^*_{\pi}(q)$ has an imaginary part itself
for $(q_0, \vec q)$ values which correspond to the excitation of {\sl p--h},
{$\Delta$--h}, {\sl 2p--2h}, etc states on the mass shell. By expanding the pion
propagator $G_{\pi}(q)$ as in Fig.~\ref{self1} and integrating Eq.~(\ref{Sigma1}) 
over $q_0$, the nuclear matter $\Lambda$ decay width of Eq.~(\ref{Gamma})
becomes \cite{Os85}:
\begin{eqnarray}
\label{Sigma2}
{\Gamma}_{\Lambda}(\vec k,\rho)&=&-6(G m_{\pi}^2)^2\int \frac{d\vec q}
{(2\pi)^3}{\theta}(\mid \vec k- \vec q \mid -k_F)
{\theta}(k_0-E_N(\vec k-\vec q)-V_N) \nonumber \\
& & \times {\rm Im}\left[{\alpha}(q)\right]_{q_0=k_0-E_N(\vec k-\vec q)-V_N} ,
\end{eqnarray}
where:
\begin{eqnarray}
\label{Alpha}
{\alpha}(q)&=&\left(S^2+\frac{P^2}{m_{\pi}^2}\vec q\,^2\right)F_{\pi}^2(q)
G_{\pi}^0(q)+\frac{\tilde{S}^2(q)U_L(q)}{1-V_L(q)U_L(q)} \nonumber \\
& & +\frac{\tilde{P}_L^2(q)U_L(q)}{1-V_L(q)U_L(q)}+
2\frac{\tilde{P}_T^2(q)U_T(q)}{1-V_T(q)U_T(q)} .
\end{eqnarray}
In Eq.~(\ref{Sigma2}) the first ${\theta}$ function forbids
intermediate nucleon momenta smaller than the 
Fermi momentum (see Fig.~\ref{self}), while
the second one requires the pion energy $q_0$ to be 
positive. Moreover, the ${\Lambda}$ energy, $k_0=E_{\Lambda}(\vec k)+V_{\Lambda}$,
contains a phenomenological binding term. 
With the exception of diagram (a), the pion lines of Fig.~\ref{self1} 
have been replaced, in Eq.~(\ref{Alpha}), by the effective interactions 
$\tilde{S}$, $\tilde{P}_L$, $\tilde{P}_T$,$V_L$, $V_T$ 
($L$ and $T$ stand for spin--longitudinal and spin--transverse, respectively),
which include $\pi$-- and $\rho$--exchange
modulated by the effect of short range repulsive correlations.
The potentials $V_L$ and $V_T$ represent the (strong) {\sl p--h} interaction and
include a Landau parameter $g^{\prime}$,
which accounts for the short range repulsion, 
while $\tilde{S}$, $\tilde{P}_L$ and
$\tilde{P}_T$ correspond to the lines connecting weak and strong 
hadronic vertices and contain another Landau parameter, $g^{\prime}_{\Lambda}$,
which is related to the strong $\Lambda N$
short range correlations. For details on these potentials see
appendix \ref{app1}. Furthermore, in Eq.~(\ref{Alpha}):
\begin{equation}
G_{\pi}^0(q)=\frac{1}{q_0^2-\vec q\,^2-m_{\pi}^2} , \nonumber
\end{equation}
is the free pion propagator,
while $U_L(q)$ and $U_T(q)$ contain the Lindhard functions for {\sl p--h} and 
{\sl ${\Delta}$-h} excitations \cite{Wa71}
and also account for the irreducible {\sl 2p--2h} polarization propagator:
\begin{equation}
\label{propU}
U_{L,T}(q)=U^{ph}(q)+U^{\Delta h}(q)+U^{2p2h}_{L,T}(q) .
\end{equation}
They appear in Eq.~(\ref{Alpha}) within the standard RPA expression.
The decay width (\ref{Sigma2}) depends both explicitly and through $U_{L,T}(q)$
on the nuclear matter density $\rho=2k_F^3/3{\pi}^2$.
The Lindhard function for the {\sl p--h} excitation is defined by
\cite{Wa71}:
\begin{equation}
U^{ph}(q)=-4i\int \frac{d^4p}{(2\pi)^3}G_N^0(p) G_N^0(p+q) , \nonumber
\end{equation}
where:
\begin{equation}
G_N^0(p)=\frac{{\theta}(\mid \vec p \mid-k_F)}{p_0-T_N(\vec p)+i{\epsilon}}+
\frac{{\theta}(k_F-\mid \vec p \mid)}{p_0-T_N(\vec p)-i{\epsilon}} , \nonumber
\end{equation}
is the free nucleon propagator. In the above equation, $T_N$ is the
nucleon kinetic energy. The Lindhard function 
$U^{\Delta h}$ is obtained from $U^{ph}$ by replacing the {\sl p--h}
propagators with the {\sl $\Delta$--h} ones. Analytical 
expressions of $U^{ph}$ and $U^{\Delta h}$ are given in Refs.~\cite{Wa71,Os90}.

For the evaluation of $U^{2p2h}_{L,T}$ we discuss two different
approaches. In Refs.~\cite{Ra95,Al99} a phenomenological parameterization was 
adopted: as we shall see in paragraph \ref{p2p2hp}, this consists in relating 
$U^{2p2h}_{L,T}$ to the available phase space for on--shell {\sl 2p--2h} 
excitations in order
to extrapolate for off--mass shell pions the experimental data of
$P$--wave absorption of real pions in pionic atoms. 
In an alternative approach~\cite{Al99a}, as we shall discuss
in detail in subsection \ref{func},  $U^{2p2h}_{L,T}$ is
evaluated microscopically, starting from a classification of the relevant 
Feynman diagrams according to the so--called bosonic loop expansion, 
which will be obtained by means of a functional approach. 

In the spin--longitudinal channel, $U(q)$ is related to the $P$--wave
pion proper self--energy through:
\begin{equation}
{\Sigma}_{\pi}^{(P)\,*}(q)=\frac{\displaystyle \vec q\,^2
\frac{f_{\pi}^2}{m_{\pi}^2} F_{\pi}^2(q)U_L(q)}
{1-\displaystyle \frac{f_{\pi}^2}{m_{\pi}^2}g_L(q)U_L(q)} , \nonumber
\end{equation}
where the Landau function $g_L(q)$ is given in appendix \ref{app1}.
The full pion (proper) self--energy:
\begin{equation}
{\Sigma}_{\pi}^*(q)={\Sigma}_{\pi}^{(S)\,*}(q)+{\Sigma}_{\pi}^{(P)\,*}(q) , \nonumber
\end{equation}
also contains an $S$--wave term, which, by using
the parameterization of Ref.~\cite{Se83}, can be written as: 
\begin{equation}
\Sigma_{\pi}^{(S)\,*}(q)=-4\pi \left(1+\frac{m_{\pi}}{m_N}\right)b_0\rho , \nonumber
\end{equation}
with $b_0=-0.0285/m_{\pi}$. The function ${\Sigma}_{\pi}^{(S)\,*}$ is real
(constant and positive), 
therefore it contributes only to the mesonic decay 
[diagram (d) in Fig.~\ref{self1} is the relative lowest order]. 
On the contrary, the $P$--wave self--energy is complex and attractive: 
${\rm Re}\;{\Sigma}_{\pi}^{(P)\,*}(q)<0$.

The propagator method provides a unified picture of the decay widths. 
A non--vanishing imaginary part in a
self--energy diagram requires placing simultaneously on--shell the particles of the
considered intermediate state. For instance, diagram (b) in Fig.~\ref{self1} has two sources
of imaginary part.  One comes from cut 1, where the nucleon and the pion are placed
on--shell. This term contributes to the mesonic channel: the final pion 
eventually interacts with the
medium through a {\sl p--h} excitation and then escapes from the nucleus. 
Diagram (b) and further iterations lead to a 
renormalization of the pion in the medium which may increase the mesonic rate 
even by one or two orders of magnitude in heavy nuclei \cite{Os85,Os93,Mo94}. 
The cut 2 in Fig.~\ref{self1}(b) places a nucleon and a {\sl p--h} pair on shell, 
so it is the lowest order contribution to the physical process 
${\Lambda}N\rightarrow NN$; analogous considerations apply to all the considered 
diagrams. 

In order to evaluate the various contributions to the width stemming 
from Eq.~(\ref{Sigma2}), it is convenient to consider all the intervening 
free meson propagators as real. Then the
imaginary part of (\ref{Alpha}) will develop the following contributions:
\begin{equation}
\label{pres}
{\rm Im}\, \frac{U_{L,T}(q)}{1-V_{L,T}(q)U_{L,T}(q)}=
\frac{{\rm Im}\,U^{ph}(q)+{\rm Im}\,U^{\Delta h}(q)+{\rm Im}\,U^{2p2h}_{L,T}(q)}
{\mid 1-V_{L,T}(q)U_{L,T}(q)\mid ^2} .
\end{equation}
The three terms in the numerator of Eq.~(\ref{pres}) can be interpreted as different
decay mechanisms of the hypernucleus. The term proportional to ${\rm Im}\,U^{ph}$
provides the one--nucleon induced non--mesonic rate, $\Gamma_1$. There is no overlap 
between ${\rm Im}\,U^{ph}(q)$ and the pole $q_0=\omega(\vec q)$ in the (dressed)
pion propagator $G_{\pi}(q)$: thus the separation of the mesonic and one--body
stimulated non--mesonic channels is unambiguous. 

Further, ${\rm Im}\,U^{\Delta h}$
accounts for the $\Delta \to \pi N$ decay width, thus representing a contribution 
to the mesonic decay. 

The third contribution of Eq.~(\ref{pres}), 
proportional to ${\rm Im}\,U^{2p2h}_{L,T}$, intervenes in a 
wide kinematical range, in which the above mentioned cuts put on the mass shell
not only the {\sl 2p--2h} lines, but possibly also the pionic line.
Indeed the renormalized pion pole in Eq.~(\ref{proppion}) is given by the
dispersion relation:
\begin{equation}
{\omega}^2(\vec q)-\vec q\,^2-m_{\pi}^2-{\rm Re}\,{\Sigma}_{\pi}^*
\left[{\omega}(\vec q),\vec q\,\right]=0 , \nonumber
\end{equation}
with the constraint:
\begin{equation}
\omega({\vec q})=k_0-E_N(\vec k -\vec q)-V_N . \nonumber
\end{equation}
At the pion pole, ${\rm Im}\,U^{2p2h}_{L,T}\neq 0$, thus the two--body induced
non--mesonic width, $\Gamma_2$, cannot be disentangled from the mesonic
width, $\Gamma_{\rm M}$. In other words, part of the decay rate calculated 
from ${\rm Im}\,U^{2p2h}_{L,T}$ is due to the excitations of the renormalized pion and
gives in fact $\Gamma_{\rm M}$, with the exception of the mesonic contribution
originating from ${\rm Im}\,U^{\Delta h}$, which is, however, only a
small fraction of $\Gamma_{\rm M}$. In order to separate $\Gamma_{\rm M}$ from 
$\Gamma_2$, in the numerical calculation it is convenient to evaluate the
mesonic width by adopting the following prescription. We start from
Eq.~(\ref{Sigma2}), setting:
\begin{equation} 
\label{alphaprop} 
{\alpha}(q)={\alpha}_M(q)\equiv
\left(S^2+\frac{P^2}{m_{\pi}^2}\vec q\,^2\right)F^2_{\pi}(q)G_{\pi}(q) ,
\end{equation} 
and omitting ${\rm Im}\,{\Sigma}_{\pi}^*$ in $G_{\pi}$ (which corresponds to 
set ${\rm Im}\, U^{ph}={\rm Im}\, U^{\Delta h}={\rm Im}\, U^{2p2h}_{L,T}=0$).
Then ${\rm Im}\,{\alpha}_M(q)$ only accounts for the (real) contribution of the
pion pole:
\begin{equation}
{\rm Im}\,G_{\pi}(q)= -\pi \delta \left[q_0^2-\vec q\,^2-m_{\pi}^2- 
{\rm Re}\,{\Sigma}_{\pi}^*(q)\right] . \nonumber 
\end{equation} 
We notice that the compact relation (\ref{alphaprop}) between ${\alpha}(q)$ and
the pion propagator is valid only for the calculation of the mesonic decay
mode. In fact in this case the following substitutions must be performed
in Eq.~(\ref{Alpha}) (see also appendix \ref{app1}):
\begin{equation} 
\begin{array}{l l l} 
\tilde{S}(p)&\rightarrow & \displaystyle\frac{f_{\pi}}{m_{\pi}}SF^2_{\pi}(q) 
G^0_{\pi}(q)\mid\vec q\mid , \nonumber \\ 
\tilde{P}_L(p)&\rightarrow & \displaystyle\frac{f_{\pi}}{m_{\pi}}\frac{P}{m_{\pi}}\vec q\,^2 
F^2_{\pi}(q)G^0_{\pi}(q) , \nonumber \\ 
\tilde{P}_T(p)&\rightarrow & 0 , \nonumber 
\end{array} 
\end{equation}
and hence the various terms in $\alpha (q)$ can be combined to give the
expression (\ref{alphaprop}). Obviously this implies that no correlation
other than the pion is active between the $\Lambda$ and the strong vertices
($g^{\prime}_{\Lambda}=0$). 

Once the mesonic decay rate is known, one can calculate the three--body
non--mesonic rate by subtracting $\Gamma_{\rm M}$ and $\Gamma_1$ from the
total rate $\Gamma_{\rm T}$, which one gets via the full expression for
$\alpha(q)$ [Eq.~(\ref{Alpha})].

\subsubsection{Finite nuclei}

Using the Polarization Propagator approach, the
decay widths in finite nuclei are obtained from the ones evaluated
in nuclear matter via the LDA: 
the Fermi momentum is made $r$--dependent (namely 
a local Fermi sea of nucleons is introduced) and related to the 
nuclear density by the same relation which holds in nuclear matter: 
\begin{equation} 
\label{local} 
k_F(\vec r)=\left\{\frac{3}{2}{\pi}^2\rho 
(\vec r)\right\}^{1/3} . 
\end{equation} 
Moreover, the nucleon binding potential $V_N$ also becomes $r$--dependent in 
LDA. In Thomas--Fermi approximation one assumes: 
\begin{equation} 
\epsilon_F(\vec r)+V_N(\vec r)\equiv \frac{k_F^2(\vec r)}{2m_N}+V_N(\vec r)=0 . \nonumber
\end{equation} 
For the ${\Lambda}$ binding energy, $V_{\Lambda}$, the experimental values
\cite{Pi91,Ha96} can be used. 
With these prescriptions one can then evaluate the decay width in finite nuclei 
by using the semiclassical approximation, through the relation: 
\begin{equation} 
\label{local1} 
{\Gamma}_{\Lambda}(\vec k)=\int d\vec r \,| {\psi}_{\Lambda}(\vec r)| ^2 
{\Gamma} _{\Lambda}\left[\vec k,\rho (\vec r)\right] , 
\end{equation} 
where ${\psi}_{\Lambda}$ is the appropriate ${\Lambda}$ wave function and 
${\Gamma} _{\Lambda}\left[\vec k,\rho (\vec r)\right]$ is given by 
Eqs.~(\ref{Sigma2}), (\ref{Alpha}). This decay rate can be regarded as the 
$\vec k$--component of the ${\Lambda}$ decay rate in the nucleus with density 
$\rho(\vec r)$. It can be used to estimate the decay rates 
by averaging over the ${\Lambda}$ momentum distribution 
$|\tilde{\psi}_{\Lambda}(\vec k)|^2$. One then obtains the
following total width: 
\begin{equation} 
\label{local2} 
{\Gamma}_{\Lambda}=\int d\vec k \,| \tilde{\psi}_{\Lambda}(\vec k)|^2{\Gamma}_{\Lambda}
(\vec k) ,
\end{equation} 
which can be compared with the experimental results.

\subsubsection{Phenomenological {\sl 2p--2h} propagator}
\label{p2p2hp}

Coming to the phenomenological evaluation of the {\sl 2p--2h}
contributions in the $\Lambda$ self--energy, we recall that the authors of 
Ref.~\cite{Ra95} employed the following equation for the imaginary part of
$U^{2p2h}_{L,T}$:
\begin{equation}
\label{pheim}
{\rm Im}\, U^{2p2h}_{L,T}(q_0,\vec q;\rho)=\frac{P(q_0,\vec q;\rho)}{P(m_{\pi},\vec 0;\rho_{\rm eff})}
{\rm Im}\, U^{2p2h}_{L,T}(m_{\pi}, \vec 0; \rho_{\rm eff}) ,
\end{equation}
where $\rho_{\rm eff}=0.75\rho$. By neglecting the energy and 
momentum dependence of the {\sl p--h} interaction,
the phase space available for on--shell {\sl 2p--2h} excitations
[calculated, for simplicity, from
diagram \ref{self1}(e)] at energy--momentum $(q_0,\vec q)$ and density $\rho$
turns out to be:
\begin{eqnarray}
P(q_0,\vec q;\rho)&\propto &\int \frac{d^4k}{(2\pi)^4}\,{\rm Im}\, U^{ph}
\left(\frac{q}{2}+k;\rho\right){\rm Im}\,U^{ph}\left(\frac{q}{2}-k;\rho\right) \nonumber \\
& &\times \theta \left(\frac{q_0}{2}+k_0\right)
\theta \left(\frac{q_0}{2}-k_0\right) . \nonumber
\end{eqnarray}

In the region of $(q_0,\vec q)$ where the {\sl p--h} and {\sl $\Delta$--h} excitations are
off--shell, the relation between $U^{2p2h}_L$ and the $P$--wave pion--nucleus 
optical potential $V_{\rm opt}$ is given by:
\begin{equation}
\label{phe1}
\frac{\displaystyle \vec q\,^2\frac{f^2_{\pi}}{m^2_{\pi}}F^2_{\pi}(q)U^{2p2h}_L(q)}
{\displaystyle 1-\frac{f^2_{\pi}}{m^2_{\pi}}g_L(q)U_L(q)}=2q_0V_{\rm opt}(q) ;
\end{equation}
at the pion threshold $V_{\rm opt}$ is usually parameterized as:
\begin{equation}
\label{phe2}
2q_0V_{\rm opt}(q_0\simeq m_{\pi},\vec q\simeq \vec 0; \rho)
=-4\pi \vec q\,^2 \rho^2 C_0 ,
\end{equation}
where $C_0$ is a complex number which can be extracted from experimental
data on pionic atoms. By combining Eqs.~(\ref{phe1}) and (\ref{phe2})
it is possible to parameterize the proper {\sl 2p--2h} excitations 
in the spin--longitudinal channel through Eq.~(\ref{pheim}), by setting:
\begin{equation}
\label{phebare}
\vec q\,^2\frac{f^2_{\pi}}{m^2_{\pi}}F^2_{\pi}(q_0\simeq m_{\pi},\vec q \simeq \vec 0)
U^{2p2h}_L(q_0\simeq m_{\pi},\vec q \simeq \vec 0;\rho)=-4\pi \vec q\,^2 \rho^2 C^*_0 .
\end{equation}
The value of $C^*_0$ also depends 
on the correlation function $g_L$. From the analysis of pionic atoms data 
made in Ref.~\cite{Ga92} and taking $g'\equiv g_L(0)=0.615$, one obtains:
\begin{equation}
C^*_0=(0.105+i0.096)/m^6_{\pi} . \nonumber
\end{equation}

The spin--transverse component of $U^{2p2h}$ is assumed to be 
equal to the spin--longitudinal one, $U^{2p2h}_T=U^{2p2h}_L$, and
the real parts of $U^{2p2h}_{L}$ and $U^{2p2h}_{T}$ 
are considered constant [by using Eq.~(\ref{phebare})]
because they are not expected to be too sensitive to variations of $q_0$ and $\vec q$.
The assumption $U^{2p2h}_{T}=U^{2p2h}_{L}$
is not {\it a priori} a good approximation, but it is the only
one which can be employed in the present phenomenological description. Yet,
the differences between $U^{2p2h}_L$ and $U^{2p2h}_T$ (which will be discussed
in subsection \ref{micres}; see, in particular, figure \ref{u_obl}) can only mildly
change the partial decay widths: in fact, $U^{2p2h}_{L,T}$ are summed to $U^{ph}$,
which gives the dominant contribution. Moreover, for $U^{2p2h}_L=U^{2p2h}_T$
the transverse contribution to $\Gamma_2$ [fourth term in the right hand side 
of Eq.~(\ref{Alpha})]  is only about 16\% of $\Gamma_2$ (namely
$2\div 3$\% of the total width) in medium--heavy hypernuclei.

\subsection{Functional approach to the $\Lambda$ self--energy}
\label{func}

In alternative to the above mentioned phenomenological approach for the 
two--body induced  decay, we discuss here a microscopic approach. 
In particular, we will show how the most relevant Feynman diagrams  
for the calculation of the $\Lambda$ self--energy 
can be obtained in the framework of a functional method: following Ref.~\cite{Al99a}
we will shortly derive a classification of the diagrams according to the
prescription of the so--called bosonic loop expansion (BLE). 

The baryon--baryon strong interactions cannot be treated with the 
standard perturbative method. Indeed, in the study of nuclear phenomena 
we always need to sum, up to infinite order, the series of pertinent
diagrams. For instance, one usually performs
the summation of the infinite classes of diagrams entailed by the RPA and
Dyson equations. However, in the above quoted 
schemes no prescription is given to evaluate the ``next--to--leading'' order. 

The functional techniques can provide a theoretically founded derivation of 
new classes of expansion in terms of powers of suitably chosen parameters. 
On the other hand, as we will see, the ring approximation (a subclass of RPA) 
automatically appears in this framework at the mean field level. 
This method has been extensively applied to the analysis of
 different processes in nuclear physics \cite{Ne82,Al87,Ce97}. 
Here it will be employed 
for the calculation of the $\Lambda$ decay rates in nuclear
matter, which can be expressed through the nuclear responses 
to pseudoscalar--isovector and vector--isovector fields. 
The polarization propagators obtained in this framework 
include ring--dressed meson propagators 
(which represent the mean field level of the theory)
and almost the whole spectrum of {\sl 2p--2h} excitations (expressed in terms of a 
one--loop expansion with respect to the ring--dressed meson propagators),
which are required for the evaluation of $\Gamma_2$. 
Actually, the semiclassical expansion
leads to the prescription of grouping the relevant Feynman diagrams in a consistent 
many--body description of the ``in medium'' meson self--energies: 
the general theorems and sum rules of the theory are preserved.

Let us first consider the polarization propagator in the
pionic (spin--longitudinal) channel. 
In order to exemplify, it is useful to start from 
a Lagrangian describing a system of 
nucleons interacting with pions through a pseudoscalar--isovector
coupling:
\begin{equation}
{\mathcal L}_{\pi N}=\overline \psi (i\dsla-m_N)\psi 
+\frac{1}{2}\partial_{\mu}{\vec \phi}\cdot{\partial^{\mu} {\vec \phi}}
-\frac{1}{2}m_{\pi}^2 \vec\phi\,^2-i\overline \psi \vec \Gamma\psi \cdot \vec \phi , \nonumber
\end{equation}
where $\psi$ ($\vec \phi$) is the nucleonic (pionic) field, and:
\begin{equation}
\vec\Gamma=g\gamma_5\vec\tau  \nonumber
\end{equation}
($g=2f_{\pi}m_N/m_{\pi}$)
is the spin--isospin matrix in the spin--longitudinal isovector channel. 
We remind the reader that in the calculation of the hypernuclear decay rates one also 
needs the polarization propagator in the transverse channel 
[see Eqs.~(\ref{Sigma2}) and (\ref{Alpha})]: hence, we will have to include in the
model another mesonic degree of freedom, 
the $\rho$ meson. This is relatively straightforward, since 
the semiclassical expansion is characterized by the
topology of the diagrams, so the same scheme can be easily 
applied to mesonic fields other than the pionic one.
In this subsection we present a relativistic formalism, its non--relativistic
reduction being trivial. 

Let us now introduce a {\it classical} external field $\vec\varphi$ 
with the quantum numbers
of the pion. The Lagrangian then becomes:
\begin{equation}
{\mathcal L}_{\pi N} \rightarrow {\mathcal L}_{\pi N}-i\overline \psi 
\vec\Gamma\psi \cdot \vec \varphi. \nonumber
\end{equation}
The corresponding generating functional in terms of Feynman path integrals 
has the form:
\begin{equation}
\label{Z}
Z[\vec \varphi\,]=\int {\mathcal D} \left[\overline \psi, \psi, \vec \phi \,\right]
\exp\left\{i \int dx \left[{\mathcal L}_{\pi N}(x)-i
\overline \psi(x) \vec\Gamma\psi(x) \cdot \vec \varphi(x)\right]\right\} 
\end{equation}
(here and in the following the coordinate integrals are 4--dimensional).
All the fields in the functional integrals have to be considered as classical
variables, but with the correct commuting properties
(hence the fermionic fields are Grassman variables). The physical quantities
of interest for the problem are deduced from the generating functional by means
of functional differentiations. 
In particular, by introducing a new functional $Z_c$ such that:
\begin{equation}
\label{connect}
Z[\vec\varphi\,]=\exp{\left\{iZ_c[\vec\varphi\,]\right\}} ,
\end{equation}
the spin--longitudinal, isovector polarization propagator turns out to 
be the second functional derivative of $Z_c$ with respect to the source 
$\vec\varphi$ of the pionic field:
\begin{equation}
\label{proppol}
\Pi_{ij}(x,y)=-\left[\displaystyle \frac{\delta^2 Z_c[\vec\varphi\,]}{\delta \varphi_i(x) 
\delta \varphi_j(y)}\right]_{\vec \varphi=0}  .
\end{equation}
We notice that the use of $Z_c$ instead of $Z$ in Eq.~(\ref{proppol}) amounts 
to cancel the disconnected diagrams of the corresponding perturbative expansion 
(linked cluster theorem). 
From the generating functional $Z$ one can obtain different 
approximation schemes according 
to the order in which the functional integrations are performed. 

By integrating Eq.~(\ref{Z}) over the mesonic degrees of freedom {\it first}, 
the generating functional
can be written in terms of a fermionic effective action $S^F_{\rm eff}$. 
Up to an irrelevant multiplicative 
constant:
\begin{equation}
Z[\vec\varphi\,]=\int {\mathcal D} \left[\overline \psi, \psi\right]
\exp\left\{i S^F_{\rm eff}\left[\overline \psi, \psi\right]\right\} . \nonumber
\end{equation}
The remaining integration variables are interpreted as 
physical fields and, beyond the kinetic term, 
$S^F_{\rm eff}$ describes a quadrilinear non--local, time-- or 
energy--dependent nucleon--nucleon interaction induced by the exchange of one pion:
\begin{eqnarray}
\label{SeffF}
S^F_{\rm eff}\left[\overline \psi, \psi \right]&=&\int dx\,dy\, 
\left[\overline\psi(x) \right. G^{-1}_N(x-y)\psi(y) \\
& & +\frac{1}{2}\sum_{i=1}^{3}\overline \psi(x) \Gamma_i\psi(x) G^0_{\pi}(x-y)
\left.\overline \psi(y) \Gamma_i\psi(y)\right] , \nonumber
\end{eqnarray}
where $G_N$ and $G^0_{\pi}$ are the nucleon and free pion propagators, 
respectively, which satisfy the following field equations: 
\begin{equation}
\left(i\dsla_x-m_N-i
\vec\Gamma \cdot \vec \varphi\right)G_N(x-y)=\delta(x-y) , \nonumber
\end{equation}
\begin{equation}
\left(\dal_{\hspace*{-0.08cm}x}+m^2_{\pi}\right)G^0_{\pi}(x-y)=-\delta(x-y) . \nonumber
\end{equation}
The pion propagator is diagonal in the isospin
indices: $\left(G^0_{\pi}\right)_{ij}=\delta_{ij}G^0_{\pi}$.
The effective action (\ref{SeffF}) can then be utilized in the framework
of ordinary perturbation theory and does not bring significant novelties 
with respect to the usual calculations;
furthermore, it cannot be correctly renormalized due to the absence of a
term proportional to $\vec\phi^4$, which is needed 
to cancel the divergence of the 4--points
fermion loops.
\subsubsection{The bosonic effective action}
\label{model2}
\begin{figure}
\begin{center}
\mbox{\epsfig{file=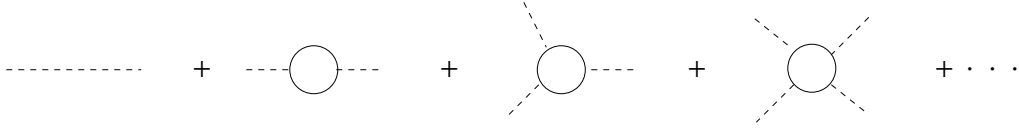,width=.9\textwidth}}
\vskip 2mm
\caption{Diagrammatic representation of the bosonic effective action 
\protect(\ref{seffb}).}
\label{vpi}
\end{center}
\end{figure}

Alternatively it is possible to eliminate, in Eq.~(\ref{Z}),
the nucleonic degrees of freedom {\it first} (without destroying 
the renormalizability of the theory \cite{Al87}). By introducing the change of
variable $\vec\phi\rightarrow \vec\phi-\vec\varphi$, Eq.~(\ref{Z}) becomes:
\begin{eqnarray}
\label{Z1}
Z\left[\vec \varphi\,\right]&=&\exp\left\{\frac{i}{2}\int dx\,dy\, \vec\varphi(x) \cdot
G^{0^{-1}}_{\pi}(x-y)\vec\varphi(y)\right\} \\ 
& & \times \int {\mathcal D} \left[\overline \psi, \psi, \vec \phi\,\right]
\exp\left\{i \int dx\,dy\, \right.\left[\overline\psi(x)G^{-1}_N(x-y)\psi(y)\right. 
\nonumber \\
& & \left. \left. +\frac{1}{2}\vec\phi(x)\cdot G^{0^{-1}}_{\pi}(x-y)
\left(\vec\phi(y)+2\vec\varphi(y)\right)\right]\right\} , \nonumber
\end{eqnarray}
where the integral over $\left[\overline\psi, \psi\right]$ is gaussian:
\begin{equation}
\int {\mathcal D} \left[\overline \psi, \psi \right] 
\exp\left\{i \int dx\,dy\, \overline\psi(x)G^{-1}_N(x-y)\psi(y)\right\}=
\left({\rm det}\,G_N \right)^{-1} . \nonumber
\end{equation}
Hence, after multiplying Eq.~(\ref{Z1}) by the unessential factor 
${\rm det}\,G^0_N$ ($G^0_N$ being the free nucleon propagator), 
which only redefines the normalization constant of the generating functional, and using the
property ${\rm det}\,X=\exp\left\{{\rm Tr}\ln X\right\}$, one obtains:
%
\begin{eqnarray}
\label{Z2}
Z[\vec \varphi\,]&=&\exp\left\{\frac{i}{2}\int dx\,dy\, \vec\varphi(x)\cdot
G^{0^{-1}}_{\pi}(x-y)\vec\varphi(y)\right\} \\
& &\quad\times
\int {\mathcal D} \left[\vec \phi \,\right]
\exp\left\{i S^B_{\rm eff}\left[\vec\phi \,\right]\right\} , \nonumber
\end{eqnarray}
with:
\begin{equation}
\label{seffb}
S^B_{\rm eff}\left[\vec\phi\,\right]=\int dx\,dy\, 
\left\{\frac{1}{2}\vec\phi(x)\cdot G^{0^{-1}}_{\pi}(x-y)
\left[\vec\phi(y)+2\vec\varphi(y)\right]+V_{\pi}\left[\vec\phi\,\right]\right\} , 
\end{equation}
\begin{eqnarray}
\label{pionpot}
V_{\pi}\left[\vec\phi\,\right]&=&i{\rm Tr}\sum_{n=1}^{\infty}\frac{1}{n}
\left(i\vec\Gamma\cdot\vec\phi G^0_N\right)^n \\
& =&\frac{1}{2}\sum_{i,j}{\rm Tr}\left(\Gamma_i\Gamma_j\right)\int dx\,dy\, \Pi^0(x,y)
\phi_i(x)\phi_j(y) \nonumber \\
& +&\frac{1}{3}\sum_{i,j,k}{\rm Tr}\left(\Gamma_i\Gamma_j\Gamma_k\right)
\int dx\,dy\,dz\, \Pi^0(x,y,z)\phi_i(x)\phi_j(y)\phi_k(z)+ 
{\mathcal O}(\vec\phi^4) . \nonumber
\end{eqnarray}
In the above\footnote{Eq.~(\ref{pionpot}) is a compact writing: for example, 
the $n=2$ term must be interpreted as:
\begin{equation}
\frac{i}{2}{\rm Tr}\left(i\vec\Gamma\cdot\vec\phi G^0_N\right)^2=
\frac{i}{2}\int dx\,dy\,{\rm Tr}\sum_{i.j}i\Gamma_i G^0_N(x-y)\,
i\Gamma_j G^0_N(y-x)\phi_i(x)\phi_j(y)  , \nonumber
\end{equation}
where the trace in the right hand side acts on the vertices $\vec \Gamma$, 
and so on.}:
\begin{eqnarray}
\label{pizero}
-i\Pi^0(x,y)&=&iG^0_N(x-y)iG^0_N(y-x) , \\
-i\Pi^0(x,y,z)&=&iG^0_N(x-y)iG^0_N(y-z)iG^0_N(z-x) ,\hspace{0.15 in} {\rm etc} .
\end{eqnarray}
With this procedure we have thus derived an effective action for the bosonic field 
$\vec\phi$. This action contains a term for the
free pion field and also a highly non--local pion self--interaction 
$V_{\pi}$, which is illustrated by the
Feynman diagrams shown in Fig.~\ref{vpi}. This effective interaction 
is given by the sum of all diagrams containing one closed fermion loop 
and an arbitrary number of pionic legs.
We note that the function in Eq.~(\ref{pizero}) is the free 
particle--hole polarization
propagator, namely the Lindhard function. Moreover, the functions 
$\Pi^0(x,y,\ldots,z)$ are symmetric for cyclic permutations of the arguments.

\subsubsection{Semiclassical expansion}
\label{model3}
The next step is the evaluation of the functional integral over the
bosonic degrees of freedom in Eq.~(\ref{Z2}). A perturbative approach
to the bosonic effective action (\ref{seffb}) does not seem to provide any
valuable results within the capabilities of the present computing tools
and we will follow here another approximation scheme, namely the
semiclassical method. 

\subsubsection*{Mean field level}
\label{model4}
The lowest order of the semiclassical expansion is the stationary phase
approximation (also called saddle point approximation in the Euclidean space): 
the bosonic effective action is required to be stationary with respect to
arbitrary variations of the fields $\phi_i$:
\begin{equation}
\displaystyle\frac{\delta S^B_{\rm eff}\left[\vec\phi\,\right]}
{\delta \phi_i(x)}=0 . \nonumber
\end{equation}
From the partial derivative of Eq.~(\ref{seffb}) one obtains the following 
equation of motion for the classical field $\vec\phi$:
\begin{equation}
\label{eqmoto}
\left(\dal +m^2_{\pi}\right)\phi_i(x)=\int dy\, 
G^{0^{-1}}_{\pi}(x-y)\varphi_i(y)+
\displaystyle \frac{\delta V_{\pi}\left[\vec\phi\,\right]}{\delta \phi_i(x)} ,
\end{equation}
whose solutions are functional of the external source $\vec\varphi$. The exact 
solution cannot be written down explicitly. However, due to the
particular form of $V_{\pi}[\vec\phi \,]$, when $\vec\varphi\rightarrow 0$
one solution is $\vec\phi=0$; the general solution of Eq.~(\ref{eqmoto}) 
can then be expressed as an expansion in powers of $\vec\varphi$:
\begin{eqnarray}
\label{soluz}
\phi_i(x)&=&\sum_{j}\int dy\, A_{ij}(x,y)\varphi_j(y) \\
& &+\frac{1}{2}\sum_{j,k}\int dy\,dz\, B_{ijk}(x,y,z)\varphi_j(y)\varphi_k(z)+
{\mathcal O}\left(\vec\varphi^3\right) . \nonumber
\end{eqnarray}
By substituting Eqs.~(\ref{soluz}) and (\ref{pionpot}) 
into (\ref{eqmoto}) and keeping only terms linear in
$\varphi_i$, one obtains the following relation for $A_{ij}$:
\begin{equation}
\label{A}
A_{ij}(x,y)-{\rm Tr}\left(\Gamma_i^2\right)\int du\,dv\, 
G^0_{\pi}(x-u)\Pi^0(u,v)A_{ij}(v,y)=\delta_{ij}\delta(x-y) .
\end{equation}
Finally, by introducing the ring--dressed pion propagator 
$G^{\rm ring}_{\pi}$, which satisfies the Dyson equation:
\begin{equation}
G^{\rm ring}_{\pi}(x-y)=G^0_{\pi}(x-y)
+{\rm Tr}\left(\Gamma_i^2\right)\int du\,dv\, G^0_{\pi}(x-u)
\Pi^0(u,v)G^{\rm ring}_{\pi}(v-y) , \nonumber
\end{equation}
or, formally:
\begin{equation}
G^{\rm ring}_{\pi}=\displaystyle \frac{G^0_{\pi}}
{1-{\rm Tr}\left(\Gamma_i^2\right)G^0_{\pi}\Pi^0} \; , \nonumber
\end{equation}
the solution of Eq.~(\ref{A}) reads:
\begin{equation}
\label{AA}
A_{ij}(x,y)=\delta_{ij}\int dz\, G^{\rm ring}_{\pi}(x-z) G^{0^{-1}}_{\pi}(z-y) .
\end{equation}
Thus, the saddle point solution of Eq.~(\ref{seffb}) at first order
in the source $\vec\varphi$ is:
\begin{eqnarray}
\label{soluzring}
\phi^{\rm ring}_i(x)&=&\int dy\,dz\, G^{\rm ring}_{\pi}(x-z) 
G^{0^{-1}}_{\pi}(z-y)\varphi_i(y) \\
& \equiv &\int dy\, \left(G^{\rm ring}_{\pi} G^{0^{-1}}_{\pi}\right)(x-y)
\varphi_i(y) , \nonumber
\end{eqnarray}
and the corresponding bosonic effective action reads:
%
\begin{eqnarray}
S^B_{\rm eff}\left[\vec\phi^{\rm ring}\right]&=&-\frac{1}{2}
\int dx\,dy\,du\,dv\, 
G^{0^{-1}}_{\pi}(x-u) \nonumber \\ 
& & \times \vec\varphi(u)\cdot G^{\rm ring}_{\pi}(x-y)
G^{0^{-1}}_{\pi}(y-v)\vec\varphi(v) . \nonumber
\end{eqnarray}
Now, the generating functional of Eq.~(\ref{Z2}) takes the form:
\begin{eqnarray}
Z\left[\vec\varphi \,\right]&=&\exp\left\{\frac{i}{2}\right. 
\int dx\,dy\,du\,dv\,
\vec\varphi(u)\cdot G^{0^{-1}}_{\pi}(x-u) \nonumber \\
& & \times \left. \left[G^0_{\pi}(x-y)-G^{\rm ring}_{\pi}(x-y)\right]
G^{0^{-1}}_{\pi}(y-v)\vec\varphi(v)\right\} , \nonumber
\end{eqnarray}
and the polarization propagator can then be evaluated by using 
Eqs.~(\ref{connect}), (\ref{proppol}).
One obtains that in the saddle point approximation it coincides with the 
well known ring expression:
\begin{eqnarray}
\Pi_{ij}(x,y)&=&\delta_{ij}\left[\Pi^0(x,y)+{\rm Tr}\left(\Gamma^2_i\right)
\int du\,dv\, \Pi^0(x,u)G^{\rm ring}_{\pi}(u-v)\Pi^0(v,y)\right] \nonumber \\
& \equiv & \delta_{ij}\Pi^{\rm ring}(x,y) , \nonumber
\end{eqnarray}
or, formally:
\begin{equation}
\Pi=\displaystyle \frac{\Pi^0}
{1-{\rm Tr}\left(\Gamma_i^2\right)G^0_{\pi}\Pi^0}
\equiv \Pi^{\rm ring} . \nonumber
\end{equation}
Hence, the ring approximation corresponds to the mean field level
of the present effective theory.

\subsubsection*{Quantum fluctuations around the mean field solution
(one--boson--loop corrections)}
\label{model5}
In the next step of the semiclassical expansion we write the bosonic 
effective action as:
\begin{eqnarray}
S^B_{\rm eff}\left[\vec\phi\,\right]&=&S^B_{\rm eff}\left[\vec\phi^0\right]+
\frac{1}{2} \sum_{ij} \int dx\,dy\, \displaystyle 
\left[\frac{\delta^2S^B_{\rm eff}\left[\vec\phi\,\right]}
{\delta\phi_i(x)\delta\phi_j(y)}
\right]_{\vec\phi=\vec\phi^0} \nonumber \\
& &\times \left[\phi_i(x)-\phi^0_i(x)\right]\left[\phi_j(y)-\phi^0_j(y)\right], \nonumber
\end{eqnarray}
where now $\vec\phi^0$ also contains the second order term in the source 
$\vec\varphi$ [see Eq.~(\ref{soluz})]. Then, after performing the
gaussian integration over
$\vec\phi$, the generating functional (\ref{Z2}) reads:
\begin{eqnarray}
\label{Z3}
Z\left[\vec\varphi \,\right]&=&\exp\left\{\frac{i}{2}\int dx\,dy\, 
\vec\varphi(x)\cdot G^{0^{-1}}_{\pi}(x-y)\vec\varphi(y)\right\} \\
& & \times \exp\left\{iS^B_{\rm eff}\left[\vec\phi^0\right]
-\frac{1}{2}{\rm Tr}\ln \left[\displaystyle
\frac{\delta^2S^B_{\rm eff}\left[\vec\phi\,\right]}
{\delta\phi_i(x)\delta\phi_j(y)}\right]_{\vec\phi=\vec\phi^0}\right\} , \nonumber
\end{eqnarray}
and the polarization propagator is:
\begin{equation}
\label{proppol1}
\Pi_{ij}(x,y)=-\left\{\frac{\delta^2}
{\delta \varphi_i(x)\delta \varphi_j(y)}\left[S^B_{\rm eff} \left[\vec\phi^0\right] 
+\frac{i}{2}{\rm Tr}\ln \left(\displaystyle \frac{\delta^2S^B_{\rm eff}\left[\vec\phi\,\right]}
{\delta\phi_k(x)\delta\phi_l(y)}\right)_{\vec\phi=\vec\phi^0}\right] \right\}_{\vec\varphi=0} . 
\end{equation}
In the above, the second derivative of the effective action (\ref{seffb}) at the
order $\vec\phi^2$ turns out to be:
\begin{eqnarray}
\label{deriv2}
&&\frac{\delta^2S^B_{\rm eff}\left[\vec\phi\,\right]}
{\delta\phi_i(x)\delta\phi_j(y)}=\delta_{ij}G^{0^{-1}}_{\pi}(x-y)
+{\rm Tr}\left(\Gamma_i\Gamma_j\right)\Pi^0(x,y) \\
& &+\sum_{k} \int du \left[{\rm Tr}\left(\Gamma_i\Gamma_j\Gamma_k\right)\Pi^0(x,y,u)
+{\rm Tr}\left(\Gamma_j\Gamma_i\Gamma_k\right)\Pi^0(y,x,u)\right]\phi_k(u) \nonumber \\
& &+\sum_{k,l}\int du\,dv\, \left[{\rm Tr}\left(\Gamma_i\Gamma_j\Gamma_k\Gamma_l\right)
\Pi^0(x,y,u,v)
+{\rm Tr}\left(\Gamma_j\Gamma_i\Gamma_k\Gamma_l\right)\Pi^0(y,x,u,v) \right. \nonumber \\
& &+\left. {\rm Tr}\left(\Gamma_i\Gamma_l\Gamma_j\Gamma_k\right)\Pi^0(x,v,y,u)\right]
\phi_k(u)\phi_l(v) . \nonumber
\end{eqnarray}
The second term in the right hand side of Eq.~(\ref{deriv2}) does not 
affect the calculation of Eq.~(\ref{proppol1}). By substituting Eq.~(\ref{soluz}) in
the equation of motion (\ref{eqmoto}), from the terms of order $\vec\varphi^2$
one gets for the $B_{ijk}$ functions the following expression:
\begin{eqnarray}
\label{BB}
B_{ijk}(x,y,z)&=&2\,{\rm Tr}\left(\Gamma_i\Gamma_j\Gamma_k\right)\int du\,dv\,dt\,
\Pi^0(u,v,t)G^{\rm ring}_{\pi}(x-u) \\
& & \times \left(G^{\rm ring}_{\pi} 
G^{0^{-1}}_{\pi}\right)(v-y)
\left(G^{\rm ring}_{\pi} G^{0^{-1}}_{\pi}\right)(t-z) . \nonumber
\end{eqnarray}
There remains now 
to calculate the logarithm in Eq.~(\ref{proppol1}) up to second 
order in $\vec\varphi$. One can multiply the generating functional (\ref{Z3}) 
by the factor $\left({\rm det} G^0_{\pi}\right)^{-1/2}$, inessential in the 
calculation of the polarization propagator (this corresponds to multiply
Eq.~(\ref{deriv2}) by $G^0_{\pi}$). Then, after calculating Eq.~(\ref{deriv2})
for $\vec\phi=\vec\phi^0$, with  $\vec\phi^0$ given by the 
Eqs.~(\ref{soluz}), (\ref{AA}), (\ref{BB}),
we expand the logarithm up to $\vec\varphi^2$ and take the trace 
to the same order. This is rather tedious, but, at the end,
the derivation with respect to the external
source provides the following total polarization propagator:
\begin{equation}
\Pi_{ij}(x,y)=\delta_{ij}\Pi (x,y), \nonumber
\end{equation}
where:
\begin{eqnarray}
\label{fluctua}
\hspace{-0.5cm} \Pi(x,y)&=&\Pi^{\rm ring}(x,y)+\sum_{kl}{\rm Tr}\left(\Gamma_k\Gamma_l\right)
\int du\,dv\, G^{\rm ring}_{\pi}(u-v) \Pi^0(x,u,y,v) \\
& & +\sum_{kl}{\rm Tr}\left(\Gamma_k\Gamma_l\right)
\int du\,dv\,G^{\rm ring}_{\pi}(u-v) \left[\Pi^0(x,u,v,y)+\Pi^0(x,y,v,u)\right] 
\nonumber \\
& &+\int du\,dv\,dw\,ds\, G^{\rm ring}_{\pi}(u-w) G^{\rm ring}_{\pi}(v-s)\Pi^0(x,u,v)
\nonumber \\
& &\times \sum_{klmn}\left[{\rm Tr}\left(\Gamma_k\Gamma_l\Gamma_m\Gamma_n\right)
\Pi^0(y,w,s)+{\rm Tr}\left(\Gamma_k\Gamma_l\Gamma_n\Gamma_m\right)
\Pi^0(y,s,w)\right] . \nonumber
\end{eqnarray}

We remind the reader that the second derivative of 
$S^B_{\rm eff}\left[\vec\phi^{\rm ring}\right]$ and 
$S^B_{\rm eff}\left[\vec\phi^0\right]$ with respect to the external source, 
with $\vec\phi^{\rm ring}$
$\left[\vec\phi^0\right]$ given by Eq.~(\ref{soluzring}) [Eqs.~(\ref{soluz}),
(\ref{AA}), (\ref{BB})], gives the same result 
(the ring polarization propagator) when evaluated at $\vec\varphi=0$.

\begin{figure}
\begin{center}
\mbox{\epsfig{file=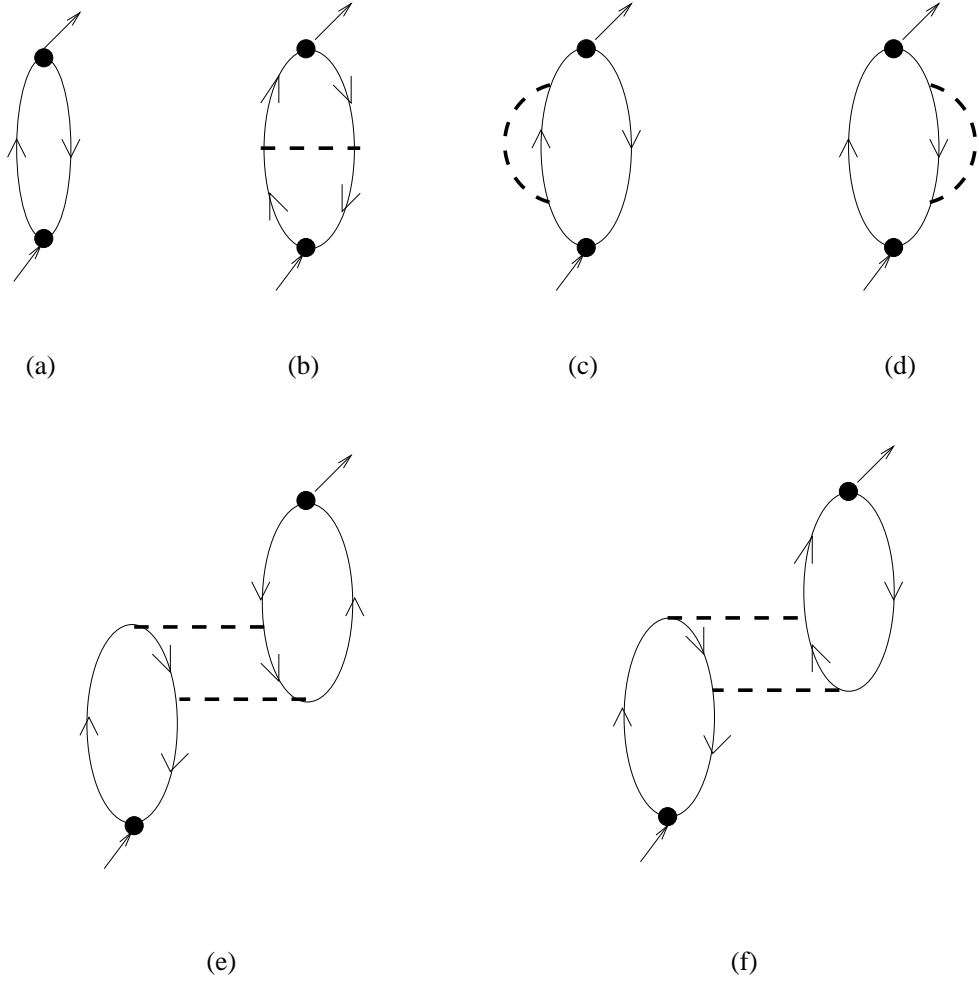,width=.85\textwidth}}
\vskip 5mm
\caption{Feynman diagrams for the polarization propagator of 
Eq.~\protect(\ref{fluctua}): (a) particle--hole;
(b) exchange; (c) and (d) self--energy--type; (e) and (f) correlation diagrams. 
Only the first contribution to the ring expansion has been drawn. The
dashed lines represent ring--dressed pion propagators.}
\label{onelooponeloop}
\end{center}
\end{figure}

\begin{figure}
\begin{center}
\mbox{\epsfig{file=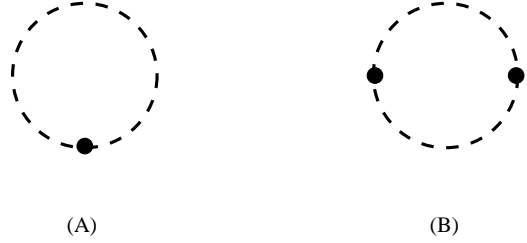,width=.45\textwidth}}
\vskip 5mm
\caption{First order diagrams in the bosonic loop expansion. Diagrams 
(b), (c) and (d) of Fig.~\protect\ref{onelooponeloop} reduce to diagram (A), 
while (e) and (f) reduce to (B).}
\label{oneloop1oneloop1}
\end{center}
\end{figure}

The Feynman diagrams corresponding to Eq.~(\ref{fluctua}) are 
depicted in Fig.~\ref{onelooponeloop}.
Diagram (a) represents the Lindhard function $\Pi^0(x,y)$, 
which is the first term of $\Pi^{\rm ring}(x,y)$. In (b) we have an 
{\it exchange} diagram (the thick dashed lines representing ring--dressed
pion propagators); (c) and (d) are {\it self--energy} diagrams, while in (e) 
and (f) we show the {\it correlation} diagrams of the present approach. 
The approximation
scheme developed here is also referred to as bosonic loop expansion
(BLE). The practical
rule to classify the Feynman diagrams according to their order in the
BLE is to reduce to a point all its fermionic lines and
to count the number of bosonic loops left out. In this case the diagrams 
(b)--(f) of Fig.~\ref{onelooponeloop} reduce to a one--boson--loop.
Diagrams (b), (c), (d) can be represented by the loop (A) of 
Fig.~\ref{oneloop1oneloop1}, while (e) and (f) correspond to 
the loop (B) of the same figure.

The polarization propagator of Eq.~(\ref{fluctua}) is the central result
of this microscopic approach, which will be used in the calculation of 
the $\Lambda$ decay width in nuclear matter. Notice that the model can 
easily include the excitation of baryonic resonances, by replacing 
the fermionic field with multiplets. The topology
of the diagrams remains the same as in Fig.~\ref{onelooponeloop} but, introducing
for example the $\Delta$ resonance,
each fermionic line represents either a nucleon or a $\Delta$, taking care of
isospin conservation. One thus obtains 15 exchange, 14 self--energy
and 98 correlation diagrams (see Ref.~\cite{Ce97} for the whole diagrammology).

Moreover, since the BLE is characterized by the topology
of the diagrams, one can include in the model additional mesonic degrees of
freedom, together with phenomenological short range correlations. In 
particular, the extension to other spin--isospin channels simply amounts to 
change the definition of the vertices $\Gamma_i$ in Eq.~(\ref{fluctua})
and the same occurs for the non--relativistic 
reduction of the theory. Accordingly, for the non--relativistic 
pion--exchange, $\Gamma_i$ becomes (apart from the coupling constant)
$(\vec \sigma\cdot {\vec q})\tau_i$, for the $\rho$--exchange
it reads $(\vec \sigma\times {\vec q})_k \tau_i$, $k$ being a 
spatial index, and for the $\omega$--exchange 
$\Gamma_i\propto (\vec \sigma\times {\vec q})_i$. 
The exchange of $\omega$--mesons
is taken into account only inside the one--boson--loop 
diagrams (b)--(f) of Fig.~\ref{onelooponeloop},
but not in the mesonic lines stemming from the $\Lambda$ decay vertex, 
where the considered exchanged meson is, necessarily, of isovector nature 
($\pi$ or $\rho$). 
Beyond $\pi$, $\rho$ and $\omega$ mesons,
the present approach also contains (partly) 
the exchange of the scalar--isoscalar $\sigma$--meson:
indeed, in the phenomenology of the Bonn $NN$ potential~\cite{Ma87}, the 
latter is described through box diagrams (which are contained
in the correlation diagrams of Fig.~\ref{onelooponeloop}), namely by the exchange of
two pions with the simultaneous excitation of one or both the intermediate
nucleons to a $\Delta$ resonance.

A further difficulty arises if one starts from a potential model rather then 
from a Lagrangian containing bosons as true degrees of freedom. However
this disease is easily overcome by means of a Hubbard--Stratonovitch 
transformation, which enables one to substitute
a potential with a two--body interaction between nucleons by a suitably 
introduced auxiliary field. 
As an example, for a scalar--isoscalar potential $V$, the relevant identity 
reads:
\begin{eqnarray}
&\exp\left\{\displaystyle \frac{i}{2}\int dx\,dy\,\overline\psi(x)
\psi(x)V(x-y)\overline\psi(y)\psi(y)\right\}=\sqrt{\rm det\, V}& \nonumber \\
&\times \displaystyle \int {\mathcal D}[\sigma]\exp\left\{\frac{i}{2}
\int dx\,dy\,\sigma(x)V^{-1}(x-y)
\sigma(y)+i\int dx\, \overline\psi(x)\psi(x)\sigma(x)\right\}& , \nonumber
\end{eqnarray}
where $\sigma$ is the auxiliary field.
Clearly, the previous derivation will remain valid, providing one substitutes
the inverse propagator of the auxiliary field with the inverse potential in
the ``free'' part of the action.

Finally, a relevant point for the feasibility of the calculations is that
all fermion loops in Fig.~\ref{onelooponeloop} can be evaluated analytically \cite{Ce92},
so that each diagram reduces to a 3--dimensional (numerical) integral.

In particular, the formalism can be applied to evaluate the functions $U_{L,T}$ of 
Eq.~(\ref{propU}), which are required in Eqs.~(\ref{Sigma2}), (\ref{Alpha}). In the
one--boson--loop (OBL) approximation of Eq.~(\ref{fluctua}) and Fig.~\ref{onelooponeloop} 
we have to replace Eq.~(\ref{Alpha}) with: 
\begin{eqnarray}
\label{Alpha1}
{\alpha}(q)&=&\left(S^2+\frac{P^2}{m_{\pi}^2}\vec q\,^2\right)F_{\pi}^2(q)
G_{\pi}^0(q)+\frac{\tilde{S}^2(q)U_1(q)}{1-V_L(q)U_1(q)} \\
& & +\frac{\tilde{P}_L^2(q)U_1(q)}{1-V_L(q)U_1(q)}+
2\frac{\tilde{P}_T^2(q)U_1(q)}{1-V_T(q)U_1(q)} \nonumber \\
& & +\left[\tilde S^2(q)+\tilde P_L^2(q)\right]U^{\rm OBL}_L(q)+2\tilde P_T^2(q)
U^{\rm OBL}_T(q) , \nonumber
\end{eqnarray}
where 
\begin{equation}
U_1=U^{ph}+U^{\Delta h} , \nonumber
\end{equation}
while $U^{\rm OBL}_{L,T}$ are evaluated from the diagrams 
\ref{onelooponeloop}(b)--\ref{onelooponeloop}(f)  
using the standard Feynman rules. The normalization of
these functions is such that $U^{ph}(x,y)=4\Pi^0(x,y)$, $\Pi^0$ being 
given by Eq.~(\ref{pizero}). One relevant difference between the
OBL formula (\ref{Alpha1}) and the RPA expression of
Eq.~(\ref{Alpha}) lies in the fact that in the former, to be consistent with 
 Eq.~(\ref{fluctua}), the {\sl 2p--2h} diagrams (which contribute to 
$U^{\rm OBL}_{L,T}$) are not RPA--iterated.

\subsection{Results of the phenomenological calculation} 
\label{pheres} 
We shall illustrate here and in the following subsection the results which can
be obtained for hypernuclear decay widths by employing the two approaches 
(phenomenological and microscopic) illustrated above. 

To start with let us consider the PPM combined with the LDA:
in order to evaluate the width from Eqs.~(\ref{local1}), (\ref{local2}) one needs to specify 
the nuclear density and the wave 
function for the ${\Lambda}$. The former is assumed to be a Fermi distribution
(normalized to the nuclear mass number $A$):
\begin{equation}
\label{density}
\rho_A(r)=\frac{A}{\displaystyle \frac{4}{3}\pi R^3(A)
\left\{1+\left[\frac{\pi a}{R(A)}\right]^2\right\}} \,
\frac{1}{\left\{\displaystyle 1+\exp\left[\frac{r-R(A)}{a}\right]\right\}} , 
\end{equation}
with radius $R(A)=1.12A^{1/3}-0.86A^{-1/3}$ fm and thickness $a=0.52$ fm.
The ${\Lambda}$ wave function is obtained from a $\Lambda$--nucleus potential of
Woods--Saxon shape, with fixed diffuseness and with radius and depth such that it
exactly reproduces the first two single particle eigenvalues ($s$ and $p$ $\Lambda$ levels)
of the hypernucleus under analysis. 
\subsubsection{Short range correlations and $\Lambda$ wave function -- $^{12}_{\Lambda}$C}
\label{resgamma}
A crucial ingredient in the calculation of the decay
widths is the short range part of the strong $NN$ and ${\Lambda}N$ interactions. 
They are expressed by the functions $g_{L,T}(q)$ and $g_{L,T}^{\Lambda}(q)$
reported in appendix \ref{app1} and contain the Landau parameters $g^{\prime}$ and
$g^{\prime}_{\Lambda}$, respectively. No experimental information is available on 
$g^{\prime}_{\Lambda}$, while many constraints have been set on $g^{\prime}$, for
example by the well known quenching of the Gamow--Teller resonance. Realistic
values of $g^{\prime}$ within the framework of the ring approximation are in the
range $0.6\div 0.7$ \cite{Os82}. However, in the present context $g^{\prime}$
correlates not only {\sl p--h} pairs but also {\sl p--h} with {\sl 2p--2h} states. 
In order to fix the correlation parameters in this new contest, in Ref.~\cite{Al99} the 
calculated non--mesonic width of $^{12}_{\Lambda}$C has been compared with the 
experimental one. 

\begin{figure}
\begin{center}
\mbox{\epsfig{file=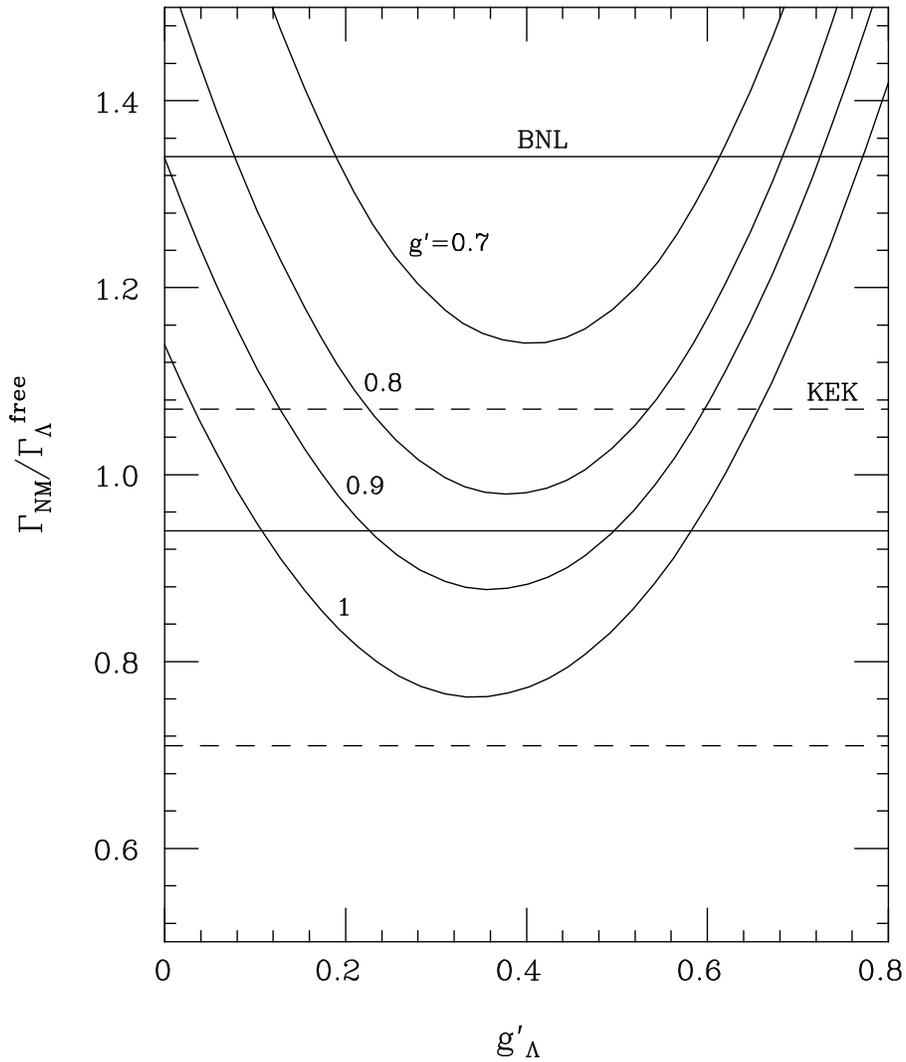,width=0.8\textwidth}}
\vskip 2mm
\caption{Dependence of the non--mesonic width on the Landau 
parameters $g^{\prime}$ and $g^{\prime}_{\Lambda}$ for $^{12}_{\Lambda}$C. 
The experimental value from BNL~\protect\cite{Sz91} (KEK~\protect\cite{No95}) 
lies in between the horizontal solid (dashed) lines 
(taken from Ref.~\cite{Al99}).}
\label{carb}
\end{center}
\end{figure}

In Fig.~\ref{carb} we see how the total non--mesonic width for carbon 
depends on the Landau parameters. 
The rate decreases as $g^{\prime}$ increases. This characteristic
is well established in RPA [see Eq.~(\ref{Alpha})]. Moreover, for fixed $g^{\prime}$, 
there is a minimum for $g^{\prime}_{\Lambda}\simeq 0.4$ (almost independent of 
the value of $g^{\prime}$). This is due to the fact 
that for $g^{\prime}_{\Lambda}\ll 0.4$ the
longitudinal $P$--wave contribution in Eq.~(\ref{Alpha})
dominates over the transverse one and 
the opposite occurs for $g^{\prime}_{\Lambda}\gg 0.4$ (we also
remind the reader that the $S$--wave interaction [Eq.~(\ref{s})] is independent 
of $g^{\prime}_{\Lambda}$). Moreover, the longitudinal
$P$--wave ${\Lambda}N \rightarrow NN$ interaction [Eq.~(\ref{pl})] contains
the pion exchange plus short range correlations, while the 
transverse $P$--wave ${\Lambda}N \rightarrow NN$ interaction [Eq.~(\ref{pt})]
only contains repulsive correlations, so with increasing $g^{\prime}_{\Lambda}$
the $P$--wave longitudinal contribution to the width decreases, 
while the $P$--wave transverse part increases. From Fig.~\ref{carb} we see that there is
a broad range of choices of $g^{\prime}$ and $g^{\prime}_{\Lambda}$ 
values which fit the ``experimental band'': 
$\Gamma^{\rm exp}_{NM}/\Gamma^{\rm free}_{\Lambda}=0.94\div 1.07$. 
The latter represents decay widths which are
compatible with both the BNL \cite{Sz91} and KEK \cite{No95} experiments. 
One should notice that the theoretical curves reported in Fig.~\ref{carb}
contain the contribution of the three--body process 
$\Lambda NN\rightarrow NNN$; should the latter be neglected
(ring approximation), then one could get equivalent results with $g^{\prime}$
values smaller than the ones reported in the figure 
(typically $\Delta g^{\prime}\simeq -0.1$). The phenomenology of the $(e,e')$ 
quasi--elastic scattering suggests, in ring approximation, $g^{\prime}$ values in
the range $0.6\div 0.7$. Here, by taking into account also {\sl 2p--2h} contributions, 
``equivalent" $g^{\prime}$ values larger than in ring 
approximation are used. From Fig.~\ref{carb}, the experimental band appears to be 
compatible with $g^{\prime}$ in the range $0.75\div 0.85$ and 
$g^{\prime}_{\Lambda}$ in the range $0.3\div 0.5$. On the other hand, the new KEK
results \cite{Sa98,Bh98,Ou00,Sa00} set an upper limit of about $1.03$ for the
non--mesonic width, which practically forced us to chose 
$g^{\prime}\gsim 0.8$ and $g^{\prime}_{\Lambda}$ in the above mentioned interval;
considering that $\Gamma_{\rm NM}$ does not change dramatically in this range,
$g^{\prime}_{\Lambda}=0.4$ is a reasonable choice. The new KEK data
are: $\Gamma_{T}/\Gamma^{\rm free}_{\Lambda}=1.14\pm0.08$ and 
$\Gamma_{\pi^-}/\Gamma^{\rm free}_{\Lambda}=0.113\pm 0.014$; 
taking for $\Gamma_{\pi^0}$ the data from \cite{Sz91}, 
$\Gamma_{\pi^0}/\Gamma^{\rm free}_{\Lambda}=0.06^{+0.08}_{-0.05}$, 
and \cite{Sa91}, $\Gamma_{\pi^0}/\Gamma_{T}=0.174\pm 0.058$, which gives
$\Gamma_{\pi^0}/\Gamma^{\rm free}_{\Lambda}=0.198\pm 0.067$ (the calculation of 
Refs.~\cite{Os93,Mo94} supply $\Gamma_{\pi^0}$ values which lie
in between the above central data), by subtraction from the total width one
obtains $\Gamma_{\rm NM}/\Gamma^{\rm free}_{\Lambda}= 0.97^{+0.11}_{-0.10}$ or 
$\Gamma_{\rm NM}/\Gamma^{\rm free}_{\Lambda}=0.83\pm 0.11$ for the two choices of $\Gamma_{\pi^0}$. 
Finally, from Fig.~\ref{carb} we see that the values compatible with both these 
intervals ($0.87\div 0.94$) require $g^{\prime}= 0.85\div 0.90$. This argument
somewhat enlarges the above considered experimental band of Fig.~\ref{carb} 
($0.94\div 1.07$) from below, giving
a new interval, $0.87\div 1.07$, whose central value is reproduced by
fixing $g^{\prime}=0.8$, $g^{\prime}_{\Lambda}=0.4$.

Using these values for the Landau parameters, 
we illustrate now the sensitivity of the calculation of Ref.~\cite{Al99}
to the ${\Lambda}$ wave function in $^{12}_{\Lambda}$C. In addition
to the Woods--Saxon potentials (New W--S) that reproduces the $s$ and $p$ 
$\Lambda$--levels, other choices have also been used. In particular: an
harmonic oscillator wave function (H.O.) with an "empirical" frequency $\omega$
\cite{Pi91,Ha96}, obtained from the $s-p$ energy shift, the 
Woods--Saxon wave function of Ref.~\cite{Do88} (Dover W--S) and
the microscopic wave function (Micr.) calculated, in Ref.~\cite{Po98},
from a non--local self--energy using a realistic $\Lambda N$ interaction.
The results are shown (in units of the free $\Lambda$ width)
in table~\ref{wf sens}, where they are compared with the
experimental data from BNL \cite{Sz91} and KEK \cite{No95,Bh98,Ou00,Sa00}.
\begin{table}
\begin{center}
\caption{Sensitivity of the decay rates to the $\Lambda$ wave function for 
$^{12}_{\Lambda}$C (taken from Ref.~\cite{Al99}).} 
\label{wf sens}
\vspace{0.5cm}
\begin{tabular}{|c|c c c c|c c c|} \hline
\mc {1}{|c|}{} &
\mc {1}{c}{Micr.} &
\mc {1}{c}{Dover} &
\mc {1}{c}{H.O.} &
\mc {1}{c}{New} &
\mc {1}{|c}{BNL} &
\mc {1}{c}{KEK} &
\mc {1}{c|}{KEK New} \\
                &      & W--S  &      & W--S  & \cite{Sz91} & \cite{No95} & \cite{Bh98,Ou00,Sa00}
 \\ \hline\hline
${\Gamma}_M$    & 0.25 & 0.25 & 0.26 & 0.25 & $0.11\pm 0.27$ & $0.36\pm 0.13$ & $0.31\pm 0.07$\\
${\Gamma}_1$    & 0.69 & 0.77 & 0.78 & 0.82 &                &                &    \\
${\Gamma}_2$    & 0.13 & 0.15 & 0.15 & 0.16 &                &                &\\
${\Gamma}_{NM}$ & 0.81 & 0.92 & 0.93 & 0.98 & $1.14\pm 0.20$ & $0.89\pm 0.18$ & $0.83\pm 0.11$ \\
${\Gamma}_{T}$  & 1.06 & 1.17 & 1.19 & 1.23 & $1.25\pm 0.18$ & $1.25\pm 0.18$ & $1.14\pm 0.08$ \\ \hline 
\end{tabular}
\end{center}
\end{table}
By construction, the chosen $g^{\prime}$ and $g^{\prime}_{\Lambda}$ reproduce the 
experimental non--mesonic width using the
W--S wave function which gives the right $s$ and $p$ hyperon levels in
$^{12}_\Lambda$C (column New W--S). 
We note that it is possible to generate the microscopic wave function of
Ref.~\cite{Po98} for carbon via a local hyperon--nucleus W--S potential with 
radius $2.92$ fm and depth $-23$ MeV. 
Although this potential reproduces fairly well the experimental $s$--level for the 
${\Lambda}$ in $^{12}_{\Lambda}$C, it does not reproduce the $p$--level.
A completely phenomenological $\Lambda$--nucleus potential, that can easily
be extended to heavier nuclei and reproduces the experimental
$\Lambda$ single particle levels as well as possible, has been preferably
adopted in Ref.~\cite{Al99}.
Except for $s$--shell hypernuclei, where the experimental data require
$\Lambda$--nucleus potentials with a repulsive core
at short distances, the $\Lambda$ binding energies have been well reproduced by W--S potentials.
The authors of Ref.~\cite{Al99}
use a W--S potential with fixed diffuseness ($a=0.6$ fm) and adjust 
the radius and depth to reproduce the $s$ and $p$ $\Lambda$--levels. The parameters 
of the potential for carbon are $R=2.27$ fm and $V_0=-32$ MeV.

To analyze the results of table~\ref{wf sens}, we note that the
microscopic wave function 
is substantially more extended than all the other wave functions used in
the present study.
The Dover's parameters \cite{Do88}, namely $R=2.71$ fm and $V_0=-28$ MeV, give rise to
a $\Lambda$ wave function that is somewhat more extended than the new W--S one but 
is very similar to the one obtained from a harmonic oscillator with an empirical 
frequency $\hbar \omega =10.9$ MeV. Consequently, the non--mesonic width from the Dover's
wave function is very similar to the one obtained from the harmonic oscillator
and slightly smaller than the new W--S one. 
The microscopic wave--function predicts the smallest non--mesonic widths due to
the more extended $\Lambda$ wave--function, which explores regions of lower 
density, where the probability of interacting 
with one or more nucleons is smaller. From table~\ref{wf sens} we also see
that, against intuition, the mesonic width is quite insensitive to 
the ${\Lambda}$ wave function. On this point we remind the reader that the more
extended is the wave function in $r$--space, the larger is the mesonic
width, since the Pauli blocking effects on the emitted nucleon are reduced.
However, the integral over the ${\Lambda}$--momenta in Eq.~(\ref{local2}) is
weighted by the momentum distribution $|\tilde{\psi}_{\Lambda}(\vec k)|$, which
correspondingly tends to cancel the above mentioned effect: as a result, 
$\Gamma_{\rm M}$ is insensitive to the different wave functions used in the calculation
and it is consistent with both the BNL and KEK data.
In summary, different (but realistic) $\Lambda$ wave functions
give rise to total decay widths which may differ at most by 15\%. 
\subsubsection{Decay widths of light to heavy $\Lambda$--hypernuclei}
\label{heavy}
Using the new W--S wave functions and the Landau parameters $g^{\prime}=0.8$
and $g^{\prime}_{\Lambda}=0.4$, in Refs.~\cite{Al99,Tesi} the 
calculation has been extended to hypernuclei from $^5_{\Lambda}$He to 
$^{208}_{\Lambda}$Pb. We note that, in order to reproduce 
the experimental $s$ and $p$ levels for the hyperon in the different 
nuclei one must use potentials with nearly 
constant depth, around $28\div 32$ MeV, in all but the lightest hypernucleus 
($^5_{\Lambda}$He).
Radii and depths of the employed W--S potentials are quoted in table~\ref{ws par}.
\begin{table}
\begin{center}
\caption{W--S parameters (taken from Ref.~\cite{Al99}).}
\label{ws par}
\vspace{0.5cm}
\begin{tabular}{|c|c c|} \hline
\mc {1}{|c|}{$ ^{A+1}_{\Lambda}Z$} &
\mc {1}{c}{$R$ (fm)} &
\mc {1}{c|}{$V_0$ (MeV)} \\ \hline\hline
$^{12}_{\Lambda}$C    & 2.27 & $-32.0$ \\
$^{28}_{\Lambda}$Si   & 3.33 & $-29.5$  \\
$^{40}_{\Lambda}$Ca   & 4.07 & $-28.0$ \\
$^{56}_{\Lambda}$Fe   & 4.21 & $-29.0$ \\
$^{89}_{\Lambda}$Y    & 5.07 & $-28.5$ \\
$^{139}_{\Lambda}$La  & 6.81 & $-27.5$ \\
$^{208}_{\Lambda}$Pb  & 5.65 & $-32.0$ \\ \hline 
\end{tabular}
\end{center}
\end{table}
In the case of helium, the $\Lambda$--nucleus mean potential
has a repulsive core. For this hypernucleus the most convenient
$\Lambda$ wave function turn out to be the one derived in
Ref.~\cite{St93}, within a quark model description of $^5_{\Lambda}$He.

\begin{figure}
\begin{center}
\mbox{\epsfig{file=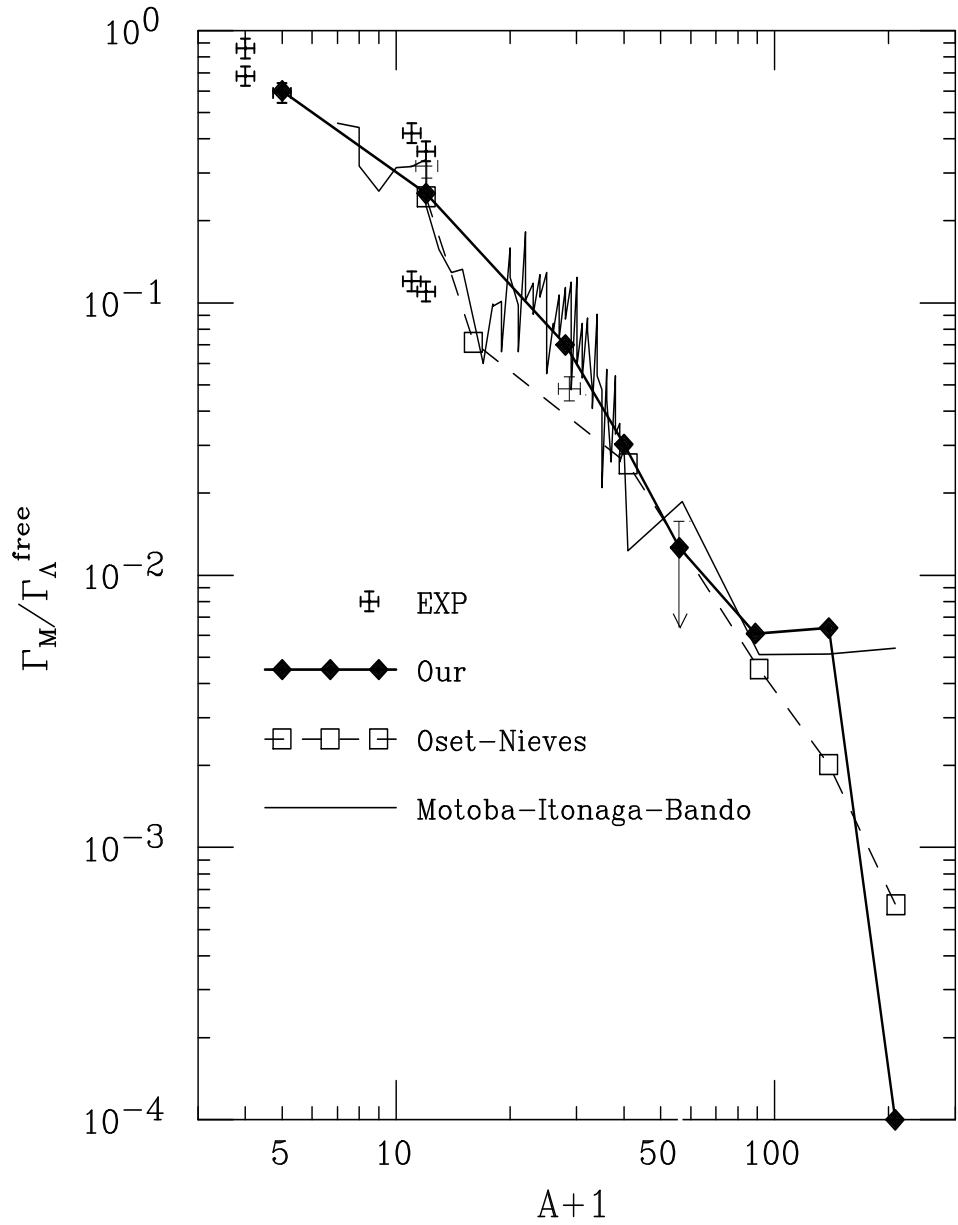,width=.88\textwidth}}
\vskip 2mm
\caption{Mesonic width as a function of the nuclear mass number $A$. The results
of Ref.~\cite{Al99,Tesi} (thick solid line) are compared with the calculations of Nieves--Oset 
\protect\cite{Os93} (dashed line) and Motoba--Itonaga--Band$\overline{\rm o}$ 
\protect\cite{It88,Mo94} (solid line). Available experimental data 
\cite{Sz91,No95,Sa01} are also shown. See text for details on data.}
\label{mesonic}
\end{center}
\end{figure}

The resulting hypernuclear decay rates are shown in table~\ref{sat} \cite{Al99,Tesi}. 
\begin{table}
\begin{center}
\caption{Mass dependence of the hypernuclear weak decay rates.}
\label{sat}
\vspace{0.5cm}
\begin{tabular}{|c|c c c c|} \hline
\mc {1}{|c|}{$ ^{A+1}_{\Lambda}Z$} &
\mc {1}{c}{${\Gamma}_M$} &
\mc {1}{c}{${\Gamma}_1$} &
\mc {1}{c}{${\Gamma}_2$} &
\mc {1}{c|}{${\Gamma}_{T}$} \\ \hline\hline
$ ^{5}_{\Lambda}$He    & 0.60             & 0.27 & 0.04 & 0.91 \\
$ ^{12}_{\Lambda}$C    & 0.25             & 0.82 & 0.16 & 1.23 \\
$ ^{28}_{\Lambda}$Si   & 0.07             & 1.02 & 0.21 & 1.30 \\
$ ^{40}_{\Lambda}$Ca   & 0.03             & 1.05 & 0.21 & 1.29 \\
$ ^{56}_{\Lambda}$Fe   & 0.01             & 1.12 & 0.21 & 1.35 \\
$ ^{89}_{\Lambda}$Y    & $6\cdot 10^{-3}$ & 1.16 & 0.22 & 1.38 \\
$ ^{139}_{\Lambda}$La  & $6\cdot 10^{-3}$ & 1.14 & 0.18 & 1.33 \\
$ ^{208}_{\Lambda}$Pb  & $1\cdot 10^{-4}$ & 1.21 & 0.19 & 1.40 \\ \hline
\end{tabular}
\end{center}
\end{table}
We observe that the mesonic rate rapidly vanishes 
by increasing the nuclear mass number $A$. 
This is well known and it is related to the decreasing phase space 
allowed for the mesonic channel, and to smaller overlaps between 
the ${\Lambda}$ wave function and the nuclear surface, as $A$ increases. 
In Fig.~\ref{mesonic} the results of Refs.~\cite{Al99,Tesi}
for $\Gamma_{\rm M}$ (thick solid line) are compared
with the ones of Nieves--Oset \cite{Os93} (dashed line)
and Motoba--Itonaga--Band$\overline{\rm o}$ \cite{It88,Mo94} (solid line), which were obtained
within a shell model framework. Also the central values of the available experimental
data \cite{Sz91,No95,Sa01} are shown. Although the wave function method (WFM) is more reliable than
the LDA for the evaluation of the mesonic rates (because of the small energies involved
in the decay, which amplify the effects of the nuclear shell structure), 
we see that this LDA calculation agrees with the WFM ones 
(apart from the case of $^{208}_{\Lambda}$Pb) and with the data. 
In particular, the results for $^{12}_{\Lambda}$C and
$^{28}_{\Lambda}$Si are in agreement with the recent KEK measurement \cite{Sa00}:
$\Gamma_{\rm M}(^{12}_{\Lambda}{\rm C})/\Gamma^{\rm free}_{\Lambda}= 
0.31\pm 0.07$, $\Gamma_{\pi^-}(^{28}_{\Lambda}{\rm Si})/\Gamma^{\rm free}_{\Lambda}=
0.047\pm 0.008$. The results for 
$^{40}_{\Lambda}$Ca, $^{56}_{\Lambda}$Fe and $^{89}_{\Lambda}$Y are in agreement 
with the old emulsion data (quoted in Ref.~\cite{Co90}), which indicates
$\Gamma_{\pi^-}/\Gamma_{\rm NM}\simeq (0.5\div 1)\cdot 10^{-2}$ 
in the region $40<A<100$. Moreover, the recent KEK experiments \cite{Sa00} obtained
the limit: $\Gamma_{\pi^-}(^{56}_{\Lambda}{\rm Fe})/\Gamma^{\rm free}_{\Lambda}
< 0.015$. It is worth noticing, in figure \ref{mesonic}, the rather pronounced
oscillations of $\Gamma_{\rm M}$ in the calculation of Refs.~\cite{It88,Mo94}, 
which are caused by shell effects.

Coming back to table~\ref{sat}, we note that, with the exception of $^5_{\Lambda}$He,
the two--body induced decay is rather independent of the hypernuclear dimension
and it is about 15\% of the total width. Previous works
\cite{Al91,Ra95} gave more emphasis to this new channel,
without, however, reproducing the experimental non--mesonic
rates. The total width does not change much with $A$,
as it is also shown by the experiment.

\begin{figure}
\begin{center}
\mbox{\epsfig{file=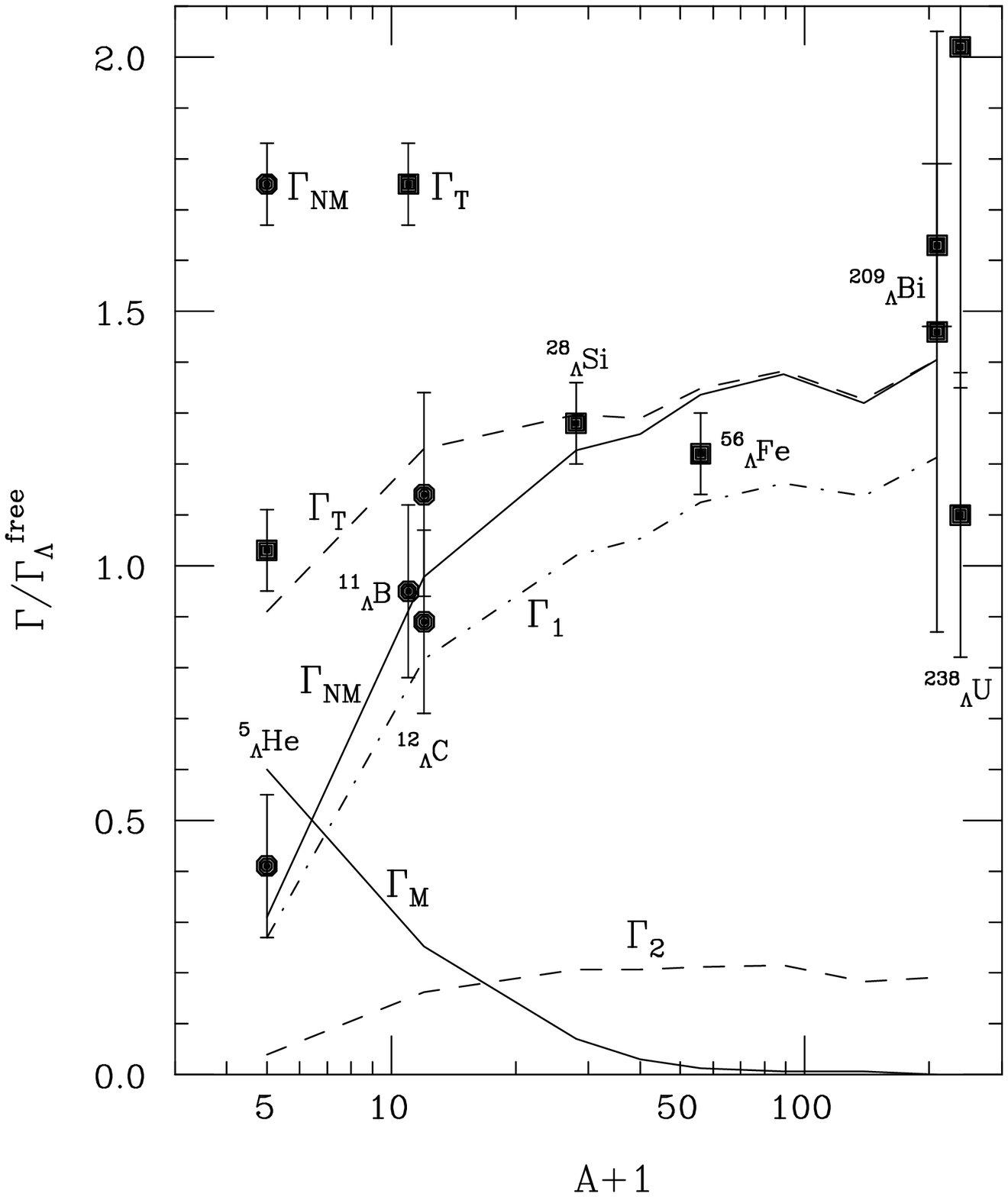,width=.9\textwidth}}
\vskip 2mm
\caption{Partial ${\Lambda}$ decay widths in finite nuclei as a function of
the nuclear mass number $A$. Experimental data are taken from Refs.~\cite{Sz91,Ar93,No95,Bh98,Ku98}}
\label{satu}
\end{center}
\end{figure}

\begin{figure}
\begin{center}
\mbox{\epsfig{file=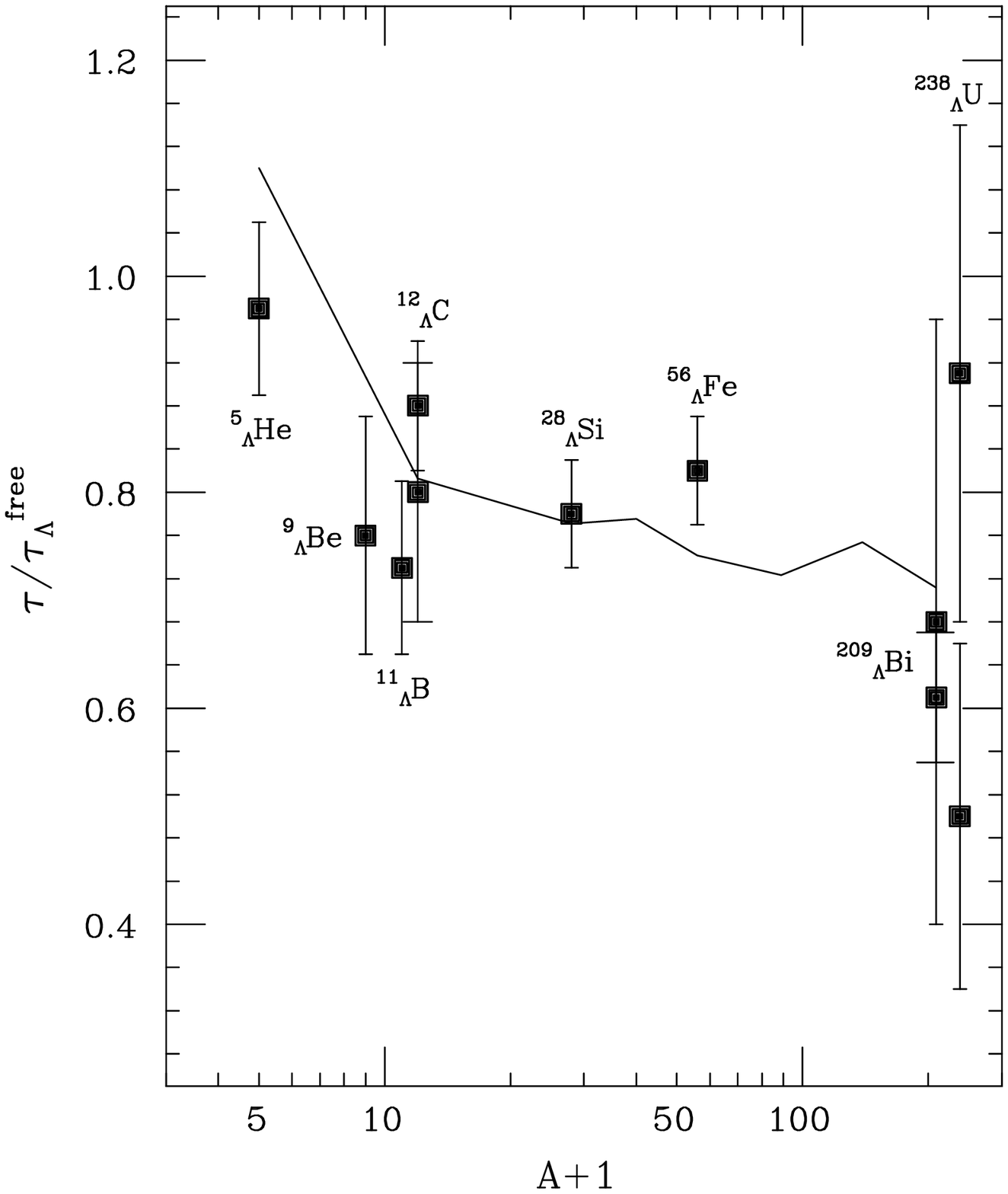,width=.9\textwidth}}
\vskip 2mm
\caption{Total ${\Lambda}$ lifetime in finite nuclei as a function of
the nuclear mass number $A$. Experimental data are taken from Refs.~\cite{Sz91,Ar93,No95,Bh98,Ku98}}
\label{life}
\end{center}
\end{figure}

In Fig.~\ref{satu} the results of table~\ref{sat} are compared
with recent (after 1990) experimental data for $\Gamma_{\rm NM}$ and 
$\Gamma_{\rm T}$ \cite{Sz91,Ar93,No95,Bh98,Ku98}, while
in figure~\ref{life} the same comparison concerns the total $\Lambda$ lifetime
$\tau=\hbar /\Gamma_{\rm T}$. The theoretical results
are in good agreement with the data over the whole hypernuclear mass range explored.
The saturation of the ${\Lambda}N\rightarrow NN$ interaction
in nuclei is well reproduced.

\subsection{Results of the microscopic calculation}
\label{micres}
The results presented in this subsection have been obtained
by applying the formalism developed in subsections \ref{pm} and \ref{func}
for nuclear matter. Although, in principle, one could extend this calculation
to finite nuclei through the local density approximation, as in previous
subsection, in practice this would require prohibitive computing times. Indeed the latter are
already quite conspicuous for the evaluation of the diagrams of
Fig.~\ref{onelooponeloop} at {\it fixed} Fermi momentum. Hence, in
order to compare the results with the experimental data in finite nuclei, different Fermi momenta,
fixed on the following basis, have been employed in the calculation for nuclear matter. 
First we remind the reader that in LDA the local
Fermi momentum $k^A_F(r)$ is related to the nuclear density (\ref{density}) 
by equation (\ref{local}). For the present purpose, the average, fixed Fermi momentum can be
obtained by weighting each local $k_F$ with the probability density of the hyperon
in the considered nucleus:
\begin{equation}
\label{kf3}
\langle k_F\rangle_A=\int d\vec r \,k^A_F(\vec r)|\psi_{\Lambda}(\vec r)|^2 .
\end{equation}
In Ref.~\cite{Al99} $\psi_{\Lambda}(\vec r)$ has been calculated 
from a $\Lambda$--nucleus Wood--Saxon
potential with thickness $a=0.6$ fm and with radius and depth which
reproduce the measured $s$ and $p$ $\Lambda$--levels.

It is possible to classify the hypernuclei, for which
experimental data on the non--mesonic decay rate 
are available, into three mass regions
(medium--light: $A\simeq 10$; medium: $A\simeq 30\div 60$;
and heavy hypernuclei: $A\gsim 200$), as shown in table~\ref{kmed}.
The experimental bands include values of the non--mesonic widths 
which are compatible with the quoted experiments. For medium and heavy 
hypernuclei the available experimental data actually refer to the {\it total}
decay rate. However, from experiments and various estimates it turns out
that the mesonic width for medium hypernuclei is at most 5\% of the total width and
rapidly decreases as $A$ increases. Therefore, because of the low precision of the
data, one can safely approximate $\Gamma^{\rm exp}_{NM}$ with $\Gamma^{\rm exp}_{T}$
for medium and heavy systems.
In the third column of table~\ref{kmed} we report the average Fermi momenta
obtained with Eq.~(\ref{kf3}).
\begin{table}
\begin{center}
\caption{Average Fermi momenta for three representative mass regions. The experimental
data are in units of the free $\Lambda$ decay rate (taken from Ref.~\cite{Al99a}).}
\vspace{0.3cm}
\label{kmed}
\begin{tabular}{|c|c c|}
\hline
\mc {1}{|c|}{} &
\mc {1}{c}{$\Gamma^{\rm exp}_{NM}$} &
\mc {1}{c|}{$\langle k_F\rangle$ (fm$^{-1}$)} \\ \hline\hline
Medium--Light: $^{11}_{\Lambda}$B - $^{12}_{\Lambda}$C & $0.94\div 1.07$ \cite{Sz91,No95} &  1.08 \\
Medium : $^{28}_{\Lambda}$Si - $^{56}_{\Lambda}$Fe    & $1.20\div 1.30$ \cite{Bh98} &  $\simeq 1.2$\\
Heavy: $^{209}_{\Lambda}$Bi - $^{238}_{\Lambda}$U     & $1.45\div 1.70$ \cite{Ar93,Ku98} &  1.36 \\ \hline
\end{tabular}
\end{center}
\end{table}
In the calculations we discuss in the following we have then used the following average
Fermi momenta: $k_F=1.1$ fm$^{-1}$ for medium--light, $k_F=1.2$ fm$^{-1}$ for medium and 
$k_F=1.36$ fm$^{-1}$ for heavy hypernuclei.

In addition to $\langle k_F\rangle$, other parameters enter into the 
microscopic calculation of hypernuclear decay widths, which are specifically
related to the baryon--meson vertices and to the short range correlations. In
Ref.~\cite{Al99a}, with the exception of the Landau parameters $g^{\prime}$ and 
$g^{\prime}_{\Lambda}$, the values of these parameters have not been
left free: rather, they have been kept fixed on the basis of the
existing phenomenology (for example in the analysis of quasi--elastic
electron--nucleus scattering, spin--isospin nuclear response functions,
etc). For the complete list of these quantities we refer to Ref.~\cite{Al99a}.

An important ingredient in the calculation of the $\Lambda$ decay rates
is the short range part of the $NN$ and $\Lambda N$ 
strong interactions: in fact, the momenta involved in the non--mesonic processes
are very large. These short range correlations can be
parameterized with the functions reported in appendix \ref{app1}. The
zero energy and momentum limits of these correlations, $g^{\prime}$ and
$g^{\prime}_{\Lambda}$, are considered
as {\it free parameters}. We remind once again the reader that no experimental 
constraint is available on $g^{\prime}_{\Lambda}$, while 
in the framework of ring approximation (namely by neglecting the 
{\sl 2p--2h} states in the $\Lambda$ self--energy), 
realistic values of $g^{\prime}$ lie in the
range $0.6\div 0.7$ \cite{Os82}. However, in the present context, $g^{\prime}$
enters into the one--boson--loop contributions;
moreover, in some diagrams [for instance (f) and (g) of Fig.~\ref{self1}] two consecutive
$g^{\prime}$ are ``connected'' to the same fermionic line, introducing a sort of 
double counting, which imposes a renormalization of $g^{\prime}$. In the picture of 
figures~\ref{self1} and \ref{onelooponeloop} 
the $\Lambda$ self--energy acquires an energy and momentum 
behaviour which cannot be explained and simulated on the basis of the
simple ring approximation.
Therefore, the physical meaning of the Landau parameters is different 
in the present scheme with respect to the customary phenomenology. Hence 
in Ref.~\cite{Al99a} $g^{\prime}$ has been used as free parameter, 
to be fixed in order to reproduce the experimental hypernuclear decay rates.

\begin{figure}  
\begin{center}
\mbox{\epsfig{file=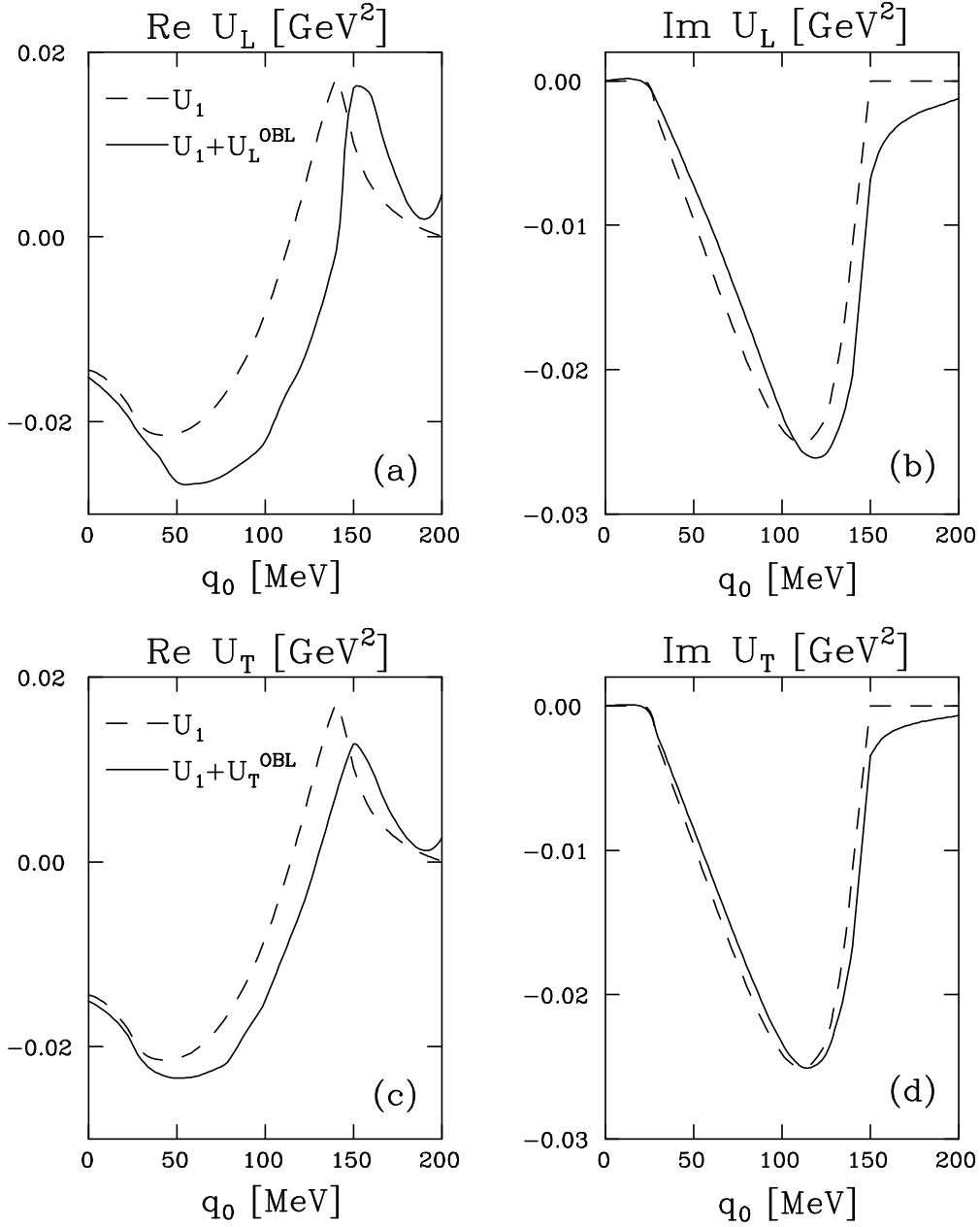,width=0.9\textwidth}}
\caption{Polarization propagators in one--boson--loop approximation
$U_{L,T}(q_0,\vec q)$ ($U_{L,T}=U_1+U^{OBL}_{L,T}$) 
of Fig.~\ref{onelooponeloop} and Eq.~(\ref{Alpha1})
as a function of $q_0$, with $q_0=k_0-E_N({\vec k}-{\vec q})-V_N$. In the
figure $g^{\prime}=0.7$ and $k_F=1.36$ fm$^{-1}$.}
\label{u_obl}
\end{center}
\end{figure}

In figure~\ref{u_obl} we report the real and
imaginary parts of the spin--longitudinal (L) and spin--transverse (T) polarization 
propagators in one--boson--loop approximation $U_{L,T}(q_0,\vec q)$
($U_{L,T}=U_1+U^{OBL}_{L,T}$), which are needed in Eq.~(\ref{Alpha1}), as
a function of $q_0$, $\vec q$ being related to $q_0$ by the 
constraint of Eq.~(\ref{Sigma2}), $q_0=k_0-E_N(\vec k - \vec q)-V_N$. 
The Landau parameter $g^{\prime}$ has been fixed to 
0.7 and the Fermi momentum to $k_F=1.36$ fm$^{-1}$.
In (a) and (b) we show real and imaginary parts of $U_1=U^{ph}+U^{\Delta h}$ 
(dashed lines) and $U_1+U^{OBL}_L$ (solid lines), respectively. The former
is the sum of the {\sl p--h} and {\sl $\Delta$--h} Lindhard functions of
Fig.~\ref{onelooponeloop}(a),
the latter has been calculated by adding the OBL diagrams 
\ref{onelooponeloop}(b)--\ref{onelooponeloop}(f). For reasons related to the
technique employed in the numerical evaluation of the Feynman diagrams,
it is not possible to separate, in the OBL contributions,
the imaginary parts which arise from placing on shell {\sl p--h} and 
{\sl 2p--2h} excitations. In (c) and (d), the above quantities are plotted for
the spin--transverse channel.

As discussed in the previous subsection, for fixed $g^{\prime}$ the
non--mesonic width (the total width in nuclear matter, where $\Gamma_{\rm M}=0$)
has a minimum as a function of $g^{\prime}_{\Lambda}$, which
is almost independent of the value of $g^{\prime}$ (see figure \ref{carb}). 
This characteristic does not
depend on the set of diagrams taken into account in the calculation, but it is simply
due to the interplay between the longitudinal and transverse parts of the 
$P$--wave $\Lambda N\rightarrow NN$ potential
[$\tilde P_L$ and $\tilde P_T$ functions of Eqs.~(\ref{Alpha}), (\ref{Alpha1})]. 
Thus, also in the microscopic calculation the minimum of $\Gamma_{\rm NM}$ is obtained for 
$g^{\prime}_{\Lambda}\simeq 0.4$. 

\begin{figure}
\begin{center}
\mbox{\epsfig{file=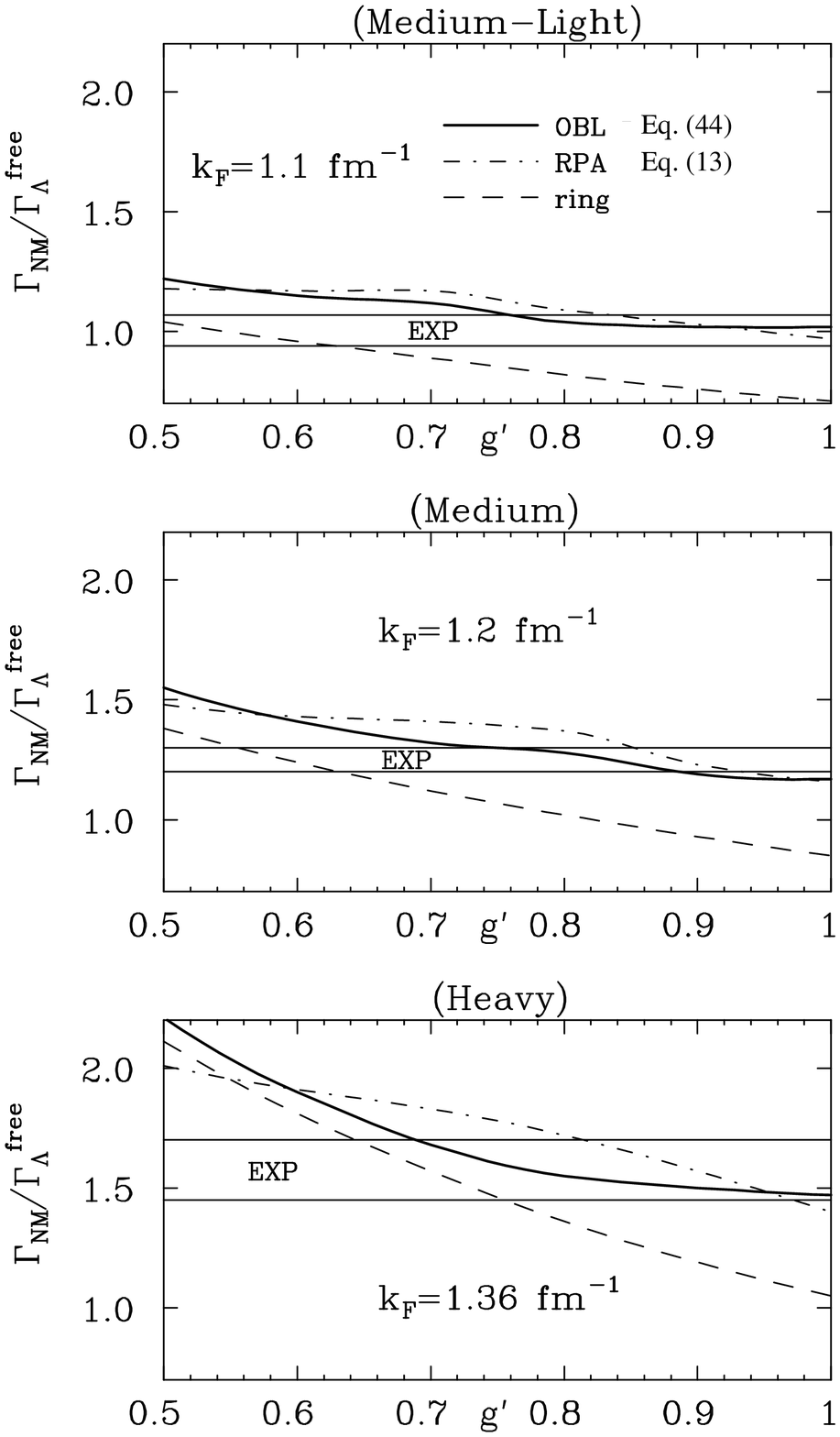,width=0.75\textwidth}}
\caption{Dependence of the non--mesonic width on the Landau 
parameter $g^{\prime}$, for $g^{\prime}_{\Lambda}=0.4$. The three plots 
correspond to the classification of
table~\ref{kmed}. The thick solid curves refer to the
one--boson--loop approximation of Eq.~(\ref{Alpha1}), 
the dot--dashed ones to the RPA calculation of Eq.~(\ref{Alpha}) and
the dashed ones to the ring approximation.
The experimental bands of table~\ref{kmed} lie in between the horizontal 
solid lines (taken from Ref.~\cite{Al99a}).}
\label{gammagprimo}
\end{center}
\end{figure}

\begin{figure}
\begin{center}
\mbox{\epsfig{file=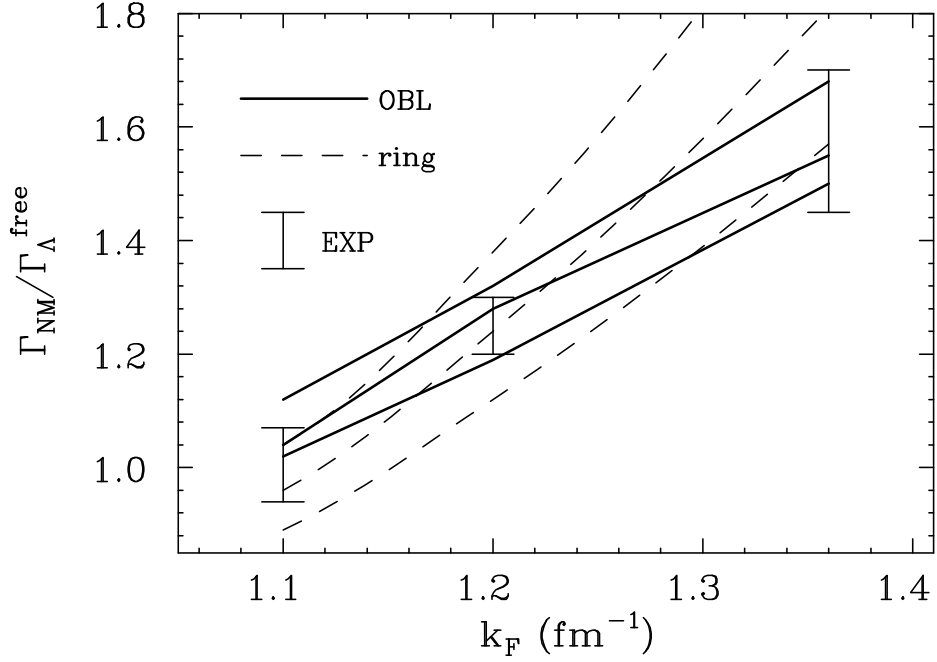,width=0.85\textwidth}}
\caption{Dependence of the non--mesonic width on the Fermi momentum of nuclear matter.
The solid curves refer to the one--boson--loop approximation
(with $g^{\prime}=0.7, 0.8, 0.9$ from the top to the bottom), 
while the dashed lines refer to the ring approximation ($g^{\prime}=0.5, 0.6, 0.7$). 
The experimental data are also shown (taken from Ref.~\cite{Al99a}).}
\label{gammakf}
\end{center}
\end{figure}

Fixing $g^{\prime}_{\Lambda}=0.4$, in ring approximation
one can reproduce the experimental decay rates
by using $g^{\prime}$ values which are compatible with the existing literature. 
In figure~\ref{gammagprimo} we show, as a function of $g^{\prime}$
(for $g^{\prime}_{\Lambda}=0.4$), the calculated non--mesonic decay widths
(in units of the free $\Lambda$ width) 
for the three mass regions of table~\ref{kmed}.
The thick solid curves refer to the one--boson--loop approximation
of Eq.~(\ref{Alpha1}) and Fig.~\ref{onelooponeloop}, while the dot--dashed curves are 
obtained through a RPA iteration of both the particle--hole and the 
one--boson--loop diagrams, namely by using Eq.~(\ref{Alpha}). 
However, we remind the reader that only the former approximation
has a theoretically founded basis, in line with the semiclassical scheme 
introduced in subsection \ref{func}; moreover, this ``inconsistent'' RPA calculation has the
tendency to overestimate, in the acceptable range of $g^{\prime}$ values, 
the experimental non--mesonic widths. The dashed lines represent the pure
ring approximation. The calculated widths are compatible with the experimental bands
for the $g^{\prime}$ values reported in table~\ref{gprimo}. 
\begin{table}
\begin{center}
\caption{$g^{\prime}$ values compatible with the experiments (taken from Ref.~\cite{Al99a}).}
\vspace{0.3cm}
\label{gprimo}
\begin{tabular}{|c|c c|}
\hline
\mc {1}{|c|}{} &
\mc {1}{c}{OBL} &
\mc {1}{c|}{ring} \\ \hline\hline
$k_F=1.1$ fm$^{-1}$  & $\gsim 0.75$     & $0.45\div 0.65$ \\
$k_F=1.2$ fm$^{-1}$  & $0.75\div 0.90$ & $0.55\div 0.65$ \\
$k_F=1.36$ fm$^{-1}$ & $0.70\div 1.00$ & $0.65\div 0.75$ \\ \hline
\end{tabular}
\end{center}
\end{table}
As we have already noticed, the intervals corresponding 
to the ring calculation are in agreement
with the phenomenology of other processes, like the $(e,e')$ quasi--elastic
scattering. However, only the full calculation (column OBL) allows for
a good description (keeping the same $g^{\prime}$ value) 
of the rates in the whole range of $k_F$ considered here. In Fig.~\ref{gammakf}
we see the dependence of the non--mesonic widths on the Fermi momentum.
The solid lines correspond to the one--loop approximation, with 
$g^{\prime}=0.7, 0.8, 0.9$ from the top to the bottom, while the dashed
lines refer to the ring approximation, with $g^{\prime}=0.5, 0.6, 0.7$, 
again from the top to the bottom. We can then conclude that for the one--loop
calculation the best choice for the Landau parameters is the following:
\begin{equation}
g^{\prime}=0.8, \hspace{0.6cm} g^{\prime}_{\Lambda}=0.4 . \nonumber
\end{equation}
This parameterization turns out to be the same that was employed in subsection \ref{pheres}.
However, we must point out that in the other calculation the
{\sl 2p--2h} contributions in the $\Lambda$ self--energy have been evaluated
by using a phenomenological parameterization of the pion--nucleus
optical potential. Here we are considering a microscopical evaluation of
all the relevant diagrams which contribute at the one--boson--loop level:
it is evident that the role played by the Landau parameters is different in the
two approaches. 

In order to compare the results of the phenomenological and the microscopical
approaches, it is appropriate to consider the former as obtained with constant
density rather than in LDA.
In table~\ref{newold} we show the comparison between the
one--boson--loop approximation (column OBL) and the 
phenomenological model (column PM) of paragraph \ref{p2p2hp} at {\it fixed} $k_F$. 
\begin{table}
\begin{center}
\caption{Comparison between the one--boson--loop approximation (column OBL) and the
phenomenological model (column PM) of paragraph \ref{p2p2hp} for $g^{\prime}=0.8$,
$g^{\prime}_{\Lambda}=0.4$. The decay rates are in units of the free $\Lambda$ width
(taken from Ref.~\cite{Al99a}).}
\vspace{0.3cm}
\label{newold}
\begin{tabular}{|c|c c| c c| c c|}
\hline
\mc {1}{|c|}{} &
\mc {2}{c|}{$k_F=1.1$ fm$^{-1}$} &
\mc {2}{c|}{$k_F=1.2$ fm$^{-1}$} &
\mc {2}{c|}{$k_F=1.36$ fm$^{-1}$} \\
\mc {1}{|c|}{} &
\mc {1}{c}{OBL} &
\mc {1}{c|}{PM} &
\mc {1}{c}{OBL} &
\mc {1}{c|}{PM} &
\mc {1}{c}{OBL} &
\mc {1}{c|}{PM} \\ \hline\hline
 $\Gamma_1$              & 0.82            & 0.81 & 1.02 & 1.00 & 1.36 & 1.33 \\
 $\Gamma_2$              & 0.22            & 0.13 & 0.26 & 0.18 & 0.19 & 0.26 \\
 $\Gamma_{\rm NM}$       & 1.04            & 0.94 & 1.28 & 1.19 & 1.55 & 1.59 \\
 \hline
\mc {1}{|c|}{$\Gamma^{\rm exp}_{NM}$} &
\mc {2}{c|}{$0.94\div 1.07$} &
\mc {2}{c|}{$1.20\div 1.30$} &
\mc {2}{c|}{$1.45\div 1.70$} \\ \hline
\end{tabular}
\end{center}
\end{table}
\begin{table}
\begin{center}
\caption{Comparison between the phenomenological model for finite nuclei
(column LDA) and at fixed $k_F$ (column PM).
The decay rates are in units of the free $\Lambda$ width and for
$g^{\prime}=0.8$, $g^{\prime}_{\Lambda}=0.4$.}
\vspace{0.3cm}
\label{newold1}
\begin{tabular}{|c|c c| c c| c c|}
\hline
\mc {1}{|c|}{} &
\mc {2}{c|}{$k_F=1.1$ fm$^{-1}$} &
\mc {2}{c|}{$k_F=1.2$ fm$^{-1}$} &
\mc {2}{c|}{$k_F=1.36$ fm$^{-1}$} \\
\mc {1}{|c|}{} &
\mc {1}{c}{LDA} &
\mc {1}{c|}{PM} &
\mc {1}{c}{LDA} &
\mc {1}{c|}{PM} &
\mc {1}{c}{LDA} &
\mc {1}{c|}{PM} \\ \hline\hline 
 $\Gamma_1$              & 0.82       & 0.81 & $1.02\div 1.12$ & 1.00 & 1.21 & 1.33 \\
 $\Gamma_2$              & 0.16       & 0.13 & 0.21            & 0.18 & 0.19 & 0.26 \\
 $\Gamma_{\rm NM}$       & 0.98       & 0.94 & $1.23\div 1.33$ & 1.19 & 1.40 & 1.59 \\
 \hline
\mc {1}{|c|}{$\Gamma^{\rm exp}_{NM}$} &
\mc {2}{c|}{$0.94\div 1.07$} &
\mc {2}{c|}{$1.20\div 1.30$} &
\mc {2}{c|}{$1.45\div 1.70$} \\ \hline
\end{tabular}
\end{center}
\end{table} 
Both calculations have been carried out with $g^{\prime}=0.8$ and
$g^{\prime}_{\Lambda}=0.4$, and reproduce with the same accuracy the data. 
For technical reasons, the OBL calculation does not allow to precisely identify
the partial rates $\Gamma_1$ and $\Gamma_2$ which contribute to the total 
$\Gamma_{\rm NM}=\Gamma_1+\Gamma_2$. In fact, one cannot separate in the imaginary parts of 
the diagrams (b)--(f) of Fig.~\ref{onelooponeloop} the contributions coming
from cuts on {\sl p--h} and {\sl 2p--2h} states, and hence the partial width 
($\Gamma_2$) stemming from the two--nucleon induced decay.
The values listed in the table for $\Gamma^{\rm OBL}_2$ have been obtained 
from the total imaginary part of the diagrams 
\ref{onelooponeloop}(b)--\ref{onelooponeloop}(f) 
[namely from the last two terms in the right hand side of Eq.~(\ref{Alpha1})]. 
In this approximation, $\Gamma^{\rm OBL}_1=\Gamma_{\rm ring}$ [second, third and fourth 
terms in the right hand side of Eq.~(\ref{Alpha1})]. 
As a matter of fact, one would expect that $\Gamma_2$
increases with $k_F$ (and this is the case for the PM calculation), but the OBL results
do not follow this statement. From table~\ref{newold} and from the study of the
$g^{\prime}$--dependence of $\Gamma^{\rm OBL}_2$, which has not been discussed here,
the only reasonable conclusion we can draw on the two--body induced processes
in OBL approximation is that for $1.1$ fm$^{-1}\lsim k_F\lsim 1.36$ fm$^{-1}$
and $g^{\prime}\simeq 0.8$, $\Gamma_2/\Gamma^{\rm free}_{\Lambda}=0.1\div 0.3$, in
agreement with the results of table \ref{sat}
obtained for finite nuclei with the phenomenological model in LDA.

We conclude by noticing that the PM results at fixed $k_F$ of table~\ref{newold} 
are consistent, when we follow the mass classification of table~\ref{kmed}, 
with the ones for finite nuclei obtained with the same model in LDA and presented in
subsection \ref{pheres} (see table~\ref{newold1}).
There is only some disagreement (at the level of 12\% on $\Gamma_{\rm NM}$) 
for $k_F=1.36$ fm$^{-1}$. This comparison provides an indication of the reliability
in using fixed Fermi momenta to simulate the $\Lambda$ decay in finite nuclei.

\newpage

\section{The $\Gamma_n/\Gamma_p$ puzzle}
\label{newpuzzle}
\subsection{Introduction}

The most relevant open problem in the study of the weak hypernuclear decay is to 
understand, theoretically, the large experimental values of the ratio $\Gamma_n/\Gamma_p$.
Actually, the large experimental uncertainties involved in the extraction of 
the ratio do not allow to reach any definitive
conclusion. The data are quite limited and not precise due to the difficulty of
detecting the products of the non--mesonic decays, especially the
neutrons. Moreover, up to now it has not been possible to distinguish between
nucleons produced by the one--body induced and the 
(non--negligible) two--body induced decay mechanism.

The Polarization Propagator Method used
to obtain the results discussed in subsections \ref{pheres} and
\ref{micres} does not distinguish between neutron-- and proton--induced
processes, but makes an ``average'' over these reactions.
However, within a $\Lambda N\to NN$ OPE model, a simple
counting of the isospin factors in the diagrams contributing to the
non--mesonic width at lowest order (low density limit) and for a $\Lambda$ at rest,
gives \cite{Os90} $\Gamma_n/\Gamma_p\simeq N/(14\,Z)$ for $N, Z \lsim 10$
($N$ and $Z$ are the number of neutrons and protons of the hypernucleus, respectively)
when only the (dominating) parity--conserving part of the $\Lambda \pi N$ 
vertex is taken into account. For heavier systems,
a nearly constant ratio ($\simeq 1/14$) is expected, as a result of the saturation of the
$\Lambda n\rightarrow nn$ and $\Lambda p\rightarrow np$ interactions. The inclusion of the
$\Lambda \pi N$ parity--violating term tends to increase the OPE ratio \cite{Os01}. 
As we have seen in section \ref{sumexpth}, more refined
calculations in OPE agree with the previous naive expectation, with values in the interval:
\begin{equation}
\label{ope-new}
\left[ \frac{\Gamma_n}{\Gamma_p} \right]^{\rm OPE} \simeq 0.05\div 0.20 ,
\end{equation}
for all the considered systems. The small OPE ratios are due to
the $\Delta I=1/2$ rule, which fixes the vertex ratio
$V_{\Lambda \pi^-p}/V_{\Lambda \pi^0n}=-\sqrt{2}$ (both in $S$-- and $P$--wave
interactions), and to the particular form of
the OPE potential, which has a strong tensor and weak central and parity--violating
components: the large tensor transition $\Lambda N(^3S_1)\rightarrow NN(^3D_1)$
requires, in fact, $I=0$ $np$ pairs in the anti--symmetric final state.
In $p$--shell and heavier hypernuclei the relative $\Lambda N$ $L=1$ state is found
to give only a small contribution to tensor transitions for the neutron--induced
decay, so it cannot improve the ratio (\ref{ope-new}).
The contribution of the $\Lambda N$ $L=1$ relative state to $\Gamma_n+\Gamma_p$
seems to be of about $5\div 15$\% in $p$--shell hypernuclei \cite{Be92,It98,Pa01}.
For these systems we expect the dominance of the $S$--wave interaction in the initial state,
due to the small $\Lambda N$ relative momentum. 

By using again a simple argument about the isospin structure of the $\Lambda N\rightarrow NN$
interaction in OPE, it is possible to estimate that for pure $\Delta I=3/2$ transitions
(for which $V_{\Lambda \pi^-p}/V_{\Lambda \pi^0n}=1/\sqrt{2}$) the OPE ratio
is increased by a factor $\simeq 2.5$ with respect to the value obtained for
pure $\Delta I=1/2$ transitions. On the other hand, the OPE model
with $\Delta I=1/2$ couplings has been able to reproduce the one--body stimulated 
non--mesonic rates $\Gamma_{1}=\Gamma_n+\Gamma_p$ for $s$-- and $p$--shell hypernuclei
\cite{It95,Pa97,Ok98,It98,Pa01}. Hence, the problem rather consists in overestimating the
proton--induced rate and underestimating the neutron--induced one.

Other ingredients beyond the OPE might be
responsible for the large experimental ratios.
A few calculations with $\Lambda N \rightarrow NN$ transition potentials including
heavy--meson--exchange or direct quark contributions
have improved the situation, without providing, nevertheless, a 
satisfactory theoretical explanation of the puzzle: 
very recent evaluations showed the importance of both $K$--meson--exchange
\cite{Ok99,Os01,Pa01} and direct quark mechanism \cite{Ok99} to obtain larger ratios.
The tensor component of $K$--exchange has opposite sign with respect to the one
for $\pi$--exchange, resulting in a reduction of $\Gamma_p$. The parity violating
$\Lambda N(^3S_1)\to NN(^3P_1)$ transition, which contributes to both the $n$--
and $p$--induced processes, is considerably enhanced by $K$--exchange 
and direct quark mechanism and tends to increase $\Gamma_n/\Gamma_p$ \cite{Ok99,Pa01}.

In table~\ref{best-ratio} we summarize the calculations
that predicted ratios considerably enhanced with respect to the 
OPE values. Experimental data are given for comparison.
\begin{table}[thb]
\caption{$\Gamma_n/\Gamma_p$ ratio.}
\label{best-ratio}
\begin{center}
\begin{tabular}{|c|c|c|c|} \hline
\mc {1}{|c|}{Ref. and Model} &
\mc {1}{c|}{$^5_{\Lambda}$He} &
\mc {1}{c|}{$^{12}_{\Lambda}$C} &
\mc {1}{c|}{Nuclear Matter} \\ \hline\hline
Itonaga {\em et al.}~1998 \cite{It98}    & & 0.36 & \\
($\pi + 2\pi/\rho + 2\pi/\sigma$) & &      & \\ \hline
Sasaki {\em et al.}~2000 \cite{Ok99} &0.701 & &0.716 \\
($\pi +K+$ DQ)               & & & \\ \hline
Jun {\em et al.}~2001 \cite{Ju01} &1.30 &1.14 & \\
(OPE + 4BPI)                  & & & \\ \hline
Jido {\em et al.}~2001 \cite{Os01} & &0.53 & \\
($\pi +K+2\pi +\omega$)      & & & \\ \hline
Parre\~{n}o--Ramos~2001 \cite{Pa01} &$0.343\div 0.457$ & $0.288\div 0.341$ & \\
($\pi +\rho +K + K^* + \omega +\eta$)                        & & & \\ \hline\hline
Exp 1974 \cite{Mo74}     &                & $0.59\pm0.15$   &    \\ \hline  
Exp BNL 1991 \cite{Sz91} &$0.93\pm0.55$ &$1.33^{+1.12}_{-0.81}$ & \\ \hline
Exp KEK 1995 \cite{No95} & &$1.87^{+0.67}_{-1.16}$ & \\ \hline
Exp KEK 1995 \cite{No95a} &$1.97\pm0.67$ & & \\ \hline
Exp KEK 2001 \cite{Sat99,Ha01} & &$1.17^{+0.22}_{-0.20}$ & $^{56}_{\Lambda}$Fe: $2.54^{+0.61}_{-0.81}$ \\ \hline
\end{tabular}
\end{center}
\end{table}
\begin{table}[thb]
\caption{Non--mesonic width $\Gamma_n+\Gamma_p$ (in units of $\Gamma^{\rm free}_{\Lambda}$).}
\label{best-nm}
\begin{center}
\begin{tabular}{|c|c|c|c|} \hline
\mc {1}{|c|}{Ref. and Model} &
\mc {1}{c|}{$^5_{\Lambda}$He} &
\mc {1}{c|}{$^{12}_{\Lambda}$C} &
\mc {1}{c|}{Nuclear Matter} \\ \hline\hline
Itonaga {\em et al.}~1998 \cite{It98}    & & 1.05 & \\
($\pi + 2\pi/\rho + 2\pi/\sigma$) & &      & \\ \hline
Sasaki {\em et al.}~2000 \cite{Ok99} &0.519 & &2.456 \\
($\pi +K+$ DQ)               & & & \\ \hline
Jun {\em et al.}~2001 \cite{Ju01} &0.426 &1.174 & \\
(OPE + 4BPI)                  & & & \\ \hline
Jido {\em et al.}~2001 \cite{Os01} & &0.769 & \\
($\pi +K+2\pi +\omega$)      & & & \\ \hline
Parre\~{n}o--Ramos~2001 \cite{Pa01} &$0.317\div 0.425$ & $0.554\div 0.726$ & \\
($\pi +\rho +K + K^* + \omega +\eta$)                        & & & \\ \hline\hline
Exp BNL 1991 \cite{Sz91} &$0.41\pm0.14$    &$1.14\pm0.20$ & \\ \hline
Exp CERN 1993 \cite{Ar93}   & & & $\bar{p}+$Bi: $1.46^{+1.83}_{-0.52}$ \\
                            & & & $\bar{p}+$U:  $2.02^{+1.74}_{-0.63}$ \\ \hline
Exp KEK 1995 \cite{No95} & &$0.89\pm0.18$ & \\ \hline
Exp KEK 1995 \cite{No95a} &$0.50\pm0.07$ & & \\ \hline
Exp COSY 1998 \cite{Ku98}    & & & $p+$Bi: $1.63^{+0.19}_{-0.14}$  \\ \hline
Exp KEK 2000 \cite{Bh98,Ou00} & &$0.83\pm 0.11$ & $^{56}_{\Lambda}$Fe: $1.22\pm 0.08$ \\ \hline
Exp COSY 2001 \cite{Ka01}   & & & $p+$Au: $2.02^{+0.56}_{-0.35}$ \\ \hline
Exp COSY 2001 \cite{Ku01}   & & & $p+$U:  $1.91^{+0.28}_{-0.22}$ \\ \hline
\end{tabular}
\end{center}
\end{table}
Almost all calculations reproduce the observed non--mesonic widths 
$\Gamma_n+\Gamma_p$, as one can see in table~\ref{best-nm}: only Parre\~{n}o and 
Ramos tends to underestimate the data for $^{12}_{\Lambda}$C, whereas 
Sasaki {\em et al.}~overestimate the most accurate experiments for very heavy hypernuclei.
Itonaga {\em et al.}~predict a too small $\Gamma_n/\Gamma_p$.
The results of Sasaki {\em et al.}~for  $\Gamma_n/\Gamma_p$ and $\Gamma_n+\Gamma_p$
in $^5_{\Lambda}$He are compatible with data, but for nuclear matter the authors 
underestimate $\Gamma_n/\Gamma_p$ and overestimate $\Gamma_n+\Gamma_p$. The phenomenological
fit of Jun {\em et al.}~reproduces $\Gamma_n/\Gamma_p$ and $\Gamma_n+\Gamma_p$
for $^5_{\Lambda}$He and $^{12}_{\Lambda}$C. However, the values of some of
the coupling constants of their 4--baryon point interaction, which are required to fit
the data, are questionable. Jido {\em et al.}~give a ratio for $^{12}_{\Lambda}$C
compatible with the lower limits of the data. Finally, Parre\~{n}o and Ramos
obtain a ratio compatible with the lower limits of the data for 
$^5_{\Lambda}$He but they underestimate the experiments for $^{12}_{\Lambda}$C.
Clearly, a variety of situations, sometimes contradictory, which give a flavour of the
difficulties inherent to $\Gamma_n/\Gamma_p$.

\subsection{Two--body induced decay and nucleon final state interactions}
\label{FSI2b}
The analysis of the ratio ${\Gamma}_n/{\Gamma}_p$ is influenced by the  
two--nucleon induced process ${\Lambda}NN\rightarrow NNN$, whose experimental 
identification is rather difficult and it is a challenge for the future.  
By assuming that the meson produced in the weak  
vertex is mainly absorbed by an isoscalar $NN$ correlated pair 
(quasi--deuteron approximation), the three--body process turns out to be
${\Lambda}np\rightarrow nnp$, so that    
a considerable fraction of the measured neutrons 
could come from this channel and not only from $\Lambda n \rightarrow nn$ 
and $\Lambda p \rightarrow np$. In this way it might be 
possible to explain the large experimental ${\Gamma}_n/{\Gamma}_p$ ratios, 
which originally have been analyzed without taking into account  
the two--body stimulated process. 
Nevertheless, the situation is far from being clear and simple, both from the  
theoretical and experimental viewpoints. 
The new non--mesonic mode was introduced in Ref.~\cite{Al91} and its calculation 
was improved in Ref.~\cite{Ra95}, where the authors found that the inclusion of the 
new channel would bring to extract from the experiment
even larger values for the ${\Gamma}_n/{\Gamma}_p$ ratios, thus worsening the 
disagreement with the theoretical estimates. However, in the 
hypothesis that only two out of the three nucleons coming from the three--body decay are  
detected, the reanalysis of the experimental data would lead back to smaller ratios 
\cite{Os98}. The above hypothesis is plausible for the following reason. 
The two--body induced decay mode takes place when the pion emitted by the  
$\Lambda$ vertex is not too far from being on its renormalized 
mass--shell (on the contrary, the particle--hole region of the in--medium 
pion excitation spectrum, which contributes to the one--body 
induced decay, would be quite far from the pionic branch in the medium). 
It occurs that the pionic branch (which is a delta function on the 
energy--momentum dispersion relation in free space) is renormalized in the medium and  
has a width associated to its  capability in exciting {\sl 2p--2h} states. Part of this strength 
overcomes the Pauli blocking, giving rise to the two--body induced decay.  
As a consequence of the emission of an ``almost on--shell'' pion, 
the nucleon coming out from the $\Lambda$ vertex will have a small kinetic energy 
($T_N\simeq 5$ MeV for a rigorously on--shell pion) and hence will be,
most probably, below the experimental detection threshold, which was around
$30\div 40$ MeV in the experiments quoted in table \ref{best-ratio}.

These observations show that ${\Gamma}_n/{\Gamma}_p$ is sensitive to the detection threshold 
and to the detailed kinematics of the process. For instance, the calculated  
energy spectra of the emitted nucleons clearly display the above statement 
about the slow nucleon emitted in the weak vertex \cite{Ra97};  
their calculation also requires a careful treatment of the nucleon final state interactions. 
In Ref.~\cite{Ra97} the nucleon energy distributions have been  
calculated by using a Monte Carlo simulation to describe nucleon's
rescattering inside the nucleus: the ratio 
$\Gamma_n/\Gamma_p$ has been taken as a free parameter and extracted  
by comparing the simulated spectra with the experimental data. The momentum distributions of 
the primary nucleons were determined within the polarization propagator scheme discussed 
in subsection \ref{pm}. In their way out of the nucleus, the nucleons, due to the collisions 
with other nucleons, continuously change energy, direction, charge, and secondary  
nucleons are emitted as well. Then, the energy distribution of the observable nucleons,  
which also loose their energy by the interactions with 
the experimental set--up, is different from the one at the level of the primary nucleons.
The shape of the proton spectrum obtained in Ref.~\cite{Ra97} is sensitive to 
the ratio $\Gamma_n/\Gamma_p$. The protons from the three--nucleon 
mechanism $\Lambda NN\rightarrow NNN$ appear mainly at low energies, while,
for $^{12}_{\Lambda}$C, those from the one--nucleon stimulated
process peak around 75 MeV. Since the experimental spectra show a 
fair amount of protons in the low energy region, they would favour a relatively larger 
two--body induced decay rate and/or a reduced number of protons from the 
one--body induced process. Consequently, for $\Gamma_2=0.27\, \Gamma^{\rm free}_{\Lambda}$
the authors of Ref.~\cite{Ra97} found that
the experimental spectra of Refs.~\cite{Mo74,Sz91} were compatible with 
values of $\Gamma_n/\Gamma_p$ around $3$ for $^{12}_{\Lambda}$C, in strong contradiction
with the theoretical predictions. However, by using available data 
on the total number of emitted neutrons and protons, the same calculation shows 
that the experimental error bars on $\Gamma_n/\Gamma_p$ are increased by the 
inclusion of the three--body channel, leading to values which, within one standard deviation, 
can be even compatible with the OPE values. In Ref.~\cite{Ra97} it was also pointed out 
the convenience of measuring the number of outgoing protons per decay event. 
This observable, which can be measured from delayed fission events in the decay of heavy 
hypernuclei, gives a more reliable neutron to proton ratio 
and it is less sensitive to the details of the intra--nuclear cascade 
calculation determining the final shape of the spectra. 

\begin{figure}  
\begin{center} 
\mbox{\epsfig{file=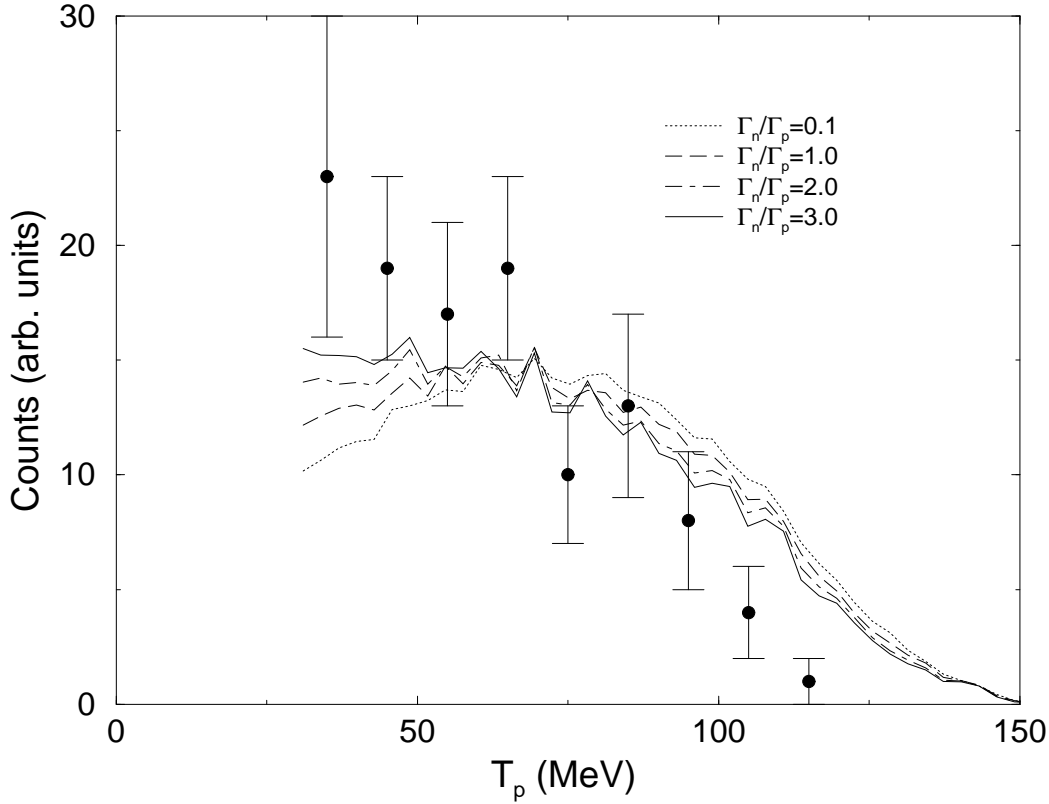,width=0.9\textwidth}}  
\vskip 2mm 
\caption{Proton spectrum from the decay of $^{12}_\Lambda$C for different values 
of $\Gamma_n/\Gamma_p$. The experimental data are from Ref.~\protect\cite{Sz91}
(taken from Ref.~\cite{Al99}).}
\label{fig:spec} 
\end{center} 
\end{figure}

\begin{figure} 
\begin{center} 
\mbox{\epsfig{file=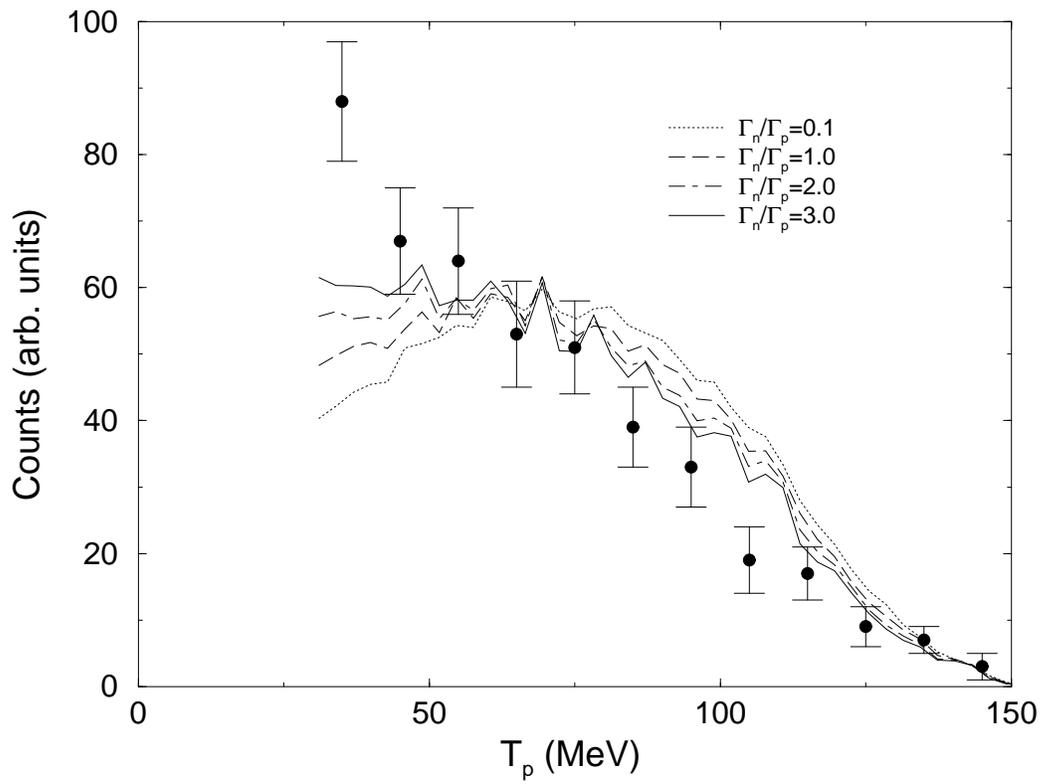,width=0.9\textwidth}} 
\vskip 2mm 
\caption{Proton spectrum from the decay of $^{12}_\Lambda$C for different values 
of $\Gamma_n/\Gamma_p$. The experimental data are taken from  
Ref.~\protect\cite{Mo74}.} 
\label{fig:mo} 
\end{center} 
\end{figure}

The excellent agreement of the calculations discussed in subsection \ref{pheres} 
for the experimental total non--mesonic decay rates
made it worth to explore again the predictions for the nucleon spectra \cite{Al99}. 
The question is whether the model used in \ref{pheres} affects the momentum distribution of 
the primary emitted nucleons strongly enough to obtain good agreement with 
the experimental proton spectra without requiring 
very large values for $\Gamma_n/\Gamma_p$.
The nucleon spectra from the decay of several hypernuclei have been thus
generated by using the Monte Carlo simulation of Ref.~\cite{Ra97}.
The spectra obtained for various values of $\Gamma_n/\Gamma_p$, used again as a 
free parameter, are compared in Fig.~\ref{fig:spec} (Fig.~\ref{fig:mo}) 
with the data from the BNL experiment of Ref.~\cite{Sz91} (from Ref.~\cite{Mo74}). 
\noindent We remark that, although the total non--mesonic widths are smaller  
than those of Ref.~\cite{Ra97} by about 35\%, the resulting nucleon 
spectra, once they are normalized to the 
same non--mesonic rate, are practically identical. 
The reason is that the ratio $\Gamma_2/\Gamma_1$ of  
two--body induced versus one--body induced decay rates is
essentially the same in  
both models (between 0.2 and 0.15 from medium to heavy hypernuclei), and the  
momentum distributions for the primary emitted protons are also very similar. 
As a consequence, the conclusions drawn in Ref.~\cite{Ra97} still hold and
the new calculation also favours very large values of $\Gamma_n/\Gamma_p$
when compared with experimental spectra.

On the basis of the above considerations, 
the origin of the discrepancy between theory and experiment for
$\Gamma_n/\Gamma_p$ is far from being resolved. On the theoretical side, 
there is still room for improving the numerical simulation of the nuclear 
final state interactions: Coulomb distortion, multiple scattering and the evaporating process 
should be incorporated in the calculation. In particular, multiple scattering 
and the evaporating process are important 
ingredients, which increase the nucleon spectra at low
energies. On the experimental side, although more recent spectra are available 
\cite{No95,Bh98a}, they have not been corrected for the energy losses inside 
the target and detector as well as for the geometry of the detector, so a direct
comparison with theoretical predictions is not possible.  

Attempts to incorporate these corrections by combining a theoretical 
model for the nucleon rescattering in the nucleus with a simulation of the 
interactions in the experimental set--up have been done at KEK \cite{Sat99,Bh01,Ha01}. 
The results reported in Refs.~\cite{Sat99,Ha01} show that $\Gamma_n/\Gamma_p$ increases
with the hypernucleus mass number, with values in the range $1\div 3$ for $^{12}_{\Lambda}$C,
$^{28}_{\Lambda}$Si and $^{56}_{\Lambda}$Fe. In the next paragraph we shall
discuss in detail these recent results.

A decisive forward step towards a
clean extraction of $\Gamma_n/\Gamma_p$ would be obtained if the nucleons from the different
non--mesonic processes, $\Lambda N \to NN $ and $\Lambda NN \to NNN$, were disentangled.
Through the measurement of coincidence spectra and angular correlations 
of the outgoing nucleons, it could be possible, in the near future, 
to split the non--mesonic decay width into its two components 
${\Gamma}_1$ and ${\Gamma}_2$ \cite{Ze98,FI98,Ou00a} and obtain a more precise
and direct measurement of $\Gamma_n/\Gamma_p$. We shall discuss this important point
more extensively in paragraph \ref{poten}. By using a simple argument about the 
detection efficiency in coincidence measurements, the authors of  
Ref.~\cite{Ze98} evaluated the influence of the final state interactions 
and of the two--body induced process on $\Gamma_{\rm NM}$ to be
of $(15\pm 15$)\% for $s$--shell hypernuclei. 
According to the calculation reported in table \ref{sat}, for $^5_{\Lambda}$He the effect of 
the two--body stimulated decay alone is $\Gamma_2/\Gamma_{\rm NM}= 13$\%. 

\subsubsection{Recent experimental spectra}
\label{recexp}
\begin{figure}
\begin{center}
\mbox{\epsfig{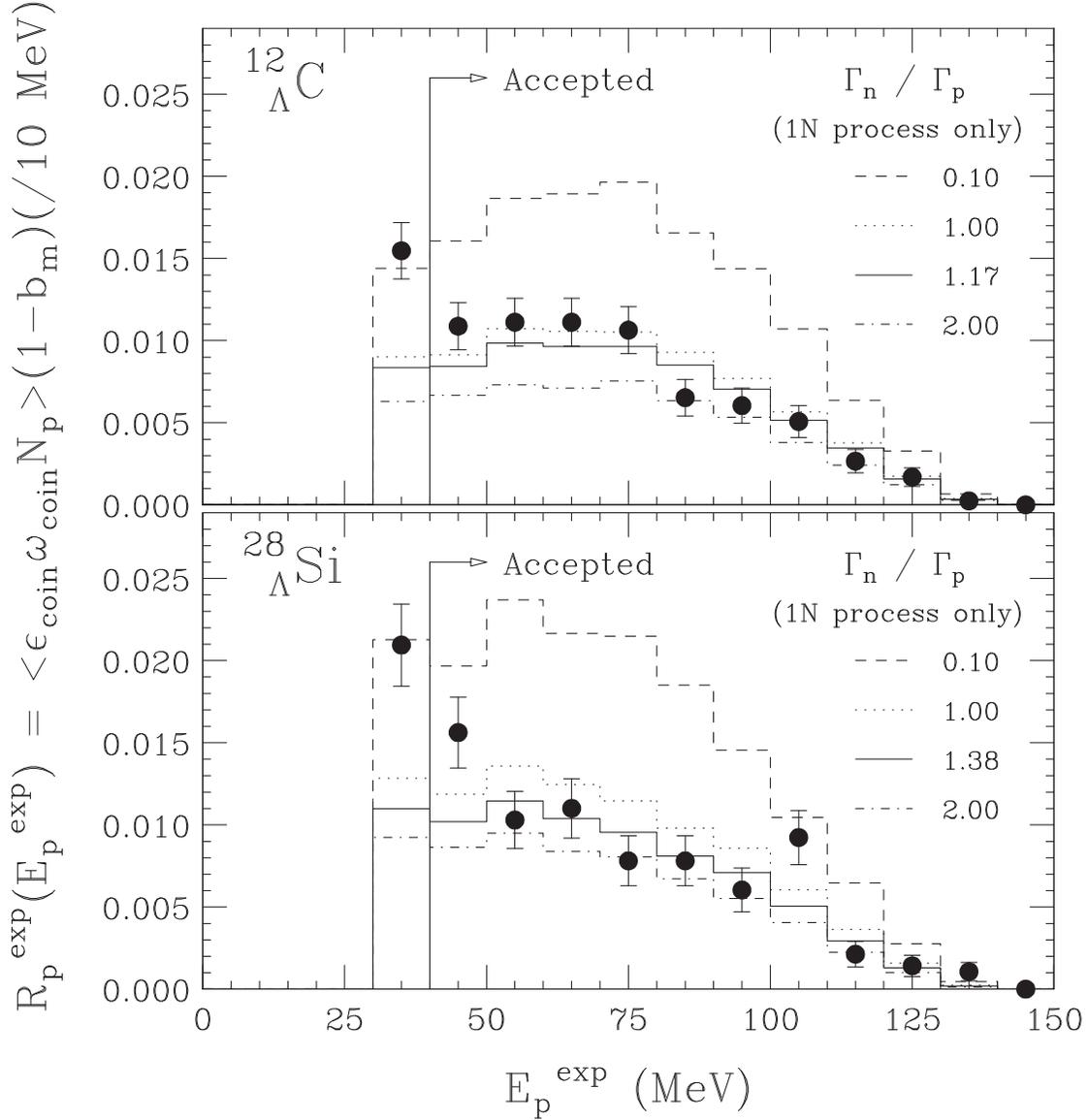}}
\vskip 2mm
\caption{Proton energy spectra measured at KEK--E307 (taken from Ref.~\cite{Ha01})}
\label{kek-307}
\end{center}
\end{figure}

Very recently, at KEK--E307 \cite{Sat99,Ha01}, the proton spectra 
for $^{12}_{\Lambda}$C, $^{28}_{\Lambda}$Si and $^{56}_{\Lambda}$Fe have been 
measured and compared with theoretical simulations of the intra--nuclear 
cascades after the weak processes, obtained with the MC code of Ref.~\cite{Ra97}. 
Corrections for the detector geometry and the
nucleonic interactions inside the target and detector materials
have also been implemented, through a GEANT MC code. The proton energy spectra 
have been measured by means of a coincidence counter system identifying
the hypernuclear production instant time through the detection of the
kaon emitted in the $n(\pi^+,K^+)\Lambda$ production reaction. 
In figure \ref{kek-307} the spectra obtained for $^{12}_{\Lambda}$C and $^{28}_{\Lambda}$Si
are shown. The vertical axes are normalized to the number of protons per hypernuclear
decay.

The results of KEK--E307 supply a $\Gamma_n/\Gamma_p$ ratio, again estimated
by fitting the proton spectra, which increases with the mass number \cite{Sat99,Ha01}:
\begin{eqnarray}
\label{newratio}
\frac{\Gamma_n}{\Gamma_p}(^{12}_{\Lambda}{\rm C})&=&1.17^{+0.22}_{-0.20} ,  \\
\frac{\Gamma_n}{\Gamma_p}(^{28}_{\Lambda}{\rm Si})&=&1.38^{+0.30}_{-0.27} , \nonumber \\
\frac{\Gamma_n}{\Gamma_p}(^{56}_{\Lambda}{\rm Fe})&=&2.54^{+0.61}_{-0.81} , \nonumber
\end{eqnarray}
the last one being preliminary.
Because of the non--zero experimental energy threshold for proton detection, 
the obtained fits for $\Gamma_n/\Gamma_p$ turn out to be slightly sensitive 
to the two--body induced process, although this mechanism gives a non--negligible 
contribution to the non--mesonic rate: for $^{12}_{\Lambda}$C and $^{28}_{\Lambda}$Si,
the central values of $\Gamma_n/\Gamma_p$ are reduced 
by about 16\% when the two--body induced process is taken into account, with  
results anyhow compatible within the error bars in both descriptions \cite{Ha01}.
The fits including the two--nucleon stimulated decay 
have been performed by using $\Gamma_2/(\Gamma_1 +\Gamma_2) \simeq 0.3$ as input.
This small effect of the two--nucleon stimulated decay is principally due
to the rather large value of the proton energy detection threshold 
($E_{\rm Th}\simeq 40$ MeV) with respect to the average energy of the protons
from the process $\Lambda NN \to NNN$.
The results reported in Eq.~(\ref{newratio}) and figure \ref{kek-307} refer to the analysis in which 
only the one--nucleon induced process is taken into account. The fits which
included the two--nucleon induced processes lead to:
${\Gamma_n}/{\Gamma_p}(^{12}_{\Lambda}{\rm C})=0.96^{+0.24}_{-0.23}$ and
${\Gamma_n}/{\Gamma_p}(^{28}_{\Lambda}{\rm Si})=1.18^{+0.33}_{-0.31}$.
At the present level of precision, the observation of signals from 
the two--body induced decay is thus impossible. 
However, the degree of accuracy of the new KEK measurements   
allowed to significantly improve the error bars with respect to the previous 
experiments (see data listed in table~\ref{best-ratio}): this leads to exclude neutron to proton 
ratios smaller than $0.73$ ($0.50$) at the $1\sigma$ ($2\sigma$) level for 
$^{12}_{\Lambda}$C, in the analysis including the two--nucleon induced process. 

We want to make the following remark on the mass dependence of the KEK--E307 results. 
The ratio $\Gamma_n/\Gamma_p$ sizeably increases in going from 
$^{28}_{\Lambda}$Si \cite{Ha01} to $^{56}_{\Lambda}$Fe \cite{Sat99}
(we remind the reader that the data for iron are only preliminary).
This is in disagreement with the well known behaviour of the $\Lambda N\rightarrow NN$ 
interaction in nuclei, namely its saturation for large mass numbers. In fact, should we 
estimate, as in paragraph \ref{heavy}, the mesonic rate for $^{28}_{\Lambda}$Si
and $^{56}_{\Lambda}$Fe to be $0.07$ and $0.01$, respectively (here and in the following the 
decay rates are in units of $\Gamma^{\rm free}_{\Lambda}$) and use the total decay 
rates measured in the same experiments \cite{Bh98}, then for the central values of the  
non--mesonic decay width we would obtain:
$\Gamma_{\rm NM}(^{28}_{\Lambda}{\rm Si})= 
\Gamma_{\rm NM}(^{56}_{\Lambda}{\rm Fe}) = 1.21$. 
This value, together with the ratios of Eq.~(\ref{newratio}), provides:   
$\Gamma_n(^{28}_{\Lambda}{\rm Si})= 0.70$, 
$\Gamma_p(^{28}_{\Lambda}{\rm Si})= 0.51$,
$\Gamma_n(^{56}_{\Lambda}{\rm Fe})= 0.87$ 
and $\Gamma_p(^{56}_{\Lambda}{\rm Fe})= 0.34$. 
As a consequence, $\Gamma_n$ and $\Gamma_p$ do not follow the saturation behaviour
(the contrary occurs for the observed total rate $\Gamma_{\rm NM}=\Gamma_n+\Gamma_p$),  
which predicts $n$-- and $p$--stimulated rates increasing with $N$ and $Z$, respectively, 
and saturating for $N, Z\simeq 10$: $\Gamma_n/\Gamma_p\simeq 
\Gamma_n^{\rm sat}/\Gamma_p^{\rm sat}$ for $N, Z \gsim 10$. 
Since one also expects a saturation for the neutron-- and proton--induced
decay rates separately, this result could
represent a signal of a systematic error in the experimental analysis
employed to extract the $\Gamma_n/\Gamma_p$ ratios.

\subsubsection{Possible improvements}
\label{improvements}
As discussed above, the $\Gamma_n/\Gamma_p$ ratios of Eq.~(\ref{newratio}), 
extracted from the recently measured proton spectra, confirm 
the results of previous experiments: the neutron--
and proton--induced decay rates are of the same order of magnitude over a large
hypernuclear mass number range. However, since the new experiments have
significantly improved the quality of the data, small values of $\Gamma_n/\Gamma_p$
(say smaller than $0.7$ for $^{12}_{\Lambda}$C) are now excluded with good precision. 

After having inspected both the experimental procedures
and the theoretical models until now developed to determine the ratio, we want to 
summarize here some interesting features which could bring to future perspectives:
\begin{itemize}
\item [1)] The proton spectra originating from neutron-- and proton--induced
processes (and, eventually, from two--nucleon stimulated decays)
are added {\it incoherently} in the Monte Carlo intra--nuclear cascade
calculations used to determine $\Gamma_n/\Gamma_p$. In this way a possible 
quantum--mechanical {\it interference} effect between the two channels 
is lost. Therefore, in an experiment like KEK--E307, in which 
only charged particles are detected, one cannot go back up to the decay
mechanism which produced an observed proton. The conclusion of 
paragraph \ref{recexp} about a possible systematic error in the experimental
analyses of $\Gamma_n/\Gamma_p$ could then be due to the incorrect procedure
of summing incoherently, i.e. in a classical picture,
the proton spectra from the $\Lambda n\to nn$ and $\Lambda p\to np$
processes.

\item[2)] The experimental spectra of figure \ref{kek-307} 
do not exhibit a peak around the energy ($\simeq 75$ MeV) which corresponds to the 
kinematical situation of back--to--back nucleon pairs coming from
one--nucleon induced decays. The shape of the spectra
just above the 30 MeV detection threshold are quite flat, or even
decreasing for increasing energy, and are not well fitted by the simulations
used to extract $\Gamma_n/\Gamma_p$. This is in principle due to different, hardly
distinguishable, effects: 1) the nucleon energy losses in the nucleus,
2) the nucleon energy losses in the target and detector materials,
and 3) the relevance of the two--nucleon stimulated non--mesonic decay.
Because of the present level of experimental accuracy,
an analysis which takes into account the two--nucleon induced
decay {\it alone} cannot improve the comparison between experiment and theory.
It would then be advisable to explore the effects of
{\it robust} nucleon final state interactions on the simulations
used to determine the ratio from the experimental spectra.

\item [3)] The direct observation of the three--body emission 
events is quite difficult and up to now no signal
has been found. The calculated nucleon spectra \cite{Ra97} for this channel present a maximum at
energies below the detection threshold, and only a fraction (about 40\% for
$\Gamma_n/\Gamma_p=1$ and $E_{\rm Th}= 40$ MeV) of nucleons from three--body emission
can be detected. Moreover, for $E>E_{\rm Th}$ the nucleon
distribution from one--body induced processes superimposes to the previous one.
The spectrum simulated for the two--nucleon emission dominates and,
for $^{12}_{\Lambda}$C, peaks at an energy ($\simeq 75$ MeV) that corresponds
to the situation in which the two nucleons come out back--to--back. This observation shows that the
separation of the nucleons from the two non--mesonic channels is only possible by
{\it angular correlation measurements}. In Ref.~\cite{Ze98} the authors studied how the back--to--back
kinematics is able to select the one--nucleon induced process, and 
$NN$ coincidence measurements (of energies and
angular distributions) are expected in the near future 
at Da$\Phi$ne \cite{FI98}, KEK \cite{Ou00a} and BNL \cite{Gi01}.

\item [4)] In order to disentangle the two--nucleon induced decay events from
the one--nucleon induced ones, the direct observation of the outgoing {\it neutrons}
is thus needed. Neutron spectra can be measured down to about 10 MeV kinetic
energy, since they are less affected than the proton ones by 
energy losses in the target and detector materials. 
The joint observation of proton and neutron spectra could then help to disentangle 
the set--up material effects from the nucleon final state interactions 
occurring inside the residual nucleus.
A very recent experiment, KEK--E369, measured neutron spectra from $^{12}_{\Lambda}$C
and $^{89}_{\Lambda}$Y non--mesonic decays \cite{Bha01}. A preliminary analysis of
data is consistent with a ratio in the range $0.5\div 1$ for $^{12}_{\Lambda}$C,
obtained through a new intra--nuclear calculation. The newly developed Monte Carlo code 
with the same range of $\Gamma_n/\Gamma_p$ values also seems to be able to 
reproduce the $^{12}_{\Lambda}$C proton spectrum observed at KEK--E307.
These analyses have been performed by neglecting the contribution of the two--nucleon
induced decay.

\item [5)] In our opinion, the key point to 
avoid the possible deficiencies of the {\it single} nucleon spectra measurements 
discussed in point 1), will be to employ coincidence detections of the final
nucleons. Only such a procedure leads to 
a {\it direct} and unambiguous determination of $\Gamma_n/\Gamma_p$.
In the experiment KEK--E462 \cite{Ou00a}, an angular and energy correlation
measurement will study the decay of $^5_{\Lambda}$He hypernuclei.
Very low detection thresholds ($\simeq 10$ MeV for neutrons and $\simeq 20$ MeV for protons)
and statistics improved with respect to previous measurements will be used.
By using a light, spin--isospin saturated hypernucleus such as $^5_{\Lambda}$He, 
one has the good point that the nuclear final state interaction are considerably reduced. 
The use of low nucleon threshold energies will make it possible to observe 
essentially all the final state interactions effects. 
Another experiment, at BNL \cite{Gi01}, will measure $\Gamma_n/\Gamma_p$ for $^4_{\Lambda}$H,
again by $nn$ and $np$ coincidence measurements.
\end{itemize}

\subsubsection{Potentialities of coincidence experiments}
\label{poten}
We concentrate here on the potentialities of future experiments employing
double coincidence nucleon detection. The purpose is to stress the importance
of this kind of measurements both for the solution of the
$\Gamma_n/\Gamma_p$ puzzle and for the observation of two--nucleon stimulated decay
events.
 
A simplistic analysis supplies the following expressions for the numbers of
detected neutrons, $N_n=N^{\rm 1B}_n+N^{\rm 2B}_n$, and protons, 
$N_p=N^{\rm 1B}_p+N^{\rm 2B}_p$, in terms of the non--mesonic partial decay widths:
\begin{equation}
\label{numeri-n}
\begin{array}{l l}
N^{\rm 1B}_n = \displaystyle 
\epsilon_n \, \Omega_n \, R_n \, \frac{2\Gamma_n+\Gamma_p}{\Gamma_{T}} \, N , 
& \hspace{0.5cm}
N^{\rm 1B}_p = \displaystyle 
\epsilon_p \, \Omega_p \, R_p \, \frac{\Gamma_p}{\Gamma_{T}} \, N , \\
 & \\
N^{\rm 2B}_n = \displaystyle 
\epsilon_n \, \Omega_n \, R_n \, \frac{2\Gamma_2}{\Gamma_{T}} \, N ,
& \hspace{0.5cm}
N^{\rm 2B}_p = \displaystyle
\epsilon_p \, \Omega_p \, R_p \, \frac{\Gamma_2}{\Gamma_{T}} \, N .
\end{array}
\end{equation}
Here, $\epsilon_n$ ($\epsilon_p$) is the neutron (proton) detection efficiency,
$\Omega_n$ ($\Omega_p$) the detector acceptance for neutrons (protons) 
and $R_n$ ($R_p$) the fraction of outgoing neutrons (protons) with kinetic
energy above the detection threshold. The quantities $R_n$ and $R_p$ take
into account the nucleon rescattering effects in the nucleus, which influence,
as previously discussed, the numbers of observed neutrons and protons. 
Moreover, in the relations for the number of neutrons and protons originating from two--body 
induced decays, $N^{\rm 2B}_n$ and $N^{\rm 2B}_p$, we have employed the 
quasi--deuteron approximation, in which the three--body processes proceeds mainly 
through the channel $\Lambda np\to nnp$.   
By imposing $\epsilon_n \, \Omega_n \, R_n=\epsilon_p \, \Omega_p \, R_p=1$,
the previous equations supply the number of nucleons at the weak decay vertex.
Finally, $\Gamma_{\rm T}$ is the total decay width 
($\Gamma_{\rm T}=\Gamma_n+\Gamma_p+\Gamma_2+\Gamma_{\rm M}$) and 
$N$ the total number of decayed hypernuclei. In an experimental analysis, 
the ratio $\Gamma_n/\Gamma_p$ can then be obtained from the measurement of $N_n$, 
$N_p$, $N$ and $\Gamma_{\rm T}$ and the theoretical evaluation of $\Gamma_2$ as follows:
\begin{equation}
\label{n/p-simple}
\frac{\Gamma_n}{\Gamma_p}=
\frac{\alpha \,  \displaystyle \frac{N_n}{N_p} -1 + \left(\alpha \, \displaystyle \frac{N_n}{N_p} 
-2\right)\displaystyle \frac{\Gamma_2}{\Gamma_n+\Gamma_p}}
{2-\left(\alpha \, \displaystyle \frac{N_n}{N_p} -2\right)\displaystyle \frac{\Gamma_2}
{\Gamma_n+\Gamma_p}} ,
\end{equation}
where:
\begin{equation}
\Gamma_n+\Gamma_p= \frac{1}{2}\left(\frac{N_n}{\epsilon_n\, \Omega_n\, R_n}
+\frac{N_p}{\epsilon_p\, \Omega_p\, R_p}\right)\frac{\Gamma_{\rm T}}{N} -\frac{3}{2}\Gamma_2 ,
\end{equation}
and:
\begin{equation}
\alpha \equiv \frac{\epsilon_p \, \Omega_p \, R_p}{\epsilon_n \, \Omega_n \, R_n} . \nonumber
\end{equation}

On the other hand, the numbers of $nn$ and $np$ coincidence detections 
for an opening angle $\theta$ between the pairs are:
\begin{eqnarray}
\label{nn-np-1b}
N^{\rm 1B}_{nn}(\theta)&=& \epsilon_n^2 \, \Omega_{NN}(\theta)\,
f^{\rm 1B}_{NN}(\theta) \, R^{\rm 1B}_{nn} \, \frac{\Gamma_n}{\Gamma_{\rm T}}\, N ,  \\
N^{\rm 1B}_{np}(\theta)&=& \epsilon_n \, \epsilon_p \, \Omega_{NN}(\theta) \,
f^{\rm 1B}_{NN}(\theta) \, R^{\rm 1B}_{np} \, \frac{\Gamma_p}{\Gamma_{\rm T}}\, N , \nonumber
\end{eqnarray}
respectively, when two--body stimulated decays are neglected.
With $\Omega_{NN}(\theta)$ we have denoted the average acceptance
for nucleon pairs detected at an opening angle $\theta$, while $f^{\rm 1B}_{NN}(\theta)$
is the $NN$ angular correlation function for $\Lambda N\to NN$. 
Finally, $R^{1B}_{nn}$ and $R^{1B}_{np}$ are, respectively, 
the fraction of $nn$ and $np$ pairs from one--body induced processes 
leaving the residual nucleus with energies above the detection thresholds. The ratio
$\Gamma_n/\Gamma_p$ can then be measured through the following relation:
\begin{equation}
\label{n/p-new}
\frac{\Gamma_n}{\Gamma_p}=\frac{N^{\rm 1B}_{nn}}{N^{\rm 1B}_{np}}\, 
\frac{\epsilon_p}{\epsilon_n} \, \frac{R^{\rm 1B}_{np}}{R^{\rm 1B}_{nn}} ,
\end{equation}
$N^{\rm 1B}_{nn}$ and $N^{\rm 1B}_{np}$ being the {\it total} numbers of detected
$nn$ and $np$ pairs from one--body induced decays, respectively.

Angular two--nucleon correlation measurements allow to disentangle the
two--body stimulated decays from the total set of data. 
The number of $nn$ and $np$ pair detected at an angle $\theta$
and originating from three--body decays are:
\begin{eqnarray}
\label{nn-np-2b}
N^{2B}_{nn}(\theta)&=&\epsilon^2_n \, \Omega_{NN}(\theta)\,
f^{\rm 2B}_{NN}(\theta) \, R^{2B}_{nn} \, \frac{\Gamma_2}{\Gamma_{\rm T}}\, N , \\
N^{2B}_{np}(\theta)&=&\epsilon_n \, \epsilon_p \, \Omega_{NN}(\theta)\, 
f^{\rm 2B}_{NN}(\theta) \, R^{2B}_{np} \, \frac{2\Gamma_2}{\Gamma_{\rm T}}\, N ,
\nonumber
\end{eqnarray}
respectively,
the factor 2 in the second equation being the number of $np$ pairs in the three--particle 
final state $nnp$. Besides, $f^{\rm 2B}_{NN}(\theta)$ is the $NN$ angular correlation function 
for three--body decays, while $R^{2B}_{nn}$ ($R^{2B}_{np}$) is the fraction of $nn$ ($np$)
pairs from two--nucleon induced decays leaving the nucleus 
with energies above the detection thresholds. 
Nucleon pairs from one--body induced decays are mainly emitted back--to--back
with $\simeq 75$ MeV kinetic energy. A detailed study \cite{Ze98} of the 
$NN$ angular correlation function in $\Lambda N\to NN$, $f^{\rm 1B}_{NN}(\theta)$, 
shows that the $NN$ opening angles are ``with great probability'' larger than $140^\circ$
for $s$--shell hypernuclei. On the contrary, the function
$f^{\rm 2B}_{NN}(\theta)$ peaks around $120^\circ$. 
By using the approximation that all pairs detected at angles $\theta > 140^\circ$
($\theta < 140^\circ$) come from one--nucleon (two--nucleon)
induced processes (this assumption is realistic only for light hypernuclei,
where a small effect of the final state interactions is expected),
one can give an estimate of the various $N$'s.

To do this, we refer to the case of the experiment KEK--E462 \cite{Ou00a}, which
will study the decay of $^5_{\Lambda}$He hypernuclei.
Let us start by assuming that $\Gamma_n=\Gamma_p=0.20$ and $\Gamma_{\rm T}=1.00$
(the widths are in units of $\Gamma^{\rm free}_{\Lambda}$).
These values agrees with the results of the 1991 BNL experiment \cite{Sz91}.
Moreover, from the calculation presented in
table \ref{sat} it follows a ratio $\Gamma_2/(\Gamma_n+\Gamma_p)=0.15$.
It is not easy to evaluate $R_n$, $R_p$, $R^{1B}_{nn}$, $R^{1B}_{np}$,
$R^{2B}_{nn}$ and $R^{2B}_{np}$:
neglecting the nucleon rescattering effects in the residual nucleus and assuming a 0 MeV
detection threshold, these quantities are equal to 1. The nuclear final state
interactions increase the total number of nucleons outgoing from the nucleus 
with respect to the number of nucleons at the weak decay vertex, but a 
non--zero energy detection thresholds decreases the number of final nucleons
which can be observed \cite{Ra97}. The $R$'s factors depend on a delicate 
balance between these two effects.
A simulation of single and coincidence nucleon spectra
fitting experimental data allows to determine these quantities.
Here, we can use as a guidance the results of Refs.~\cite{Sz91,Ze98,Ra97} to estimate that, 
very roughly, for the detection thresholds which will be used at
KEK--E462 (around $10\div 20$ MeV), $R_n$, $R_p$, $R^{1B}_{nn}$ and $R^{1B}_{np}$ 
are sufficiently close to 1 and $R^{2B}_{nn}\simeq R^{2B}_{np}\simeq 0.8$ for $^5_{\Lambda}$He.
Further, the following parameters of this experiment are needed \cite{Ou00a}:
\begin{equation}
\begin{array}{l l}
\epsilon_n=0.23 ,             & \hspace{0.3mm} \epsilon_p=0.85 , \\
\Omega_n=0.27 ,               & \hspace{0.3mm} \Omega_p=0.18 , \\
\Omega^{\rm 1B}_{NN}=0.143 ,  & \hspace{0.3mm} \Omega^{\rm 2B}_{NN}=0.05 ,
\end{array} \nonumber
\end{equation}
where $\Omega^{\rm 1B}_{NN}$ and $\Omega^{\rm 2B}_{NN}$ are the total detector acceptances 
for $NN$ pairs coming from one--body and two--body induced processes,
respectively. The latter quantities 
are obtained by averaging the functions $\Omega_{NN}(\theta) \, f^{\rm 1B}_{NN}(\theta)$ 
and $\Omega_{NN}(\theta) \, f^{\rm 2B}_{NN}(\theta)$ of Eqs.~(\ref{nn-np-1b}), 
(\ref{nn-np-2b}) over the
intervals $\theta > 140^\circ$ and $\theta < 140^\circ$, respectively. 
By observing $N=100000$ decays of $^5_{\Lambda}$He hypernuclei, we thus expect the
following total number of counts:
\begin{equation}
\label{num-exp}
\begin{array}{l l}
N^{\rm 1B}_n=3726 ,   & N^{\rm 1B}_p= 3060 , \\
N^{\rm 2B}_n=745 ,    & N^{\rm 2B}_p=918   ,  \\
N^{\rm 1B}_{nn}=151 , & N^{\rm 1B}_{np}=559 , \\
N^{\rm 2B}_{nn}=13,   & N^{\rm 2B}_{np}=94 .
\end{array}
\end{equation}
If, on the contrary, one assumes $\Gamma_n+\Gamma_p=0.3$ (this value agrees with the
calculation presented in table~\ref{sat}) and $\Gamma_n/\Gamma_p=0.5$, the number of
counts are:
\begin{equation}
\label{num-exp2}
\begin{array}{l l}
N^{\rm 1B}_n=2484 ,   & N^{\rm 1B}_p= 3060 , \\
N^{\rm 2B}_n=559 ,    & N^{\rm 2B}_p=689   ,  \\
N^{\rm 1B}_{nn}=76 , & N^{\rm 1B}_{np}=559 , \\
N^{\rm 2B}_{nn}=10,   & N^{\rm 2B}_{np}=70 ,
\end{array}
\end{equation}
respectively. From these estimates one reaches the following important
conclusion: if the two--nucleon induced decay rate
is about 15\% of the total non--mesonic width, from existing data 
and calculations on $\Gamma_n$ and $\Gamma_p$ one expects a non--negligible 
number of $NN$ coincidence counts (of the order of 100) coming from two--body 
induced processes for an ensemble of $N=100000$ hypernuclear decays. 

In an experiment which would measure the quantities of Eqs.~(\ref{num-exp}), (\ref{num-exp2}) 
with sufficient statistics, one has two independent ways to determine the ratio $\Gamma_n/\Gamma_p$
[by using Eqs.~(\ref{n/p-simple}) and (\ref{n/p-new})] and,
once $\Gamma_{\rm T}$ is measured, two independent ways
to determine $\Gamma_2$ [Eqs.~(\ref{nn-np-2b})].
Careful analyses of the nucleon final state interactions must be done,
in order to estimate the different factors $R$'s. We must also note
that the use of Eq.~(\ref{n/p-simple}) to obtain the ratio could be affected
by problems related to interference effects between neutron-- and 
proton--stimulated decays, as mentioned in point 1) of paragraph 
\ref{improvements}. A determination of the ratio with both Eqs.~(\ref{n/p-simple}) 
and (\ref{n/p-new}) will thus be able to quantify these interference effects. 

To study the effect of the two--nucleon stimulated decay on the determination of
$\Gamma_n/\Gamma_p$ for experiments which do not detect the nucleons in 
double coincidence, let us consider the data from
the BNL experiment of Ref.~\cite{Sz91} for $^5_{\Lambda}$He:
\begin{equation}
\begin{array}{l l}
\displaystyle \frac{N_n}{\epsilon_n \, \Omega_n}\left(^5_{\Lambda}{\rm He}\right)
=3000\pm 1300 ,  & \hspace{0.4cm} 
\displaystyle \frac{N_p}{\epsilon_p \, \Omega_p}\left(^5_{\Lambda}{\rm He}\right)
=1730\pm 260 . \nonumber
\end{array}
\end{equation}
In order to calculate $\Gamma_n/\Gamma_p$ with Eq.~(\ref{n/p-simple}),
an estimate of the nuclear final state interaction effects for the outgoing nucleons
is required. By using as a guidance the analyses of Refs.~\cite{Sz91,Ra97,Ze98,Bha01},
one has that, very roughly, $R_p/R_n\simeq 1\div 1.1$ for the energy thresholds of the
BNL experiment ($\simeq 30\div 40$ MeV). By assuming $\Gamma_2/(\Gamma_n+\Gamma_p)=0.15$, 
Eq.~(\ref{n/p-simple}) then supplies:
\begin{equation}
\label{con}
\frac{\Gamma_n}{\Gamma_p}\left(^5_{\Lambda}{\rm He}\right)=0.44^{+0.53}_{-0.44} 
\hspace{1cm} ({\rm 1B+2B},\, R_p/R_n=1.1) ,
\end{equation}
while neglecting the two--nucleon induced channel:
\begin{equation}
\label{senza}
\frac{\Gamma_n}{\Gamma_p}\left(^5_{\Lambda}{\rm He}\right)=0.45\pm 0.44 
\hspace{1cm} ({\rm 1B \, \, only},\, R_p/R_n=1.1) . 
\end{equation}
For $R_p/R_n=1$, the ratios of Eqs.~(\ref{con}) and (\ref{senza}) become
slightly smaller, namely:
\begin{equation}
\label{con1}
\frac{\Gamma_n}{\Gamma_p}\left(^5_{\Lambda}{\rm He}\right)=0.34^{+0.47}_{-0.34}
\hspace{1cm} ({\rm 1B+2B},\, R_p/R_n=1) , 
\end{equation}
\begin{equation}
\label{senza1}
\frac{\Gamma_n}{\Gamma_p}\left(^5_{\Lambda}{\rm He}\right)=0.37^{+0.40}_{-0.37}
\hspace{1cm} ({\rm 1B \, \, only},\, R_p/R_n=1) . 
\end{equation}
respectively. 

A similar analysis can be performed on $^{12}_{\Lambda}$C data,
again from the BNL experiment of Ref.~\cite{Sz91}:
\begin{equation}
\begin{array}{l l}
\displaystyle \frac{N_n}{\epsilon_n \, \Omega_n}\left(^{12}_{\Lambda}{\rm C}\right)
=3400\pm 1100 ,  & \hspace{0.4cm}
\displaystyle \frac{N_p}{\epsilon_p \, \Omega_p}\left(^{12}_{\Lambda}{\rm C}\right)
=1410\pm 200 . \nonumber
\end{array}
\end{equation}
The calculation of table \ref{sat} supplies a ratio $\Gamma_2/(\Gamma_n+\Gamma_p)=0.2$
for $^{12}_{\Lambda}$C. Thus, from Eq.~(\ref{n/p-simple}) one obtains:
\begin{equation}
\label{con2}
\frac{\Gamma_n}{\Gamma_p}\left(^{12}_{\Lambda}{\rm C}\right)=1.14\pm 0.80
\hspace{1cm} ({\rm 1B+2B},\, R_p/R_n=1.2) ,
\end{equation}
\begin{equation}
\label{senza2}
\frac{\Gamma_n}{\Gamma_p}\left(^{12}_{\Lambda}{\rm C}\right)=0.95\pm 0.51
\hspace{1cm} ({\rm 1B \, \, only},\, R_p/R_n=1.2) ,
\end{equation}
\begin{equation}
\label{con3}
\frac{\Gamma_n}{\Gamma_p}\left(^{12}_{\Lambda}{\rm C}\right)=0.78\pm 0.60
\hspace{1cm} ({\rm 1B+2B},\, R_p/R_n=1) ,
\end{equation}
\begin{equation}
\label{senza3}
\frac{\Gamma_n}{\Gamma_p}\left(^{12}_{\Lambda}{\rm C}\right)=0.71\pm 0.43
\hspace{1cm} ({\rm 1B \, \, only},\, R_p/R_n=1) .
\end{equation}
Again, because of the big error bars, the effect of $\Gamma_2$ on the determination
of $\Gamma_n/\Gamma_p$ with Eq.~(\ref{n/p-simple}) is negligible [this also occurs 
for very large (and unrealistic) values of $\Gamma_2$]. 
In order to have a determination of the ratios of Eqs.~(\ref{con}) and (\ref{con2}) 
with relative errors of about $20$\%, $N_n$ and $N_p$ must be measured with
very small errors: $\Delta N_n/N_n\simeq 2\, \Delta N_p/N_p \simeq 8$\%.
Thus, to determine $\Gamma_2$ one must resort to $NN$ correlation measurements,
as previously discussed. It is worth noticing that 
the ratios of Eqs.~(\ref{con})--(\ref{senza1})
are considerably smaller than the result published in Ref.~\cite{Sz91}
for $^5_{\Lambda}$He ($\Gamma_n/\Gamma_p=0.93\pm 0.55$) and compatible with the OPE calculations.
Only sizeable final state interactions ($R_p/R_n\simeq 1.7$) can give ratios around 1 by employing
Eq.~(\ref{n/p-simple}) for $^5_{\Lambda}$He. The values of Eqs.~(\ref{con2})--(\ref{senza3}),
instead, are closer to the published result for $^{12}_{\Lambda}$C 
($\Gamma_n/\Gamma_p=1.33^{+1.12}_{-0.81}$) 
and disagree with the OPE calculations within one standard deviation. Interestingly, 
the ratio of Eq.~(\ref{n/p-simple}) for $^{12}_{\Lambda}$C
becomes equal to the central data point ($1.33$) when $R_p/R_n \simeq 1.3$.

\subsection{Phenomenological analysis of $s$--shell hypernuclei}
\label{passh}
The analysis of the non--mesonic decays in $s$--shell hypernuclei offers an important 
tool both for the solution of the $\Gamma_n/\Gamma_p$ puzzle and for testing the 
validity of the related $\Delta I= 1/2$ rule. Since in these hypernuclei the $\Lambda N$   
pair is necessarily in the $L=0$ relative state, the only possible 
$\Lambda N\rightarrow NN$ transitions are the following ones 
(we use the spectroscopic notation $^{2S+1}L_J$):  
\begin{eqnarray}
\label{partial}  
{^1S_0} &\rightarrow &{^1S_0} \hspace{0.3cm}(I_f=1)  \\  
        &\rightarrow &{^3P_0} \hspace{0.3cm}(I_f=1) \nonumber \\
{^3S_1} &\rightarrow &{^3S_1} \hspace{0.3cm}(I_f=0) \nonumber \\ 
        &\rightarrow &{^1P_1} \hspace{0.3cm}(I_f=0) \nonumber \\ 
        &\rightarrow &{^3P_1} \hspace{0.3cm}(I_f=1) \nonumber \\
        &\rightarrow &{^3D_1} \hspace{0.3cm}(I_f=0) . \nonumber   
\end{eqnarray} 
The $\Lambda n\rightarrow nn$ process has final states with isospin  
$I_f=1$ only, while for $\Lambda p\rightarrow np$ both $I_f=1$ and  
$I_f=0$ are allowed. 

We discuss in this subsection an analysis performed by 
the authors of the present review \cite{Al99b} in order to explore the validity of 
the $\Delta I= 1/2$ rule in the one--nucleon induced $\Lambda$--decay.
This analysis is based on the phenomenological model of Block and Dalitz
\cite{Bl62,Bl63}, which we briefly outline now.

The interaction probability of a particle which crosses an infinite
homogeneous system of thickness $ds$ is, classically, $dP=ds/\lambda$, where 
$\lambda=1/(\sigma \rho)$ is the mean free path of the projectile, 
$\sigma$ is the relevant cross section and $\rho$ is the
density of the system. Then, if we refer to the process $\Lambda N\to NN$, the width
$\Gamma_{\rm NM}=dP_{\Lambda N\to NN}/dt$ can be written as:  
\begin{equation} 
\Gamma_{\rm NM}=v\sigma \rho , \nonumber
\end{equation}
$v=ds/dt$ being the $\Lambda$ velocity in the rest frame of the homogeneous system.  For a
finite nucleus of density $\rho(\vec r)$, by introducing a local Fermi sea of nucleons,
one can write, within the semiclassical approximation:  
\begin{equation} 
\Gamma_{\rm NM}=\langle v\sigma \rangle \int d{\vec r} \rho({\vec r}) \mid 
\psi_{\Lambda}({\vec r})\mid^2 , \nonumber
\end{equation}
where $\psi_{\Lambda}({\vec r})$ is the $\Lambda$ wave function in the  
hypernucleus and $\langle \rangle$ denotes an average over spin and isospin
states. In the above equation the nuclear density is normalized to the 
mass number $A=N+Z$, hence the integral gives the average nucleon density $\rho_A$ 
at the position of the $\Lambda$ particle. In this scheme, the non--mesonic width 
$\Gamma_{\rm NM}=\Gamma_n+\Gamma_p$ of the hypernucleus $^{A+1}_{\Lambda}Z$ turns
out to be:  
\begin{equation}  
\Gamma_{\rm NM}(^{A+1}_{\Lambda}Z)= 
\frac{N\overline{R}_n(^{A+1}_{\Lambda}Z)+Z\overline{R}_p(^{A+1}_{\Lambda}Z)}{A}\rho_A , \nonumber
\end{equation}
where $\overline{R}_n$ ($\overline{R}_p$) denotes the spin--averaged rate for   
the neutron--induced (proton--induced) process appropriate for the 
considered hypernucleus. 

Furthermore, by introducing the rates $R_{NJ}$ for the spin--singlet ($R_{n0}$, $R_{p0}$) and
spin--triplet ($R_{n1}$, $R_{p1}$) elementary $\Lambda N\to NN$ 
interactions, the non--mesonic decay widths of
$s$--shell hypernuclei are \cite{Bl62,Bl63}:  
\begin{eqnarray} 
\label{phen}
\Gamma_{\rm NM}(^3_{\Lambda}{\rm H})&=&
\left(3R_{n0}+R_{n1}+3R_{p0}+R_{p1}\right)\frac{\rho_2}{8} ,\\
\Gamma_{\rm NM}(^4_{\Lambda}{\rm H})&=& 
\left(R_{n0}+3R_{n1}+2R_{p0}\right)\frac{\rho_3}{6},\nonumber \\ 
\Gamma_{\rm NM}(^4_{\Lambda}{\rm He})&=&
\left(2R_{n0}+R_{p0}+3R_{p1}\right)\frac{\rho_3}{6} ,\nonumber \\
\Gamma_{\rm NM}(^5_{\Lambda}{\rm He})&=&
\left(R_{n0}+3R_{n1}+R_{p0}+3R_{p1}\right)\frac{\rho_4}{8} .
\nonumber 
\end{eqnarray}
These relations take into account that the total hypernuclear angular momentum is 0 for
$^4_{\Lambda}$H and $^4_{\Lambda}$He and 1/2 for $^3_{\Lambda}$H and $^5_{\Lambda}$He.
In terms of the rates associated to the partial--wave transitions (\ref{partial}), the
$R_{NJ}$'s of Eqs.~(\ref{phen}) read:  
\begin{eqnarray} 
R_{n0}&=&R_n(^1S_0)+R_n(^3P_0) , \nonumber \\ 
R_{p0}&=&R_p(^1S_0)+R_p(^3P_0) , \nonumber \\
R_{n1}&=&R_n(^3P_1) , \nonumber \\ 
R_{p1}&=&R_p(^3S_1)+R_p(^1P_1)+R_p(^3P_1)+R_p(^3D_1), \nonumber 
\end{eqnarray} 
the quantum numbers of the $NN$ final state being reported
in brackets.  If one assumes that the $\Lambda N\to NN$ weak interaction occurs with a
change $\Delta I=1/2$ of the isospin, the following relations (simply derived by
angular momentum coupling coefficients)  hold among the rates for transitions to $I=1$
final states:  
\begin{equation} 
\label{uno1} 
R_n(^1S_0)=2R_p(^1S_0) , \hspace{0.2cm}
R_n(^3P_0)=2R_p(^3P_0) , \hspace{0.2cm} R_n(^3P_1)=2R_p(^3P_1) .
\end{equation}
Hence: 
\begin{equation}
\label{uno2} 
\frac{R_{n1}}{R_{p1}}\leq \frac{R_{n0}}{R_{p0}}=2 .  
\end{equation}
For pure $\Delta I=3/2$ transitions, the factors 2 in 
Eqs.~(\ref{uno1}) are replaced by 1/2. Hence, by further introducing the ratio:  
\begin{equation}
r=\frac{\langle I_f=1||A_{1/2}||I_i=1/2\rangle} {\langle
I_f=1||A_{3/2}||I_i=1/2\rangle} \nonumber
\end{equation}
between the $\Delta I=1/2$ and $\Delta I=3/2$
$\Lambda N\to NN$ transition amplitudes for isospin 1 final states ($r$ being real, as
required by time reversal invariance), for a general $\Delta I=1/2\: -\: \Delta I=3/2$
mixture one gets:  
\begin{equation} 
\label{mixt1} 
\frac{R_{n1}}{R_{p1}}=
\frac{4r^2-4r+1}{2r^2+4r+2+6\lambda^2}\leq
\frac{R_{n0}}{R_{p0}}=\frac{4r^2-4r+1}{2r^2+4r+2} , 
\end{equation}
where:  
\begin{equation} 
\label{lam}
\lambda=\frac{\langle I_f=0||A_{1/2}||I_i=1/2\rangle} 
{\langle I_f=1||A_{3/2}||I_i=1/2\rangle} .  
\end{equation}
The partial rates of Eq.~(\ref{mixt1}) supply
the $\Gamma_n/\Gamma_p$ ratios for $s$--shell hypernuclei through Eqs.~(\ref{phen}),
which provide the sum $\Gamma_n +\Gamma_p$ for each considered hypernucleus. 
For example, for $^5_{\Lambda}$He one has:  
\begin{equation} 
\label{ggtot}
\frac{\Gamma_n}{\Gamma_p}(^5_{\Lambda}{\rm He})= 
\frac{R_{n0}+3R_{n1}}{R_{p0}+3R_{p1}} .
\end{equation}                                                             

By using Eqs.~(\ref{phen}) and (\ref{mixt1}) together with the available
experimental data it is possible to extract the spin and isospin behaviour
of the $\Lambda N\to NN$ interaction without resorting to a detailed knowledge 
of the interaction mechanism. This reasoning was applied for
the first time by Block and Dalitz \cite{Bl62,Bl63}.
Unfortunately, up to now, the large experimental error bars have not allowed to draw 
definitive conclusions about the validity of the 
$\Delta I=1/2$ rule in non--mesonic decays by employing the previous  
model. There are indications for a sizeable 
violation of this rule \cite{Co90a,Sc92,Ru99}, but more precise measurements are  
needed, especially for $^3_{\Lambda}$H and $^4_{\Lambda}$H. 
If confirmed, this would represent the first evidence of such a violation  
in non--leptonic strangeness changing processes.
By using the phenomenological model of Block and Dalitz, in the next paragraph 
we shall discuss, through a new analysis \cite{Al99b} which employs recent data, 
the validity of the $\Delta I=1/2$ rule in the process $\Lambda N\to NN$.

Before proceeding, we note that Eqs.~(\ref{phen}) make use  
of several assumptions, which cannot be easily tested: 
the decays are treated incoherently on the stimulating  
nucleons within a simple 4--baryons point interaction model, thus  
interference effects originating from antisymmetrization of the two--nucleon final state 
as well as final state interactions are neglected. Moreover, the calculation requires the 
average nuclear density at the $\Lambda$ position and does not
take into account non--mesonic decays induced by more than one nucleon.  
However, given the high momentum of the outgoing nucleons and
the present level of accuracy of the data, the above  
approximations can be considered as satisfactory.

\subsubsection{Experimental data and $\Delta I=1/2$ rule} 
\label{blda} 
In Ref.~\cite{Al99b} a phenomenological analysis
of experimental data on non--mesonic decay of $s$--shell hypernuclei 
is employed to study the possible violation of the $\Delta I=1/2$ rule 
in the $\Lambda N\to NN$ interaction. 
In that paper we have analyzed recent data (which are summarized in table~\ref{data}) 
by using a quite different method with respect to the previous works 
of Refs.~\cite{Do87,Co90a,Sc92}.  
\begin{table}  
\begin{center} 
\caption{Experimental data (in units of $\Gamma^{\rm free}_{\Lambda}$) 
for $s$--shell hypernuclei (taken from Ref.~\protect\cite{Al99b}).} 
\vspace{0.5cm} 
\label{data} 
\begin{tabular}{|c|c c c c c|} 
\hline
\mc {1}{|c|}{} & 
\mc {1}{c}{$\Gamma_n$} &  
\mc {1}{c}{$\Gamma_p$} & 
\mc {1}{c}{$\Gamma_{\rm NM}$} &
\mc {1}{c}{$\Gamma_n/\Gamma_p$} & 
\mc {1}{c|}{Ref.} \\ \hline\hline  
$^4_{\Lambda}$H      &                       &                & $0.22\pm 0.09$ & &  
reference value \\
                     &                       &                & $0.17\pm 0.11$ & 
& 
KEK \cite{Ou98}\\ 
                     &                       &                & $0.29\pm 0.14$ & 
& 
\cite{Bl63}\\ \hline 
$^4_{\Lambda}$He     &$0.04\pm 0.02$         & $0.16\pm 0.02$ & $0.20\pm 0.03$ & $0.25\pm  
0.13$         &   
BNL \cite{Ze98} \\ \hline
$^5_{\Lambda}$He     &$0.20\pm 0.11$         & $0.21\pm 0.07$ & $0.41\pm 0.14$ & $0.93\pm  
0.55$         & 
BNL \cite{Sz91} \\ \hline  
\end{tabular} 
\end{center} 
\end{table} 

Unfortunately, no data are available on the non--mesonic decay of hypertriton
and on $\Gamma_n/\Gamma_p$ for $^4_{\Lambda}$H. Indeed, we shall see in the following 
that the future measurement of $\Gamma_n/\Gamma_p$ for $^4_{\Lambda}$H at BNL 
\cite{Gi01} will be of great importance for a test of the $\Delta I=1/2$ rule.
The BNL data \cite{Ze98,Sz91} for $^4_{\Lambda}$He and $^5_{\Lambda}$He of table~\ref{data} 
together with the {\it reference value} for $^4_{\Lambda}$H have been used
in our analysis. This last number is the weighted
average of the previous estimates \cite{Ou98,Bl63},
which have not been obtained from direct measurements but rather by using 
theoretical constraints. One has then 5 independent data which allow to fix,  
from Eqs.~(\ref{phen}), the 4 rates $R_{N,J}$ and $\rho_3$. Indeed, the 
average nucleon density $\rho_4$ at the $\Lambda$ position for $^5_{\Lambda}$He, 
also entering into Eqs.~(\ref{phen}), has been estimated to be 
$\rho_4=0.045$ fm$^{-3}$ by employing the 
$\Lambda$ wave function of Ref.~\cite{St93} (which was obtained through a quark model
description of the $\Lambda N$ interaction) and the gaussian density for $^4$He that  
reproduces the experimental mean square radius of the nucleus.    
For $^4_{\Lambda}$H and $^4_{\Lambda}$He, instead, no realistic hyperon wave 
function is available and we can obtain the value  
$\rho_3=0.026$ fm$^{-3}$ from the data of table~\ref{data},  
by imposing that [see Eqs.~(\ref{phen})]:  
\begin{equation} 
\frac{\Gamma_p(^5_{\Lambda}{\rm He})}{\Gamma_p(^4_{\Lambda}{\rm He})}= 
\frac{3}{4} \frac{\rho_4}{\rho_3} .  \nonumber
\end{equation}

The best choice to determine the rates $R_{N,J}$ by fitting experimental data
corresponds to use the relations for the observables:  
\begin{equation}
\Gamma_{\rm NM}(^4_{\Lambda}{\rm H}) , \hspace{0.3cm} \Gamma_{\rm NM}(^4_{\Lambda}{\rm He}) ,
\hspace{0.3cm} \Gamma_{\rm NM}(^5_{\Lambda}{\rm He}) , \hspace{0.3cm} 
\frac{\Gamma_n}{\Gamma_p}(^4_{\Lambda}{\rm He}) ,  \nonumber
\end{equation}
which have the smallest experimental uncertainties. 
After solving these equations we obtained the following partial rates 
(as usual, the decay widths of Eqs.~(\ref{phen}) are considered in units of the 
free $\Lambda$ decay width):
\begin{eqnarray}
\label{results} 
R_{n0}&=&(4.7\pm 2.1)\:{\rm fm}^3 , \\ 
R_{p0}&=&(7.9^{+16.5}_{-7.9})\:{\rm fm}^3 , \\
R_{n1}&=&(10.3\pm 8.6)\:{\rm fm}^3 , \\   
\label{results2}  
R_{p1}&=&(9.8\pm 5.5)\:{\rm fm}^3 , \\  
\overline{R}_n(^5_{\Lambda}{\rm He})\equiv   
\frac{1}{4}\left(R_{n0}+3R_{n1}\right)&=&(8.9\pm 6.5)\:{\rm fm}^3,  \nonumber \\  
\overline{R}_p(^5_{\Lambda}{\rm He})\equiv 
\frac{1}{4}\left(R_{p0}+3R_{p1}\right)&=&(9.3\pm 5.8)\:{\rm fm}^3 , \nonumber
\end{eqnarray} 
The errors have been obtained with the standard formula: 
\begin{equation}
\delta[O(r_1,\ldots,r_N)]=\sqrt{\sum_{i=1}^{N} 
\left(\frac{\partial O}{\partial r_i}\delta r_i\right)^2} , \nonumber
\end{equation}
namely by treating the data as independent and uncorrelated.
Due to the large relative errors [especially in the 
measures of $\Gamma_{\rm NM}(^4_{\Lambda}{\rm H})$ 
and $\Gamma_{\rm NM}(^5_{\Lambda}{\rm He})$]  
implied in the extraction of the above rates, 
the gaussian propagation of the uncertainties has to be regarded as a   
poor approximation. 

For the ratios of Eq.~(\ref{mixt1}) we have then:  
\begin{equation}  
\label{gen1} 
\frac{R_{n0}}{R_{p0}}=0.6^{+1.3}_{-0.6} , 
\end{equation}
\begin{equation}
\label{gen2}
\frac{R_{n1}}{R_{p1}}=1.0^{+1.1}_{-1.0} . 
\end{equation} 
while the ratios of the spin--triplet to the spin-singlet interaction 
rates are:  
\begin{equation} 
\frac{R_{n1}}{R_{n0}}=2.2\pm 2.1 , \nonumber
\end{equation}
\begin{equation} 
\frac{R_{p1}}{R_{p0}}=1.2^{+2.7}_{-1.2} . \nonumber
\end{equation}
The large uncertainties do not allow 
to draw definite conclusions about the possible violation of the 
$\Delta I=1/2$ rule and the spin--dependence of the transition rates. 
Eqs.~(\ref{gen1}) and (\ref{gen2}) are still compatible 
with Eq.~(\ref{uno2}), namely with the $\Delta I=1/2$ rule, although 
the central value in Eq.~({\ref{gen1}}) is more in 
agreement either with a pure $\Delta I=3/2$ transition ($r\simeq 0$) or
with $r\simeq 2$ [see Eq.~(\ref{mixt1})].
Actually, Eq.~(\ref{gen1}) is compatible with $r$   
in the range $-1/4\div 40$, while the ratio $\lambda$ of 
Eqs.~(\ref{mixt1}) and (\ref{lam}) is completely undetermined. 

By using the results of Eqs.~(\ref{results})--(\ref{results2}) we  
can predict the neutron to proton ratio for $^3_{\Lambda}$H, $^4_{\Lambda}$H and 
$^5_{\Lambda}$He, which turn out to be:
\begin{eqnarray} 
\frac{\Gamma_n}{\Gamma_p}(^3_{\Lambda}{\rm H})&=&0.7^{+1.1}_{-0.7} , \nonumber \\
\frac{\Gamma_n}{\Gamma_p}(^4_{\Lambda}{\rm H})&=&2.3^{+5.0}_{-2.3} , \nonumber \\
\frac{\Gamma_n}{\Gamma_p}(^5_{\Lambda}{\rm He})&=&0.95\pm 0.92 , \nonumber
\end{eqnarray}
and, by using $\rho_2=0.001$ fm$^{-3}$ \cite{Bl63},
\begin{equation} 
\Gamma_{\rm NM}(^3_{\Lambda}{\rm H})=0.007\pm 0.006 . \nonumber
\end{equation}
The latter is of the same order of magnitude of the 
detailed 3--body calculation of Ref.~\cite{Go97},
which provides a non--mesonic width equal to 
1.7\% of the free $\Lambda$ width. The ratio obtained for $^5_{\Lambda}$He 
is in good agreement with the data of table \ref{data}. An accurate measurement of
$\Gamma_{\rm NM}(^3_{\Lambda}{\rm H})$ and  $\Gamma_n/\Gamma_p$ for $^3_{\Lambda}{\rm H}$
and $^4_{\Lambda}{\rm H}$ would then provide a test of the weak 
decay model of Eqs.~(\ref{phen}) if the rates of Eqs.~(\ref{results})--(\ref{results2})
could be extracted with less uncertainty from data.

The compatibility of the data with the $\Delta I=1/2$ 
rule can be discussed in a different way: by {\it assuming} this rule, 
we fix $R_{n0}/R_{p0}=2$. Then, by using the observables:  
\begin{equation}  
\Gamma_{\rm NM}(^4_{\Lambda}{\rm He}) , 
\hspace{0.3cm} \Gamma_{\rm NM}(^5_{\Lambda}{\rm He}) , \hspace{0.3cm}
\frac{\Gamma_n}{\Gamma_p}(^4_{\Lambda}{\rm He}) , \nonumber
\end{equation}
the extracted partial rates ($R_{n0}$, $R_{n1}$, $\overline{R}_n$ and $\overline{R}_p$ 
are unchanged with respect to the above derivation) are: 
\begin{eqnarray} 
R_{n0}&=&(4.7\pm 2.1)\:{\rm fm}^3 , \nonumber \\  
R_{p0}\equiv R_{n0}/2&=&(2.3\pm 1.0)\:{\rm fm}^3 , \nonumber \\ 
R_{n1}&=&(10.3\pm 8.6)\:{\rm fm}^3 , \nonumber \\ 
R_{p1}&=&(11.7\pm 2.4)\:{\rm fm}^3 . \nonumber
\end{eqnarray} 
These values are compatible with the ones in 
Eqs.~(\ref{results})--(\ref{results2}). 
For pure $\Delta I=1/2$ transitions the spin--triplet interactions 
seem to dominate over the spin--singlet ones:   
\begin{equation}
\frac{R_{n1}}{R_{n0}}=2.2\pm 2.1 , \nonumber
\end{equation}
\begin{equation}  
\frac{R_{p1}}{R_{p0}}=5.0\pm 2.4 . \nonumber
\end{equation}
Moreover, since: 
\begin{equation} 
\frac{R_{n1}}{R_{p1}}=0.9\pm 0.8 , \nonumber
\end{equation}
from Eq.~(\ref{mixt1}) one obtains the following estimate   
for the ratio between the $\Delta I=1/2$ amplitudes: 
\begin{equation} 
\left|\frac{\langle I_f=0||A_{1/2}||I_i=1/2\rangle} 
{\langle I_f=1||A_{1/2}||I_i=1/2\rangle}\right| \simeq \frac{1}{3.7}\div 2.3 . \nonumber
\end{equation}
The other independent observables which here have not been utilized are then predicted to be:   
\begin{equation}
\Gamma_{\rm NM}(^4_{\Lambda}{\rm H})=0.17\pm 0.11 , \nonumber
\end{equation}
and:
\begin{equation}
\frac{\Gamma_n}{\Gamma_p}(^5_{\Lambda}{\rm He})=0.95\pm 0.72 , \nonumber 
\end{equation} 
in good agreement with the values of table~\ref{data}, with a $\chi^2$ for 
one degree of freedom of 0.31 (corresponding to a $0.56\, \sigma$ deviation).
This means that the data are consistent with the hypothesis of validity  
of the $\Delta I=1/2$ rule at the level of 60\%. In other words, 
the $\Delta I=1/2$ rule is excluded at the 40\% confidence level. 

The observables for which experimental data are not available at present
are predicted to be:
\begin{eqnarray}
\frac{\Gamma_n}{\Gamma_p}(^3_{\Lambda}{\rm H})=1.3\pm 0.6 , \nonumber  \\
\frac{\Gamma_n}{\Gamma_p}(^4_{\Lambda}{\rm H})=7.6\pm 6.2 , \nonumber 
\end{eqnarray}
and, for $\rho_2=0.001$ fm$^{-3}$,
\begin{equation}
\Gamma_{\rm NM}(^3_{\Lambda}{\rm H})=0.005\pm 0.003 . \nonumber
\end{equation}
We note that the central value of $\Gamma_n/\Gamma_p$ for $^4_{\Lambda}{\rm H}$
in the analysis which enforces the $\Delta I=1/2$ rule is considerably 
larger than the central value obtained in the general analysis previously
discussed. Thus, the future measurement \cite{Gi01} of this quantity
will represent an important test of the $\Delta I=1/2$ rule.

We conclude this subsection by considering a simple extension to hypernuclei 
of the $p$--shell. In table~\ref{data2} the data on weak non--mesonic decay of 
$^{12}_{\Lambda}$C are quoted.   
\begin{table} 
\begin{center} 
\caption{Experimental data for $^{12}_{\Lambda}$C (taken from Ref.~\cite{Al99b}).} 
\vspace{0.5cm}  
\label{data2} 
\begin{tabular}{|c|c|c|} 
\hline 
\mc {1}{|c|}{$\Gamma_{\rm NM}$} & 
\mc {1}{c|}{$\Gamma_n/\Gamma_p$} & 
\mc {1}{c|}{Ref.} \\ \hline\hline    
$1.14\pm 0.20$ & $1.33^{+1.12}_{-0.81}$ & BNL \cite{Sz91} \\ 
$0.89\pm 0.18$ & $1.87^{+0.67}_{-1.16}$ & KEK \cite{No95}  \\ 
$1.01\pm 0.13$ & $1.61^{+0.57}_{-0.66}$ & average \\ \hline 
\end{tabular} 
\end{center}
\end{table}
The relevant decay rate can be written in the following form:
\begin{equation}  
\label{carbonio} 
\Gamma_{\rm NM}(^{12}_{\Lambda}{\rm C})=\frac{\rho^s_{11}}{\rho_4} 
\Gamma_{\rm NM}(^5_{\Lambda}{\rm He})+ 
\frac{\rho^p_{11}}{7}\left    
[3\overline{R}_n(p)+4\overline{R}_p(p)\right] , 
\end{equation}
where $\rho^s_{11}$ ($\rho^p_{11}$) is the mean $s$--shell 
($p$--shell) nucleon density at the hyperon position, while 
$\overline{R}_n(p)$ [$\overline{R}_p(p)$] is the spin--averaged  
$p$--shell neutron--induced (proton--induced) rate.  
By using the previous results from $s$--shell hypernuclei 
and the weighted average values in table~\ref{data2}, we obtain:  
\begin{eqnarray}
\overline{R}_n(p)&=&(18.3\pm 10.7)\:{\rm fm}^3 , \nonumber \\ 
\overline{R}_p(p)&=&(3.6^{+12.6}_{-3.6})\:{\rm fm}^3 . \nonumber
\end{eqnarray} 
The densities $\rho^s_{11}$ ($=0.064$ fm$^{-3}$) and 
$\rho^p_{11}$ ($=0.043$ fm$^{-3}$) have been calculated from the 
appropriate nucleon $s$-- and $p$--shell Woods--Saxon wave functions. 
The $s$-- and $p$--shell contributions in Eq.~(\ref{carbonio})  
are $0.58\pm 0.20$ and $0.43\pm 0.24$, respectively.   
The contribution of the $\Lambda N$ $P$ partial waves to $\Gamma_{\rm NM}$ 
is estimated to be only $5\div 15$\% in $p$--shell hypernuclei \cite{Be92,It98,Pa01}.
Thus, about $10\div 30$\% of the $^{12}_{\Lambda}$C $p$--shell contribution is
expected to be originated by $\Lambda N$ relative states with $L=1$.


\newpage
\section{Non--mesonic decay of polarized $\Lambda$--hypernuclei: the asymmetry puzzle}
\label{polarized}

\subsection{Introduction}
\label{intro-pol}
Lambda hypernuclear states can be produced with a sizeable amount of polarization 
\cite{Ba89}. The development of angular distribution measurements of decay 
particles (photons, pions and protons) from polarized hypernuclei is of crucial importance 
in order to extract new information on hypernuclear production, structure and decay.

A new open problem, of very recent origin, in the study of the weak hypernuclear decay
concerns large discrepancies among the results of two experiments \cite{Aj92,Aj00},
performed at KEK, which observed
the asymmetric emission of non--mesonic decay protons from polarized hypernuclei.
Theoretical predictions are able to reproduce, although not very accurately, the older measurement, 
but very recent observations have completely changed the situation, leading to a
puzzling status. We analyze the problem in this section.

Thanks to the large momentum transfer involved, the $n(\pi^+,K^+)\Lambda$ reaction has been used,
at $p_{\pi}=1.05$ GeV and small $K^+$ laboratory scattering angles 
($2^\circ\lsim \theta_{K}\lsim 15^\circ$), 
to produce hypernuclear states with a substantial amount of spin--polarization,
preferentially aligned along the axis normal to the reaction plane \cite{Aj92,Aj00}.
The origin of hypernuclear polarization is twofold \cite{Ba89}. 
It is known that the distortions (absorptions) of the initial ($\pi^+$) 
and final ($K^+$) meson--waves produce a small polarization of the
hypernuclear orbital angular momentum up to laboratory scattering angles $\theta_{K}\simeq 15^\circ$
(at larger scattering angles, the orbital polarization increases with a negative sign). 
At small but non--zero angles, the main source of polarization is due to 
an appreciable spin--flip interaction term in the elementary reaction $\pi^+ n\to \Lambda K^+$,
which interferes with the spin--nonflip amplitude.
In a typical experimental situation with $p_{\pi}=1.05$ GeV and $\theta_{K}\simeq 15^\circ$,
the polarization of the hyperon spin in the free $\pi^+ n\to \Lambda K^+$ process is about 0.75. 

The KEK experiment of Ref.~\cite{Aj92} measured for the first time 
the asymmetry of the angular distribution of protons produced in 
the non--mesonic decay, $\vec{\Lambda} p\to np$,
of polarized $p$--shell hypernuclei, produced on $^{12}$C target.
The difference between the number of protons emitted along the
polarization axis and the number of protons outgoing in the opposite
direction must be determined. As we shall briefly discuss in the next subsection,
this proton asymmetry is related to the {\it interference} 
between the parity--violating and parity--conserving transition amplitudes 
with different values of the $NN$ isospin \cite{Ba90}. Due to the 
antisymmetry of the $NN$ state, the $\Lambda N\to NN$
parity--violating and parity--conserving amplitudes correspond to
$S+I=$ even and $S+I=$ odd final states, respectively ($S=$ spin, $I=$ isospin). 
This means that the interference terms contributing to the proton asymmetry
occur between amplitudes with the same $NN$ intrinsic spin $S$.
The non--mesonic partial rates are dominated by the parity--conserving amplitudes.
Thanks to the information on the spin--parity structure of the process which
can be obtained with the study of the asymmetric emission of 
protons from polarized hypernuclei, new constraints can then be imposed on the
$\Lambda N\to NN$ decay mechanism.

\subsection{Spin--polarization observables}
\label{pol-obs}
In this subsection we briefly outline the formal derivation of the
proton asymmetry parameter and its relation with the other spin observables.
More details can be found in Ref.~\cite{Ra92}.
The intensity of protons emitted in the non--mesonic decay of a
polarized hypernucleus along a direction forming an angle $\Theta$ 
with the polarization axis is defined by:
\begin{eqnarray}
\label{int-prot}
I(\Theta, J)&\equiv &{\rm Tr}[\mathcal{M}\rho(J)\mathcal{M}^{\dagger}](\Theta) \\
\nonumber \\
&=&\sum_{F,M,M^{\prime}}\langle F; \Theta|\mathcal{M}|I; J, M\rangle \langle I; J, M|\rho(J)|
I; J, M^{\prime}\rangle \langle I; J, M^{\prime}|\mathcal{M}^{\dagger}|F; \Theta \rangle . \nonumber
\end{eqnarray}
Here, $\mathcal{M}$ is the operator describing the 
$\vec{\Lambda} p\to np$ transition, $|I; J, M\rangle$ is the initial hypernuclear state,
$M$ denoting the third component of the hypernuclear total spin $J$, $|F; \Theta \rangle$ 
the many--body final state (given by the residual nucleus and the outgoing nucleons,
with a proton emerging at an angle $\Theta$) and $\rho$
is the density matrix of the polarized hypernucleus. 

With reference to the $(\pi^+,K^+)$ production reaction, the density matrix
for pure vector polarization along $\vec{k}_{\pi}\times \vec{k}_K$ is given by \cite{Ra92}:
\begin{equation}
\label{vec-pol}
\rho(J)= \frac{1}{2J+1}\left[1+P_y(J)\, S_y\, \frac{3}{J+1}\right] ,
\end{equation}
in the Madison frame, in which the $z_M$--axis is along the direction of the incoming
pion and the $y_M$--axis is along 
$\vec{k}_{\pi}\times \vec{k}_K$. In Eq.~(\ref{vec-pol}) $P_y$ is the hypernuclear
polarization and $S_y$ the projection along the $y_M$--axis of the spin operator $J$. 
From Eq.~(\ref{int-prot}) one then obtains the proton distribution in the form:
\begin{equation}
I(\Theta, J)=I_0(J)\left[1+\mathcal{A}(\Theta, J)\right] , \nonumber
\end{equation}
where:
\begin{equation}
I_0(J)=\frac{{\rm Tr}(\mathcal{M}\mathcal{M}^{\dagger})}{2J+1}  \nonumber
\end{equation} 
is the (isotropic) intensity for an unpolarized hypernucleus.
The asymmetry of the angular distribution for the outgoing protons is expressed by:
\begin{equation} 
\mathcal{A}(\Theta, J)=P_y(J)\, \frac{3}{J+1}\, \frac{{\rm Tr}
(\mathcal{M}S_y\mathcal{M}^{\dagger})(\Theta)}
{{\rm Tr} \left(\mathcal{M}\mathcal{M}^{\dagger}\right)} . \nonumber
\end{equation} 
One easily obtains that this {\it proton asymmetry parameter} is proportional
to ${\rm cos}\, \Theta$ \cite{Ra92}:
\begin{equation} 
\label{asymm-a-vec}
\mathcal{A}(\Theta, J)=P_y(J)\, A_y(J)\, {\rm cos}\, \Theta .
\end{equation}
Here, the quantity:
\begin{equation}
\label{asymm-y} 
A_y(J) = \frac{3}{J+1} \, \frac{\sum_{M} M\, \sigma(J, M)}{\sum_{M} \sigma(J, M)} ,
\end{equation}
which is a property of the hypernuclear non--mesonic decay only, is usually referred to
as the {\it hypernuclear asymmetry parameter}. The hypernuclear polarization $P_y$
depends both on the kinematics ($p_{\pi}$ and $\theta_{K}$) and dynamics of the production
reaction. In Eq.~(\ref{asymm-y}):
\begin{equation} 
\label{inten-prot}
\sigma(J, M)=\sum_{F} \left|\langle F|\mathcal{M}|I; J, M\rangle \right| ^2 
\end{equation}
is the intensity of protons emitted along the quantization axis $z$
for a projection $M$ of the hypernuclear total spin. The transition amplitudes
appearing in Eqs.~(\ref{int-prot}) and (\ref{inten-prot}) are evaluated in the proton
helicity frame, whose $z$--axis is along the direction of the outgoing proton.

In the shell model weak--coupling scheme with the $1s$ $\Lambda$--hyperon
coupled to the nuclear core ground state, $P_y$ is directly related to the
polarization $p_{\Lambda}$ of the $\Lambda$ spin in the hypernucleus as follows:
\begin{equation} 
\label{p-lambda}
p_{\Lambda}(J)=
\begin{cases}
\displaystyle -\frac{J}{J+1}P_y(J) & \text{if}\, \, J=J_C-\frac{1}{2}  \\
P_y(J) & \text{if}\, \, J=J_C+\frac{1}{2} , 
\end{cases} 
\end{equation} 
$J_C$ being the total spin of the nuclear core.
It is useful to introduce an {\it intrinsic lambda asymmetry parameter}
$a_{\Lambda}$, which is characteristic of the elementary process $\vec{\Lambda} p\to np$
and should be independent of the hypernucleus, such that:
\begin{equation}
\label{asymm-a-vec1} 
\mathcal{A}(\Theta, J)=p_{\Lambda}(J)\, a_{\Lambda}\, {\rm cos}\, \Theta . 
\end{equation} 
From Eqs.~(\ref{asymm-a-vec}) and (\ref{p-lambda}) it follows then:
\begin{equation}
\label{a-lambda}
a_{\Lambda}= 
\begin{cases} 
\displaystyle -\frac{J+1}{J}A_y(J) & \text{if}\, \, J=J_C-\frac{1}{2}  \\ 
A_y(J) & \text{if}\, \, J=J_C+\frac{1}{2} , 
\end{cases}
\end{equation}
and $a_{\Lambda}=A_y(J)=0$ if $J=0$.
In the case of $^5_{\Lambda}\vec{\rm H}$e, $J_C=0$ and $J=1/2$, thus:
\begin{equation}
a_{\Lambda}\equiv A_y(^5_{\Lambda}\vec{\rm H}{\rm e})=
\frac{\sigma \left(^5_{\Lambda}\vec{\rm H}{\rm e}, +1/2\right)-
\sigma \left(^5_{\Lambda}\vec{\rm H}{\rm e}, -1/2\right)}
{\sigma \left(^5_{\Lambda}\vec{\rm H}{\rm e}, +1/2\right) + 
\sigma \left(^5_{\Lambda}\vec{\rm H}{\rm e}, -1/2\right)} , \nonumber
\end{equation} 
and $-1\leq a_{\Lambda}\leq 1$.

\subsection{Experiments}
\label{expts}

Experimentally, the proton asymmetry parameter is obtained by comparing the
number of protons emerging parallel and antiparallel to the $y_M$--axis:
\begin{equation}
\mathcal{A}(0^{\circ})=\frac{I(0^{\circ})-I(180^{\circ})}
{I(0^{\circ})+I(180^{\circ})} .
\end{equation}

The asymmetry $\mathcal{A}(0^{\circ})$ measured by the
KEK experiments \cite{Aj92,Aj00,Bh01} suffered 
from large uncertainties, principally due to limited statistics, final state interaction
effects (which attenuate the weak decay vertex proton asymmetry)
and to the poor knowledge of the hypernuclear polarization. Moreover, two--nucleon 
induced decays, not taken into account in the experimental analyses, 
are expected to contribute. 

In the first experiment \cite{Aj92}, KEK--E160, $^{11}_{\Lambda}\vec{\rm B}$, 
$^{12}_{\Lambda}\vec{\rm C}$ and other $p$--shell
hypernuclei were produced by the $(\pi^+,K^+)$ reaction on $^{12}$C.
At about 10 MeV excitation energy with respect to the
$^{12}_{\Lambda}$C$(1^-)$ ground state, the reaction can create proton--unbound states, 
which then populate the $^{11}_{\Lambda}$B$(\frac{1}{2}^+)$ ground state by proton and photon 
emissions. The high excitation energy region,
around 20 MeV, is called quasi--bound region since, even if here the $\Lambda$
has a finite escape probability, deexcitations via the emission of
one or more nucleons are also possible, and lead to a light hyperfragment (LH)
with $A\leq 10$: for example, the emission of
a $p$, $n$, $d$, $^3$He or $\alpha$ particle produces a final
$^{11}_{\Lambda}$B, $^{11}_{\Lambda}$C, $^{10}_{\Lambda}$B,
$^9_{\Lambda}$Be or $^8_{\Lambda}$Be hypernucleus, respectively.
The statistics and  energy resolution ($5\div 7$ MeV) of the kaon spectrometer 
were limited at KEK--E160; moreover, the polarization of the produced hypernuclei, 
whose decay protons were observed, had to be evaluated theoretically in order to determine
the intrinsic $\Lambda$ asymmetry, $a_{\Lambda}$, from the measured
$\mathcal{A}$. Such a calculation requires a delicate analysis of 
1) the polarization of the hypernuclear states directly produced in the
$(\pi^+,K^+)$ reaction and 2) the depolarization
effects due to strong and electromagnetic transitions of the populated excited states,
which take place before the weak decay.
In table~\ref{pol-92} we list the observed asymmetries.
\begin{table}
\begin{center}
\caption{Asymmetries observed at KEK--E160 \cite{Aj92}.}
\vspace{0.5cm}
\label{pol-92}
\begin{tabular}{|c|c|c|c|}
\hline
\mc {1}{|c|}{} &
\mc {1}{c|}{$^{12}_{\Lambda}\vec{\rm C}$} &
\mc {1}{c|}{$^{11}_{\Lambda}\vec{\rm B}$} &
\mc {1}{c|}{LH ($A\leq 10$)} \\ \hline\hline
$\mathcal{A}(0^\circ)$ & $-0.01\pm 0.11$ & $-0.19\pm 0.10$ & $-0.24\pm 0.09$  \\ \hline
$a_{\Lambda}=\displaystyle \frac{\mathcal{A}(0^\circ)}{k\, p_{\Lambda}}$   
                       & $-0.17\pm 1.83$ & $-1.33\pm 0.72$ & $-1.50\pm 0.68$  \\ 
                       & $(p_{\Lambda}=0.06\div 0.09)$ & $(p_{\Lambda}=0.16\div 0.21)$ & 
$(p_{\Lambda}=0.15\div 0.26)$  \\ 
                       & $-0.13\pm 1.45$ & $-0.77\pm 0.41$ &  \\ 
                       & $(p_{\Lambda}=0.095)$   & $(p_{\Lambda}=0.31)$   &  \\  \hline
\end{tabular}
\end{center}
\end{table} 
According to Eq.~(\ref{asymm-a-vec1}),
the proton asymmetry $\mathcal{A}(0^\circ)$ should depend linearly on the polarization 
$p_{\Lambda}$ of the hyperon in the nucleus (see second and fourth line of the table),
which is always positive, reflecting the positive sign of the $\Lambda$ polarization in the
elementary $\pi^+ n\to \Lambda K^+$ reaction. 
The values for the intrinsic $\Lambda$ asymmetry, 
$a_{\Lambda}=\mathcal{A}(0^\circ)/(k\, p_{\Lambda})$, of the third line 
are obtained by using the theoretical evaluations of $P_y$ originally employed in the analysis
of Ref.~\cite{Aj92}. The attenuation factor $k$, estimated to be 
around 0.8, is due to the $\Lambda$ Fermi motion and the rescattering of the
emitted protons. The main reason of the attenuation in the observed asymmetry is the 
detection of secondary protons, emitted as a consequence of the scattering of decay
neutrons and protons with the nucleons of the residual nucleus.
By assuming that $a_{\Lambda}$ is independent of the hypernucleus,
the weighted average of the three results supplies a very large and negative asymmetry:
$a_{\Lambda}=-1.3\pm 0.4$, namely in the physically acceptable range
between $-0.9$ and $-1$. In the fifth line of the table, more realistic evaluations of 
the polarization, extracted from Refs.~\cite{Au83,Ej89,It94}, are used to obtain $a_{\Lambda}$. 
A weighted average among the improved results for $^{11}_{\Lambda}\vec{{\rm B}}$
and $^{12}_{\Lambda}\vec{{\rm C}}$ and the original one for lighter hyperfragments 
gives a smaller asymmetry value: $a_{\Lambda}=-0.9\pm 0.3$. 

More recently \cite{Aj98}, it has been possible to measure the polarization
of $^5_{\Lambda}\vec{\rm H}$e hypernuclei, which, from Eq.~(\ref{p-lambda}),
coincides with the $\Lambda$ polarization: 
$P_y(^5_{\Lambda}\vec{\rm H}{\rm e})=p_{\Lambda}(^5_{\Lambda}\vec{\rm H}{\rm e})$. 
The $^6$Li$(\pi^+,K^+)^6_{\Lambda}$Li reaction is used to produce a polarized 
$^6_{\Lambda}$Li hypernucleus. The ground state of $^6_{\Lambda}$Li lies above the
$^5_{\Lambda}{\rm He}+p$ threshold, thus an $^5_{\Lambda}$He  
hypernucleus in the $0^+$ ground state is exclusively produced by the emission of a proton. 
The polarization of $^5_{\Lambda}$He is measured
by observing the asymmetric emission of negative pions in its mesonic decay,
$\mathcal{A}^{\pi^-}=P_y\, A_y^{\pi^-}$. To obtain the polarization from the
observed $\mathcal{A}^{\pi^-}$, $A_y^{\pi^-}$ was assumed \cite{Mot94} to be equal
to the value for the free $\Lambda \to \pi^- p$ decay: $\alpha_{\pi^-}=-0.642\pm 0.013$
\cite{pdb}.
This approximation is reasonable, since in $^5_{\Lambda}$He the hyperon is coupled
to a spin--parity $0^+$ $^4$He core.
Unfortunately, the small branching ratio and asymmetry parameter for the 
mesonic decay of $p$--shell hypernuclei makes such a measurement very difficult
for these systems.
The distorted wave impulse approximation of Ref.~\cite{Mot94} reproduces
quite well the measured values of $P_y(^5_{\Lambda}\vec{\rm H}{\rm e})$
\cite{Aj98}. However, it is not clear whether 
such a model is able to account for the polarization mechanism
of $p$--shell hypernuclei. 

The experimental values of $P_y(^5_{\Lambda}\vec{\rm H}{\rm e})$ have been
employed, very recently, to determine $a_{\Lambda}\equiv A_y(^5_{\Lambda}\vec{\rm H}{\rm e})$ 
from a measurement of the proton asymmetry in $^5_{\Lambda}\vec{{\rm H}}$e 
(KEK--E278) \cite{Aj00}. Again, $^5_{\Lambda}\vec{\rm H}$e hypernuclei have been produced by the 
$^6$Li$(\pi^+,K^+)^6_{\Lambda}$Li reaction. When compared with an experiment employing
a $p$--shell hypernucleus, the use of $^5_{\Lambda}\vec{\rm H}$e has evident virtues:
the measured hypernuclear polarization is larger and approximatively equal to that of
the $\Lambda$--hyperon, since $J^P(^4{\rm He})=0^+$; the
nuclear effects on the observed asymmetry $\mathcal{A}$ are smaller; finally, only the relative
$S$--wave in the initial $\Lambda p$ system is active. All these features help the
theoretical interpretation of data. In table~\ref{pol-00} the obtained results are quoted.
\begin{table} 
\begin{center} 
\caption{Asymmetries observed at KEK--E278 \cite{Aj00} for $^5_{\Lambda}\vec{\rm H}$e.
The values of $p_{\Lambda}$ in the third line are taken from Ref.~\cite{Aj98}.}
\vspace{0.5cm} 
\label{pol-00}
\begin{tabular}{|c|c|c|}   \hline
\mc {1}{|c|}{$K^+$ scattering--angle} & 
\mc {1}{c|}{$2^\circ < |\theta_K | < 7^\circ$} &  
\mc {1}{c|}{$7^\circ < |\theta_K | < 15^\circ$} \\ \hline\hline 
$\mathcal{A}(0^\circ)$ & $0.082\pm 0.060$ & $0.035\pm 0.080$ \\ \hline 
$p_{\Lambda}$          & $0.247\pm 0.082$ & $0.393\pm 0.094$ \\ \hline
$a_{\Lambda}=\displaystyle\frac{\mathcal{A}(0^\circ)}{\epsilon\, k\, p_{\Lambda}}$
          & $0.441\pm 0.356$ & $0.120\pm 0.271$  \\  
\mc {1}{|c|}{weighted average} &
\mc {2}{c|}{$0.24\pm 0.22$} \\ \hline
\end{tabular}
\end{center}
\end{table}
The proton asymmetry $\mathcal{A}(0^\circ)$ has been measured for two 
$K^+$ scattering--angle regions, $2^\circ < |\theta_K | < 7^\circ$ and 
$7^\circ < |\theta_K | < 15^\circ$.
The reduction factor, $\epsilon=0.804$, is due to the finite acceptance of the
decay counter system, while the attenuation factor, $k=0.935$, is again due to 
nuclear effects. Both these quantities, estimated through
Monte Carlo simulations, and the $\Lambda$ polarization in $^5_{\Lambda}\vec{{\rm H}}$e
are required in order to derive the intrinsic $\Lambda$ asymmetry. A statistical
fluctuation caused a remarkable difference between the values of $a_{\Lambda}$ in the two 
scattering--angle regions. However, in the hypothesis that this observable depends
on the one--body induced non--mesonic decay only, a weighted average is permitted and leads to a 
relatively small, positive value, within 2 standard deviations.

The experiments thus revealed an opposite sign of
$a_{\Lambda}$ for $^5_{\Lambda}\vec{\rm H}$e \cite{Aj00} and $p$--shell hypernuclei 
\cite{Aj92}. This is puzzling, since from its definition 
one expects $a_{\Lambda}$ to be not much sensitive to the nuclear structure effects:
Ref.~\cite{Ra92} (\cite{Pa01}) demonstrated that this is true
within 25\% (6\%) in a calculation for $^5_{\Lambda}\vec{\rm H}$e and 
$^{12}_{\Lambda}\vec{\rm C}$ (see next subsection). 
The weak coupling scheme is known to be a good approximation for describing the
ground states of $\Lambda$ hypernuclei. However, one must note that, in the experiment on 
$p$--shell hypernuclei, due to the low energy resolution, several excited hypernuclear states
enter into the game. The procedure used to calculate the hypernuclear polarization
in this case is complicated and could have led to an unrealistic value of $a_{\Lambda}$.
For example, in Ref.~\cite{Ra92}, a sizeable reduction
(increase) of the hypernuclear polarization $P_y$ has been found for
$^{12}_{\Lambda}\vec{\rm C}$ ($^{11}_{\Lambda}\vec{\rm B}$) once the spin
depolarization of possibly populated excited states of these hypernuclei
are taken into account. 
It is difficult, however, to think that the sign difference is only due to this effect. 
Also a statistical fluctuation can hardly cause such a difference between the
two experiments. Another possible explanation, suggested in Ref.~\cite{Aj00},
could arise from a dominance of the $L=1$ $\Lambda N$ interaction 
in $p$--shell hypernuclei. However, this hypothesis is incompatible with calculations
\cite{Be92,It98,Pa01} which proved how the $\Lambda N$ $L=0$ relative state is the dominant
one in the non--mesonic decay of those hypernuclei, giving about $85\div 95$\% of the   
total non--mesonic rate. 

The preliminary results of KEK--E307 \cite{Bh01}, which employs carbon, silicon and iron 
targets, show a large and positive value of $a_{\Lambda}$ for $^{12}_{\Lambda}\vec{\rm C}$
within 2 standard deviations: they find
$a_{\Lambda}=0.85\pm 0.39$, in complete disagreement with 
the outcome of KEK--E160. To improve statistics, in the E307 
experiment an analysis of $a_{\Lambda}$ including all the non--mesonic decay events
gated to both the bound and continuum regions of $^{12}_{\Lambda}\vec{\rm C}$, 
$^{27}_{\Lambda}\vec{\rm A}$l, $^{28}_{\Lambda}\vec{\rm S}$i and 
$_{\Lambda}\vec{\rm F}$e is in progress. A preliminary result confirms a positive
value of $a_{\Lambda}$ \cite{Bh01} for hypernuclei beyond the $s$--shell.
However, future data analysis as well as improved
statistical and systematic uncertainties are needed before this conclusion can be ensured. 

\subsection{Theory {\it vs} experiment}
\label{th-exp-asy}

Within the model of Block and Dalitz \cite{Bl62,Bl63}, discussed in subsection
\ref{passh}, the intrinsic lambda asymmetry parameter of Eqs.~(\ref{asymm-y}),
(\ref{a-lambda}) is evaluated through the following formula \cite{Na99}:
\begin{equation}
\label{bl-asy}
a_{\Lambda}\equiv A_y(^5_{\Lambda}\vec{\rm H}{\rm e})=
\frac{\displaystyle 2\sqrt{3}\, {\mathcal{R}}e \left[a_p e^*_p - 
\frac{1}{\sqrt{3}}b_p (c^*_p-\sqrt{2}d^*_p)+f_p (\sqrt{2}c^*_p+d^*_p)\right]}
{|a_p|^2 + |b_p|^2  + 3(|c_p|^2  + |d_p|^2  +|e_p|^2  + |f_p|^2 )} ,
\end{equation}
where:
\begin{eqnarray}
a_p &=& \langle np; {^1S_0}| t |\Lambda p; {^1S_0} \rangle , \nonumber \\
b_p &=& \langle np; {^3P_0}| t |\Lambda p; {^1S_0} \rangle ,  \nonumber \\
c_p &=& \langle np; {^3S_1}| t |\Lambda p; {^3S_1} \rangle ,  \nonumber \\ 
d_p &=& \langle np; {^3D_1}| t |\Lambda p; {^3S_1} \rangle ,  \nonumber \\ 
e_p &=& \langle np; {^1P_1}| t |\Lambda p; {^3S_1} \rangle ,  \nonumber \\ 
f_p &=& \langle np; {^3P_1}| t |\Lambda p; {^3S_1} \rangle ,  \nonumber
\end{eqnarray} 
are the elementary $\Lambda p\to np$ transition amplitudes. The use of Eq.~(\ref{bl-asy})
to estimate $a_{\Lambda}$ only provides approximate results. Indeed, within the 
model of Block and Dalitz, interference effects as well as final state interactions of the two 
outgoing nucleons with the residual nucleus are neglected. However, it is evident from
Eq.~(\ref{bl-asy}) that the asymmetry is due to the interference
between parity--conserving ($a_p$, $c_p$ and $d_p$) and parity--violating 
($b_p$, $e_p$ and $f_p$) $\Lambda p\to np$ amplitudes with the same value of
the $np$ intrinsic spin $S$. Hence, interference terms between
spin--singlet ($J=0$) and spin--triplet ($J=1$)
amplitudes (terms in $a_p e^*_p$, $b_p c^*_p$ and $b_p d^*_p$) enter $a_\Lambda$.

In table~\ref{asy-th} we summarize the calculations of the intrinsic $\Lambda$ 
asymmetry. Previously discussed experimental data are reported for comparison.
\begin{table} 
\begin{center}
\caption{Calculations of the intrinsic lambda parameter $a_{\Lambda}$. Note that
$A_y(^5_{\Lambda}\vec{\rm H}{\rm e})=a_{\Lambda}$ and
$A_y(^{12}_{\Lambda}\vec{\rm C})=-a_{\Lambda}/2$.}
\vspace{0.5cm}
\label{asy-th}
\begin{tabular}{|c|c|c|c|}
\hline
\mc {1}{|c|}{Ref. and Model} &
\mc {1}{c|}{$^5_{\Lambda}\vec{\rm H}$e} &
\mc {1}{c|}{$^{12}_{\Lambda}\vec{\rm C}$} &
\mc {1}{c|}{NM} \\ \hline\hline
Ramos {\em et al.}~1992 \cite{Ra92}           &           &            & \\ 
                               OPE      & $-0.524$  & $-0.397$   &  \\
                            $\pi+K$     & $-0.509$  & $-0.375$   & \\ \hline
Dubach {\em et al.}~1996 \cite{Du96}          &           &            & \\
                               OPE      &           &            & $-0.192$ \\
                               OME      &           &            & $-0.443$  \\ \hline
Sasaki {\em et al.}~2001 \cite{Sa01}          &           &            & \\
                                    OPE & $-0.441$  &            &  \\
                                $\pi+K$ & $-0.362$  &            &  \\
                            $\pi+K+$DQ  & $-0.678$  &            &  \\ \hline
Parre\~{n}o {\em et al.}~2001 \cite{Pa01}     &           &            &  \\ 
                                    OPE & $-0.252$  & $-0.340$ & \\
                                $\pi+K$ & $-0.572\div -0.606$ & $-0.626\div -0.640$ & \\
                                  OME   & $-0.675\div -0.682$ & $-0.716\div -0.734$ & \\ \hline\hline
Exp KEK--E160 1992 \cite{Aj92}                &                     & $-0.9\pm 0.3$       & \\ \hline
Exp KEK--E278 2000 \cite{Aj00}                & $0.24\pm 0.22$      &                     & \\ \hline
Exp KEK--E307 2001 \cite{Bh01}                &                     & $0.85\pm 0.39$ (prel.)   & \\ \hline
\end{tabular}
\end{center}
\end{table}
All evaluations provide a negative asymmetry, between $-0.38$ and $-0.73$ for the complete
results, in fair agreement with the old KEK result of 1992, but 
in strong disagreement with the positive sign revealed by the recent experiments.
As expected, the calculations show a moderate sensitivity of the asymmetry 
to the details of nuclear structure.
The work of Ramos {\em et al.}~\cite{Ra92} has been performed in a relativistic nuclear model
by applying formula (\ref{asymm-y}), which defines, through Eq.~(\ref{a-lambda}), 
the intrinsic asymmetry.
The nuclear matter calculation of Dubach {\em et al.}~\cite{Du96} refers to a OME model including the exchange 
of $\pi, \rho, K, K^*, \omega$, and $\eta$ mesons. In this case, only relative $S$--wave
interactions are considered in the initial $\Lambda p$ state; moreover, $a_{\Lambda}$
has been calculated through Eq.~(\ref{bl-asy}) by neglecting the above
mentioned interference terms between the $J=0$ and $J=1$ $\Lambda p\to np$ transitions. 
These terms must be included in the calculation, and are quantitatively 
important: for example, in the OPE calculation of Sasaki {\em et al.}~\cite{Sa01},
the complete formula supplies an asymmetry equal to $-0.441$, to compare with the result,
$-0.159$ \cite{Sa-pr}, obtained by disregarding the $J=0$ -- $J=1$ interference
terms. Incidentally, this approximation is allowed only when the
$\Lambda p\to np$ process occurs in free space: more precisely, in free space only
a spin--triplet $\Lambda p$ initial state contributes to the asymmetry
[see Eq.~(\ref{bl-asy})]. Finally, Parre\~{n}o {\em et al.}~\cite{Pa01} applied Eq.~(\ref{asymm-y}) to
$^5_{\Lambda}\vec{\rm H}$e and $^{12}_{\Lambda}\vec{\rm C}$ hypernuclei within
a shell model framework with a OME transition potential including the exchange of 
$\pi, \rho, K, K^*, \omega$, and $\eta$ mesons. We note that these authors find a considerable 
increase of the asymmetry when the $K$--meson is added to the pion. On the contrary, Sasaki
{\em et al.}~\cite{Sa01} obtained a lower asymmetry in the $\pi+K$ calculation
with respect to the pure OPE value. However the OME and $\pi+K+$DQ results of the two 
calculations agree with each another, with values around $-0.7$. At variance with the above 
discussed results, the calculation of Ramos {\em et al.}~\cite{Ra92} supplies practically the same asymmetry
in the OPE and $\pi+K$ models. The origin of these discrepancies is unknown:
the difference among the various OPE calculations are due to the use of different $\pi \Lambda N$
form factors and short range correlations for the initial $\Lambda p$ and final $np$ states.

In conclusion, further investigations are required to clarify the situation:
on the theoretical side there seems to be no way (even by forcing the model parameters to  
unrealistic values) to obtain positive asymmetry values \cite{Pa-pr}; 
on the experimental side the present anomalous
discrepancy between different data needs to be resolved.
It has been advanced the hypothesis that the
asymmetry puzzle could have the same origin of the previously discussed
puzzle on the $\Gamma_n/\Gamma_p$ ratio \cite{Aj00}. At present there is no firm
evidence of this relation. Indeed, the situation is even more confused for the asymmetry
than for the $\Gamma_n/\Gamma_p$ ratio: in the former case, the experiments
cannot provide any guidance for new theoretical speculations.
We hope that future experimental studies of the
inverse reaction $\vec{p} n\to p\Lambda $ in free space could help in disentangling the puzzling 
situation. Indeed, the weak production of the $\Lambda$--hyperon through the scattering of
longitudinally polarized protons on neutron targets can give a richer and cleaner 
(with respect to the non--mesonic hypernuclear decay) piece of
information on the $\Lambda$ polarization--observables \cite{Na99}.

\newpage
\section{Summary and perspective}
\label{concl}

In this review we have discussed the present status of hypernuclear physics. 
Beyond an extensive and updated description of our present understanding
of weak hypernuclear decay processes, which are the main topic of the 
paper, we have also illustrated some phenomenological aspects
of the $YN$, $YY$ interaction and the hypernuclear structure
and reviewed the reactions which are used to produce hypernuclei.  

Measurements of the $YN$ and $YY$ cross sections are very difficult to perform,
because of the very short lifetimes of hyperons. As a consequence, the various 
phenomenological models developed to describe
these interactions are not completely satisfactory. One of the major
 reason of interest on hypernuclear phenomena lies thus in
the information which can be extracted about the $YN$ and $YY$ interactions
(both of strong and weak nature, the former being relevant for
hypernuclear structure studies and the latter for hypernuclear weak decays). 

Further, we have introduced the weak decay modes of $\Lambda$--hypernuclei:
beyond the mesonic channel, which is observed also for a free $\Lambda$,
the hypernuclear decay proceeds through non--mesonic processes, mainly induced
by one nucleon or by a pair of correlated nucleons. This channel is the 
dominant one in medium--heavy hypernuclei, where the Pauli principle strongly
suppresses the mesonic decay.

The results obtained within the various models proposed to describe the
mesonic and non--mesonic decay rates as well as the asymmetry parameters
in the decay of the $\Lambda$--hyperon in nuclei have been thoroughly discussed.
The mesonic rates have been reproduced quite well by calculations performed in different
frameworks. The non--mesonic rates have been considered within a variety of 
phenomenological and microscopic models, most of them being based on the exchange of
a pion between the decaying $\Lambda$ and the nucleon(s). 
More complex meson exchange potentials, as well as direct quark models have also
been considered for the evaluation of non--mesonic decay rates. In this context, 
particular interest has been devoted to the partial rates  $\Gamma_n$ and $\Gamma_p$ 
and to their ratio.

In spite of the fact that several calculations have been able to reproduce, already at 
the OPE level, the total non--mesonic width,  $\Gamma_{NM}=\Gamma_n+\Gamma_p (+\Gamma_2)$,
the values therewith obtained for $\Gamma_n/\Gamma_p$ reveal a strong disagreement with
the measured central data. Actually, due to the large experimental uncertainties
involved in the extraction of $\Gamma_n/\Gamma_p$, at present one cannot draw
definite conclusions, and different and more refined experimental analysis are required
to correct for eventual deficiencies of the models.                

Notably, the non--mesonic partial rates $\Gamma_n$ and $\Gamma_p$
are dominated by parity--conserving transition amplitudes. The asymmetric emission of 
protons from proton--induced non--mesonic decays of polarized hypernuclei is related 
to the interference between the parity--conserving and parity--violating transition 
amplitudes to $NN$ states with the same intrinsic spin $S$. 
Therefore, the study of the decay asymmetries complements the one of the non--mesonic 
partial rates, providing, at least in principle, new constraints on the 
$\Lambda N\to NN$ decay  mechanism.

Nuclear structure uncertainties are under control, and cannot influence
very much the calculation of the hypernuclear observables for the non--mesonic decay.
The total non--mesonic widths turn out to be relatively insensitive to the
details of the weak interaction model. On the contrary, the ratio $\Gamma_n/\Gamma_p$
strongly depends on the decay mechanism. Nevertheless,
the OME calculations, including the exchange of mesons more massive than the pion
such as the $\rho$, $K$, $K^*$, $\omega$ and $\eta$, as well as the (correlated or 
uncorrelated) two--pion exchange models cannot improve the comparison with the 
experimental $\Gamma_n/\Gamma_p$ ratios and decay asymmetries. These evaluations
are rather sensitive to the models used for the required
meson--baryon--baryon strong and weak vertexes. However, only by using rather 
unrealistic coupling constants it is possible to fit, simultaneously, the data
on $\Gamma_n +\Gamma_p$ and $\Gamma_n/\Gamma_p$ for different hypernuclei 
\cite{Pa-pr}.

The OPE mechanism alone is able to reproduce the observed total non--mesonic
widths, but strongly underestimates (by about one order of magnitude) the central data 
for the ratio. Only the $K$--meson--exchange turned out to be important to obtain
considerably larger $\Gamma_n/\Gamma_p$ ratios, but the central data remains underestimated. 
The inclusion in the non--mesonic transition potential of quark degrees
of freedom suffers from large theoretical uncertainties. The models that
implemented direct quark interactions in OME calculations found
$\Gamma_n/\Gamma_p$ values considerably larger than the OPE estimates,
also as a result of the $K$--meson exchange, but problems
remains in reproducing both the ratio and $\Gamma_n +\Gamma_p$ for the
considered systems.

Although some of the discussed improvements could represent a step forward in the solution
of the ${\Gamma}_n/{\Gamma}_p$ puzzle, further efforts (especially on the
experimental side) must be invested in order to understand the detailed
dynamics of the non--mesonic decay. From the theoretical point of view,
it is not easy to imagine new mechanisms as responsible for the large observed ratios.
Very recent experiments at KEK have considerably reduced the error bars on
 ${\Gamma}_n/{\Gamma}_p$, by means of {\it single} nucleon spectra measurements. 
The new experiments confirmed previous data,  with improved accuracy. 
However, in order to avoid possible deficiencies of this kind of observations, a direct 
and unambiguous extraction of the ratio is compulsory. 
As widely discussed in the present review, for such a determination 
good statistics coincidence measurements of the $nn$ and $np$ emitted pairs are required.
These correlation measurements will also allow for the identification of
the nucleons which come out from the different one-- and two--nucleon induced processes.

As far as the asymmetry parameters are concerned,
the situation is even more puzzling. Indeed, strong inconsistencies already appear at the
experimental level: the two existing experiments revealed an opposite sign of the
intrinsic asymmetry parameter, $a_{\Lambda}$, for
$^5_{\Lambda}\vec{\rm H}$e and $p$--shell hypernuclei. This is in strong contradiction
with the theoretical expectation of an intrinsic asymmetry which should be, in principle,
rather insensitive to nuclear structure effects. Some calculations reproduced the first
measurement of $a_\Lambda$, which found a large and negative value for $p$--shell
hypernuclei, but no calculation could obtain a positive value for $^5_{\Lambda}\vec{\rm H}$e.
The experiments thus cannot provide any guidance for further theoretical evaluations.
Improved experiments, establishing with certainty the sign and magnitude of $a_\Lambda$
for $s$-- and $p$--shell hypernuclei, are then strongly awaited.
                                                                
We conclude this work by reminding the reader that hypernuclear physics is 49 years old,
yet a lot of efforts remain to be done, both experimentally and theoretically, in order 
to fully understand the hyperon dynamics and decay inside the nuclear medium.
The impressive progress experienced in the last few years is promising and we hope
that it deserves a definite answer to the intriguing open questions which we have
illustrated here.

\section*{Acknowledgments} 
Fruitful and friendly discussions with H. C. Bhang, R. H. Dalitz, O. Hashimoto, A. Molinari,
E. Oset, H. Outa, A. Parre\~{n}o, K. Sasaki and Y. Sato are acknowledged.
We are especially grateful to the members of KEK for providing us with a paper of theirs prior
to publication; also we thank E. Botta and A. Feliciello for technical support and assistance 
during and after the {\em VII International Conference on Hypernuclear and Strange Particle Physics}.
We also take the opportunity to warmly acknowledge our colleagues
A. De Pace, R. Cenni and A. Ramos, who collaborated with
us in obtaining some of the results discussed in the review.
A special recognition goes to A. Parre\~{n}o, for her suggestions and comments after reading 
part of the manuscript.
The work has been partially supported by the EEC through TMR Contract CEE--0169. 

\newpage

\appendix
\section{Spin--isospin $NN\to NN$ and ${\Lambda}N\to NN$ interactions}
\label{app1}

In this appendix we show how the repulsive $NN$ and ${\Lambda}N$ strong
correlations at short distances are implemented in the $NN\rightarrow NN$
and ${\Lambda}N\rightarrow NN$ interactions and then in the
hypernuclear decay width calculated within the Polarization Propagator
Method of subsection \ref{pm}. 
The $NN\rightarrow NN$ interaction can be described through an effective
potential
given by:
\begin{equation}
\label{gmat}
G(r)=g(r)V(r) .
\end{equation}
Here $g(r)$ is a two--body correlation function, which vanishes as
$r\rightarrow 0$
and goes to 1 as $r\rightarrow \infty$, while $V(r)$ is the meson exchange
potential, which in our case 
contains $\pi$ and $\rho$ exchange: $V=V_{\pi}+V_{\rho}$.
A practical and realistic form for $g(r)$ is \cite{Os82}:
\begin{equation}
\label{pract}
g(r)=1-j_0(q_cr) ,
\end{equation}
where $j_0$ is the Bessel spherical function of order 0.
With $q_c=m_{\omega}\simeq 780$ MeV one gets a good reproduction of
realistic
$NN$ correlation functions obtained from $G$--matrix calculations.
The inverse of $q_c$ is indicative of the hard core radius of the
interaction.
Since there are no experimental indications,
the same correlation momentum is generally used for the strong ${\Lambda}N$
interaction. On the other hand, we remind the reader that $q_c$
is not necessarily the same in the two cases, given the different nature
of the repulsive forces involved. Using the correlation function (\ref{pract}) it is
easy to get the effective interaction, Eq.~(\ref{gmat}), in momentum space.
It reads:
\begin{equation}
G_{NN\to NN}(q)=V_{\pi}(q)+V_{\rho}(q)+
\frac{f_{\pi}^2}{m_{\pi}^2}\left\{g_L(q)\hat{q}_i \hat{q}_j+
g_T(q)({\delta}_{ij}-\hat{q}_i \hat{q}_j)\right\}{\sigma}_i{\sigma}_j
\vec{\tau}\cdot \vec{\tau} , \nonumber
\end{equation}
where the correlations are embodied in the functions $g_L$ and $g_T$.
Then, the spin--isospin $NN\rightarrow NN$ interaction can be separated
into a spin--longitudinal and a spin--transverse parts, as follows:
\begin{equation}
G_{NN\to NN}(q)=\left\{V_L(q)\hat{q}_i \hat{q}_j+V_T(q)({\delta}_{ij}-
\hat{q}_i \hat{q}_j)\right\}{\sigma}_i{\sigma}_j\vec{\tau}\cdot \vec{\tau}
\hspace{0.8cm} (\hat{q}_i=q_i/\mid \vec q\mid) , \nonumber
\end{equation}
where:
\begin{equation}
V_L(q)=\frac{f_{\pi}^2}{m_{\pi}^2}\left\{\vec
q\,^2F_{\pi}^2(q)G_{\pi}^0(q)+
g_L(q)\right\} , \nonumber
\end{equation}
\begin{equation}
V_T(q)=\frac{f_{\pi}^2}{m_{\pi}^2}\left\{\vec q\,^2C_{\rho}F_{\rho}^2(q)
G_{\rho}^0(q)+g_T(q)\right\} . \nonumber
\end{equation}
In the above, $F_{\pi}$ and $F_{\rho}$ are the $\pi NN$ and
$\rho NN$ form factors, respectively, while $G_{\pi}$ and $G_{\rho}$ are
the corresponding free meson propagators:
$G_m^0=1/(q_0^2-\vec q\,^2-m_m^2)$.

The ${\Lambda}N\rightarrow NN$ transition potential, modified by the
effect of the strong $\Lambda N$ correlations, splits into a
$P$--wave (again spin--longitudinal and spin--transverse) part:
\begin{equation}
G_{{\Lambda}N\rightarrow NN}(q)=
\left\{\tilde{P}_L(q)\hat{q}_i \hat{q}_j+\tilde{P}_T(q)({\delta}_{ij}-
\hat{q}_i \hat{q}_j)\right\} {\sigma}_i{\sigma}_j\vec{\tau}\cdot
\vec{\tau} , \nonumber 
\end{equation}
with:
\begin{equation}
\label{pl}
\tilde{P}_L(q)=\frac{f_{\pi}}{m_{\pi}}\frac{P}{m_{\pi}}
\left\{\vec q\,^2F_{\pi}^2(q)G_{\pi}^0(q)+g_L^{\Lambda}(q)\right\} ,
\end{equation}
\begin{equation}
\label{pt}
\tilde{P}_T(q)=\frac{f_{\pi}}{m_{\pi}}\frac{P}{m_{\pi}}g_T^{\Lambda}(q) ,
\end{equation}
and an $S$--wave part:
\begin{equation}
\label{s}
\tilde{S}(q)=\frac{f_{\pi}}{m_{\pi}}S\left\{F_{\pi}^2(q)G_{\pi}^0(q)-
\tilde{F}_{\pi}^2(q)\tilde{G}_{\pi}^0(q)\right\}\mid \vec q \mid .
\end{equation}
Form factors and propagators with a tilde imply that they are calculated
by replacing $\vec q\,^2\rightarrow \vec q\,^2+q_c^2$, while $C_{\rho}$ is
given by:
\begin{equation}
\label{rhocoup}
C_{\rho}={\frac{f_{\rho}^2}{m_{\rho}^2}}
\left[\frac{f_{\pi}^2}{m_{\pi}^2}\right]^{-1} .
\end{equation}
The expressions for the correlation functions are the following ones:
\begin{equation}
g_L(q)=-\left\{\vec q\,^2+\frac{1}{3}q_c^2\right\}\tilde{F}_{\pi}^2(q)
\tilde{G}_{\pi}^0(q)-\frac{2}{3}q_c^2C_{\rho}\tilde{F}_{\rho}^2(q)
\tilde{G}_{\rho}^0(q) , \nonumber
\end{equation}
\begin{equation}
g_T(q)=-\frac{1}{3}q_c^2\tilde{F}_{\pi}^2(q)\tilde{G}_{\pi}^0(q)
-\left\{\vec q\,^2+\frac{2}{3}q_c^2\right\}C_{\rho}\tilde{F}_{\rho}^2(q)
\tilde{G}_{\rho}^0(q) , \nonumber
\end{equation}
\vspace{0.08in}
\begin{equation}
g_L^{\Lambda}(q)=-\left\{\vec
q\,^2+\frac{1}{3}q_c^2\right\}\tilde{F}_{\pi}^2(q)
\tilde{G}_{\pi}^0(q) , \nonumber
\end{equation}
\begin{equation}
g_T^{\Lambda}(q)=-\frac{1}{3}q_c^2\tilde{F}_{\pi}^2(q)\tilde{G}_{\pi}^0(q) . \nonumber
\end{equation}
The functions $g_L$ and $g_T$ [$g^{\Lambda}_L$ and $g^{\Lambda}_T$]   
have been obtained from Eqs.~(\ref{gmat}) and (\ref{pract}) with  
$V=V_{\pi}+V_{\rho}$ [$V=V_{\pi}$]. Using the set of parameters:
\begin{equation}
q_c=780\, {\rm MeV}, \hspace{0.1in} {\Lambda}_{\pi}=1.2\,
{\rm GeV}, \hspace{0.1in}
{\Lambda}_{\rho}=2.5\, {\rm GeV}, \hspace{0.1in} f^2_{\pi}/4\pi= 0.08,  
\hspace{0.1in} C_{\rho}=2 , \nonumber
\end{equation}
at zero energy and momentum we have:
\begin{equation}
g_L(0)=g_T(0)=0.615, \hspace{0.4in}
g_L^{\Lambda}(0)=g_T^{\Lambda}(0)=0.155 , \nonumber
\end{equation}
which can be identified with the customary Landau--Migdal parameters.
However, if one wishes to keep the zero energy and momentum limit of $g_{L,T}$
and $g_{L,T}^{\Lambda}$ as free parameters, a replacement of the previous functions by:
\begin{equation}
g_{L,T}(q)\rightarrow g^{\prime}\frac{g_{L,T}(q)}{g_{L,T}(0)},
\hspace{0.4in}
g_{L,T}^{\Lambda}(q)\rightarrow
g^{\prime}_{\Lambda}\frac{g_{L,T}^{\Lambda}(q)}
{g_{L,T}^{\Lambda}(0)} , \nonumber
\end{equation}
is required.

Now we come to the implementation of the above spin--isospin effective
potentials in the nuclear matter $\Lambda$ self--energy of Eq.~(\ref{Sigma1}). 
From the graphs of Fig.~\ref{self1},
by applying the Feynman rules we obtain:
\begin{eqnarray}
\label{Sigma3}
&&{\Sigma}_{\Lambda}(k)=3i(G m_{\pi}^2)^2\int \frac{d^4q}{(2\pi)^4}
G_N(k-q)\left(S^2+\frac{P^2}{m_{\pi}^2}\vec q\,^2\right)F_{\pi}^2(q)
\left\{G^0_{\pi}(q) \right. \\
& &+\,{G^0}^2_{\pi}(q)\frac{f^2_{\pi}}{m^2_{\pi}}F^2_{\pi}(q)U(q)q_iq_j
\left[\delta_{ij}+\left\{V_L(q)\hat{q}_i \hat{q}_j+V_T(q)({\delta}_{ij}-
\hat{q}_i \hat{q}_j)\right\}U(q) \right. \nonumber \\
& &+\left. \left. \left\{V_L(q)\hat{q}_j \hat{q}_k+V_T(q)({\delta}_{jk}-
\hat{q}_j \hat{q}_k)\right\}\left\{V_L(q)\hat{q}_k
\hat{q}_i+V_T(q)({\delta}_{ki}-
\hat{q}_k \hat{q}_i)\right\}U^2(q)
+ . . .^{\,}\right] \right\} , \nonumber 
\end{eqnarray}
where only the $NN$ short range correlations
are taken into account. The function $U=U^{ph}+U^{\Delta h}+U^{2p2h}$ contains the
{\sl p--h}, {\sl $\Delta$--h} and irreducible {\sl 2p--2h} proper polarization propagators.
Now we must include in the previous equation the repulsive correlations
in the lines connecting weak and strong vertices.
For the $P$--wave interaction this corresponds to perform the replacement:
\begin{equation}
\frac{f_{\pi}}{m_{\pi}}\frac{P}{m_{\pi}}\vec q\,^2F^2_{\pi}(q)G^0_{\pi}(q)
\hat{q}_i\hat{q}_j\rightarrow
\tilde{P}_L(q)\hat{q}_i \hat{q}_j+\tilde{P}_T(q)({\delta}_{ij}-
\hat{q}_i \hat{q}_j) , \nonumber
\end{equation}
while the interaction which connects the $S$--wave weak vertex and the
$P$--wave strong vertex becomes:
\begin{equation}
\frac{f_{\pi}}{m_{\pi}}SF^2_{\pi}(q)G^0_{\pi}(q)\mid\vec q\mid
\hat{q}_i\rightarrow
\tilde{S}(q)\hat{q}_i . \nonumber
\end{equation}
The functions $\tilde{P}_L$, $\tilde{P}_T$ and $\tilde{S}$ are given by
equations (\ref{pl}), (\ref{pt}) and (\ref{s}).
Moreover, the polarization propagator $U$ in the modified
Eq.~(\ref{Sigma3})
has to be understood as $U_L$ when multiplied by a spin--longitudinal potential
($V_L$, $\tilde{P}_L$, $\tilde{S}$), while it is $U_T$ when multiplied
by $V_T$ or $\tilde{P}_T$.
By introducing these prescriptions in Eq.~(\ref{Sigma3}) and summing the two
geometric series (there is no interference between longitudinal and
transverse modes) one obtains:
\begin{equation}
{\Sigma}_{\Lambda}(k)=3i(G m_{\pi}^2)^2\int \frac{d^4q}{(2\pi)^4}
G_N(k-q)\alpha(q) , \nonumber
\end{equation}
with $\alpha(q)$ given by Eq.~(\ref{Alpha}). Then, integrating over $q_0$,
for the $\Lambda$ decay width one finally obtains Eq.~(\ref{Sigma2}).

\vfill\eject

\newpage
\begin{thebibliography}{999}
\label{biblio}
\bibitem{Da53} M. Danysz and J. Pniewski, {\em Philos. Mag.} {\bf 44} (1953) 348.
\bibitem{Ha95} J. Haidenbuaer, K. Holinde, K. Kilian And T. Sefzick,
{\em Phys. Rev.} {\bf C 52} (1995) 3496.
\bibitem{Ki98} T. Kishimoto, {\em Nucl. Phys.} {\bf A 629} (1998) 369c;
T. Kishimoto {\sl et al.}, {\em Nucl. Phys.} {\bf A 663} (2000) 509. 
\bibitem{Pa99} A Parre\~{n}o, A. Ramos, N. G. Kelkar and C. Bennhold,
{\em Phys. Rev.} {\bf C 59} (1999) 2122. 
\bibitem{Re94} B. Holzenkamp, K. Holinde and J. Speth, {\em Nucl. Phys.} {\bf A 500}
(1989) 485; \\ A. Reuber, K. Holinde and J. Speth, {\em Nucl. Phys.} {\bf A 570} (1994) 543.
\bibitem{Ri96}  M. M. Nagels, Th. A. Rijken and J. J. de Swart, 
{\em Phys. Rev.} {\bf D 20} (1979) 1633; 
P. M. M. Maessen, Th. A. Rijken and J. J. de Swart, {\em Phys. Rev.} {\bf C 40} (1989) 2226; 
Th. A. Rijken and V. G. J. Stoks, {\em Phys. Rev.} {\bf C 54} (1996) 2851.
\bibitem{Ri99} Th. A. Rijken, V. G. J. Stoks and Y. Yamamoto, {\em Phys. Rev.} {\bf C 59} (1999) 21.
\bibitem{Ri99a} V. G. J. Stoks and Th. A. Rijken, {\em Phys. Rev.} {\bf C 59} (1999) 3009.
\bibitem{Ya92} M. Oka, K. Shimizu and K. Yazaki, {\em Nucl. Phys.} {\bf A 464} 
(1987) 700; K. Yazaki, {\em Perspectives of Meson Science}, T. Yamazaki, K. Nakai and
K. Nagamine eds (Elsevier Science Publisher, 1992) p. 795.
\bibitem{St90}U. Straub, Z. Y. Zhang, K. Br\"{a}uer, A. Faessler, 
S. B. Khadkikar and G. L\"{u}beck, {\em Nucl. Phys.} {\bf A 483} (1988) 686;
ibid. {\bf A 508} (1990) 385c.
\bibitem{Fu96} Y. Fujiwara, C. Nakamoto and Y. Suzuki, {\em Phys. Rev.} {\bf C 54} (1996) 2180.
\bibitem{Us98} Q. N. Usmani and A. R. Bodmer, {\em Nucl. Phys.} {\bf A 639} (1998) 147c.
Q. N. Usmani and A. R. Bodmer, {\em Phys. Rev.} {\bf C 60} (1999) 055215.
\bibitem{Sh99} A. R. Bodmer and Q. N. Usmani, {\em Nucl. Phys.} {\bf A 477} (1988) 621;
M. Shoeb, N. Neelofer, Q. N. Usmani and M. Z. Rahman Khan,
{\em Phys. Rev.} {\bf C 59} (1999) 2807.
\bibitem{Ak00} Y. Akaishi {\sl et al.}, {\em Phys. Rev. Lett.} {\bf 84} (2000) 3539.
\bibitem{Mo86} T. Motoba and H. Band$\overline{\rm o}$, {\em Prog. Theor. Phys.} {\bf 76} (1986) 1321.
\bibitem{Gi88} B. F. Gibson, {\em Nucl. Phys.} {\bf A 479} (1988) 115c.
\bibitem{Gi94} B. F. Gibson, I. R. Afnan, J. A. Carlson and D. R. Lehman, 
{\em Progr. Theor. Phys. Suppl.} {\bf 117} (1994) 339.
\bibitem{Gi95} B. F. Gibson, E. V. Hungerford, {\em Phys. Rep.} {\bf 257} (1995) 349.
\bibitem{Hi01} E. Hiyama, M. Kamimura, T. Motoba, T. Yamada and Y. Yamamoto, 
e--print archive {\bf nucl--th/0106070}.
\bibitem{Po96} M. Hjorth-Jensen, A. Polls, A. Ramos and H. M\"{u}ther,
{\em Nucl. Phys.} {\bf A 605} (1996) 458.
\bibitem{Mill85} D. J. Millener, C. B. Dover, A. Gal and R. H. Dalitz, {\em Phys. Rev.}
{\bf C 31} (1985) 499.
\bibitem{Ta98} H. Tamura {\sl et al.}, {\em Nucl. Phys.} {\bf A 639} (1998) 83c;
Proceedings of the APCTP workshop {\em SNP99}, Il--T. Cheon, S. W. Hong
and T. Motoba eds, World Scientific Singapore (2000) p. 411;
{\em Phys. Rev. Lett.} {\bf 84} (2000) 5963.
\bibitem{Tam01} H. Akikawa {\sl et al.}, {\em Nucl. Phys.} {\bf A 691} (2001) 134c;
H. Tamura, {\em Hirschegg 2001, Structure of Hadrons} (2001) p. 290. 
\bibitem{Mill01} D. J. Millener, {\em Nucl. Phys.} {\bf A 691} (2001) 93c.
\bibitem{Has95} T. Hasegawa  {\sl et al.}, {\em Phys. Rev. Lett.} {\bf 74} (1995) 224.
\bibitem{Ba86} H. Band$\overline{\rm o}$ and T. Motoba, {\em Prog. Theor. Phys.} {\bf 76} (1986) 1321.
\bibitem{Ba90} H. Band$\overline{\rm o}$, T. Motoba and J. \v{Z}ofka, {\em Int. J. Mod.
Phys.} {\bf A 5} (1990) 4021.
\bibitem{Br78} W. Br\"{u}ckner {\sl et al.}, {\em Phys. Lett.} {\bf 79 B} (1978) 157.
\bibitem{Bo80} A. Bouyssy, {\em Phys. Lett.} {\bf 84 B} (1979) 41;
ibid. {\bf 91 B} (1980) 15.
\bibitem{Ma81} M. May {\sl et al.}, {\em Phys. Rev. Lett.} {\bf 47} (1981) 1106,
ibid. {\bf 51} (1983) 2085.
\bibitem{Je90} B. K. Jennings, {\em Phys. Lett.} {\bf B 246} (1990) 325.
\bibitem{Ha99} O. Hashimoto,
Proceedings of the APCTP workshop {\em SNP99}, Il--T. Cheon, S. W. Hong
and T. Motoba eds, World Scientific Singapore (2000) p. 116.
\bibitem{Sa99} A. Sakaguchi {\sl et al.},
Proceedings of the APCTP workshop {\em SNP99}, Il--T. Cheon, S. W. Hong
and T. Motoba eds, World Scientific Singapore (2000) p. 231.
\bibitem{Hi98} E. Hiyama, M. Kamimura, T. Motoba, T. Yamada and Y. Yamamoto,
{\em Nucl. Phys.} {\bf A 639} (1998) 173c, {\em Phys. Rev. Lett.} {\bf 85} (2000) 270.
\bibitem{Ho01} H. Hotchi {\sl et al.}, {\em Phys. Rev.} {\bf C 64} (2001) 044302. 
\bibitem{Do88} C. B. Dover, A. Gal and D. J. Millener, {\em Phys. Rev.} 
{\bf C 38} (1988) 2700.
\bibitem{Pi91} P. H. Pile {\sl et al.}, {\em Phys. Rev. Lett.} {\bf 66} (1991) 2585.
\bibitem{Ha96} T. Hasegawa {\sl et al.}, {\em Phys. Rev.} {\bf C 53} (1996) 1210.
\bibitem{Tan01} K. Tanida {\sl et al.}, {\em Phys. Rev. Lett.} {\bf 86} (2001) 1982. 
\bibitem{Hi99} E. Hiyama, M. Kamimura, K. Miyazaki and T, Motoba, 
{\em Phys. Rev.} {\bf C 59} (1999) 2351.
\bibitem{Ar93} T. A. Armstrong {\sl et al.}, {\em Phys. Rev.} {\bf C 47} (1993) 1957.
\bibitem{Ku98} H. Ohm {\sl et al.}, {\em Phys. Rev.} {\bf C 55} (1997) 3062;
P. Kulessa {\sl et al.}, {\em Phys. Lett.} {\bf B 427} (1998) 403.
\bibitem{Su99} H. Nemura, Y. Suzuki, Y. Fujiwara and C. Nakamoto, 
{\em Prog. Theor. Phys.} {\bf 101} (1999) 981. 
\bibitem{Mo91} T. Motoba, H. Band$\overline{\rm o}$, T. Fukuda and J. \v{Z}ofka, 
{\em Nucl. Phys.} {\bf A 534} (1991) 597.
\bibitem{St93} U. Straub, J. Nieves, A. Faessler and E. Oset,
{\em Nucl. Phys.} {\bf A 556} (1993) 531.
\bibitem{Ou98} H. Outa {\sl et al.}, {\em Nucl. Phys.} {\bf A 639} (1998) 251c.
\bibitem{Do89} C. B. Dover, D. J. Millener and A. Gal, 
{\em Phys. Rep.} {\bf 184} (1989) 1.
\bibitem{Ha90} T. Harada, S. Shinmura, Y. Akaishi and H. Tanaka,
{\em Nucl. Phys.} {\bf A507} (1990) 715.
\bibitem{Ak97} Y. Akaishi and T. Yamazaki, {\em Prog. Part. Nucl. Phys.} {\bf 39} (1997) 565.
\bibitem{Da81} R. H. Dalitz, {\em Nucl. Phys.} {\bf A 354} (1981) 101c.
\bibitem{Be85} R. Bertini {\sl et al.}, {\em Phys. Lett.} {\bf 90 B} (1980) 375; ibid.
{\bf 136 B} (1984) 29; ibid. {\bf 158 B} (1985) 19.
\bibitem{Ba99} S. Bart {\sl et al.}, {\em Phys. Rev. Lett.} {\bf 83} (1999) 5238.
\bibitem{Br82} R. Brockmann and E. Oset, {\em Phys. Lett.} {\bf 118 B} (1982) 33.
\bibitem{Os90} E. Oset, P. Fern\'andez de C\'ordoba, L. L. Salcedo and
R. Brockmann, {\em Phys. Rep.} {\bf 188} (1990) 79.
\bibitem{Ya94} T. Yamada and K. Ikeda, {\em Prog. Theor. Phys. Suppl.}
{\bf 177} (1994) 201.
\bibitem{Mar95} J. Mare\v{s} {\sl et al.}, {\em Nucl. Phys.} {\bf A 594} (1995) 311.
\bibitem{Ha89} R. S. Hayano {\sl et al.}, {\em Phys. Lett.} {\bf B 231} (1989) 355.
\bibitem{Ou94} H. Outa, T. Yamazaki, M. Iwasaki and R. S. Hayano, 
{\em Prog. Theor. Phys. Suppl.} {\bf 117} (1994) 177.
\bibitem{Na98} T. Nagae {\sl et al.}, {\em Phys. Rev. Lett.} {\bf 80} (1998) 1605.
\bibitem{Ak86} Y. Akaishi, {\em Few--body Syst. Suppl.} {\bf 1} (1986) 120.
\bibitem{Ba94} C. J. Batty, E. Friedman and A. Gal, {\em Phys. Lett.} {\bf B 335}
(1994) 273.
\bibitem{Do83} C. B. Dover and A. Gal, {\em Ann. Phys.} {\bf 146} (1983) 309.
\bibitem{Nak98} K. Nakazawa, {\em Nucl. Phys.} {\bf A 639} (1998) 345c.
\bibitem{Ma98} M. May, {\em Nucl. Phys.} {\bf A 639} (1998) 363c.
\bibitem{Fuk98} T. Fukuda, {\em Nucl. Phys.} {\bf A 639} (1998) 355c;
{\em Phys. Rev.} {\bf C 58} (1998) 1306. 
\bibitem{Ka00} P. Khaustov {\sl et al.}, {\em Phys. Rev.} {\bf C 61} (2000) 054603. 
\bibitem{Mo01} T. Motoba, {\em Nucl. Phys.} {\bf A 691} (2001) 213c. 
\bibitem{Ta95} S. Tadokoro, H. Kobayashi and Y. Akaishi, {\em Nucl Phys.}
{\bf A 585} (1995) 225c.
\bibitem{Ao91} S. Aoki {\sl et al.}, {\em Prog. Theor. Phys.} {\bf 85} (1991) 1287.
\bibitem{Ic01} A. Ichikawa {\sl et al.}, {\em Phys. Lett.} {\bf B 500} (2001) 37.
\bibitem{Ah00} J. K. Ahn {\sl et al.}, {\em Phys. Rev.} {\bf C 62} (2000) 055201. 
\bibitem{Ku95} I. Kumagai-Fuse and Y. Akaishi, {\em Prog. Theor. Phys.} {\bf 94}
(1995) 151. 
\bibitem{Ja77} R. L. Jaffe, {\em Phys. Rev. Lett.} {\bf 38} (1977) 195.
\bibitem{Ya98} K. Yamamoto {\sl et al.}, {\em Nucl. Phys.} {\bf A 639} (1998) 371c.
\bibitem{Ch98} R. E. Chrien, {\em Nucl. Phys.} {\bf A 629} (1998) 388c.
\bibitem{Ya00} K. Yamamoto {\sl et al.}, {\em Phys. Lett.} {\bf B 478} (2000) 401.
\bibitem{Al00} A. Alavi--Harati {\sl et al.}, {\em Phys. Rev. Lett.} {\bf 84} (2000) 2593.  
\bibitem{Tak01} H. Takahashi {\sl et al.}, {\em Phys. Rev. Lett.} {\bf 87} (2001) 212503.
\bibitem{Ah01} J. K. Ahn, {\sl et al.}, {\em Phys. Rev. Lett.} {\bf 87} (2001) 132504. 
\bibitem{La98} D. E. Lanskoy, {\em Phys. Rev.} {\bf C 58} (1998) 3351.
\bibitem{Af00} I. R. Afnan, 
Proceedings of the APCTP workshop {\em SNP99}, Il--T. Cheon, S. W. Hong
and T. Motoba eds, World Scientific Singapore (2000) p. 149. 
\bibitem{It01} K. Itonaga, T. Ueda and T. Motoba, {\em Nucl. Phys.} {\bf A 691} (2001) 197c. 
\bibitem{Pa01a} A. Parre\~{n}o, A. Ramos and C. Bennhold, 
e--print archive {\bf nucl--th/0106054}, in press on {\em Phys. Rev.} {\bf C}.
\bibitem{Be81} R. Bertini {\sl et al.}, {\em Nucl. Phys.} {\bf A 360} (1981) 315;
ibid. {\bf A 368} (1981) 365.
\bibitem{Aj01} S. Ajimura {\sl et al.}, {\em Phys. Rev. Lett.} {\bf 86} (2001) 4255;
H. Kohri {\sl et al.}, e--print archive {\bf nucl--ex/0110007}. 
\bibitem{Da97} R. H. Dalitz, D. H. Davis, T. Motoba and D. N. Tovee, {\em Nucl. Phys.}
{\bf A 625} (1997) 71.
\bibitem{Mo89} T. Motoba, {\em Nuovo Cim.} {\bf 102 A} (1989) 345.
\bibitem{Ta01} H. Tamura {\sl et al.}, {\em Nucl. Phys.} {\bf A 670} (2000) 249c;
K. Tanida {\sl et al.}, {\em Nucl. Phys.} {\bf A 684} (2001) 560c; ibid. {\bf A 691} (2001) 115c. 
\bibitem{Gr85} R. Grace {\sl et al.}, {\em Phys. Rev. Lett.} {\bf 55} (1985) 1055.
\bibitem{Sz91}J. J. Szymanski {\sl et al.}, {\em Phys. Rev.} {\bf C 43} (1991) 849.
\bibitem{Ya86} T. Yamazaki {\sl et al.}, {\em Nucl. Phys.} {\bf 450} (1986) 1c.
\bibitem{Sa91} A. Sakaguchi {\sl et al.}, {\em Phys. Rev.} {\bf C 43} (1991) 73.
\bibitem{FI98} V. Lucherini {\sl et al.}, {\em Nucl. Phys.} {\bf A 639} (1998) 529c;
V. Filippini {\sl et al.}, {\em Nucl. Phys.} {\bf A 639} (1998)  537c;
T. Bressani {\sl et al}, 
Proceedings of the APCTP workshop {\em SNP99}, Il--T. Cheon, S. W. Hong
and T. Motoba eds, World Scientific Singapore (2000) p. 383;
A. Feliciello, {\em Nucl. Phys.} {\bf A 691} (2001) 170c;
P. Gianotti, {\em Nucl. Phys.} {\bf A 691} (2001) 483c.
\bibitem{Ou95} H. Outa {\sl et al.}, {\em Nucl. Phys.} {\bf A 585} (1995) 109c.
\bibitem{Ha98} O. Hashimoto, {\em Nucl. Phys.} {\bf A 639} (1998) 93c.
\bibitem{Mi85} C. Milner {\sl et al.}, {\em Phys. Rev. Lett.} {\bf 54} (1985) 1237.
\bibitem{Pe86} J. C. Peng, {\em Nucl. Phys.} {\bf A 450} (1986) 129c.
\bibitem{No95} H. Noumi {\sl et al.}, {\em Phys. Rev.} {\bf C 52} (1995) 2936.
\bibitem{Bh98} H. C. Bhang {\sl et al.}, {\em Phys. Rev. Lett.} {\bf 81} (1998) 4321;
H. Park {\sl et al.}, {\em Phys. Rev.} {\bf C 61} (2000) 054004. 
\bibitem{Na01} N. Nagae, {\em Nucl. Phys.} {\bf A 691} (2001) 76c. 
\bibitem{Sa98} Y. Sato {\sl et al.}, {\em Nucl. Phys.} {\bf A 639} (1998) 279c.
\bibitem{No98} H. Noumi, {\em Nucl. Phys.} {\bf A 639} (1998) 121c; K. Imai, 
Proceedings of the APCTP workshop {\em SNP99}, Il--T. Cheon, S. W. Hong
and T. Motoba eds, World Scientific Singapore (2000) p. 419. 
\bibitem{Hu00} E. V. Hungerford,
Proceedings of the APCTP workshop {\em SNP99}, Il--T. Cheon, S. W. Hong
and T. Motoba eds, World Scientific Singapore (2000) p. 356.
\bibitem{Ta99} L. Tang,
Proceedings of the APCTP workshop {\em SNP99}, Il--T. Cheon, S. W. Hong
and T. Motoba eds, World Scientific Singapore (2000) p. 393. 
\bibitem{Gi79} F. J. Gillman and M. B. Wise, {\em Phys. Rev.} {\bf D 20} (1979) 2392.
\bibitem{Pa90} E. A. Paschos, T. Schneider and Y. L. Wu, {\em Nucl. Phys.}
{\bf B 332} (1990) 285.
\bibitem{Ok98a} K. Takayama and M. Oka, {\bf hep-ph/9809388}; {\bf hep-ph/9811435}.
\bibitem{Os93} J. Nieves and E. Oset {\em Phys. Rev.} {\bf C 47} (1993) 1478.
\bibitem{Mo94} T. Motoba and K. Itonaga, {\em Prog. Theor Phys. Suppl.} {\bf 117}
(1994) 477.
\bibitem{It88} K. Itonaga, T. Motoba and H. Band$\overline{\rm o}$, {\em Z. Phys.} {\bf A 330}
(1988) 209; {\em Nucl. Phys.} {\bf A 489} (1988) 683.
\bibitem{Co90} J. Cohen, {\em Prog. Part. Nucl. Phys.} {\bf 25} (1990) 139.
\bibitem{Ku95a}I. Kumagai-Fuse, S. Okabe and Y. Akaishi, {\em Phys. Lett.} 
{\bf B 345} (1995) 386.
\bibitem{Da59} R. H. Dalitz, {\em Phys. Rev.} {\bf 112} (1958) 605;
R. H. Dalitz and L. Liu, {\em Phys. Rev.} {\bf 116} (1959) 1312.
\bibitem{Os98} E. Oset and A. Ramos, {\em Prog. Part. Nucl. Phys.} {\bf 41} 
(1998) 191.
\bibitem{Go97} J. Golak, K. Miyagawa, H. Kamada, H. Witala, W. Gl\"{o}ckle, 
A. Parre\~{n}o, A. Ramos and C. Bennhold, {\em Phys. Rev.} {\bf C 55} (1997) 2196;
Erratum: ibid. {\bf C 56} (1997) 2892;
H. Kamada, J. Golak, K. Miyagawa, H. Witala, W. Glockle, {\em Phys. Rev.} 
{\bf C 57} (1998) 1595.
\bibitem{Be92} C. Bennhold and A. Ramos, {\em Phys. Rev.} {\bf C 45} (1992) 3017.
\bibitem{It98} K. Itonaga, T. Ueda and T. Motoba, {\em Nucl. Phys.}  
{\bf A 639} (1998) 329c.
\bibitem{Pa01}  A. Parre\~{n}o and A. Ramos, e--print archive {\bf nucl--th/0104080},
in press on {\em Phys. Rev.} {\bf C}.
\bibitem{It95} K. Itonaga, T. Ueda and T. Motoba, {\em Nucl. Phys.} {\bf A 577}
(1994) 301c ; ibid. {\bf A 585} (1995) 331c.  
\bibitem{Pa97} A. Parre\~{n}o, A. Ramos and C. Bennhold,
{\em Phys. Rev.} {\bf C 56} (1997) 339.
\bibitem{Ok98} T. Inoue, M. Oka, T. Motoba and K. Itonaga,
{\em Nucl. Phys.} {\bf A 633} (1998) 312. 
\bibitem{Du86} J. F. Dubach, G. B. Feldman, B. R. Holstein and L. de la Torre,
{\em Nucl. Phys.} {\bf A 450} (1986) 71c. 
\bibitem{Du96} J. F. Dubach, G. B. Feldman, and B. R. Holstein,
{\em Ann. Phys.} {\bf 249} (1996) 146.
\bibitem{Sh94} M. Shmatikov, {\em Nucl Phys.} {\bf A 580} (1994) 538.
\bibitem{Os01} D. Jido, E. Oset and J. E. Palomar, {\em Nucl. Phys.} {\bf A 694} (2001) 525.  
\bibitem{Ma94} K. Maltman and M. Shmatikov, {\em Phys. Lett.} {\bf B 331} (1994) 1; 
{\em Nucl. Phys.} {\bf A 585} (1995) 343c.  
\bibitem{Ma95} K. Maltman and M. Shmatikov, {\em Phys. Rev.} {\bf C 51} (1995) 1576.  
\bibitem{Pa98} A. Parre\~{n}o, A. Ramos, C. Bennhold and K. Maltman, {\em Phys. Lett.}  
{\bf B 435} (1998) 1.  
\bibitem{Ch83} C.-Y. Cheung, D. P. Heddle and L. S. Kisslinger,  
{\em Phys. Rev.} {\bf C 27} (1983) 335; 
D. P. Heddle and L. S. Kisslinger, {\em Phys. Rev.} {\bf C 33} (1986) 608. 
\bibitem{Ok96} T. Inoue, S. Takeuki and M. Oka, {\em Nucl. Phys.} {\bf A 597} (1996) 563.
\bibitem{Ok99} K. Sasaki, T. Inoue and M. Oka, 
{\em Nucl. Phys.} {\bf A 669} (2000) 331; Erratum: ibid. {\bf A 678} (2000) 455. 
\bibitem{Ju01} J.-H. Jun and H. C. Bhang, {\em Nuovo Cim.} {\bf 112 A} (1999) 649;
J.-H. Jun, {\em Phys. Rev.} {\bf C 63} (2001) 044012. 
\bibitem{Te62} V. L. Telegdi, {\em Sci. Am.} {\bf 206} (1962) 50.
\bibitem{Po76} B. Povh, {\em Rep. Prog. Phys.} {\bf 39} (1976) 823.
\bibitem{Mo74} A. Montwill {\sl et al.}, {\em Nucl. Phys.} {\bf A 234} (1974) 413.  
\bibitem{Ze98} V. J. Zeps, {\em Nucl. Phys.} {\bf A 639} (1998) 261c. 
\bibitem{Bo87} J. P. Bocquet {\sl et al.}, {\em Phys. Lett.} {\bf B 182} (1986) 146; 
ibid. {\bf B 192} (1987) 312.   
\bibitem{Ka01} B. Kamys {\sl et al.}, {\em Eur. Phys. J.} {\bf A 11} (2001) 1.
\bibitem{Ku01} P. Kulessa {\sl et al.}, e--print archive {\bf nucl--ex/0108027}.
\bibitem{Has89} A. Hashimoto {\sl et al.}, {\em Nuovo Cim.} {\bf 102 A} (1989) 679. 
\bibitem{Ru98} A. Rusek, {\em Nucl. Phys.} {\bf A 639} (1998) 111c. 
\bibitem{Ch53} W. Cheston and H. Primakov, {\em Phys. Rev.} {\bf 92} (1953) 1537. 
\bibitem{Bl62} R. H. Dalitz and G. Rajasekaran, {\em Phys. Lett.} {\bf 1} (1962) 58.  
\bibitem{Bl63} M. M. Block and R. H. Dalitz, {\em Phys. Rev. Lett.} {\bf 11} (1963) 96.
\bibitem{Da73} R. H. Dalitz, Proceedings of the {\em Summer Study Meeting on Nuclear 
and Hypernuclear Physics with Kaon Beams}, BNL Report No. 18335 (1973) p. 41. 
\bibitem{Do87} C. B. Dover, {\em Few--Body Systems Suppl.} {\bf 2} (1987) 77.  
\bibitem{Sc92} R. A. Schumacher, {\em Nucl. Phys.} {\bf A 547} (1992) 143c. 
\bibitem{Al99b} W. M. Alberico and G. Garbarino,
{\em Phys. Lett.} {\bf B 486} (2000) 362.
\bibitem{Co90a} J. Cohen, {\em Phys. Rev.} {\bf C 42} (1990) 2724. 
\bibitem{Ru99} Z. Rudy {\sl et al.}, {\em Eur. Phys. J.} {\bf A 5} (1999) 127;
W. Cassing {\sl et al.}, e--print archive {\bf nucl--ex/0109012}. 
\bibitem{Ad67} J. B. Adams, {\em Phys. Rev.} {\bf 156} (1967) 1611. 
\bibitem{Mc84} B. H. J. McKellar and B. F. Gibson, {\em Phys. Rev.} {\bf C 30} (1984) 322. 
\bibitem{Na88} G. Nardulli, {\em Phys. Rev.} {\bf C 38} (1988) 832. 
\bibitem{Pa95-96} A. Parre\~no, A. Ramos and C. Bennhold, {Phys. Rev.} {\bf C 52} (1995)
R1768; Erratum: ibid {\bf C 54} (1996) 1500. 
\bibitem{Ta85} K. Takeuki, H. Takaki and
H. Band$\overline{\rm o}$, {\em Prog. Theor. Phys.} {\bf 73} (1985) 841. 
\bibitem{Ba85} H. Band$\overline{\rm o}$, {\em Prog. Theor. Phys. Suppl.} {\bf 81} (1985) 181. 
\bibitem{Ra94} A. Ramos and C. Bennhold, {\em Nucl. Phys.} {\bf A 577} (1994) 287c. 
\bibitem{HP01} K. Hagino and A. Parre\~no, {\em Phys. Rev.} {\bf C 63} (2001) 044318.
\bibitem{Ba88} H. Band$\overline{\rm o}$, Y. Shono and H. Takaki, {\em Int. J. Mod. Phys.}
{\bf A 3} (1988) 1581. 
\bibitem{Sh94a} M. Shmatikov, {\em Phys. Lett.} {\bf B 322} (1994) 311.
\bibitem{Sh95} S. Shinmura, {\em Nucl. Phys.} {\bf A 585} (1995) 357c. 
\bibitem{Ok99a} M. Oka, {\em Nucl. Phys.} {\bf A 647} (1999) 97.
\bibitem{Os85} E. Oset and L. L. Salcedo, {\em Nucl. Phys.} {\bf A 443} (1985) 704.  
\bibitem{Al91} W. M. Alberico, A. De Pace, M. Ericson and A. Molinari,   
{\em Phys. Lett.} {\bf B 256} (1991) 134.
\bibitem{Ra95} A. Ramos, E. Oset and L. L. Salcedo,
{\em Phys. Rev.} {\bf C 50} (1995) 2314.
\bibitem{Al99} W. M. Alberico, A De Pace, G. Garbarino and A. Ramos,
{\em Phys. Rev.} {\bf C 61} (2000) 044314.
\bibitem{Al99a} W. M. Alberico, A. De Pace, G. Garbarino, R. Cenni, 
{\em Nucl. Phys.} {\bf A 668} (2000) 113. 
\bibitem{Zh99} L. Zhou and J. Piekarewicz, {\em Phys. Rev.} {\bf C 60} 
(1999) 024306. 
\bibitem{Sh93} S. Shinmura, {\em Prog. Theor. Phys.} {\bf 89} (1993) 1115.
\bibitem{Pa95} A. Parre\~no, A. Ramos and E. Oset, {\em Phys. Rev.}
{\bf C 51} (1995) 2477.  
\bibitem{Ou00} H. Outa {\sl et al.}, {\em Nucl. Phys.} {\bf A 670} (2000) 281c. 
\bibitem{Os86a} E. Oset, L. L. Salcedo and Q. N. Usmani, {\em Nucl. Phys.}
{\bf A 450} (1986) 67c. 
\bibitem{No95a} H. Noumi {\sl et al.}, Proceedings of the {\em IV International
Symposium on Weak and Electromagnetic Interactions in Nuclei},
Ed. H. Ejiri, T. Kishimoto and T. Sato (World Scientific, 1995) p. 550.
\bibitem{Sa00} Y. Sato {\sl et al.},
{\em Nucl. Phys.} {\bf A 691} (2001) 189c. 
\bibitem{Er90} M. Ericson and H. Band$\overline{\rm o}$, {\em Phys. Lett.} {\bf B 237} (1990) 169.
\bibitem{Mo92} T. Motoba, {\em Nucl. Phys.} {\bf A 547} (1992) 115c.  
\bibitem{Ha01} O. Hashimoto {\sl et al.},
http://lambda.phys.tohoku.ac.jp/$\tilde{\, \,}$hashimot/E307/,
submitted to {\em Phys. Rev. Lett.} 
\bibitem{Sat99} Y. Sato {\sl et al.}, {\em Ph.D. Thesis}, Tohoku University, 1999;
Proceedings of the workshop on {\em Hypernuclear Physics with Electromagnetic Probes},
Dec. 2--4, 1999, Hampton, Virginia, USA.
\bibitem{Os82} E. Oset, H. Toki and W. Weise, {\em Phys. Rep.} {\bf 83} (1982) 281.
\bibitem{Os94} E. Oset, P. Fern\'andez de C\'ordoba, J. Nieves, A. Ramos and
L. L. Salcedo, {\em Prog. Theor. Phys. Suppl.} {\bf 117} (1994) 461. 
\bibitem{Os91} P. Fern\'{a}ndez De C\'{o}rdoba and E. Oset,
{\em Nucl. Phys.} {\bf A 528} (1991) 736.
\bibitem{Wa71} A. L. Fetter and J. D. Walecka, {\em Quantum Theory of Many
Particle Systems} (McGraw--Hill, New York, 1971).
\bibitem{Se83} R. Seki and K. Masutani, {\em Phys. Rev.} {\bf C 27} (1983) 2799.
\bibitem{Ga92} C. Garcia-Recio, J. Nieves and E. Oset,
{\em Nucl. Phys.} {\bf A 547} (1992) 473.
\bibitem{Ne82} J. W. Negele, {\em Rev. Mod. Phys.} {\bf 54} (1982) 813.
\bibitem{Al87} W. M. Alberico, R. Cenni, A. Molinari and P. Saracco,
{\em Ann. Phys.} {\bf 174} (1987) 131.
\bibitem{Ce97} R. Cenni, F. Conte and P. Saracco,
{\em Nucl. Phys.} {\bf A 623} (1997) 391.
\bibitem{Ma87} R.Machleidt, K.Holinde, and Ch. Elster, {\em Phys. Rep.} {\bf 149} (1987) 1.
\bibitem{Ce92}R. Cenni and P. Saracco, {\em Nucl. Phys.} {\bf A 487} (1988) 279;
R. Cenni, F. Conte, A. Cornacchia and P. Saracco,
{\em Nuovo Cim.} vol. 15 n. 12 (1992).
\bibitem{Po98} J. Vida\~na, A. Polls, A. Ramos and M. Hjorth-Jensen, {\em Nucl. Phys.}
{\bf A 644} (1998) 201.
\bibitem{Tesi} G. Garbarino, {\em Ph.D. Thesis}, University of Turin, 2000;
{\em Nucl. Phys.} {\bf A 691} (2001) 193c. 
\bibitem{Ra97} A. Ramos, M. J. Vicente--Vacas and E. Oset,
{\em Phys. Rev.} {\bf C 55} (1997) 735. 
\bibitem{Bh98a} H. C. Bhang {\sl et al.}, {\em Nucl. Phys.} {\bf A 629} (1998) 412c.
\bibitem{Bh01} H. C. Bhang {\sl et al.}, {\em Nucl. Phys.} {\bf A 691} (2001) 156c;
H. C. Bhang, invited talk presented at the
{\em VII International Conference On Hypernuclear and Strange Particle Physics},
Oct. 23--27, 2000, Torino, Italy.
\bibitem{Ou00a} H. Outa {\sl et al.}, Proposal of KEK--PS E462 (2000). 
\bibitem{Gi01} R. L. Gill, {\em Nucl. Phys.} {\bf A 691} (2001) 180c.
\bibitem{Bha01} T. Nagae, Proceedings of the III International Workshop on  
{\em Physics and detectors for DAPHNE}, Eds. S. Bianco et al., 
Nov. 16--19, 1999, Frascati, Italy, p. 701; 
H. Bhang, in International Symposium on {\em Hadrons and Nuclei}, Feb. 20--22, 2001, Seoul, Korea;
T. Nagae et al., to appear in the Proceedings of the {\em INPC 2001},
Jul. 30 -- Aug. 3, 2001, Berkeley, USA.
\bibitem{Ba89} H. Band$\overline{\rm o}$, T. Motoba M. Sotona and J. \v{Z}ofka, 
{\em Phys. Rev.} {\bf C 39} (1989) 587;
T. Kishimoto, H. Ejiri and H. Band$\overline{\rm o}$, {\em Phys. Lett.} {\bf B 232} (1989) 24.
\bibitem{Aj92} S. Ajimura {\sl et al.}, {\em Phys. Lett.} {\bf B 282} (1992) 293. 
\bibitem{Aj00} S. Ajimura {\sl et al.}, {\em Phys. Rev. Lett.} {\bf 84} (2000) 4052.
\bibitem{Ra92} A. Ramos, E. van Meijgaard, C. Bennhold and B. K. Jennings,
{\em Nucl. Phys.} {\bf A 544} (1992) 703.  
\bibitem{Au83} E. H. Auerbach {\sl et al.}, {\em Ann. Phys.} {\bf 148} (1983) 381.
\bibitem{Ej89} H. Ejiri, T. Kishimoto and H. Noumi, {\em Phys. Lett.} {\bf B 225} (1989) 35.
\bibitem{It94} K. Itonaga, T. Motoba, O. Richter and M. Sotona, {\em Phys. Rev.}
{\bf C 49} (1994) 1045.
\bibitem{Aj98} S. Ajimura {\sl et al.}, {\em Phys. Rev. Lett.} {\bf 80} (1998) 3471. 
\bibitem{Mot94} T. Motoba and K. Itonaga, {\em Nucl. Phys.} {\bf A 577} (1994) 293c.
\bibitem{pdb} D. E. Gromm {\sl et al.}, Review of Particle Physics,
{\em Eur. Phys. J.} {\bf C 15} (2000) 1. 
\bibitem{Na99} H. Nabetani, T. Ogaito, T. Sato and T. Kishimoto, 
{\em Phys. Rev.} {\bf C 60} (1999) 017001. 
\bibitem{Sa01} K. Sasaki, T. Inoue and M. Oka, {\em Nucl. Phys.} {\bf A 691} (2001) 201c.
\bibitem{Sa-pr} K. Sasaki, private communication.
\bibitem{Pa-pr} A. Parre\~{n}o, private communication.  
\end{thebibliography}
\end{document}